\documentclass{elsart}

\usepackage{epsf}
\usepackage{epsfig}
\usepackage{psfig}
\usepackage{latexsym}
\usepackage{amssymb}
\usepackage{amsfonts,amsbsy}
\usepackage{amsmath}

\newcommand{\btau}{\boldsymbol{\tau}}
\newcommand{\bttau}{\boldsymbol{\tilde{\tau}}}
\newcommand{\pauli}{\boldsymbol{{\sigma}}}

\begin{document}
\runauthor{Chowdhury, Santen and Schadschneider}
\begin{frontmatter}
\title{Statistical Physics of Vehicular Traffic and Some Related 
Systems\thanksref{SFB}}
\author[Cologne]{Debashish Chowdhury\thanksref{IITK}}
\author[Cologne]{Ludger Santen\thanksref{Email1}}
\author[Cologne]{Andreas Schadschneider\thanksref{Email2}}
\thanks[SFB]{Partially supported by SFB 341 (K\"oln-Aachen-J\"ulich)}

\address[Cologne]{Institut f\"ur Theoretische Physik, Universit\"at zu 
K\"oln, D-50923 K\"oln, Germany}
\thanks[IITK]{On leave from Physics Department, I.I.T., Kanpur 208016, 
India; E-mail: debch@iitk.ac.in}
\thanks[Email1]{present address: CNRS-Laboratoire de Physique Statistique,
Ecole Normale Sup{\'{e}}rieure, 24, rue Lhomond, 75231 Paris Cedex 05, 
France ; E-mail: Ludger.Santen@lps.ens.fr}
\thanks[Email2]{E-mail: as@thp.uni-koeln.de}
\begin{abstract} 

  In the so-called "microscopic" models of
  vehicular traffic, attention is paid explicitly to each individual
  vehicle each of which is represented by a "particle"; the nature of
  the "interactions" among these particles is determined by the way
  the vehicles influence each others' movement. Therefore, vehicular
  traffic, modeled as a system of {\it interacting} "particles" {\it
    driven far from equilibrium}, offers the possibility to study
  various fundamental aspects of truly nonequilibrium systems which
  are of current interest in statistical physics. Analytical as well
  as numerical techniques of statistical physics are being used to
  study these models to understand rich variety of physical phenomena
  exhibited by vehicular traffic. Some of these phenomena, observed in
  vehicular traffic under different circumstances, include transitions
  from one dynamical phase to another, criticality and self-organized
  criticality, metastability and hysteresis, phase-segregation, etc.
  In this critical review, written from the {\it perspective of
    statistical physics}, we explain the guiding principles behind all
  the main theoretical approaches. But we present detailed discussions
  on the results obtained mainly from the so-called "particle-hopping"
  models, particularly emphasizing those which have been formulated in
  recent years using the language of cellular automata.

\end{abstract}
\vspace{1cm}
\begin{center}
Physics Reports {\bf 329}, 199 (2000)
\end{center}
\end{frontmatter}
\newpage
\tableofcontents
\newpage

\section{Introduction}

The {\it concepts and techniques of statistical physics} are being
used nowadays to study several aspects of {\it complex systems}
\cite{complex} many of which, till a few decades ago, used to fall
outside the traditional domain of physical systems \cite{montroll}.
{Physical-,} chemical-, earth-, biological- and
social--sciences as well as technology meet at this frontier area of
inter-disciplinary research. Flow of vehicular traffic and granular
matter \cite{proc1,proc2}, folding of proteins \cite{protein},
formation and growth of bacterial colonies \cite{bacteria},
biological evolution of species
\cite{species} and transactions in financial markets \cite{econo} are
just a few examples of exotic phenomena in such systems.  Most of
these systems are interesting not only from the point of view of
Natural Sciences for fundamental understanding of how Nature works but
also from the points of view of applied sciences and engineering for
the potential practical use of the results of the investigations. Our
review of the current status and future trends of research on the
theory of vehicular traffic (and some related systems) will, we hope,
convince you that, indeed, the results of recent studies of
complex systems have been "a conceptual revolution, a paradigm shift
that has far reaching consequences for the very definition of physics"
\cite{parisi}.

For almost half a century physicists have been trying to understand
the fundamental principles governing the flow of vehicular traffic
using theoretical approaches based on statistical physics
\cite{proc1,proc2,hg,gazis,ph,helbook}.  The approach of a physicist
is usually quite different from that of a traffic engineer. A
physicist would like to develop a model of traffic by incorporating
only the most essential ingredients which are absolutely necessary to
describe the {\it general features} of typical real traffic. The
theoretical analysis and computer simulation of these models not only
provide deep insight into the properties of the model but also help in
better understanding of the complex phenomena observed in real
traffic. {\it Our aim in this review} is to present a critical survey
of the progress made so far towards understanding the {\it
  fundamental} aspects of traffic phenomena from the {\it perspective
  of statistical physics}.

There are two different conceptual frameworks for modeling traffic. In
the "coarse-grained" fluid-dynamical description, the traffic is
viewed as a compressible fluid formed by the vehicles but these
individual vehicles do not appear explicitly in the theory. In
contrast, in the so-called "microscopic" models of vehicular traffic
attention is explicitly focused on individual vehicles each of which
is represented by a "particle"; the nature of the interactions among
these particles is determined by the way the vehicles influence each
others' movement. In other words, in the "microscopic" theories
vehicular traffic is treated as a system of {\it interacting}
"particles" {\it driven far from equilibrium}. Thus, vehicular traffic
offers the possibility to study various fundamental aspects of the
dynamics of truly nonequilibrium systems which are of current
interest in statistical physics \cite{sz,gs,spohn,vp}.

In order to provide a broad perspective, we describe both
"macroscopic" and "microscopic" approaches although we put more
emphasis on the latter. Sometimes the phenomenological equations 
of traffic flow in the "macroscopic" models can be obtained
from microscopic considerations in the same spirit in which 
macroscopic or phenomenological theories of matter are derived 
from their molecular-theoretic description.

At present, even within the conceptual framework of "microscopic"
approach, there are several different types of mathematical
formulations of the dynamical evolution of the system. For example,
the probabilistic description of vehicular traffic in the {\it kinetic
  theory} is developed by appropriately modifying the kinetic theory
of gases. On the other hand, a deterministic description of the motion
of individual vehicles is provided by the so-called {\it car-following
  theories} which are based on the basic principles of classical
Newtonian dynamics. In contrast, the so-called {\it particle-hopping
  models} describe traffic in terms of a stochastic dynamics of
individual vehicles.  We explain the guiding principles behind all
these formulations. But we discuss in detail the results obtained
mainly from the investigations of the recently developed
"particle-hopping" models which are usually formulated using the
language of cellular automata (CA) \cite{wolfram}. At present,
there is no traffic model yet which can account for all aspects of
vehicular traffic. In this review we consider a wide variety of CA
models which describe various different types of traffic phenomena.

We map the particle-hopping models of vehicular traffic onto some
other model systems; these mappings indicate the possibility of
exploiting powerful techniques, used earlier for other systems, to
study traffic models and, sometimes, enable us to obtain results for
traffic models directly from the known results for models of other
systems. We present pedagogical summaries of the {\it statistical
  mechanical treatments} of the CA models of traffic. We critically
examine the regimes of validity of the {\it approximation} schemes of
{\it analytical} calculations which we illustrate with explicit
calculations in those limiting cases where these usually yield {\it
  exact} results. The results of the theoretical analysis of these 
models are compared with those obtained from computer simulations and,
wherever possible, with the corresponding {\it empirical} results from
real traffic.  We also compare vehicular traffic with many other
similar physical systems to show the ubiquity of some physical
phenomena.

{\it Computer simulations} are known to provide sufficiently accurate
{\it quantitative} results when analytical treatments require
approximations which are too crude to yield results of comparable
accuracy. In this review we demonstrate how computer simulations often
help in getting deep insight into various phenomena involved in
traffic and in {\it qualitative} understanding of the basic principles
governing them thereby avoiding potentially hazardous experiments with
real traffic.  Computer simulations of the "microscopic" models of
traffic have not only attracted the attention of a growing number of
statistical physicists in the recent years, but have also been
received positively by many traffic engineers. The ongoing research
efforts to utilize computer simulations of the microscopic models for
{\it practical applications} in planning and design of transportation
networks have been reviewed very recently by Nagel et al.~\cite{nagel99}.

Our review is complimentary to those published in the recent years by
Helbing \cite{helbook} and by Nagel et al.~\cite{nagel99}.  A large
number of important papers on traffic published in recent years are
based on the particle-hopping models. But these works have received
very little attention in \cite{helbook}. We discuss the methods and
results for the particle-hopping models in great detail in this review
after explaining the basic principles of all the theoretical
approaches. Moreover, we focus almost exclusively on the {\it
  fundamental principles} from the point of view of statistical
physics while Nagel et al.~\cite{nagel99} emphasize {\it practical
  applications} which are directly relevant for traffic engineering.

At this point a skeptic may raise a serious question: "can 
we ever predict traffic phenomena with statistical mechanical 
theories without taking into account effects arising from 
widely different human temperaments and driving habits of 
the individual drivers?" We admit that, unlike the particles 
in a gas, a driver is an intelligent agent who can "think", 
make individual decisions and "learn" from experience. 
Besides, the action of the driver may also depend on his/her 
physical and mental states (e.g., sorrow, happiness, etc.). 
It is also true that the behavior of each individual driver 
does not enter explicitly into the "microscopic" models of 
traffic. Nevertheless, as we shall show in this review, many 
general features of traffic can be explained in general terms 
with these models provided the different possible behavioural 
effects  are captured collectively through a probabilistic 
description which requires only a few phenomenological 
parameters. Similar strategies have been suggested also for 
capturing the behavioural effects of some individual traders 
in financial markets collectively through probabilistic 
descriptions \cite{complex}. These probabilistic descriptions 
make the dynamics of the models intrinsically stochastic.

Throughout this article the terms vehicle (or, car) and driver 
are used interchangeably, although each of these terms usually 
refers to the composite unit consisting of the vehicle and 
the driver.  The main questions addressed by physicists are 
posed as problems in Section \ref{section_pract}. As a motivation 
we present some relevant empirically observed general features of 
real traffic as well as their plausible phenomenological explanations 
in Section \ref{secion_empir}. The following sections review the 
different theoretical approaches. The classification of the models
into different classes is not unique. Mostly we choose a classification
according to the use of discrete or continuous space, time and state
variables. The conceptual basis of the older 
theoretical approaches, namely, the fluid-dynamical theories, the kinetic 
theories and the car-following theories, are explained, in the 
Sections \ref{section_fluid},\ref{section_kinetic} and \ref{sec_carfoll}, 
respectively, where the corresponding recent developments are also 
summarized. These model classes are continuous in space, time and state
variables. Some coupled map lattice models of traffic are considered in 
Section \ref{Sec_coupmap}. They are discrete in time. CA models
are discrete in space, time and state variables. The model 
suggested by Nagel and Schreckenberg (NaSch) \cite{ns} is the 
minimal model of traffic on highways; the theoretical results 
on various aspects of this model are discussed in Section \ref{sec_NaSch}  
where the nature of the spatio-temporal organization of vehicles 
are also investigated and the fundamental question of the 
(im-)possibility of any dynamical phase transition in the NaSch 
model is addressed.  Various generalizations and extensions of 
the NaSch model (including those for multi-lane traffic) are 
reviewed in Section \ref{Sec_gener}. The occurrence of self-organized 
criticality in the so-called cruise-control limit of the 
NaSch model is pointed out. It is demonstrated how additional 
"slow-to-start" rules of CA can give rise to {\it metastability},
hysteresis and {\it phase-separation} in the generalized NaSch 
models, in qualitative agreement with empirical observations. 
In Section \ref{Sec_disorder} the formation and {\it "coarsening"} 
of platoons of vehicles are investigated in an appropriate generalization 
of the NaSch model with one type of quenched randomness; your 
attention is drawn to the formal analogy between this 
phenomenon and the Bose-Einstein condensation. In this section 
the effects of other kinds of {\it quenched disorder} on the 
nature of the steady-states of the NaSch model are also considered.
In Section \ref{Sec_other} we present some other CA models of 
highway traffic which are not directly related to the NaSch model.
The Biham-Middleton-Levine (BML) ~\cite{bml} model is the earliest
CA model of traffic in idealized networks of streets in cities; it
exhibits a first order {\it phase transition}. A critical review of
this model is presented in Section \ref{Sec_city}  together with a list 
of its generalizations which have been reported so far. Furthermore
a marriage of the NaSch description of traffic and the BML model, 
which has led to the development of a novel model of city traffic, 
is explained. A brief status report of the ongoing efforts to make 
practical use of the theoretical models for traffic engineering is 
also presented. The similarities between various particle-hopping 
models of traffic and some other systems far from equilibrium are 
pointed out in Section \ref{Sec_related} followed by the concluding 
Section \ref{Sec_summ} where your attention is also drawn towards 
challenging open questions. Several Appendices deal mostly with
more technical aspects of some important calculations, but
are not necessary for an understanding of the main text.
 
\section{Fundamental and practical questions}
\label{section_pract}

The aim of {\it basic research} in {\it traffic science} is to 
discover the {\it fundamental laws} governing traffic systems. The 
main aim of {\it traffic engineering} is on planning, design and 
implementation of transportation network and traffic control systems. 
Statistical physicists have been contributing to traffic science by 
developing models of traffic and drawing general conclusions about 
the basic principles governing traffic phenomena by studying these 
models using the tools of statistical physics. Moreover, using these 
models, statistical physicists have also been calculating several 
quantities which may find {\it practical applications} in traffic 
engineering. Furthermore, several groups of statistical physicists 
are currently also engaged in developing strategies for fast on-line 
simulation and traffic control so as to optimize traffic flow; 
significant contributions to this traditional domain of traffic 
engineering can reduce the financial burden on the governments. 

\subsection{Some fundamental questions}

Because of the apparent similarities between the "microscopic" 
models of traffic and macroscopic samples of ionic conductors in 
the presence of external electric field, the tools of statistical 
mechanics seem to be the natural choice for studying these models.
However, the actual calculation of even the steady-state properties 
of traffic from the "microscopic" models is a highly difficult 
problem because  (apart from the human element involved) (a) the 
vehicles {\it interact} with each other and (b) the system is {\it 
driven} far from equilibrium, although it may attain a nonequilibrium 
steady-state. 

In principle, the time-independent observable properties of large
pieces of matter can be calculated within the general framework of
equilibrium statistical mechanics, pioneered by Maxwell, Boltzmann and
Gibbs, provided the system is in thermal equilibrium. Of course, in
practice, it may not be possible to carry out the calculations without
making approximations because of the interactions among the
constituents of the system. Some time-dependent phenomena, e.g.,
fluctuation and relaxation, can also be investigated using the Linear
Response Theory provided the system is not too far from equilibrium.
Unfortunately, so far there is no general theoretical formalism for
dealing with systems far from equilibrium. Moreover, the condition of
detailed balance does not hold \cite{derr_review} although a condition
of pairwise balance \cite{schutz96} holds for some special
systems driven far from equilibrium.

The dynamical phases of systems driven far from equilibrium are 
counterparts of the stable phases of systems in equilibrium. 
Some of the fundamental questions related to the nature of these 
phases are as follows.\\
(i) What are the various {\it dynamical phases} of traffic? Does   
traffic exhibit phase-coexistence, phase transition, criticality 
\cite{stanley,goldenfeld} or self-organized criticality 
\cite{bak,ddhar} and, if so, under which circumstances?\\
(ii) What is the nature of {\it fluctuations around the steady-states} 
of traffic? Analogous phenomenon of the fluctuations around stable 
states in equilibrium is by now quite well understood.\\ 
(iii) If the initial state is far from a stationary state of the 
driven system, how does it {\it evolve with time} to reach a 
truly steady-state? Analogous phenomena of equilibration of systems 
evolving from metastable or unstable initial states through 
nucleation (for example, in a supersaturated vapour) or spinodal 
decomposition (for example, in a binary alloy) have also been 
studied earlier extensively \cite{gunton}.\\
(iv) What are the effects of quenched (static or time-independent) 
disorder on the answers of the questions posed in (i)-(iii) above? 

\subsection{Some practical questions} 

Let is first define some characteristic quantitative features of
vehicular traffic. The {\em flux} $J$, which is sometimes also
called {\em flow} or {\em current}, is defined as the number of
vehicles crossing a detector site per unit time \cite{may90}. The
distance from a selected point on the leading vehicle to the same
point on the following vehicle is defined as the {\em
 distance-headway} \cite{may90}. The {\em time-headway} is defined
as the time interval between the departures (or arrivals) of two
successive vehicles recorded by a detector placed at a fixed position
on the highway \cite{may90}. The distributions of distance-headways
and time-headways are regarded as important characteristic of traffic
flow. For example, larger headways provide greater margins of safety
whereas higher capacities of the highway require smaller headways.

Let us now pose some questions which are of {\em practical interest} 
in traffic engineering. \\
(a) What is the relation between density $c$ and flux $J$? In traffic 
engineering, this relation is usually referred to as the {\em 
fundamental diagram}.\\ 
(b) What are the distributions of the distance-headway and time-headway?\\
(c) How should on- and off-ramps be designed?\\
(d) Does an additional lane really lead to an improvement?\\
(e) What are the effects of a new road on the performance of the
road network?\\ 
(f) What type of signaling strategy should be adopted to optimize 
the traffic flow on a given network of streets and highways?\\
(g) The generalized travelling salesman problem: Is the shortest
trip also the fastest?\\

\section{Some empirical facts and phenomenological explanations} 
\label{secion_empir}

For several reasons, it is difficult to obtain very reliable 
(and reproducible) detailed empirical data on real traffic.  
First of all, unlike controlled experiments performed in the 
conventional fields of research in physical sciences, it is 
not possible to perform such laboratory experiments on vehicular 
traffic. In other words, empirical data are to be collected 
through passive observations rather than active experiments. 
Secondly, unambiguous interpretation of the collected data is 
also often a subtle exercise because traffic states depend on 
several external influences, e.g. the weather conditions.
The systematic investigation of traffic flow has a quite long 
history \cite{greenshield,greenberg,hallTGF}. 
Although we now have a clear understanding of many aspects of 
real traffic several other controversial aspects still remain  
intellectual challenges for traffic scientists. In this section 
we give an overview of some of the well understood experimental 
findings, which are relevant for our theoretical analysis in 
the following sections. Moreover, wherever possible, we provide 
phenomenological explanations of these empirically observed 
traffic phenomena. Furthermore, we shall also mention some of 
the more recent empirical observations for which, at present, 
there are no generally accepted explanations.

\subsection{Acceleration noise} 

In general, because of the different human temperaments and 
driving habits, different drivers react slightly differently 
to the same conditions on a highway, even when no other 
vehicle influences its motion. Consequently, even on an empty 
stretch of a highway, a driver can neither maintain a constant 
desired speed nor accelerate in a smooth fashion. In addition 
to the type of the highway (i.e., the surface conditions, 
frequency of the curves, etc.) the {\it driver-to-driver 
fluctuation} of the acceleration also depends on the density 
of vehicles on the highway. The root-mean-square deviation 
of the acceleration of the vehicles is a measure of the 
so-called {\it acceleration noise}. The distributions of the 
accelerations have been measured since mid nineteen fifties 
and are well documented \cite{winzer,hoefs}. 

\subsection{Formation and characterization of traffic jams}

Traffic jam is the most extensively studied traffic phenomenon. 
Traffic jams can emerge because of various different reasons. 
Most often traffic jams are observed at bottlenecks, e.g. 
lane-reductions or crossings of highways \cite{Daganzo99}. At 
bottlenecks the capacity of the road is locally reduced thereby 
leading to the formation of jams upstream traffic. Downstream
the bottleneck, typically, a free-flow region is observed. 
In addition, traffic accidents, which also lead to a local 
reduction of the capacity of the highway, can give rise to 
traffic jams.

\begin{figure}[ht]
\begin{center}
\epsfig{file=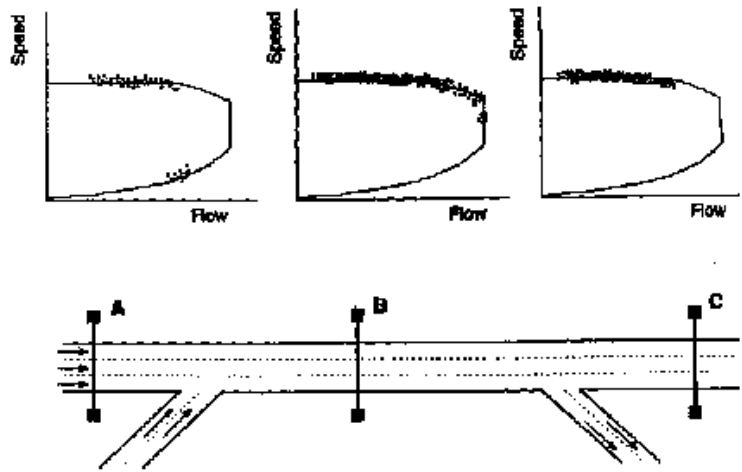,width=10cm}
  \caption{Experimental flow speed-diagrams near on- and off ramps of
    a three lane highway. The upper part of the figure shows the
    empirical results. Each dot represents a five-minute average of
    the local measurement. The lines serve merely as guide to the
    eyes. The lower part of the figure shows the location of the
    detectors (from \cite{hallTGF}).}
\label{hallbottle}
\end{center}
\end{figure}

Fig.~\ref{hallbottle} shows as an example empirical data of the
velocity at a three-lane highway close to an on- and off-ramp 
\cite{hallTGF}. The data show that downstream the bottleneck 
(at detector C) no slow vehicle has been recorded. In contrast,
in the merging regime near detector A vehicles often have to 
move slowly. In between the on- and off-ramps the vehicles move 
with larger velocities compared to those in location A although 
the number of vehicles passing detector B is maximal. Therefore, 
the on-ramp causes a local reduction of the capacity of the highway.

\begin{figure}[ht]
 \centerline{\psfig{figure=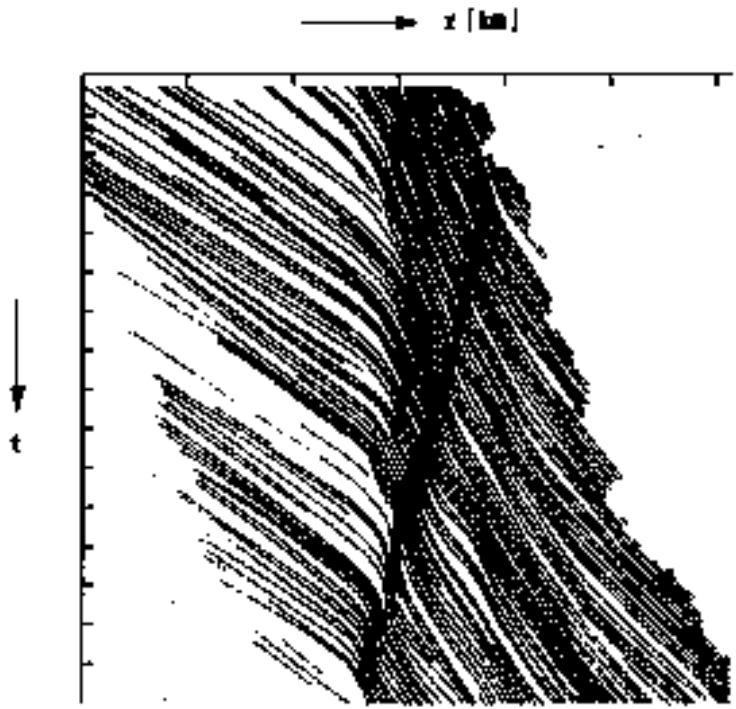,width=10cm}}
  \caption{Trajectories of individual vehicles on a single lane of a 
multi-lane highway. The trajectories were drawn from aerial photographs. 
During the analyzed time-interval the spontaneous formation, propagation 
and dissolution of a jam has been observed (from \cite{Treiterer75}).}
\label{aerial}
\end{figure}

Perhaps, what makes the study of traffic jams so interesting is 
that jams often appear, as if, from nowhere (apparently without 
obvious reasons) suddenly on crowded highways; these so-called 
"phantom jams"\footnote{"Stau aus dem Nichts" in german} are formed by 
{\em spontaneous} fluctuations in an otherwise streamlined flow. 
Direct empirical evidence for this {\em spontaneous formation
of jams} was presented by Treiterer \cite{Treiterer75} by analyzing 
a series of aerial photographs of a multi-lane highway. In 
Fig.~\ref{aerial} the picture from \cite{Treiterer75} is redrawn.  
Each line represents the trajectory of an individual vehicle on one 
lane of the highway\footnote{The discontinuous trajectories correspond 
to vehicles changing the lane.}. The space-time plot (i.e., the 
trajectories $x(t)$ of the vehicles) shows the formation and 
propagation of a traffic jam. In the beginning of the analysed time
vehicles are well separated from each other. Then, due to 
fluctuations, a dense region appears which, finally, leads to 
the formation of a jam. The jam remains stable for a 
certain period of time but, then, disappears again without any 
obvious reason. This figure clearly establishes not only the 
spontaneous formation of traffic jam but also shows that such 
jams can propagate upstream (opposite to the direction of flow 
of the vehicles). Moreover, it is possible that two or more jams 
coexist on a highway.  

A more detailed analysis of traffic jams in absence of hindrances has
been given by Kerner and Rehborn \cite{kerner,Kerner961,KernerTGF} who
pointed out the following characteristic features of wide jams.  They
found that the upstream velocity and, therefore, the outflow from a
jam is approximately constant. The outflow from a jam and the velocity
of the jam fronts are now regarded as two important empirical
parameters of highway traffic which can be used for calibrating
theoretical models.

\subsection{Flux-density relation}

Obviously, traffic flow phenomena strongly depend on the 
occupancy of the road. What type of variation of flux and 
average velocity $\langle v\rangle$ with density $c$ can one expect 
on the basis of intuitive arguments? So long as $c$ is sufficiently 
small, the average speed $\langle v \rangle$ is practically 
independent of $c$ as the vehicles are too far apart to 
interact mutually. Therefore, at sufficiently low density 
of vehicles, practically "free flow" takes place. However, 
from the practical experience that vehicles have to move 
slower with increasing density, one expects that at intermediate 
densities, 
\begin{equation}
  \label{vel_dens}
  \frac{d\langle v \rangle}{d c} \leq 0,
\end{equation}
when the forward movement of the vehicles is strongly hindred 
by others because of the reduction in the average separation 
between them. A faster-than-linear monotonic decrease of 
$\langle v \rangle$ with increasing $c$ can lead to a maximum 
\cite{may90} in the flux $\langle J \rangle = \langle c v \rangle$ 
at $c = c_m$; for $c < c_m$, increasing $c$ would lead to 
increasing $\langle J \rangle$ whereas for $c > c_m$ sharp 
decrease of $\langle v \rangle$ with increase of $c$ would 
lead to the overall decrease of $\langle J \rangle$. However, 
contrary to this naive expectation, in recent years some 
nontrivial variation of flux with density have been observed. 
The nature of the variation of the flux with the density is 
still not clearly understood \cite{Hall86} since the details 
of the complex experimental setup can strongly influence the 
empirical results.

\begin{figure}[h]
\begin{center}
\epsfig{file=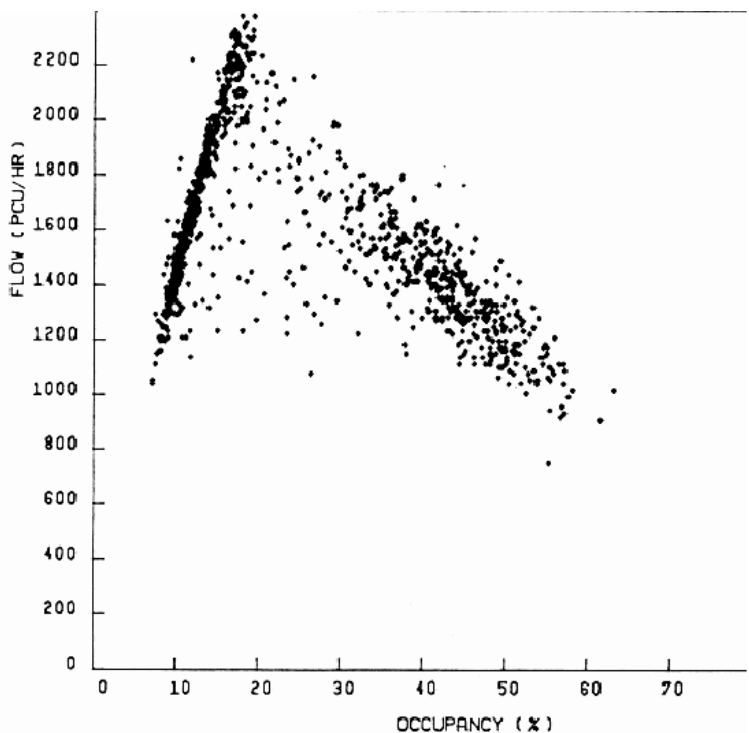,width=10cm}
  \caption{Empirical data for flow and occupancy. The data have been collected
    by counting loops on a Canadian highway. Both the occupancy and the flow
    have been directly measured by the detector. Each point in the diagram
    corresponds to an average over a time interval of five minutes (from
    \cite{Hall86}).}
\label{fig:scatter}
\end{center}
\end{figure}

Fig.~\ref{fig:scatter} shows typical time averaged local measurements of the
density and flow which have been obtained from the Queen Elizabeth Way
in Ontario (Canada) \cite{Hall86}. At low densities the
data indicate a linear dependence of the flow on the density. In
contrast strong fluctuations of the flow at large densities
exist which prevents a direct evaluation of the functional form at high
densities.

\begin{figure}[h]
\begin{center}
\epsfig{file=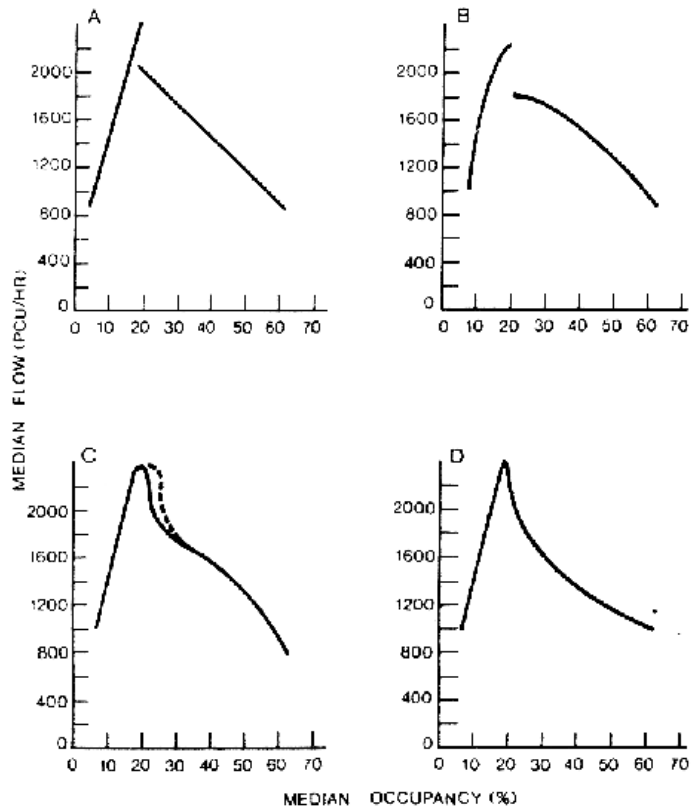,width=10cm}
  \caption{Schematic representation of fundamental diagrams consistent
    with empirical investigations (from \cite{Hall86}).}
\label{fig:hallfund}
\end{center}
\end{figure}

In order to use the empirical results for a theoretical analysis 
it is often more convenient to use the mean-values of the flow 
at a given density. Fig.~\ref{fig:hallfund} shows a collection of 
possible forms of averaged fundamental diagrams consistent with 
empirical data \cite{Hall86}. While the discontinuity of the 
fundamental diagram now seems to be well established \cite{Neubert99}  
no clear answer can be given to the question on the {\em form} of 
the diagram in the free-flow or high-density regime. In the low 
density regime linear as well as non-linear functional forms of 
the fundamental diagrams have been suggested.  For the high 
density branch no consistent picture for the high density branch 
exists. Here the results strongly depend on the specific road 
network. 

\begin{figure}[h]
  \begin{center}
     \epsfig{file=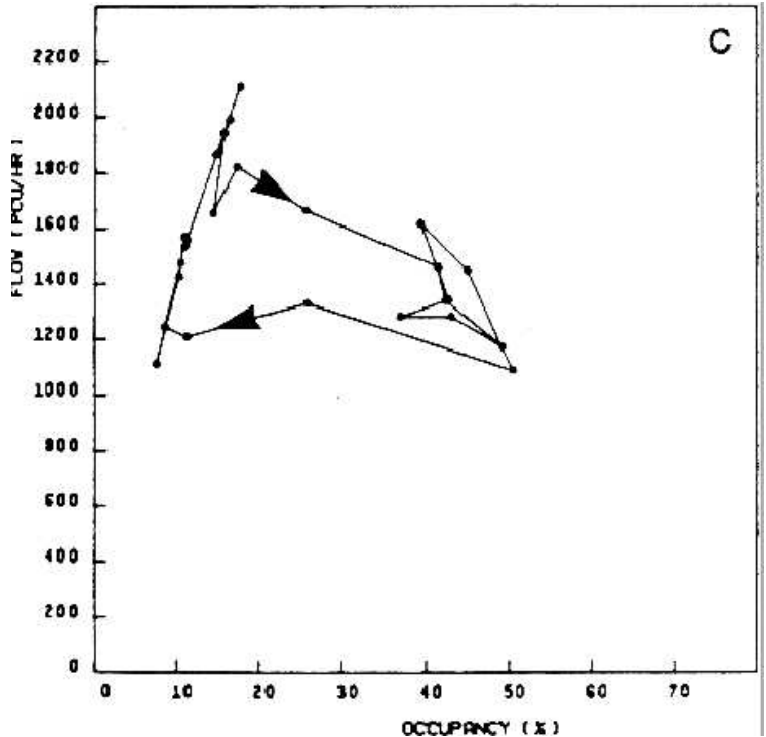,width=10cm}
     \caption{Time-traced  measurements of the flow (from
       \cite{Hall86}). Each data point corresponds to a time-averaged value
       of the flow and density obtained from a local measurement. The
       average was performed over five minute intervals.}
       \label{fig:exphys}
  \end{center}
\end{figure}

In several situations it has been observed that $J$ does not 
depend uniquely on $c$ in an intermediate regime of density; it 
indicates the existence of {\it hysteresis effects and meta-stable 
states}. In the context of traffic flow, hysteresis effects have 
the following meaning: if a measurement starts in the free-flow 
regime, an increase of the density leads to an increase of the 
flow. However, beyond a certain density, a further increase of 
the density leads to a {\it discontinuous} reduction of the stationary
flow ("capacity drop") and jams emerge. The corresponding fundamental
diagram  (Fig.~\ref{fig:hallfund} A, B) then has the so-called
'inverse-$\lambda$ form'.
Figure~\ref{fig:exphys} shows
an experimental verification of a hysteresis loop \cite{Hall86} at a
transition from a free flow to a congested state.

Recent empirical observations which have been
obtained near a crossing of highways exhibit \cite{Neubert99} 
a flat plateau (i.e., a density-indepedent flux) over an intermediate
regime of density  of the vehicles. 

\subsection{Microscopic states of traffic flow and phase transitions}

The results for the flux-density relation already suggest the
existence of at least two different dynamical phases of vehicular
traffic on highways, namely a free-flow phase and a congested phase. 
In the free-flow regime all vehicles can move with high speed 
close to the speed limit. The nature of the congested traffic is 
still under debate. Careful empirical observations in the recent 
years indicate the existence of two different congested phases, 
namely, the synchronized phase and the stop-and-go traffic phase 
\cite{Kerner961,KernerTGF,Kerner972,Kerner981,kernerPW}. 
Vehicles move rather slowly in the synchronized states, as compared to 
the free-flow states, but the flux in the synchronized states can take 
a values close to the optimum value because of relatively small 
headways. Besides, non-trivial strong
correlations between the density on different lanes exist in 
the synchronized state \cite{Kerner961,kernerPW,Koshi83} which actually
motivates the notation synchronized traffic. 
The stop-and-go traffic differs from the synchronized states in 
the sense that every vehicle inside the jams come to a complete 
halt for a certain period of time. 

\begin{figure}[h]
  \begin{center}
    \epsfig{file=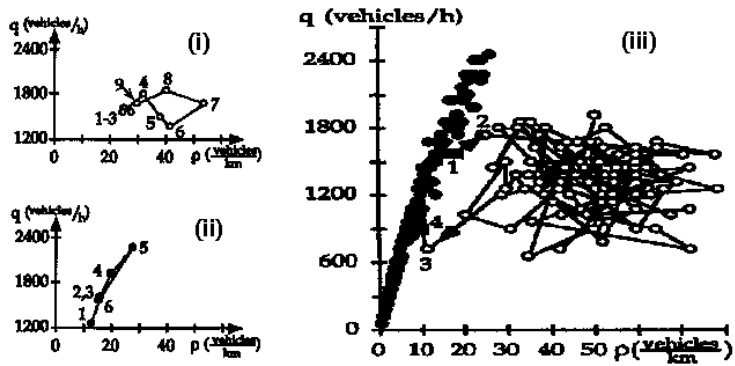,width=10cm}
    \caption{Fundamental diagrams corresponding to different types of
      synchronized flow. Measurements which correspond to synchronized
      traffic of type (i) are localised in a narrow region of the flow
      density plane, while synchronized states of type (ii) are
      similar to the results of free-flow states. In contrast for
      synchronized traffic of type (iii) the time-traced measurements
      show an unsystematic behaviour.}
   \label{fig:kernerfund}
  \end{center}
\end{figure}

Following Kerner three different types of synchronized traffic can be
distringuished by the time-dependent behaviour of the density and
flow \cite{Kerner961,KernerTGF,kernerPW}. In synchronized traffic of 
type (i) constant values of density 
and flow can be observed during a long period of time. In synchronized
traffic of type (ii) patterns of density and flow quite similar to free
flow states have been observed. The differences between
synchronized states of type (ii) and free-flow are given by the reduced
average velocities and the alignment of the speeds on different lanes
in synchronized traffic. Moreover irregular patterns of time-traced
measurements of the flow have been found in synchronized traffic (see
Fig.~\ref{fig:kernerfund}) of type (iii).

\subsection{Time- and distance-headways}

The flux $J$ can be written as $J = N/T$ where $T = \sum_{i=1}^N t_i$ 
is the sum of the time-headways recorded for all the $N$ vehicles. 
Hence, $J = 1/T_{av}$ where $T_{av} = (1/N)\sum_i t_i $ is the 
average TH. Therefore, the TH distribution contains more 
detailed informations on traffic flow than that available from 
the flux alone. With the variation of density $c$ of the vehicles, 
$T_{av}$ exhibits a minimum at $c = c_m$ where the flux is maximum 
\cite{may90}. 

  \begin{figure}
    \begin{center}
    \epsfig{file=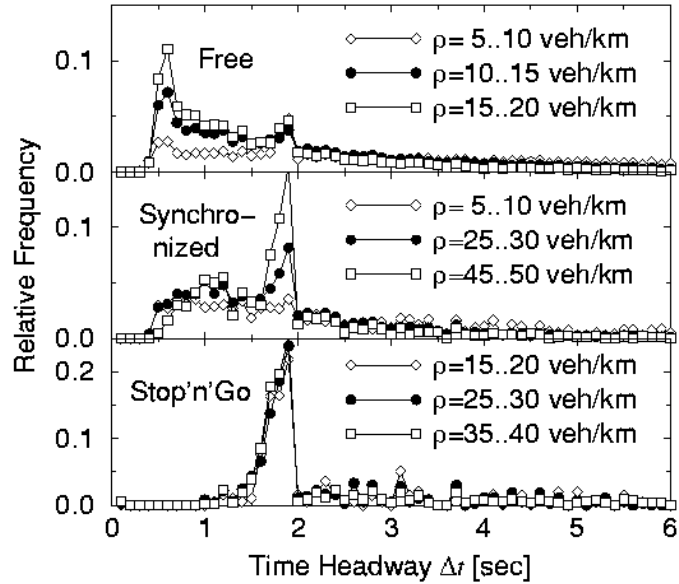,width=0.7\linewidth}
    \caption{Time-headway distribution for different density
      regimes. Top: In free-flow traffic the $\Delta t$-distribution
      is dominated by two peaks at $0.8\sec$ and $1.8\sec$. Middle: In
      synchronized traffic cars with narrow time gaps are to find as
      well as a dominating peak. Bottom: In stop-and-go traffic short
      time headways are surpressed. The peak at $1.8\sec$ remains
      since vehicles are leaving the jam with a typical temporal headway
      of approximately $2\sec$.}
  \label{thfree}
 \end{center}
\end{figure}

The results discussed in the preceeding subsections are based on time 
averaged local measurements. But it is also very useful to analyze the 
single-vehicle data directly \cite{Helbing9703,Helbing9701,Wagner,Neubert99}. 
The single-vehicle data allows calculation of the time-headway 
distributions \cite{Neubert99}.  All the time-headway distributions 
in the free-flow regime show a two peak structure. The first peak at
$\Delta t=0.8\sec$ represents the global maximum of the distribution.
On a microscopic level these short time-headways correspond to platoons
of some vehicles traveling very fast -- their drivers are taking the
risk of driving "bumper-to-bumper" with a rather high speed. These
platoons are the reason for the occurrence of high-flow states in free
traffic. The corresponding states exhibit metastability, i.e. a
perturbation of finite magnitude and duration is able to destroy such
a high-flow state \cite{KernerTGF}. Additionally, a second peak
emerges at $\Delta t=1.8\sec$ which can be associated with a typical 
drivers' urge to maintain a temporal headway of $\approx 2\sec$ 
(which is the safe distance recommended in driving schools).

Surprisingly, the small time headways have much less weight in
congested traffic. Only the peak at $\Delta t=1.8\,\sec$ is recovered,
where the time headway distribution now takes the maximum value. But
nevertheless, the small time headways ($\Delta t<1.8\,\sec$)
contribute significantly in synchronized traffic. In stop-and-go
traffic only the $1.8\,\sec$-peak remains and short time-headways are
surpressed.  The asymptotic behavior is rather unsystematic and
reflects the dynamics of vehicles inside the jams.

\begin{figure}[h]
  \begin{center}
    \epsfig{file=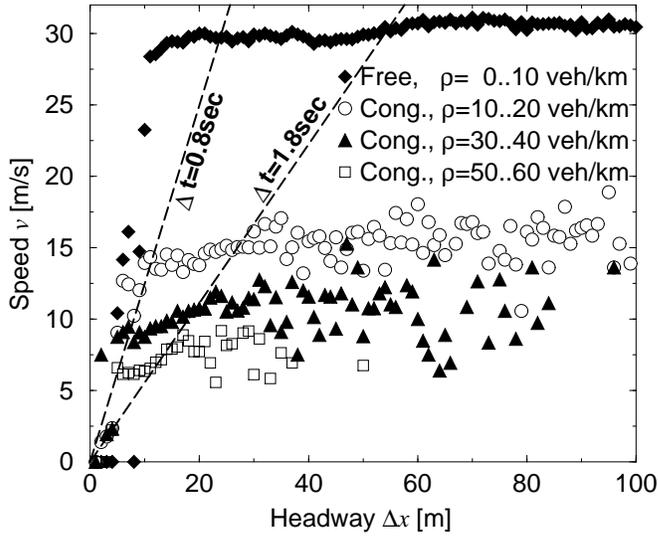,width=0.7\linewidth}
    \caption{Speed-distance relation in free and congested traffic. It
      turns out that the typical velocities at large distance headways
      strongly depend on the present traffic state (from
      \cite{Neubert99}).}
    \label{vxcomfort}
  \end{center}
\end{figure}

Another important result characterizing the microscopic states is the
dependence of the velocity of individual vehicles on the distance
headway (see Fig.~\ref{vxcomfort}). This function is also of great
importance for theoretical
approaches, e.g. it is used as input for the so-called optimal
velocity model. In the free-flow regime it is evident that the
asymptotic velocity is reached already for small distance-headways. The
slope of the velocity function is much lower than that in the free-flow
regime. Surprisingly, the asymptotic velocity depends strongly 
on the microscopic state as well as on the density, e.g., in dense 
traffic low velocities of the vehicles are also 
observed even when large distance headways are available.

\section{Fluid-dynamical Theories of vehicular traffic} 
\label{section_fluid}

When viewed from a long distance, say, an aircraft, flow of fairly 
heavy traffic appears like a stream of a fluid. Therefore, a 
"macroscopic" theory of traffic can be developed, in analogy with the 
hydrodynamic theory of fluids, by treating traffic as an effectively 
one-dimensional compressible fluid (a continuum) \cite{kuhnerev}. 
We follow the convention that traffic is flowing {\it from left to right}. 

Suppose, $c(x;t)$ and $J(x;t)$ are the "coarse-grained" density 
and flux at an arbitrary location $x$ at an arbitrary instant of 
time $t$. The equation of continuity for the fluid representing 
traffic is 
\begin{equation}
\frac{\partial c(x;t)}{\partial t} + \frac{\partial J(x;t)}{\partial x} 
= \sum_{i=1}^{J_{in}} \alpha_i(x-x_i;t) - \sum_{j=1}^{J_{out}} 
\beta_j(x-x_j;t)
\label{full-cont}
\end{equation}
where the first and the second terms on the right hand side take 
care of the sources and sinks, respectively, at the $J_{in}$ on-ramps 
situated at $x_i$ ($i=1,2,...,J_{in}$) and $J_{out}$ off-ramps 
situated at $x_j$ ($j=1,2,...,J_{out}$). We can write $\alpha_i(x-x_i;t)$ 
and $\beta_j(x-x_j;t)$ as 
\begin{equation}
\alpha_i(x-x_i;t) = \alpha_i^0(t) \phi_i(x-x_i) \quad \text{and}
\quad \beta_j(x-x_j;t) = \beta_j^0(t) \phi_j(x-x_j) 
\label{space_flux}
\end{equation}
where $\phi_i(x-x_i)$ and $\phi_j(x-x_j)$ describe the spatial 
distribution of the incoming and outgoing flux, respectively, 
while $\alpha_i^0(t)$ and $\beta_j^0(t)$ account for the 
corresponding temporal variations. 

In the following subsections, for simplicity, we shall consider a 
given stretch of highway with no entries or exits. In such special 
situations the equation of continuity reduces to the simpler 
form \cite{lw} 
\begin{equation}
\frac{\partial c(x;t)}{\partial t} 
+ \frac{\partial J(x;t)}{\partial x} = 0.
\label{eq-continuity} 
\end{equation}

One cannot get two unknowns, namely, $c(x;t)$ and $J(x;t)$ (or, 
equivalently, $v(x;t)$) by solving only one equation, namely 
(\ref{eq-continuity}), unless they are related to each other. In 
order to proceed further, one needs another independent equation, 
say, for $v(x;t)$; we shall write down such an equation later in 
subsection \ref{sec_Nav}. An alternative possibility, which Lighthill 
and Whitham \cite{lw} adopted in their pioneering work, is to assume 
that $J(x;t)$ is determined primarily by the local density $c(x;t)$ 
so that $J(x;t)$ can be treated as a function of only $c(x;t)$. 
Consequently, the number of unknown variables is reduced to one 
as, according to this assumption, the two unknowns $c(x;t)$ and 
$J(x;t)$ are not independent of each other. 

\subsection{Lighthill-Whitham theory and kinematic waves} 

As a first approximation, let us begin with Lighthill-Whitham 
assumption that 
\begin{equation}
J(x;t) = j(c(x;t)) 
\label{j-c}
\end{equation}
where $j(c)$ is a function of $c$. The functional relation (\ref{j-c}) 
between density and flux cannot be calculated within the framework 
of the fluid-dynamical theory; this must be either taken as a 
phenomenological relation extracted from empirical data or derived 
from more microscopic considerations. In general, the flux-density 
curve implied by equation (\ref{j-c}) need not be identical with 
the fundamental diagram in the steady-state. 

Under the assumption (\ref{j-c}), the $x$-dependence of the local 
flux $J(x;t)$ arises only from the $x$-dependence of $c(x;t)$. 
Alternatively, since $J(x;t) = c(x;t) v(x;t)$, assuming $v(x;t)$ 
to depend only on $c(x;t)$ the $x$-dependence of $v(x;t)$ arises 
only from the $x$-dependence of $c(x;t)$. Using (\ref{j-c}) the 
equation of continuity (\ref{eq-continuity}) can be expressed as 
\begin{equation}
\frac{\partial c(x;t)}{\partial t} + \frac{\partial c(x;t)}{\partial x}
\left[v(x;t) + c(x;t)\frac{dv}{dc}\right]  
= \frac{\partial c(x;t)}{\partial t} 
+ v_g\frac{\partial c(x;t)}{\partial x} = 0
\label{eq-cont2}
\end{equation} 
where $v_g = dJ/dc$. Note that the equations (\ref{j-c}) and 
(\ref{eq-cont2}) form the complete system of dynamical equations 
governing traffic flow in this first approximation. However, the 
equation (\ref{eq-cont2}) is {\it non-linear} because, in general, 
$v_g = dJ/dc = v(c) + c \frac{dv(c)}{dc}$ depends on $c$.  
If $v_g$ were a constant $v_0$, independent of $c$, equation 
(\ref{eq-cont2}) would become linear and 
the general solution would be of the form $c(x;t) = f(x-v_0t)$ 
where $f$ is an arbitrary function of its argument. In that case, 
the solution of any particular problem would be found by merely 
matching the function $f$ to the corresponding given initial 
and boundary conditions. Such a solution describes a {\it density 
wave} motion as an initial density profile would get translated 
by a distance $v_0t$ in a time interval $t$ {\it without any change 
in its shape}.  However, the non-linearity of the equation 
(\ref{eq-cont2}) gives rise to subtleties which are essential to 
capture at least some aspects of real traffic. 

The solution of the nonlinear equation (\ref{eq-cont2}) is of the 
general form 
\begin{equation}
c(x;t) = F(x - v_g t), 
\label{nlwave}
\end{equation}
where $F$ is an arbitrary function of its arguments. If we define 
a wave to be "recognizable signal that is transferred from one part 
of a medium to another with a recognizable velocity of propagation" 
\cite{whit} then the solutions of the form (\ref{nlwave}) can be 
regarded as a "density wave". There are several similarities between 
the density wave and the more commonly encountered waves like, for 
example, acoustic or elastic waves, But, the acoustic or elastic 
waves are solutions of linearized partial differential equations 
whereas the equation (\ref{eq-cont2}) is nonlinear and, hence, $v_g$ 
is $c$-dependent. Besides, the waves of the type (\ref{nlwave}) are 
called {\it kinematic waves} \cite{lw,whit,richards} to emphasize 
their purely kinematic origin, in contrast to the dynamic origin of 
the accoustic and elastic waves. From the initial given {\it density 
profile} $c(x;0)$ the profile $c(x;\Delta t)$ at time $\Delta t$ 
can be obtained by moving each point on the initial profile a 
distance $v_g(c) t$ to the right; obviously, the distance moved is 
different for different values of $c$. The time-evolution of the 
density profile can be shown graphically \cite{lw,whit} on the 
space-time diagram (i.e., the $x-t$ plane) where an arbitrary point 
$x_0$ on the $t = 0$ axis moves along a straight line of slope 
$v_g(c)$ if the initial density at $x_0$ is $c$. These straight 
lines are referred to as {\it characteristics}; different 
characteristics corresponding to different $c$ have different 
slopes $v_g(c)$. 

The speed $v_g(c)$ of the density wave should not be confused with 
$v(c)$, the actual speed of the continuum fluid representing traffic. 
In fact, at any instant of time $v(x;t)$ can be obtained from the 
corresponding density profile $c(x;t)$ by using the relation 
$v(x;t) = j(c(x;t))/c(x;t)$. Moreover, since 
$v_g = v(c) + c \frac{dv(c)}{dc}$ and since $\frac{dv(c)}{dc} < 0$, 
the speed of the density wave is less than that of the fluid. 
Therefore, the density wave propagates backward {\it relative to the 
traffic} and the drivers are thereby warned of density fluctuations 
ahead downstream. Furthermore, the density wave moves forward or 
backward {\it relative to the road}, depending on whether $c < c_m$ or 
$c > c_m$ where $c_m$ corresponds to the maximum in the function $j(c)$.
 
When $J(c)$ is convex, i.e., $d^2J/dc^2 < 0$, we have $dv_g/dc < 0$; 
consequently, higher values of $c$ propagate slower than lower 
values of $c$ thereby distorting the initial density profile. On the 
other hand, when $dv_g/dc > 0$ higher values of $c$ propagate faster 
and the distortion has the opposite tendency as compared to the 
case of $dv_g/dc < 0$. In both the situations the distortion of the 
initial density profile is caused by the $c$-dependence of $v_g$ 
which arises from the nonlinearity of the equation (\ref{eq-cont2}). 
The distortion of the density profile with time can also be followed 
on the space-time diagram. If $dv_g/dc < 0$, in regions of decreasing 
density (i.e., $c(x_1) > c(x_2)$ for $x_1 < x_2$) the characteristics 
move away from each other whereas, in regions of increasing density, 
the characteristics move towards each other. 

When two characteristic lines on the space-time diagram intersect 
the density would be double-valued at the point of intersection. We 
can avoid this apparently impossible scenario by the following 
interpretation: When two characteristic lines intersect a {\it shock 
wave} is generated. By definition, a shock represents a mathematical 
discontinuity in $c$ and, hence, also in $v$. 
The speed of a shock wave is given by 
\begin{equation}
v_s = \frac{J(c^+) - J(c^-)}{c^+ - c^-} 
\end{equation}
where $c^+$ and $c^-$ are, respectively, the densities immediately in 
front (downstream) and behind (upstream) the shock while $J(c^+)$ and 
$J(c^-)$ represent the corresponding downstream and upstream fluxes, 
respectively. Note that the shock wave moves downstream (upstream) if 
$v_s$ is positive (negative). Often the shock is weak in the sense that 
the relative discontinuity $(c^+-c^-)/c^-$ is small and in such cases 
the shock wave speed tends to $v_g = dJ/dc$. 
As a shock separates a section of high an low densities of the model,
it corresponds to a section of a highway where a free-flow and a
congested regime is present. In particular for large differences
between  $c^+$ and $c^-$ the velocity of the shock can be interpreted 
as the velocity of a backwards moving jam. 

One advantage of the kinematic approach outlined above over any
dynamic approach is that the dynamical equation, which will be given
in Sec.~\ref{sec_Nav}, is difficult to derive from basic first
principles and usually involve quite a few phenomenological parameters
and even a phenomenological function. On the other hand, the only
input needed for the kinematic approach is the phenomenological
function $J(c)$ which can be obtained from empirical data.

\subsection{Diffusion term in Lighthill-Whitham theory and its effects}

Let us now make improvement over the original Lighthill-Whitham theory,  
which is based on the first approximation (\ref{j-c}). We now assume 
that the local flux $J(x;t)$ is determined not only by the local density 
$c(x;t)$ but also by the gradient of the density. In other words, we 
replace the assumption (\ref{j-c}) by  
\begin{equation}
J(c) = j(c) - D \frac{\partial c}{\partial x} 
\label{j-cdiff}
\end{equation}
where $D$ is a positive constant. Note that, for fixed $c(x;t)$ (and, 
hence, fixed $j(c)$), a positive (negative) density gradient leads to 
a lower (higher) flux as the drivers are expected to reduce (increase) 
the speed of their vehicles depending on whether approaching a more 
(less) congested region. Using the relation (\ref{j-cdiff}) in the 
equation of continuity (\ref{eq-continuity}) we now get 
\begin{equation}  
\frac{\partial c(x;t)}{\partial t} 
+ v_g \frac{\partial c(x;t)}{\partial x} 
= D \frac{\partial^2 c(x;t)}{\partial x^2} 
\label{diffusion}
\end{equation}
where $v_g(c) = \frac{dj(c)}{dc}$. The equation (\ref{diffusion}) 
reduces to the equation (\ref{eq-cont2}) when $D = 0$. The nonlinearity 
and diffusion have opposite effects: the term 
$v_g(c)\frac{\partial c}{\partial x}$ 
leads to "steepening" and ultimate "breaking" of the wave whereas 
the term $D\partial^2 c/\partial x^2$ smoothens out the profile.
Nonvanishing $D$ also leads to a non-zero width of the shock wave. 

\subsection{Greenshields model and Burgers equation}

So far in the preceding subsections we have not considered any 
specific form of the function $j(c)$ relating flux with density.
One can start with the simplest (differentiable) approximation capturing
the basic form of the fundamental diagram,
\begin{equation}
J = v_{max} c (1-c). 
\label{green}
\end{equation}
Note that $v_{max}$ in (\ref{green}) is a phenomenological parameter 
and it is interpreted to be the maximum average speed for 
$c \rightarrow 0$. In traffic science and engineering, one usually 
uses $1- c/c_{jam}$ instead of $1-c$ in the equation (\ref{green}) 
and the corresponding form of the relation between $J$ and $c$ is 
known as the Greenshields \cite{greenshield} model. Substituting 
(\ref{green}) into the equation (\ref{diffusion}) we get 
\begin{equation}  
\frac{\partial c(x;t)}{\partial t} 
+ v_{max} \frac{\partial c(x;t)}{\partial x} 
- 2v_{max} c~\frac{\partial c(x;t)}{\partial x} 
= D \frac{\partial^2 c}{\partial x^2} 
\label{burg1}
\end{equation}
Introducing the linear transformation of variables 
\begin{equation}
x = v_{max} t' - x'; \quad t = t' 
\label{transform} 
\end{equation}
one gets 
\begin{equation}  
\frac{\partial c(x;t)}{\partial t'} 
+ 2v_{max} c~\frac{\partial c(x;t)}{\partial x'} 
= D \frac{\partial^2 c}{\partial {x'}^2} 
\label{burgers}
\end{equation}
which is the (deterministic) Burgers equation~\cite{whit,burgersbook}. 
Note that the transformation (\ref{transform}) takes one from the 
space-fixed coordinate system $(x,t)$ to a coordinate system $(x',t')$ 
that moves with uniform speed $v_{max}$; so, vehicles moving with 
speed $v_{max}$ with respect to the coordinate system $(x,t)$ do 
not move at all with respect to the coordinate system $(x',t')$.

The advantage of this route to the theory of traffic flow is that 
the Burgers equation (\ref{burgers}) can be transformed further 
into a diffusion equation, thereby getting rid of the nonlinearity, 
through a nonlinear transformation called the Cole-Hopf transformation 
\cite{whit}. Since it is straightforward to write down the formal 
solution to the diffusion equation, one can see clearly the role 
of the coefficient $D$ and the nature of the solutions in the limit 
$D \rightarrow 0$.

If equation (\ref{eq-cont2}) is assumed to be the only equation 
governing traffic flow then an inhomogeneous initial state can 
lead to a shock wave but the amplitude of the shock wave decreases 
with time and eventually the shock wave fades out leading to a 
homogeneous steady state in the limit $t \rightarrow \infty$. 
Leibig \cite{leibig} has studied how a random initial distribution 
of steps in the density profile evolves with time in this theory. No 
traffic jam forms {\it spontaneously} from a state of uniform
density at this level of sophistication of the fluid-dynamical approach. 

\subsection{Navier-Stokes-like momentum equation and consequences} 
\label{sec_Nav}

Corresponding to the assumption (\ref{j-cdiff}) we can write 
a velocity equation 
\begin{equation}
v(x;t) = v(c(x;t)) - \frac{D}{c} \frac{\partial c(x;t)}{\partial x} 
\label{vel1}
\end{equation}
where $v(c) = \frac{j(c(x;t))}{c(x;t)}$. In the kinematic approach 
discussed so far in the preceding subsections it is implicitly 
assumed that, following any change in the local density (and density 
gradient) is followed by an immediate response (without delay) of the
velocity field.
For a more realistic description, the local speed 
should be allowed to relax after a non-zero {\it delay time} $t_d$. 
So, it seems natural to treat the right hand side in (\ref{vel1}) 
as a {\it desired local velocity} at $x$ and write the total 
derivative $dv/dt$ of $v$ with respect to time as 
\cite{payne,payne2,kuhne1,kuhne2}
\begin{equation}
c \frac{\partial v}{\partial t} + v \frac{\partial v}{\partial x} 
= \frac{c}{t_d}\biggl[v_{safe}(c) - v\biggr]  
- D \frac{\partial c(x;t)}{\partial x}  
\label{vel3}
\end{equation}
where the function $v_{safe}(c)$ is identical to $v(c)$. Note that 
$v_{safe}(c)$ is a monotonically decreasing function of $c$, i.e, 
$dv_{safe}/dc < 0$. Equation (\ref{vel3}) is an additional dynamical 
equation describing the time-dependence of the velocity $v(x;t)$. 

Now let us interpret the two terms on the right hand side of
(\ref{vel3}).  The phenomenological function $v_{safe}(c)$ gives the
safe speed, corresponding to the vehicle density $c$, that is achieved
in time-independent and homogeneous traffic flow and $t_d$ is the
corresponding average relaxation time. Next, note that the term $D
\frac{\partial c(x;t)}{\partial x}$ takes into account the natural
tendency of the drivers to accelerate (decelerate) if the density
gradient is negative (positive), i.e.\ if the density in front
becomes smaller (larger); therefore, it can be interpreted as
proportional to the pressure gradient in the fluid describing traffic.
In addition to these terms, another term proportional to
$\frac{\partial^2 v}{\partial x^2}$ is also added to the right hand
side of the velocity equation; this tends to reduce spatial
inhomogeneities of the velocity field and is usually interpreted as
the analogue of the viscous dissipation term in the Navier-Stokes
equation.

Thus, finally, in the fluid-dynamical approach, a complete 
mathematical description of the vehicular traffic on highways
is provided by two equations, namely, the  equation of continuity 
(\ref{full-cont}) and the Navier-Stokes-like velocity equation 
\cite{kernetal,hongetal,choi,leeetal} 
\begin{eqnarray}
c \left[\frac{\partial v}{\partial t} 
+ v \frac{\partial v}{\partial x} \right] 
= - D \frac{\partial c}{\partial x} 
+ \mu \frac{\partial^2v}{\partial x^2} + \frac{c}{t_d}[v_{safe}(c)-v] 
\label{navier}
\end{eqnarray}
where $D, \mu$ and $t_d$ are phenomenological constants.

\subsection{Fluid-dynamical theories for multi-lane highways and 
city traffic}

One can describe the traffic on two-lane highways \cite{leeproc} 
by two equations each of the same form (\ref{full-cont}) 
and where the source term in the equation for lane $1$ (lane $2$) 
takes into account the vehicles which enter into it from the lane $2$ 
(lane $1$) while the sink term takes into account those vehicles 
entering the lane $2$ (lane $1$) from the lane $1$ (lane $2$). 

A lattice hydrodynamic theory for city traffic has been formulated 
recently \cite{latt_hydro}. This fluid-dynamical model is 
motivated by the CA model, developed by Biham et al.~\cite{bml}, 
which will be discussed in detail later in this review. Instead 
of generalizing the Navier-Stokes equation (\ref{navier}) a 
simpler form of the velocity equation has been assumed.

\subsection{Some recent results of the fluid-dynamical theories and 
their physical implications} 

The fluid-dynamical model of vehicular traffic has been studied 
numerically by discretizing the partial differential equations 
(\ref{full-cont}) and (\ref{navier}) together with 
appropriate initial and boundary conditions. Both periodic 
boundary conditions and open boundary conditions with time-independent 
external flux $\alpha_i^0(t) = \beta_i^0(t) = \gamma$ have been 
considered. 

\begin{figure}[ht]
\centerline{\epsfig{figure=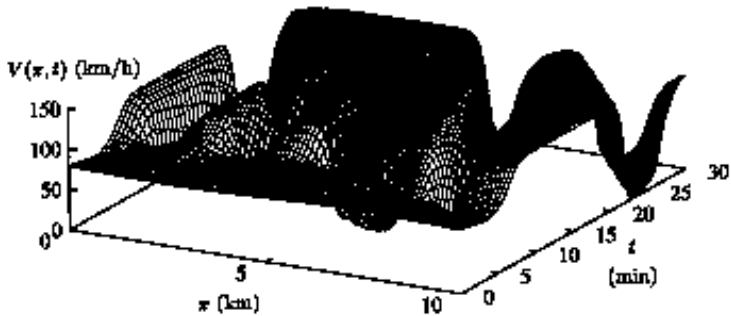,height=5cm}}
\caption{Time evolution of a density profile in the 
         K\"uhne-Kerner-Konh\"auser model \cite{kuhne1,kuhne2,kernetal}
         consisting of equations
         (\ref{full-cont}) and (\ref{navier}) with a special choice
         of $v_{safe}(c)$. Initially an approximately
         constant density profile is assumed. The figure shows that 
         the homogeneous state is not stable, already for small 
         deviations from a flat profile (from \cite{helbook}).}
\label{profile}
\end{figure}

In the fluid-dynamical theory, based on the equation of continuity 
(\ref{full-cont}) and the Navier-Stokes-like equation 
(\ref{navier}) traffic jams can appear {\it spontaneously}, 
even if the initial density profile $c(x;0)$ deviates very little 
from the homogeneous state (fig.~\ref{profile}). In order to 
understand the physical mechanism of the formation of local cluster 
of vehicles, let us consider a local increase of the density 
$\Delta c(x) > 0$ at some location $x$. Since $dv_{safe}/dc < 0$, 
the local increase of the density leads to a decrease of $v_{safe}$. 
This decrease in the safe velocity forces drivers to reduce their 
average velocity $v$ sharply if $|dv_{safe}/dc|$ is large enough. 
On the other hand, it follows from the equation of continuity that 
the local decrease of $v$ gives rise to further increase of $c$ 
around $x$ and, consequently, further subsequent decrease of $v(x)$ 
in this location. This avalanche-like process, which tends to 
increase the amplitude of the local fluctuation of the density 
around the homogeneous state, competes against other processes, 
like diffusion and viscous dissipation, which tend to decrease 
inhomogeneities.

To our knowledge, the first attempt to understand the physical 
mechanism of {\it synchronized} traffic within the framework 
of the fluid-dynamical formalism was made by Lee et 
al.~\cite{leeetal}\footnote{A brief summary of that work can be
found in \cite{helltreib}.}.
On a finite stretch of highway, of length $L$, they installed an 
on-ramp and an off-ramp on the model highway with a separation 
of $L/2$ between them. They chose the spatial distribution of the 
external flux $\phi(x)$ in equation (\ref{space_flux}) as 
\begin{equation}
\phi(x) = (2 \pi \sigma^2)^{-1/2} \exp(-x^2/2 \sigma^2) 
\end{equation}
with $\sigma = 56.7m$. They also assumed the form $v_{safe}(c) =
v_0(1-c/\hat{c})/[1+E(c/\hat{c})^{\theta}]$ for the safe velocity with
adjustable parameters $v_0, E, \theta,\hat{c}$. Lee et al.\ 
\cite{leeetal} first allowed the system to reach the steady state
after applying a weak time-independent flux $\alpha_i^0(t) =
\beta_i^0(t) = \gamma$ and simulating the time-evolution of the
traffic by solving simultaneously the equation of continuity
(\ref{full-cont}) and the Navier-Stokes-like equation (\ref{navier})
with a specific set of chosen values for the parameters $t_d, D, \mu,
v_0, E, \theta,\hat{c}$, etc.  Since they chose $\gamma < \gamma_c$,
an initially homogeneous traffic reaches a steady "free-flow" where
homogeneous regions with different densities are separated from each
other by narrow "transition layers" near the ramps ( no stable
"free-flow" exists if $\gamma \geq \gamma_c$).  Then, they applied a
pulse of additional flux $\delta q$ at the on-ramp for a short
duration $\delta t$. After a transient period, which depends on the
parameters of the model, the system was found to settle in a limit
cycle in which the local density and local flux oscillate periodically
and the oscillations are localized near the on-ramp. The {\it
  discontinuous} change of the spatio-temporally averaged velocity
induced by the localized perturbations of finite amplitude, associated
hysteresis effects and the stability of the limit cycle were found to
be qualitatively similar to some of the empirically observed
characteristics of synchronized flow in real traffic; therefore, Lee
et al.~\cite{leeetal} identified the limit cycle observed in their
theoretical investigation as the synchronized state of vehicular
traffic.  They drew analogy between this state and a "self-excited
oscillator" \cite{leeetal}. However, this mechanism of the
synchronized state is not yet accepted as the true and only possible
explanation of the phenomena associated with the synchronized state
observed empirically.
 
Meanwhile similar results have been obtained for a gas-kinetic based
traffic model \cite{hellprl1}, also using on- and off-ramps in order 
to explain the transition from free-flow to synchronised states. This 
work was completed in a recent paper, where a phase diagram was
calculated, which dependends on the on-ramp activity and the flow on
the highway \cite{hellprl2}. Summarising the recent results of the
macroscopic traffic models, there seems to be evidence that on-and
off-ramps play an important role for a theoretical explanation of
synchronised traffic. Nevertheless some experimental features are  
still not captured by these approaches. E.g. the empirical results
show that for synchonised traffic of type (iii) no correlations
between density and flow exist \cite{Neubert99}, 
in contrast to the regular patterns 
of the oscillating states found in simulations of the macroscopic
models. 

Despite its success in capturing many aspects of traffic flow
the fluid-dynamical approach has its limitations; 
for example, viscosity of traffic is not a directly measurable 
quantity. Nevertheless, the fluid dynamical approach is being 
pursued not only by some physicists but also by several members 
of the traffic engineering community \cite{newel93s,daganzo}.

\section{Kinetic theories of vehicular traffic} 
\label{section_kinetic}

In the kinetic theory, traffic is treated as a gas of 
interacting particles where each particle represents a 
vehicle. The various different versions of the kinetic 
theory of vehicular traffic 
\cite{ph,prigo,fontana,lehmann,wagkin1,wagkin2,helbkin,nagakin} 
have been developed by modifying the kinetic theory of 
gases. 

Recall that in the kinetic theory of gases \cite{huang}
$f(\vec r, \vec p;t) d^3r d^3p$ denotes the number of molecules 
which, at time $t$, have positions lying within a volume element 
$d^3r$ about $\vec r$ and momenta lying within the momentum-space 
element $d^3p$ about $\vec p$. The Boltzmann equation, which 
describes the time-evolution of the distribution $f(x,v;t)$, is 
given by 
\begin{equation}
\biggl[\frac{\partial f}{\partial t} 
+ \frac{\vec p}{m} \cdot \nabla_r 
+ \vec F \cdot \nabla_p \biggr] f(\vec r, \vec p; t) 
= \biggl(\frac{\partial f}{\partial t}\biggr)_{coll} 
\label{boltzmann_molec}
\end{equation}
where the symbols $\nabla_r$ and $\nabla_p$ denote gradient 
operators with respect to $\vec r$ and $\vec p$, respectively, 
while $\vec F$ is the external force. The term 
$\left(\frac{\partial f}{\partial t}\right)_{coll}$   
represents the rate of change of $f$, with time, which is 
caused by the mutual collisions of the molecules.

In the first of the following two subsections we present the 
earliest version of the kinetic theory of vehicular traffic 
which was introduced by Prigogine and coworkers \cite{prigo,ph} 
by modifying some of the key concepts in the kinetic theory 
of gases and by writing down an equation analogous to the 
Boltzmann equation (\ref{boltzmann_molec}). In the subsequent 
subsection we discuss the kinetic theory developed later by 
Paveri-Fontana \cite{fontana} to cure the defects from which 
the Prigogine theory was found to suffer. 

\subsection{Prigogine model}

Suppose $f(x,v;t)dx dv$ denotes the number of vehicles, at time
$t$, between $x$ and $x + dx$, having {\it actual} velocity between
$v$ and $v + dv$. In addition, Prigogine and coworkers \cite{prigo,ph}
introduced a {\it desired distribution} $f_{des}(x,v)$ which is a
mathematical idealization of the goals that the population of the
drivers {\it collectively} strives to achieve. The actual distribution
may deviate from the desired distribution because of various possible
influences, e.g., road conditions, weather conditions or interaction
with other vehicles, etc. They also argued that some of these
influences cease after some time while the interactions with the other
vehicles persist for ever.  For example, only a short stretch of the
road surface may be icy and strong winds or rain may stop after a
short duration; in such situations $f$ can relax to $f_{des}$ over a
relaxation time $\tau_{rel}$ provided mutual interactions of the
vehicles is negligibly small. On the basis of these arguments,
Prigogine and coworkers \cite{prigo,ph} suggested that the analogue of
the Boltzmann equation for the traffic should have the form
\begin{equation}
\frac{\partial f}{\partial t} + v \frac{\partial f}{\partial x} 
= \biggl(\frac{\partial f}{\partial t}\biggr)_{rel} 
+ \biggl(\frac{\partial f}{\partial t}\biggr)_{int} 
\label{boltzmann_ph}
\end{equation}
where $\left(\frac{\partial f}{\partial t}\right)_{rel}$ 
accounts for the relaxation of $f$ towards $f_{des}$ in the 
absence of mutual interactions of the vehicles while
$\left(\frac{\partial f}{\partial t}\right)_{int}$
accounts for the changes of $f$ arising from  mutual {\it 
interactions} among the vehicles. Note that the term 
$\left(\frac{\partial f}{\partial t}\right)_{int}$ on the 
right hand side of (\ref{boltzmann_ph}) may be interpreted 
as the analogue of the term 
$\left(\frac{\partial f}{\partial t}\right)_{coll}$ in 
the equation (\ref{boltzmann_molec}) whereas the term   
$\left(\frac{\partial f}{\partial t}\right)_{rel}$ in 
equation (\ref{boltzmann_ph}) may be interpreted as the 
counterpart of the term 
$\vec F \cdot \nabla_p f(\vec r, \vec p; t)$ 
in the equation (\ref{boltzmann_molec}).
 
Prigogine and coworkers wrote down an explicit form for the 
term $\left(\frac{\partial f}{\partial t}\right)_{int}$ 
by generalizing that for the term 
$\left(\frac{\partial f}{\partial t}\right)_{coll}$ 
in the kinetic theory of gases. We shall consider this 
term in the next subsection. In order to write down a simple 
explicit form of the {\it relaxation} term in the equation 
(\ref{boltzmann_ph}) they assumed that\\ 
(i) the {\it collective} relaxation, which would cause the 
actual distribution to tend towards the desired distribution, 
involves only a single relaxation time $\tau_{rel}$ so that 
\begin{equation}
\biggl(\frac{\partial f}{\partial t}\biggr)_{rel} =  
- \frac{f - f_{des}}{\tau_{rel}} 
\label{exporel}
\end{equation}
and\\ 
(ii) the desired {\it speed} distribution $F_{des}(v)$ 
remains independent of the local concentration $c(x;t)$ so 
that 
\begin{equation}
f_{des}(x,v;t) = c(x;t) F_{des}(v) 
\label{nonadap}
\end{equation}
Therefore, a  more explicit form of the Boltzmann-like equation 
(\ref{boltzmann_ph}) in the Prigogine theory is given by 
\begin{equation}
\frac{\partial f}{\partial t} + v \frac{\partial f}{\partial x} 
= - \frac{f(x,v;t) - c(x;t)F_{des}(v)}{\tau_{rel}} 
+ \biggl( \frac{\partial f}{\partial t}\biggr)_{int}
\label{fullprigo}
\end{equation}

Note that, in the absence of mutual interactions of the vehicles, 
the distribution $f(x,v;t)$ would relax exponentially with time. 
The concept of desired {\it distribution} $f_{des}(x,v;t)$ and 
this scenario of {\it collective} relaxation of $f$ towards 
$f_{des}$ has subsequently come under severe criticism \cite{fontana}. 
Analyzing a set of "ideal experiments" in the light of the 
Prigogine theory, Paveri-Fontana \cite{fontana} showed 
that the results obtained from the Boltzmann-like equation 
(\ref{fullprigo}) are physically unsatisfactory. 

More recently, Lehmann \cite{lehmann} has attempted to revive 
the Prigogine approach by reformulating it as a {\it semi-
phenomenological} theory where the distribution $f(x,v;t)$ is 
assumed to follow the simpler form 
\begin{equation}
\frac{\partial f}{\partial t} + v \frac{\partial f}{\partial x} 
= - \frac{f - f_{des}(v,c)}{\tau_{rel}}
\end{equation}
and the effects of the interactions are taken into account 
implicitly through a density-dependent desired distribution 
function $f_{des}(v,c)$ which has to determined empirically.

\subsection{Paveri-Fontana model} 

In order to remove the conceptual as well as mathematical drawbacks 
of the Prigogine model of the kinetic theory of vehicular traffic, 
Paveri-Fontana \cite{fontana} argued that each vehicle, in contrast 
to the molecules in a gas, has a {\it desired} velocity towards 
which its {\it actual} velocity tends to "relax" in the absence of 
"interaction" with other vehicles. Thus, Paveri-Fontana model is 
based on a scenario of relaxation of the velocities of the {\it 
individual} vehicles rather than a collective relaxation of the 
distribution of the velocities. 

In mathematical language, Paveri-Fontana introduced an additional
phase-space coordinate, namely, the desired velocity.  Suppose,
$g(x,v,v_{des};t) dx dv dv_{des}$ denotes the number of vehicles at
time $t$ between $x$ and $x + dx$, having {\it actual} velocity
between $v$ and $v + dv$ and {\it desired} velocity between $v_{des}$
and $v_{des} + dv_{des}$. The one-vehicle {\it actual} velocity
distribution function
\begin{eqnarray}
f(x,v;t) = \int dv_{des}  g(x,v,v_{des};t) 
\end{eqnarray}
describes the probability of finding a vehicle between $x$ and 
$x+dx$ having {\it actual} velocity between $v$ and $v+dv$ at 
time $t$. Similarly, the one-vehicle {\it desired} velocity 
distribution function 
\begin{eqnarray}
f_0(x,v_{des};t) = \int dv  g(x,v,v_{des};t) 
\end{eqnarray}
describes the probability of finding a vehicle between $x$ and 
$x+dx$ having {\it desired} velocity between $v_{des}$ and 
$v_{des}+dv_{des}$. The local density of the vehicles $c(x;t)$ 
at the position $x$ at time $t$ can be obtained from 
$$ 
c(x;t) = \int_0^{\infty} dv_{des} \int_0^{\infty} dv g(x,v,v_{des};t).
$$ 
Similarly, the corresponding average {\it actual} speed $\langle
v(x;t)\rangle$ and the average {\it desired} speed $\langle
v_{des}(x;t)\rangle$ are defined as
$$ 
\langle v(x;t)\rangle = \frac{\int_0^{\infty} dv_{des} 
\int_0^{\infty} dv v  g(x,v,v_{des};t)}{c(x;t)}. 
$$
$$ 
\langle v_{des}(x;t)\rangle = \frac{\int_0^{\infty} dv_{des} 
\int_0^{\infty} dv v_{des} g(x,v,v_{des};t)}{c(x;t)}. 
$$
Finally, the local flux $J(x;t)$ is defined as $J(x;t) = c(x;t) 
\langle v(x;t)\rangle$.

Now let us {\it assume} that the desired velocity of each 
individual driver is independent of time, i.e., $dv_{des}/dt = 0$. 
Of course, the drivers may also adapt to the changing traffic 
environment and their desired velocities may  change accordingly. 
In principle, these features can be incorporated into the kinetic 
theory at the cost of increasing complexity of the formalism. 

Next, let us also {\it assume} that, in the absence of interaction 
with other vehicles, an arbitrary vehicle reaches the desired 
velocity exponentially with time, i.e., $dv/dt = (v_{des}-v)/\tau$ 
where $\tau$ is a relaxation time. The Boltzmann-like kinetic 
equation for $g(x,v,v_{des};t)$ can be written as 
\begin{eqnarray}
\left[\frac{\partial}{\partial t} + v \frac{\partial}{\partial x}
\right] g + \frac{\partial}{\partial v}
\left[\frac{v_{des}-v}{\tau}g \right]
= \left(\frac{\partial g}{\partial t} \right)_{int} 
\label{boltzmann}
\end{eqnarray}

In order to write down an explicit form of the "interaction term" we  
have to model the interactions among the vehicles. First of all, we 
model the vehicles as point-like objects. We consider the scenario 
where a fast vehicle, when hindered by a slow leading vehicle, either 
passes or slows down to the velocity of the lead vehicle. Let us now 
make some further simplifying assumptions:\\ 
(i) The slowing down takes place with a probability $1-P_{pass}$ 
where $P_{pass}$ is the probability of passing.\\
(ii) If the fast vehicle passes the slower leading vehicle, its own 
velocity remains unchanged.\\
(iii) The velocity of the slower leading vehicle remains unchanged, 
irrespective of whether the faster following vehicle passes or slows 
down.\\
(iv) The slowing down process is instantaneous, i.e., the braking time 
is negligibly small.\\
(v) It is adequate to consider only two-vehicle interactions; there 
is no need to consider three-vehicle (or multi-vehicle) interactions.\\ 
(vi) The postulate of "vehicular chaos", which is the analogue of 
the postulate of "molecular chaos" in the kinetic theory of gases, 
holds, so that the two-vehicle distribution function 
$g_2(x,v,v_{des},x',v',v_{des}';t)$ 
can be approximated as a product of two one-particle distributions 
$g(x,v,v_{des};t)$ and $g(x',v',v_{des}';t)$, i.e., 
$g_2(x,v,v_{des},x',v',v_{des}';t) \simeq g(x,v,v_{des};t) 
g(x',v',v_{des}';t)$. Thus, the 
equation (\ref{boltzmann}) can be written explicitly as  
\begin{eqnarray}
\left[\frac{\partial}{\partial t}+v\frac{\partial}{\partial x}\right]g 
&+& \frac{\partial}{\partial v}\left[\frac{v_{des}-v}{\tau} g \right]
\nonumber\\
&=&  f(x,v;t) \int_v^{\infty} dv' (1-P_{pass}) (v' - v) 
g(x,v',v_{des};t) \nonumber \\ 
&& - g(x,v,v_{des};t) \int_0^v dv' (1-P_{pass}) (v - v') f(x,v';t)  
\label{fontanaeq}
\end{eqnarray}
where the form of the "interaction term" on the right-hand side of 
the equation (\ref{fontanaeq}) follows from the assumptions (i)-(vi) 
above. 
The first term on the right hand side of (\ref{fontanaeq}) describes the 
"gain" of probability $g(x,v,v_{des};t)$ from the interaction of vehicles 
of actual velocity $v'$ with slower leading vehicle of actual velocity 
$v$ while the second term describes the loss of the probability 
$g(x,v,v_{des};t)$ arising from the interaction of vehicles of actual 
velocity $v$ with even slower leading vehicle of actual velocity $v'$. 

The stationary homogeneous solution $g(v,v_{des})$ is, by definition, 
independent of $x$ and $t$. But, to our knowledge, so far it has 
not been possible to get even this solution of the Boltzmann-like 
integro-differential equation (\ref{fontanaeq}) by solving it 
analytically even for the simplest possible choice of the desired 
distribution function although numerical solutions  
\cite{wagkin2,nagakin} provide some insights into the regimes of 
validity of the equation (\ref{fontanaeq}) and gives indications as 
to the directions of further improvements of the Paveri-Fontana 
model. For example, the finite sizes of the vehicles must be taken 
into account at high densities. Besides, the assumption (iv) of 
instantaneous relaxation has also been relaxed in a more recent 
extension \cite{wagkin2}. 

Normally passing would require more than one lane on the highway. 
Therefore, the models discussed so far in the context of the 
kinetic theory may be regarded, more appropriately, as {\it 
quasi-one-dimensional}. These neither deal {\it explicitly} with 
$g_i(x,v,v_{des};t)$ for the individual lanes (labeled by $i$) 
nor take into account the process of lane-changing. 
Besides, all the vehicles were assumed to be of the same 
type. Now, in principle, we can generalize the formalism of the 
kinetic theory of traffic to deal with {\it different types of 
vehicles} on {\it multi-lane highways}. Suppose $g_i^a(x,v,v_{des};t)$ 
is the distribution for vehicles of type $a$ on the $i$-th lane 
of the highway. Obviously, the Boltzmann-like equations for 
the different lanes are coupled to each other. However, one 
needs additional postulates to model the lane-changing rules 
\cite{naga_kin,multi_kin}. 

Very little work has been done so far on developing kinetic 
theories of two-dimensional traffic flow which would represent,
for example, traffic in cities. Suppose, for simplicity, that 
the network of the streets consists of north-south and east-west 
streets and that east-west streets allow only east-bound traffic 
while only north-bound traffic flow takes place along the 
north-south streets. Let $P_{xx}$ ($P_{xy}$) denote the probabilities 
of an east-bound vehicle passing another east-bound (north-bound) 
vehicle. Similarly, suppose, $P_{yx}$ ($P_{yy}$) denote the 
probabilities of a north-bound vehicle passing an east-bound 
(north-bound) vehicle. Let $g_x(x,y,u,u_{des};t)$ and 
$g_y(x,y,v,v_{des};t)$ represent the distributions for the 
east-bound and north-bound vehicles, respectively, where $u$ 
and $v$ refer to the {\it actual} their actual velocities 
whereas $u_{des}$ and $v_{des}$ refer to the corresponding 
{\it desired} velocities. The Boltzmann-like equations governing 
the time evolutions of these distributions are given by \cite{city_kin}  
\begin{equation}
\left[\frac{\partial}{\partial t}+ u \frac{\partial}{\partial x}
+ v \frac{\partial}{\partial y} \right] g_x 
+ \frac{\partial}{\partial u} \left[
\frac{u_{des}-u}{\tau}g_x \right]
= \left(\frac{\partial g_x}{\partial t}\right)_{x,coll} 
+ \left(\frac{\partial g_x}{\partial t}\right)_{y,coll} 
\label{kin_city1}
\end{equation}
where 
\begin{eqnarray}
\left(\frac{\partial g_x}{\partial t}\right)_{x,coll} 
& = & f_x(x,y,u;t) \int_u^{\infty} du' (1-P_{xx}) (u' - u) 
g_x(x,y,u',u_{des};t) \nonumber \\ 
& - & g_x(x,y,u,u_{des};t) \int_0^u du' (1-P_{xx})(u - u')f_x(x,y,u';t)  
\label{kin_city2}
\end{eqnarray} 
and 
\begin{eqnarray}
\left(\frac{\partial g_x}{\partial t}\right)_{y,coll} 
& = & \delta_{u,0} \int_0^{\infty} dv' f_y(x,y,v';t) 
\int_0^{\infty} du' (1-P_{xy}) u' g_x(x,y,u',u_{des};t) \nonumber \\ 
& - & g_x(x,y,u,u_{des};t) \int_0^{\infty} dv' (1-P_{xy}) u f_y(x,y,v';t)  
\label{kin_city3}
\end{eqnarray} 
The first term on the right hand side of the equation 
(\ref{kin_city2}) describes gain of population of east-bound 
vehicles with velocity $u$ because of interaction with 
other east-bound vehicles with velocity $u' \geq u$ while 
the second term describes the loss of population of east-bound 
vehicles because of interaction with east-bound vehicles of 
velocity $u' < u$. The right hand side of the equation 
(\ref{kin_city3}) is based on the {\it assumption} that when an 
east-bound vehicle interacts with a north-bound vehicle at a 
crossing, it either passes or stops.

\subsection{Derivation of the phenomenological equations of 
the macroscopic fluid-dynamical theories from the microscopic 
gas-kinetic models} 

In this section we discuss the results of the attempts to derive 
the phenomenological equations of traffic flow in the macroscopic
fluid-dynamical theories from the microscopic gas-kinetic models. 
Several attempts have been made so far to derive the equation of 
continuity and the Navier-Stokes-like equation for traffic from 
the corresponding Boltzmann-like equation in the same spirit in 
which the derivations of the equation of continuity and Navier-Stokes 
equation for viscous fluids from the Boltzmann-equation have been 
carried out. However, because of the postulate of "vehicular chaos", 
the equation (\ref{fontanaeq}) is expected to be valid only at very 
low densities where the correlations between the vehicles is 
negligibly small whereas traffic is better approximated as a 
continuum fluid at higher densities! 

Let us define the moments
\begin{equation}
m_{k,\ell} (x;t) = \int dv ~ \int dv_{des} v^k v_{des}^{\ell} 
g(x,v,v_{des};t)
\label{moment}
\end{equation}
Note that $c = m_{0,0}$, $\langle v\rangle = m_{1,0}$.
Integrating the Boltzmann-like equation (\ref{fontanaeq}) over the
actual velocities we get
\begin{equation}
\frac{\partial}{\partial t} f_0(x,v_{des};t) 
+ \frac{\partial}{\partial x} [\bar
{v}(x,v_{des};t) f_0(x,v_{des};t)] = 0
\label{kin-eqcont}
\end{equation}
where $\bar{v}(x,v_{des};t)$ is defined as
\begin{equation}
{\bar v}(x,v_{des};t) = \frac{\int dv ~v 
~g(x,v,v_{des};t)}{f_0(x,v_{des};t)}
\end{equation}
The equation (\ref{kin-eqcont}) is an equation of continuity
for each desired speed $v_{des}$ separately; it is consequence of
the assumption that $dv_{des}/dt = 0$, i.e., no driver changes the
desired speed.

Using the Boltzmann-like equation (\ref{fontanaeq}) and the definition
(\ref{moment}) we can get separate partial differential equations
for the moments of $v$, moments of $v_{des}$ and the mixed moments
of $v$ and $v_{des}$. Unfortunately, these lead to a {\it hierarchy}
of moment equations where each evolution equation for moment
of a given order involves also moments of the next higher order.
In order to close this system of equations, one needs to
make appropriate justifiable assumptions.


\section{Car-following theories of vehicular traffic} 
\label{sec_carfoll}

In the car-following theories \cite{hg,gazis,rotheryrev} one writes, 
for each individual vehicle, an equation of motion which is the 
analogue of the Newton's equation for each individual particle in a 
system of interacting classical particles. In Newtonian mechanics, the 
acceleration may be regarded as the {\it response} of the particle 
to the {\it stimulus} it receives in the form of force which includes 
both the external force as well as those arising from its interaction 
with all the other particles in the system. Therefore, the basic 
philosophy of the car-following theories \cite{hg,gazis,rotheryrev} 
can be summarized by the equation 
\begin{equation}
[Response]_n \propto [Stimulus]_n
\end{equation}
for the $n$-th vehicle ($n = 1,2,...$). Each driver can respond to the 
surrounding traffic conditions only by accelerating or decelerating the 
vehicle. Different forms of the equations of motion of the vehicles in 
the different versions of the car-following models arise from the 
differences in their postulates regarding the nature of the stimulus 
(i.e., "behavioural force" or a "generalized force" \cite{tilch98}). 
The stimulus may be composed of the speed of the vehicle, the difference 
in the speeds of the vehicle under consideration and its lead vehicle, 
the distance-headway, etc., and, therefore, in general, 
\begin{equation}
\ddot{x}_n = f_{sti}(v_n, \Delta x_n, \Delta v_n)
\end{equation} 
where the function $f_{sti}$ represents the stimulus received by the 
$n$-th vehicle. Different versions of the car-following models 
model the function $f_{sti}$ differently. In the next two subsections 
we discuss two different conceptual frameworks for modeling $f_{sti}$. 

\subsection{Follow-the-leader model} 
\label{sec_follow}

In the earliest car-following models \cite{reuschel,pipes} 
the difference in the velocities of the $n$-th and $(n+1)$-th 
vehicles was assumed to be the stimulus for the $n$-th 
vehicle\footnote{In the following we label the vehicles in driving
direction such that the $(n+1)$-th vehicle is in front of the
$n$-th vehicle.}. In other words, it was assumed that every driver 
tends to move with the same speed as that of the corresponding 
leading vehicle so that 
\begin{equation}
\ddot{x}_{n}(t) = \frac{1}{\tau}\left[
\dot{x}_{n+1}(t) - \dot{x}_{n}(t)\right]
\label{leader}
\end{equation}
where $\tau$ is a parameter that sets the time scale of the model. 
Note that $1/\tau$ in the equation (\ref{leader}) can be interpreted 
as a measure of the sensitivity coefficient ${\mathcal{S}}$ of the driver; 
it indicates how strongly the driver responds responds to unit stimulus. 
According to such models (and their generalizations proposed in the 
fifties and sixties) the driving strategy is to follow the leader and, 
therefore, such car-following models are collectively referred to as 
the {\it follow-the-leader} model.  

Pipes \cite{pipes} derived the equation (\ref{leader}) by differentiating, 
with respect to time, both sides of the equation 
\begin{equation}
\Delta x_{n}(t) = x_{n+1}(t) - x_{n}(t) = (\Delta x)_{safe} 
+ \tau \dot{x}_{n}(t)
\end{equation}
which encapsulates his basic assumption that (a) the higher is the
speed of the vehicle the larger should be the distance-headway, and, (b)
in order to avoid collision with the leading vehicle, each driver must
maintain a {\it safe distance} $(\Delta x)_{safe}$ from the leading
vehicle.

It has been argued \cite{chandler} that, for a more realistic
description, the strength of the response of a driver at time $t$
should depend on the stimulus received from the other vehicles at time
$t - T$ where $T$ is a response time lag. Therefore, generalizing the
equation (\ref{leader}) one would get \cite{chandler}
\begin{equation}
\ddot{x}_{n}(t+T) = {\mathcal{S}}[\dot{x}_{n+1}(t)-\dot{x}_{n}(t)]
\label{chand}
\end{equation}
where the sensitivity coefficient ${\mathcal{S}}$ is a constant
independent of $n$.

According to the equations (\ref{leader}) and (\ref{chand}), 
a vehicle would accelerate or decelerate to acquire the same speed 
as that of its leading vehicle. This implies that, as if, slower 
following vehicle are dragged by their faster leading vehicle. In 
these {\it linear} dynamical models the acceleration response of 
a driver is completely independent of the distance-headway. 
Therefore, this oversimplified equation, fails to account for the 
clustering of the vehicles observed in real traffic. Moreover, 
since there is no density-dependence in this dynamical equation, 
the fundamental relation cannot be derived from this dynamics. 
In order to make the model more realistic, we now assume \cite{potts} 
that the closer is the $n$-th vehicle to the $(n+1)$-th the higher 
is the sensitivity of the driver of the $n$-th car. In this case, 
the dynamical equation (\ref{chand}) is further generalized to  
\begin{equation}
\ddot{x}_{n}(t+T) = \frac{\kappa}{[x_{n+1}(t)-x_{n}(t)]}
[\dot{x}_{n+1}(t) - \dot{x}_{n}(t)]
\end{equation}
where $\kappa$ is a constant. An even further generalization of 
the the model can be achieved \cite{rothery,gips} by expressing the 
sensitivity factor for the $n$-th driver as 
\begin{equation}
{\mathcal{S}}_{n} = 
\frac{\kappa [v_{n}(t+\tau)]^m}{[x_{n+1}(t)-x_{n}(t)]^{\ell}} 
\end{equation}
where $\ell$ and $m$ are phenomenological parameters to be fixed by 
comparison with empirical data. These generalized follow-the-leader 
models lead to coupled {\it non-linear} differential equations for 
$x_{n}$. Thus, in this "microscopic" theoretical approach, the 
problem of traffic flow reduces to problems of nonlinear dynamics. 

So far as the stability analysis is concerned, there are two types 
of analysis that are usually carried out. The {\it local} stability 
analysis gives information on the nature of the response offered by 
the following vehicle to a fluctuation in the motion of its leading 
vehicle. On the other hand, the manner in which a fluctuation in the 
motion of any vehicle is propagated over a long distance through a 
sequence of vehicles can be obtained from an {\it asymptotic} 
stability analysis.

From experience with real traffic we know that drivers often 
observe not only the leading vehicle but also a few other 
vehicles ahead of the leading vehicle. For example, the effect 
of the leading vehicle of the leading vehicle can be incorporated 
in the same spirit as the effect of "next-nearest-neighbour" 
in various lattice models in statistical mechanics. A linear 
dynamical equation, which takes into account this  
"next-nearest-neighbour" within the framework of the 
follow-the-leader model, can be written as \cite{rotheryrev}
\begin{equation}
\ddot{x}_{n}(t+T) = {\mathcal{S}}^{(1)} [\dot{x}_{n+1}(t) 
- \dot{x}_{n}(t)] + {\mathcal{S}}^{(2)} [\dot{x}_{n+2}(t) 
- \dot{x}_{n}(t)]
\label{nextneighb}
\end{equation}
where ${\mathcal{S}}^{(1)}$ and ${\mathcal{S}}^{(2)}$ are two 
phenomenological response coefficients. 

The weakest point of these theories is that these involve 
several phenomenological parameters which are determined 
through "calibration", i.e., by fitting some predictions 
of the theory with the corresponding empirical data. 
Besides, extension of these models to multi-lane traffic 
is difficult since every driver is satisfied if he/she can 
attain the desired speed! 

\subsection{Optimal velocity models} 
\label{sec_OVM}

We can express the driving strategy of the driver in the car-following 
models in terms of mathematical symbols by writing 
\begin{equation}
\ddot{x}_{n}(t) = \frac{1}{\tau}\left[V^{desired}_{n}(t) 
- v_{n}(t)\right]
\end{equation}
where $V^{desired}_{n}(t)$ is the desired speed of the $n$-th 
driver at time $t$. In all follow-the-leader models mentioned above 
the driver maintains a safe {\it distance} from the leading vehicle 
by choosing the {\it speed} of the leading vehicle as his/her own 
desired speed, i.e., $V^{desired}_{n} (t) = \dot{x}_{n+1}$.  

An alternative possibility has been explored in recent works 
based on the car-following approach
\cite{bandoetal,sugiyama95,sugiyama,sugyamTGF,komatsu,nagaov,nagaov2,naka,mason,nagaov3,muramatsu,hayanaka,wagner,lenz,mitarai,nagaemme}. 
This formulation is based on the assumption that $V^{desired}_{n}$ 
depends on the {\it distance}-headway of the $n$-th vehicle, 
i.e., $V^{desired}_{n}(t) = V^{opt}(\Delta x_{n}(t))$ 
so that
\begin{equation}
\ddot{x}_{n}(t) = \frac{1}{\tau}\left[
V^{opt}(\Delta x_{n}(t)) - v_{n}(t)\right]
\label{ov}
\end{equation}
where the so-called optimal velocity function 
$V^{opt}(\Delta x_{n})$ depends on the corresponding 
instantaneous distance-headway $\Delta x_{n}(t) = x_{n+1}(t)- x_{n}(t)$. 
In other words, according to this alternative driving strategy, 
the $n$-th vehicle tends to maintain a {\it safe speed} that 
depends on the relative position, rather than relative velocity, 
of the $n$-th vehicle. In general, $V^{opt}(\Delta x) \rightarrow 0$ 
as $\Delta x \rightarrow 0$ and must be bounded for 
$\Delta x \rightarrow \infty$. For explicit calculations, one has to 
postulate a specific functional form of $V^{opt}(\Delta x)$. 
Car-following models along this line of approach have been
introduced by Bando et al. \cite{bandoetal,sugiyama95}.
For obvious reasons, these models are usually referred to as 
{\it optimal velocity} (OV) models. 

Since the equations of motion in the follow-the-leader models 
involve only the velocities, and not positions, of the vehicles 
these can be formulated as essentially {\it first order} differential 
equations (for velocities) with respect to time. In contrast, since 
the equations of motion in the OV model involve the positions of the 
vehicles explicitly, the theoretical problems of this model are   
formulated mathematically in terms of {\it second order} differential 
equations (for the positions of the vehicles) with respect to time 
\cite{bandoetal,sugiyama95}.

The simplest choice for $V^{opt}(\Delta x)$ is 
\cite{sugiyama,sugyamTGF}
\begin{equation}
V^{opt}(\Delta x) = v_{max} \Theta(\Delta x-d), 
\label{theta}
\end{equation}
where $d$ is a constant and $\Theta$ is the Heavyside step function. 
According this form of $V^{opt}(\Delta x)$, a vehicle should 
stop if the corresponding distance-headway is less than $d$; 
otherwise, it can accelerate so as to reach the maximum allowed 
velocity $v_{max}$. A somewhat more realistic choice 
\cite{sugyamTGF,naka} is 
\begin{equation}
V^{opt}(\Delta x) = \begin{cases} 
                 0 &{\rm for} \quad \Delta x < \Delta x_A,  \\
                 f\Delta x &{\rm for} \quad \Delta x_A \leq \Delta x 
                 \leq \Delta x_B,  \\
                 v_{max} &{\rm for} \quad \Delta x_B < \Delta x.  \\ 
               \end{cases}
\label{linear}
\end{equation}
The main advantage of the forms (\ref{theta}) and (\ref{linear}) 
of the OV function is that exact analytical calculations, e.g.\ in the
jammed region, are possible \cite{sugyamTGF}. 
Although (\ref{theta}) and (\ref{linear}) may not appear very 
realistic, they capture several key features of more realistic 
forms of OV function \cite{bandoetal}, e.g., 
\begin{equation}
V^{opt}(\Delta x) = \tanh[\Delta x- \Delta x_c] + \tanh[\Delta x_c].
\label{ovfunction}
\end{equation} 
for which analytical calculations are very difficult.
For the convenience of numerical investigation, the dynamical 
equation (\ref{ov}) for the vehicles in the OV model has been 
discretized and, then, rewritten as a {\it difference} equation 
\cite{nagaemme}. 

\begin{figure}[ht]
\begin{center}
\epsfig{file=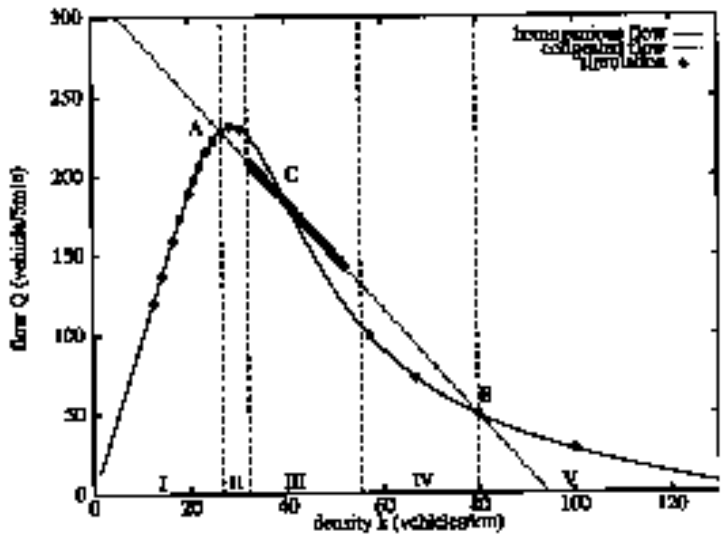,width=10cm}
  \caption{Fundamental diagram of the OV model. The solid line shows
the OV function and the dots simulation data. One can distinguish
five different density regimes with respect to the stable stationary
state (from \cite{sugyamTGF}).}
\label{ov-fund}
\end{center}
\end{figure}

The main question addressed by the OV model is the following: 
what is the {\it condition} for the stability of the homogeneous 
solution 
\begin{equation}
x_n^{h} = b n + c t 
\end{equation}
where $b = (\Delta x)_{av} = L/N$ is the constant average spacing 
between the vehicles and $c$ is the constant velocity. It is 
not difficult to argue that, in general, in the OV models the 
homogeneous flow becomes unstable when 
$\frac{\partial V^{opt}}{\partial \Delta x}\bigr|_{\Delta x=b} 
> \frac{2}{\tau}$ \cite{bandoetal}.

One can distinguish five different density regimes with respect
to the stability of microscopic states (see Fig.~\ref{ov-fund}). 
At low and at high densities the homogeneous states are stable.
For intermediate densities one can distinguish three regimes where
jammed states exist. In region III the jammed state is stable
whereas in regions II and IV both homogeneous and jammed states
form stable structures. Beyond the formation of jams also hysteresis
effects have been observed. Thus the OV model is able to reproduce
many aspects of experimental findings.

Modified Korteweg-de Vries (KdV) equation has been derived from the 
equation (\ref{ov}) in a special regime of the parameters 
\cite{komatsu} and the relations between its kink solutions and 
traffic congestion have been elucidated \cite{muramatsu}. A
generalization to two-lane traffic can be found in \cite{Nag2OV}.
In order to account for traffic consisting of two different 
types of vehicles, say, cars and trucks, Mason et al.~\cite{mason} 
generalized the formulation of Bando et al.~\cite{bandoetal} 
by replacing the constant $\tau$ by $\tau_n$ so that 
\begin{equation}
\ddot{x}_{n}(t) = \frac{1}{\tau_n}[V^{opt}_{n}(t) - v_{n}(t)]
\end{equation}
where $\tau_n$ now depends on whether the $n$-th vehicle is 
a car or a truck. Since a truck is expected to take longer to 
respond than a car we should assign larger $\tau$ to trucks 
and smaller $\tau$ to cars. Some other mathematically motivated 
generalizations of the OV model have also been considered 
\cite{nagaov2,nagaov3,hayanaka}. 

As mentioned earlier, drivers often receive stimulus not only 
from the leading vehicle but also a few other vehicles ahead of 
the leading vehicle. One possible way to generalize the OV models 
for taking into account such {\it multi-vehicle} interactions 
\cite{lenz} is to write the dynamical equations as 
\begin{equation}
\ddot{x}_{n} = \sum_{j=1}^m {\mathcal{S}}_j \left[V^{opt}
\left(\frac{x_{n+j}-x_n}{j}\right) - v_{n}\right] 
\label{multi_car}
\end{equation}
where ${\mathcal{S}}_j$ are sensitivity coefficients; one of the 
commonly used explicit forms of the optimal velocity function, 
for example (\ref{ovfunction}) can be chosen for that of the  
function $V^{opt}$ in (\ref{multi_car}). 


Before concluding our discussions on the OV models let us mention 
a very simple model \cite{nagaov} where one assumes that  
\begin{equation}
\ddot{x}_n(t) = \begin{cases} 
                 a &{\rm for} \quad \Delta x_n \geq \Delta x_c\\
                 -a & {\rm for} \quad \Delta x_n < \Delta x_c .
                 \end{cases}
\label{eq:simpleOV}
\end{equation}
with $a > 0$. Obviously, $\Delta x_c$ may be interpreted as the
safety distance. Moreover, a restriction $v_{min} \leq v_n \leq
v_{max}$ is imposed on the allowed velocities of the vehicles by
introducing the allowed minimum and maximum velocities $v_{min}$ and
$v_{max}$, respectively. Note that in this oversimplified model
$\ddot{x}_n$ depends on the corresponding distance-headway and, 
therefore, has some apparent similarities with the OV models. But, 
unlike the more general OV models, in this model $\ddot{x}_n$ does 
not depend on its instantaneous velocity. The main reason for 
considering such a oversimplified scenario is that velocity of 
the propagation of jams can be calculated analytically.


Some ideas of the OV model has been utilized by Mahnke and 
Pieret \cite{mahnke} in their master equation approach to 
the study of jam dynamics. They assumed that, at a time, 
only one vehicle can go into or come out of a jam; this, 
naturally, does not take into the merging or splitting of 
jams. Under this assumption, the master equation (see e.g.\
App.\ \ref{App_MPA}, equation (\ref{master})) governing 
the probability distribution $P(n;t)$ of the jam sizes $n$ 
is given by 
\begin{eqnarray}
\frac{dP(n;t)}{dt} &=& W_+ (n-1) P(n-1;t) + W_- (n+1) P(n+1;t)
\nonumber\\ 
&-& [W_+(n) + W_-(n)] P(n;t) 
\end{eqnarray}
where $W_+$ and $W_-$, are the "growth" and "decay" transition 
rates, respectively. It has been argued \cite{mahnke,kapusz} 
that $W_-(n) = 1/\tau = constant$ since a vehicle would require 
a constant average time $\tau$ to come out of a jam. However, 
$W_+(n)$ would depend on $V^{opt}(\Delta x)$ since the 
time taken by a free-flowing vehicle immediately behind a jam 
to get into the jam would depend on $\Delta x$ as well as on 
$V^{opt}(\Delta x)$ although the actual expression of 
$W_+(n)$ may be complicated in a reasonable ansatz \cite{kapusz}.  
Moreover, several assumptions of the model will have to be 
relaxed before the results of this approach can be compared with 
those from real traffic. 

Before concluding this section we would like to emphasize that 
while formulating the dynamical equations for updating the 
velocities and positions of the vehicles in any "microscopic" 
theory the following points should be considered:\\
(i) in the absence of any disturbance from the road conditions 
and interactions with other vehicles, a driver tends to drive 
with a {\it desired velocity} $v_{des}$; if the actual current 
velocity of the vehicle $v(t)$ is smaller (larger) than $v_{des}$, 
the vehicle accelerates (decelerates) so as to approach $v_{des}$.\\
(ii) In a freely-flowing traffic, even when a driver succeeds in 
attaining the desired velocity $v_{des}$, the velocity of the 
vehicle fluctuates around $v_{des}$ rather than remaining constant 
in time.\\
(iii) The interactions between a pair of successive vehicles in a 
lane cannot be neglected if the gap between them is shorter than 
$v_{des}$; in such situations the following vehicle must decelerate 
so as to avoid collision with the leading vehicle. 

Clearly, the reliability of the predictions of the OV model depends 
on the appropriate choice of the optimal velocity function.

\section{Coupled-map lattice models of vehicular traffic} 
\label{Sec_coupmap}

Recall that, in the car-following models, space is assumed to be a 
continuum and time is represented by a continuous variable $t$. 
Besides, velocity and acceleration of the individual vehicles are 
also real variables. However, most often, for numerical manipulations 
of the differential equations of the car-following models, one needs 
to discretize the continuous variables with appropriately chosen 
grids. In contrast, in the coupled-map lattice approach 
\cite{kaneko}, one starts with a discrete time variable and, the 
dynamical equations for the individual vehicles are formulated as 
discrete dynamical maps that relate the state variables at time $t$ 
with those at time $t+1$, although position, velocity and acceleration 
are {\it not} restricted to discrete integer values. The unit of time 
in this scheme (i.e., one time step) may be interpreted as the reaction 
time of the individual drivers as the velocity of of a vehicle at the 
time step $t$ depends on the traffic conditions at the preceding time 
step $t-1$.

Keeping in mind the general points raised at the end of the 
preceding section regarding the formulation of the dynamical 
equations for updating the velocities and positions of the 
vehicles, the general forms of the dynamical maps in the 
coupled-map lattice models can be expressed as: 
\begin{eqnarray}
v_n (t+1) &=& Map_n[v_n(t), v_{des}, \Delta x_n(t)],\label{map1}\\
x_n(t+1) &=& v_n(t) + x_n(t) 
\label{map2}
\end{eqnarray}
where $v_{des}$ is a {\it desired} velocity. In general, the dynamical
map $Map[v_n(t), v_{des}, \Delta x_n(t)]$ takes into account the
velocity $v_n(t)$ and the distance-headway $\Delta x_n(t)$ of the
$n$-th vehicle at time $t$ for deciding the velocity $v_n(t+1)$ at
time $t+1$. The effects of the interactions among the vehicles enter
into the dynamical updating rules (\ref{map1}), (\ref{map2}) only
through the distance-headway $\Delta x_n$.

\subsection{The model of Yukawa and Kikuchi}
\label{sub_Kikuchi}

Yukawa and Kikuchi \cite{yukiku,yukiku2,yukitgf} 
have studied coupled-map models based on the map
\begin{equation}
v(t+1)=F(v(t)):=\gamma v(t)+\beta \tanh\left(\frac{v^F-v(t)}{\gamma}
\right)+\epsilon
\label{cm_free}
\end{equation}
for the uninfluenced motion of a single vehicle. $v^F$ is the
preferred velocity of the vehicle and $\beta$, $\gamma$, $\delta$
and $\epsilon$ are parameters. For $\gamma$ close to 1 the map
becomes chaotic, but acceleration and deceleration are approximately
constant far from $v^F$. Their magnitude is determined by the 
parameter $\beta$. $\epsilon$ controls the difference of the
acceleration and deceleration capabilities. 
Although the model is deterministic, fluctuations in the velocity are
introduced through deterministic chaos. These fluctuations around
$v^F$ are determined by the parameter $\delta$.

If there is more than one vehicle on the road one needs an additional
deceleration mechanism to avoid collisions. This can be achieved
by introducing a deceleration map. Assuming that deceleration is 
dominated by the headway, two models have been studied in \cite{yukiku}.
In model A, the deceleration map describes a sudden braking process.
If the front-bumper to front-bumper distance $\Delta x_n$ to the next
vehicle ahead is less than the current velocity $v_n(t)$ of the
following vehicle, then the velocity is reduced to $\Delta x_n-l$ 
where $l$ is the length of the vehicles. The corresponding map
is $B(\Delta x_n(t))=\Delta x_n(t)-l$.

Model B has a more complex deceleration map:
\begin{eqnarray}
v_n(t+1)&=&G(\Delta x_n(t),v_n(t)):=
\frac{F(v_n(t))-v_n(t)}{(\alpha -1)v_n(t)}\left[\Delta x_n(t)-l
-v_n(t)\right]\nonumber\\
& &\qquad\qquad \qquad\qquad
   ({\rm for\ }v_n(t)\leq \Delta x_n(t)-l\leq\alpha v_n(t)).
\label{cm_decel}
\end{eqnarray}
The parameter $\alpha$ determines the range within which the
deceleration map $G(\Delta x,v)$ is used. For headways less than
$\alpha v_n(t)$ the map $G$ is used instead of $F(v)$. Note
that $G(\Delta x,v=\Delta x-l)=\Delta x-l$ and
$G(\Delta x,v=(\Delta x-l)/\alpha)=F(v=(\Delta x-l)/\alpha)$, i.e.\
$G$ interpolates between the free-motion map $F$ and the sudden
braking map of model A. The full velocity map of Model B is thus 
given by
\begin{equation}
Map_n(v_n(t),\Delta x_n(t))=
\begin{cases}
F(v_n(t),v_n^F)         
          & \text{for\ \ } \alpha v_n(t)\leq \Delta x_n(t),\\
G(\Delta x_n(t),v_n(t))  
          & \text{for\ \ } v_n(t)\leq \Delta x_n(t)\leq \alpha v_n(t)),\\
B(\Delta x_n(t))         
          & \text{for\ \ } \Delta x_n(t)\leq v_n(t)).
\end{cases}
\label{cm_modelB}
\end{equation}

Local measurements of the flow for a system of vehicles with different
preferred velocities $v_n^F$ produce a fundamental diagram of 
inverse-$\lambda$ shape (see Fig.~\ref{fig:hallfund}A) \cite{yukitgf}.
Here the non-uniqueness of the flow has a simple explanation.
Due to the different $v_n^F$, platoons form behind the slowest
vehicles. Whenever such a platoon passes the measurement region, a
flow value on the lower branch is recorded. Otherwise, the flow
corresponds to the upper branch.

Measurements of the power spectral density of temporal density
fluctuations, i.e.\ the Fourier transform of the time-series of
local densities, show a $1/f^\alpha$-behaviour with $\alpha\approx 1.8$
in the free-flow regime. Due to the deterministic dynamics, the
system evolves into a state with power-law fluctuations. In
\cite{yukiku2,yukitgf} it has been suggested that the origin of the
$1/f^\alpha$-fluctuations is the power-law distribution ($\propto
1/(\Delta x)^{3.0}$) of the headways $\Delta x$, since these are
related to density waves. The occurence of jams destroys long-time
correlations since vehicles loose their memory of current fluctuations
when they are forced to stop in a jam \cite{yukiku2}. Therefore
no $1/f^\alpha$-behaviour can be observed in the jammed regime.

In \cite{CMOVM} a coupled-map model based on optimal-velocity functions
has been introduced by discretizing the time variable of the
OV model (see Sec.\ \ref{sec_OVM}). This allows to study systems
with open boundaries and multilane systems. Furthermore a 
multiplicative random noise can be imposed in the velocity update
so that the velocity map is given by
\begin{equation}
v(t+1)=\left[v(t)+\alpha\left(V_{opt}(\Delta x)-v(t)\right)
\right](1+f_{noise}\xi)
\end{equation}
where $\xi \in [-1/2,1/2]$ is a uniform random variable and $f_{noise}$
the noise level.

\subsection{The model of Nagel and Herrmann}
\label{sub_NH}

Nagel and Herrmann (NH) \cite{nh} have introduced a coupled-map model 
which is related to the continuum limit of the Nagel-Schreckenberg 
cellular automata model (see Sec.~\ref{sec_NaSch}). A generalization 
of the NH model has later been presented in \cite{Sauer}.
Vehicles are characterized by a maximal velocity $v_{max}$ and
a safety distance $\alpha$. The velocity map for the NH model 
is given by

\begin{eqnarray}
v_n(t+1)=\begin{cases}
\max(\Delta x_n(t)-\delta,0) 
       &\text{for\ } v_n(t) > \Delta x_n(t)-\alpha,\\
\min(v_n(t)+a,v_{max})
       &\text{for\ } v_n(t) < \Delta x_n(t)-\beta.
\end{cases}
\end{eqnarray}

In the velocity update step, vehicles which have a headway $\Delta x$
smaller than the safety distance $\alpha$ decelerate. The headway
distance after deceleration is determined by the parameter $\delta$.
Vehicles which have a large enough headway, on the other hand,
accelerate. The acceleration coefficient $a$ is determined by
$a=a_{max}\max(1,\Delta x_n(t)/\gamma)$.

Since the dynamics of the model is deterministic, the behaviour depends
strongly on fluctuations of the initial state \cite{Sauer}. For 
equidistant starting positions of the vehicles the fundamental diagram 
consists of two linear branches with maximum flow $f(c_m)=v_{max}c_m$ 
at density $c_m=1/(v_{max}+\beta)$. For homogeneous starting positions 
the system is free-flowing up to a critical density $c_{crit}$.
Beyond this density free-flowing and congested areas coexist.

\subsection{The model of Krauss, Wagner and Gawron}
\label{sub_KWG}

Krauss et al. \cite{krauss,krauss2} introduced a whole class of
stochastic models by considering necessery conditions for the
collision-free motion of vehicles. The models are continuous in space
and discrete in time.

The vehicles are characterized by a maximum velocity $v_{max}$, their 
acceleration and deceleration capabilities $a(v)$  and $b(v)$,
respectively, and their length $l$ which will be taken to be $l=1$
in the following.
Then the update rules for the velocity $v$ and the space coordinate $x$
of each vehicle are as follows:

\noindent {\it Step 1: Determine desired velocity.} 
$$
v_{des}=\min\left[v_{max},v+a(v),v_{safe}\right]
$$

\noindent {\it Step 2: Randomization.} 
$$
v=\max\left[0,rand(v_{des}-a,v_{des})\right]
$$

\noindent{\it Step 3: Vehicle movement.}
$$
x \to x+v
$$
Here $rand(v_1,v_2)$ denotes a random number uniformly distributed
in the interval $[v_1,v_2)$ and $v_{safe}$ is a velocity which
guarantees collision motion of the vehicles. It is given
explicitly by 
\begin{equation}
v_{safe}=v_p + b(\hat{v})\frac{g-v_p}{\hat{v}+b(\hat{v})}.
\end{equation}
where $v_p$ is the velocity of the preceding vehicle located at $x_p$
and $g=x_p-x-1$ is the headway, i.e.\ the distance to the preceding 
vehicle.

\begin{figure}[ht]
\begin{center}
\epsfig{file=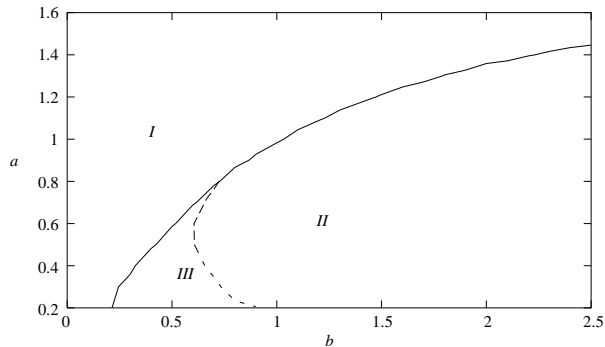,width=8cm}
  \caption{Schematic sketch of the three different classes of models
in the KWG model.}
\label{fig_KWG}
\end{center}
\end{figure}

In the simplest case the acceleration and deceleration capabilities
do not depend on the velocity, i.e.\ $a(v)=a=const$ and $b(v)=b=const$.
The behaviour of the model can be classified in three different
families (see Fig.\ \ref{fig_KWG}).

The three families of models sketched schematically in 
Fig.\ \ref{fig_KWG} can be characterized as follows:

\noindent{\em Class I: High acceleration}\\
Here no spontaneous jamming exists.  For $a\to v_{max}$ and $b\gg 1$
the behaviour is similar to that of a cellular automata model without
velocity memory introduced in \cite{brankov} which is closely related
to the Kasteleyn model \cite{kasteleyn} of statistical physics.  It
can also be interpreted as 5-vertex model \cite{brankov2}. Another
model belonging to this class has been introduced by Fukui and
Ishibashi \cite{fukuishi} (see Sec.\ \ref{sec_FI}).

\noindent{\em Class II: High acceleration -- low deceleration}\\
The outflow from a jam is identical to the maximal possible flow.
The jamming transition is not a true phase transition, but rather
a crossover. The limit $b\to\infty$, $a=1$ corresponds to a continuum 
version of the Nagel-Schreckenberg cellular automata model
\cite{ns} which will be introduced in the next section. 

\noindent{\em Class III:  Low acceleration -- low deceleration}\\
These models exhibit phase separation and metastability.  The jamming
transition is of first order. The outflow from a jam is not maximal.
For $a\ll v_{max}$ and $b\ll v_{max}$ the model is closely related to
the Gipps model \cite{gips} discussed in Sec.\ \ref{sec_follow}. Other
models belonging to this class are the Kerner-Konh\"auser model (see
Sec.\ \ref{sec_Nav}), the optimal-velocity model (Sec.\ \ref{sec_OVM})
and the models with slow-to-start rules which will be introduced in
Sec.\ \ref{Sec_s2s}.

On a macroscopic level, classes I, II and III can be distinguished by
the ordering of the densities $c_f$ and $c_c$, where $c_f$ is
the density of the outflow from a jam and $c_c$ is the density where
homogeneous flow becomes unstable \cite{krauss2}.
For $c_c > c_f$ the outflow from a jam is stable and the system
phase-separates into free-flow and jammed regions. Furthermore,
metastable states can be found. This is the type of the behaviour
found in class III. For $c_c < c_f$, on the other hand, the outflow 
from a jam is unstable and no metastable states or phase separation
can be found. This is the typical behaviour of classes I and II.
These classes can further be distinguished since in class I one
does not find any structure formation, like spontaneous jamming,
in contrast to class II.

A related model has been studied before by Migowsky et al. \cite{MWR}.
In this model vehicles are also characterized by a maximum velocity
$v_{max}$ and a bounded acceleration capability ($-b_{max}< \ddot x
<a_{max}$) which determines the safety distance $d_s$ necessary
to avoid accidents. The investigations in \cite{MWR} focussed on the
effect of so-called driving strategies. These strategies are characterized
by a vector $(f_v,f_a,f_s)$, where $f_v$, $f_a$ and $f_s$ are the
fraction of the vehicle's maximal velocity, acceleration and safety
distance actually used\footnote{In other words, the maximum velocity
is given by $v_{max}^{(n)}=f_v^{(n)}v_{max}$ etc.}, respectively.
This can lead to the possibility of accidents and allows to study the
number of crashes as a function of the driving strategies.

In \cite{MWR} different strategies have been compared. Furthermore,
dynamical changes of strategies can be introduced which allow the
drivers to adapt to the local traffic conditions. In general this
leads to a decrease in the number of accidents, jams and fuel
consumption, but at high densities the flow is reduced compared
to the case of fixed strategies.


\section{Nagel-Schreckenberg cellular automata model of vehicular 
traffic on highways}
\label{sec_NaSch}

In general, CA are idealization of physical systems in which 
both space and time are assumed to be discrete and each of 
the interacting units can have only a finite number of discrete 
states. Note that for a discretization of differential equations,
e.g.\ those of the hydrodynamic approach, space and time variables
are discrete, but the state variable still is continuous.
The concept of CA was introduced in nineteen fifties 
by von Neumann while formulating  an abstract theory of 
self-replicating computing machines \cite{codd}. 
However, it received the attention of a wider audience in the 
nineteen seventies through Conway's {\it game of life} \cite{conway}. 
The family of one-dimensional CA was studied systematically, in 
the nineteen eighties, from the point of view of dynamical 
systems and popularized by Wolfram \cite{wolfram,stauca}. Since 
then the concept of CA has been applied to model a wide variety 
of systems \cite{preston,droz,marro}.  To our knowledge, the 
first CA model for vehicular traffic was introduced by Cremer 
and Ludwig \cite{cremerca}.

\begin{figure}[ht]
\centerline{\psfig{figure=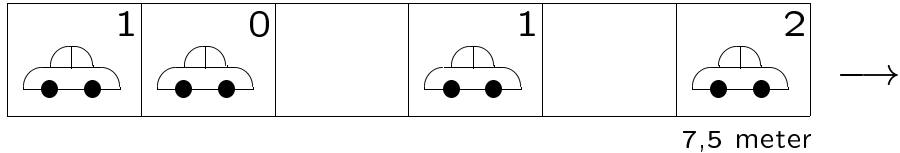,bbllx=50pt,bblly=420pt,bburx=550pt,bbury=540pt,height=3cm}}
\caption{\protect{ A typical configuration in the NaSch model. The 
number in the upper right corner is the speed of the vehicle. }}
\label{carpic}
\end{figure}

In the CA models of traffic the position, speed, acceleration as 
well as time are treated as {\it discrete} variables. In this 
approach, a lane is represented by a one-dimensional lattice. Each 
of the lattice sites represents a "cell" which can be either empty 
or occupied by at most one "vehicle" at a given instant of time 
(see Fig.~\ref{carpic}). At each {\it discrete time} step 
$t \rightarrow t+1$, the state of the system is updated following 
a well defined prescription (a summary of various possible different 
schemes of updating is given in Appendix \ref{App_updates}). The 
computational efficiency of the discrete CA models is the main 
advantage of this approach over the car-following and coupled-map 
lattice approaches. 

In the NaSch model, the speed $v$ of each vehicle can take one of 
the $v_{max}+1$ allowed {\it integer} values $v=0,1,...,v_{max}$. 
Suppose, $x_n$ and $v_n$ denote the position and speed, respectively, 
of the $n$-th vehicle. Then, $d_n = x_{n+1}-x_n$, is the gap in 
between the $n$-th vehicle and the vehicle in front of it at time $t$. 
At each time step $t \rightarrow t+1$, the arrangement of the $N$ 
vehicles on a finite lattice of length $L$ is updated {\it in 
parallel} according to the following "rules":

\noindent {\it Step 1: Acceleration.} If $ v_n < v_{max}$, the 
speed of the $n$-th vehicle is increased by one, but $v_n$ remains 
unaltered if $v_n = v_{max}$, i.e.\\
$$
v_n \rightarrow \min(v_n+1,v_{max}) \eqno ({\rm U}1). 
$$ 

\noindent{\it Step 2: Deceleration (due to other vehicles).} If 
$d_n \le v_n$, the speed of the $n$-th vehicle is reduced to $d_n-1$, 
i.e.,\\
$$v_n \rightarrow \min(v_n,d_n-1) \eqno ({\rm U}2). $$  

\noindent{\it Step 3: Randomization.} If $v_n > 0$, the speed of the $n$-th
vehicle is decreased randomly by unity with probability $p$ but $v_n$ 
does not change if $v_n = 0$, i.e., \\
$$
v_n \rightarrow \max(v_n-1,0) \quad {\rm with\ probability\ }p 
\eqno ({\rm U}3). 
$$  

\noindent{\it Step 4: Vehicle movement.} Each vehicle is moved forward 
according to its new velocity determined in Steps 1--3, i.e.
$$
x_n \rightarrow  x_n + v_n \eqno ({\rm U}4). 
$$

The NaSch model is a minimal model in the sense that all the four 
steps are necessary to reproduce the basic features of real traffic; 
however, additional rules need to be formulated to capture more 
complex situations. 
The step 1 reflects the general tendency of the drivers to drive 
as fast as possible, if allowed to do so, without crossing the 
maximum speed limit. The step 2 is intended to avoid collision 
between the vehicles. The randomization in step 3 takes into 
account the different behavioural patterns of the individual drivers, 
especially, nondeterministic acceleration as well as overreaction 
while slowing down; this is crucially important for the spontaneous 
formation of traffic jams. 
Even changing the precise order of the steps of the update rules stated 
above would change the properties of the model. E.g.\ after changing
the order of steps 2 and 3 there will be no overreactions at braking
and thus no spontaneous formation of jams.
The NaSch model may be regarded as stochastic 
CA~\cite{wolfram}. In the special case $v_{max} = 1$ the deterministic 
limit of the NaSch model is equivalent to the CA rule $184$ in Wolfram's 
notation \cite{wolfram} and some abstract extensions of this 
CA-184 rules~\cite{awazu} have been studied in the more general 
context of complex dynamics and particle flow. 

Why should the updating be done in {\it parallel}, rather than in 
random sequential manner, in traffic models like the NaSch model? 
In contrast to a random sequential update, parallel update can lead 
to a chain of overreactions. Suppose, a vehicle slows down due the 
randomization step. If the density of vehicles is large enough this 
might force the following vehicle also to brake in the deceleration 
step. In addition, if $p$ is larger than zero, it might brake even 
further in Step 3. Eventually this can lead to the stopping of a 
vehicle, thus creating a jam. This mechanism of spontaneous jam 
formation is rather realistic and cannot be modeled by the random 
sequential update.

\begin{figure}[ht]
\begin{center}
\psfig{file=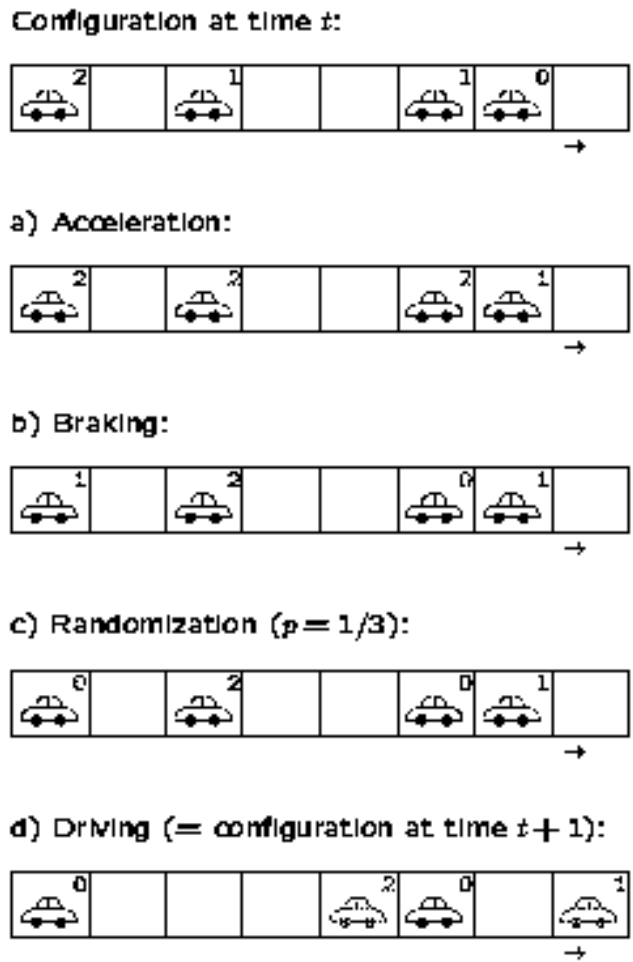,width=10cm}
\caption{\protect{Step-by-step example for the application of the
update rules. We have assumed $v_{max} = 2$ and $p = 1/3$. Therefore
on average one third of the cars qualifying will slow down in the
randomization step.}}
\label{update}
\end{center}
\end{figure}

The update scheme of the NaSch model is illustrated with a simple 
example in Fig.~\ref{update}.

\begin{figure}[ht]
 \centerline{\psfig{file=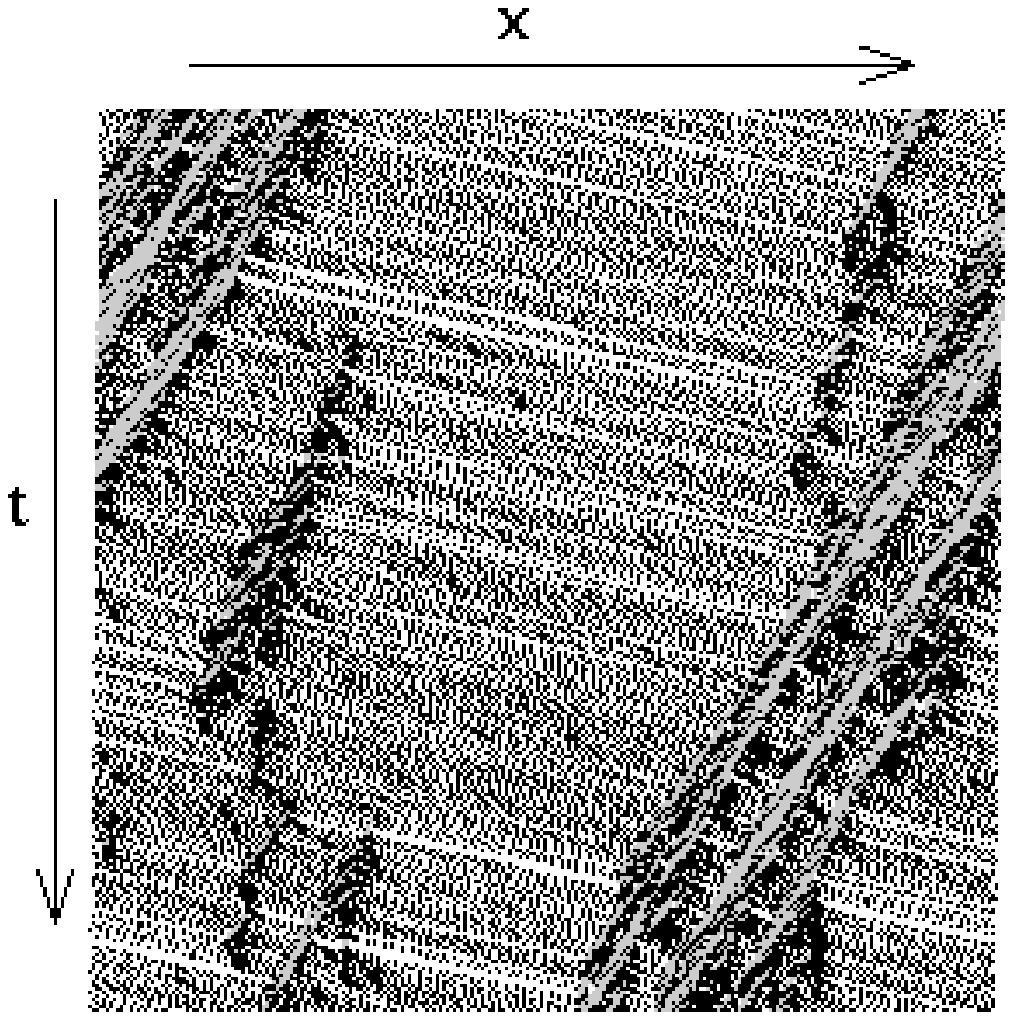,width=5cm}
\psfig{file=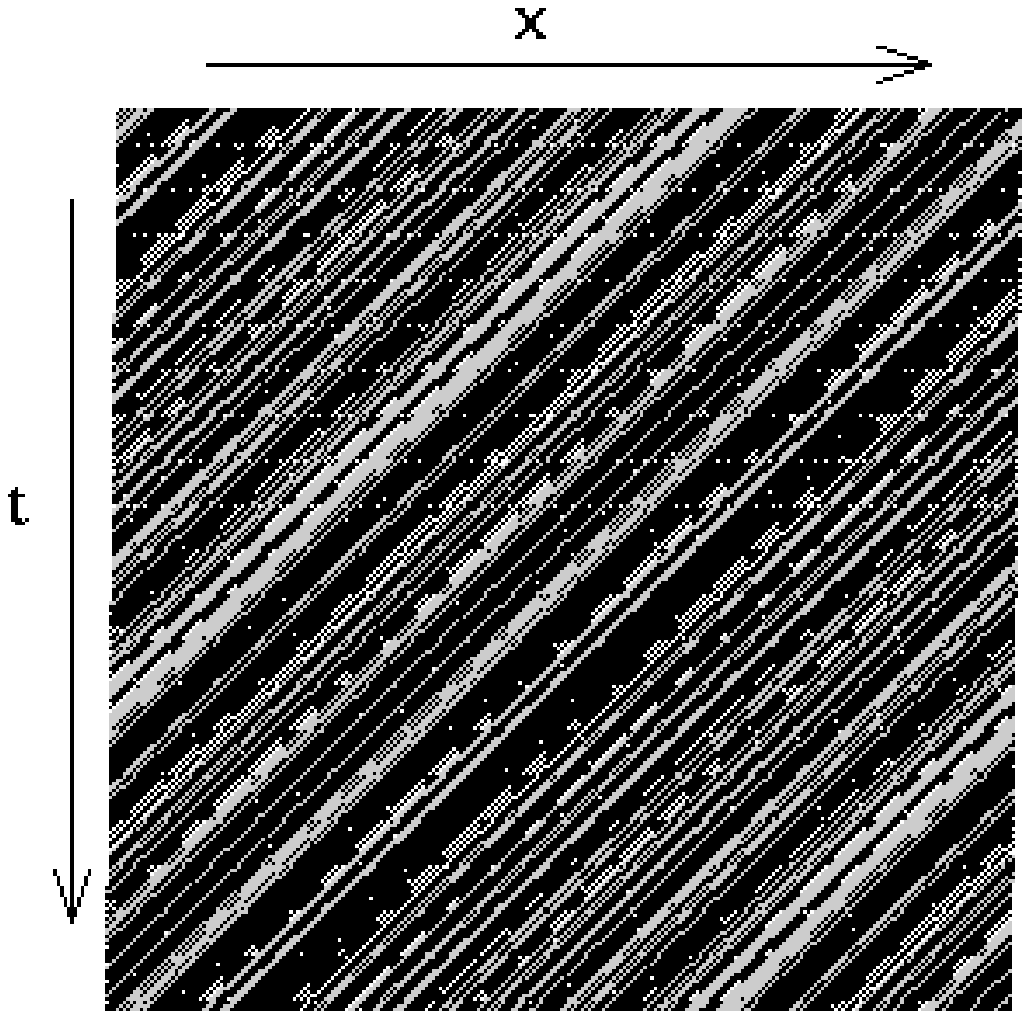,width=5cm}}
\caption{\protect{Typical space-time diagrams of the NaSch model 
with $v_{max} = 5$ and (a) $p = 0.25, c = 0.20$, (b) $p = 0, c = 0.5$. 
Each horizontal row of dots represents the instantaneous positions of 
the vehicles moving towards right while the successive rows of dots 
represent the positions of the same vehicles at the successive time steps. }}
\label{nsxt}
\end{figure}

Space-time diagrams showing the time evolutions of the NaSch model 
demonstrate that no jam is present at sufficiently low densities, 
but spontaneous fluctuations give rise to traffic jams at higher 
densities (Fig.~\ref{nsxt}(a)). From the Fig.~\ref{nsxt}(b) it should 
be obvious that the {\it intrinsic stochasticity} of the dynamics 
\cite{ns}, arising from non-zero $p$, is essential for triggering 
the jams ~\cite{ns,nh}. For a realistic description of highway traffic 
\cite{ns}, the typical length of each cell should be about $7.5$m which
is the space occupied by a vehicle in a dense jam. 
When $v_{max} = 5$ each time step should correspond to approximately 
$1$ sec of real time which is of the order of the shortest relevant
timescale in real traffic, namely the reaction time of the drivers.

Almost all the models of traffic considered in this review, 
including the NaSch model, have been formulated in such a way 
that no accident between successive vehicles is possible. 
However, accident of the vehicles is possible if the condition 
for safe driving is relaxed. For example, Boccara et al.~\cite{boccara} 
replaced the update rule of the NaSch model by the rule 
\begin{equation}
{\rm if\ \ } v_{n+1}(t) > 0 {\rm \ \ then\ \ } 
x_n(t+1) = x_n(t) + v_n(t+1) + \Delta v 
\end{equation}
where $\Delta v$ is a Bernoulli random variable which takes 
the value $1$ with probability $p_{careless}$ and zero with 
the probability $1-p_{careless}$. The probability $P_{ac}$ 
of accident per vehicle per time step is a non-monotonic 
function of the vehicle density $c$ \cite{boccara,dwhuang}. 

\subsection{Relation with other models}


\subsubsection{Relation with totally asymmetric simple exclusion process}
\label{Sec_asep}

\begin{figure}[ht]
\centerline{\epsfig{figure=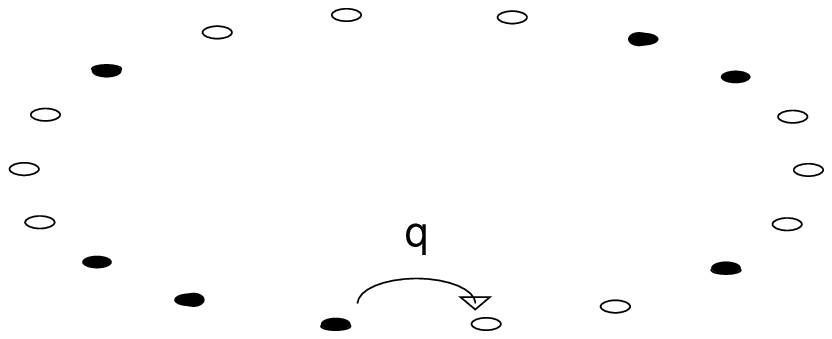,height=4cm}}
\caption{ASEP with periodic boundary conditions.}
\label{asep}
\end{figure}

Now we point out the similarities and differences between the $v_{max}
= 1$ limit of the NaSch model and the totally asymmetric simple
exclusion process (TASEP) which is the simplest prototype model
of interacting systems driven far from equilibrium \cite{spohn,sz,gs}.
In the TASEP (Fig.~\ref{asep}) a randomly chosen particle can move
forward, by one lattice spacing, with probability $q$ if the lattice
site immediately in front of it is empty. It corresponds to a Kawasaki
dynamics~\cite{kawasaki} for exchange of a charged particle and hole
on nearest-neighbour lattice sites at infinite temperature and in the
presence of an infinite electric field \cite{katzetal} (see Appendix
\ref{App_TASEP} for some technical aspects of TASEP). Several
different generalized variants of the TASEP have been considered. For
example, in the $k$-hop model \cite{binderetal} a particle can
exchange its position with the nearest-hole on its right with
probability $q$, provided the separation of the two sites under
consideration is not more than $k$ lattice spacings. The $k$-hop model
reduces to the TASEP in the special case $k = 1$.

Note that in the NaSch model
with $v_{max} = 1$ every vehicle moves forward with probability
$q = 1-p$ in the time step $t+1$ if the site immediately in front
of it were empty at the time step $t$; this, is similar to TASEP.
But, updating is done in parallel in the NaSch model whereas that
in the TASEP is done in a random sequential manner. Nevertheless,
the special case of $v_{max} = 1$ for the NaSch model achieves
special importance from the fact that so far it has been possible
to derive exact analytical results for the NaSch model only
in the special limits  (a) $p = 0$ and arbitrary $v_{max}$ (which
we have already considered), and (b) $v_{max} = 1$ and arbitrary $p$.

\subsubsection{Relation with surface growth models and the 
phenomenological fluid-dynamical theories of traffic}
\label{sub_growth}

\begin{figure}[hbt]
 \centerline{\epsfig{figure=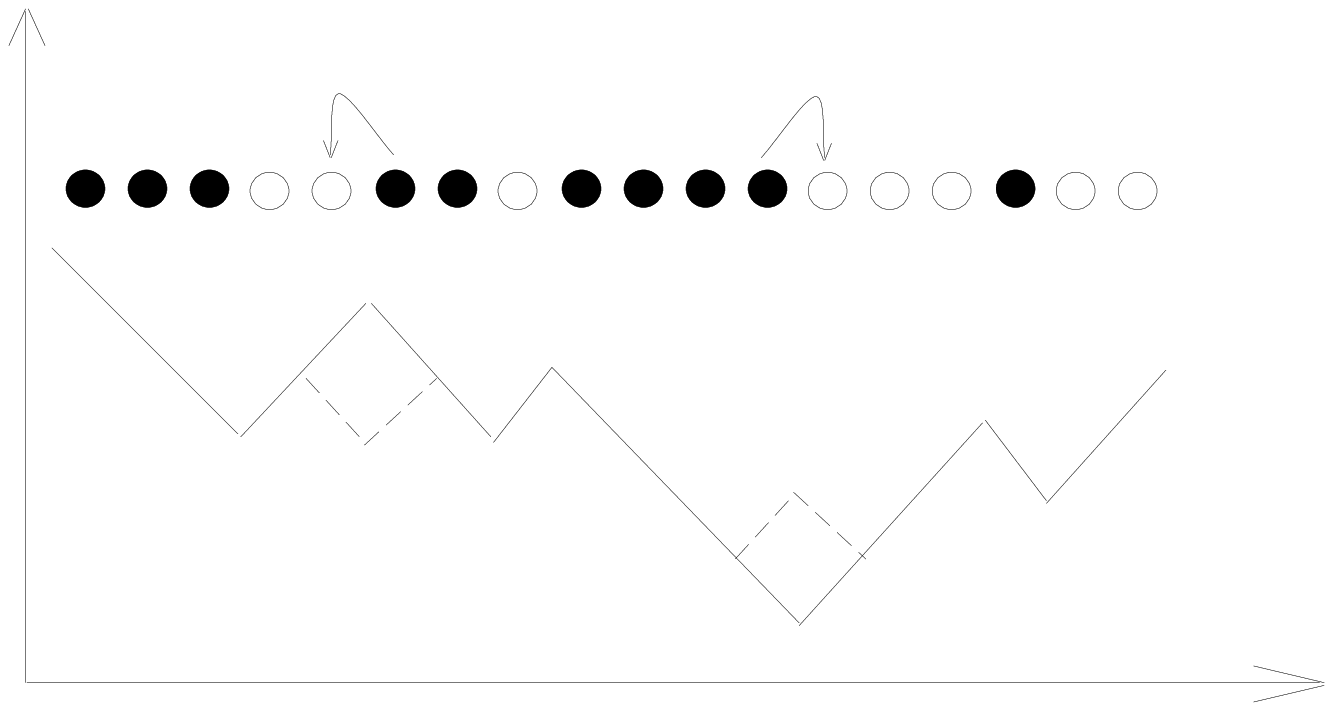,height=5.3cm}}
\caption{The schematic pictorial representation
of the mapping of the NaSch model with $v_{max} = 1$ onto a
stochastic model of surface growth.}
\label{fig9}
\end{figure}

The NaSch model with $v_{max} = 1$ can be mapped onto stochastic
growth models of one-dimensional surfaces in a two-dimensional
medium, the single-step model \cite{barabasi}. Corresponding to 
each configuration $\{\sigma_j\}$ of the NaSch model in the 
site-oriented description, one can obtain a unique surface 
profile $\{H_j\}$ through the relation 
$H_j = \half \sum_{j \leq k} (1 - 2 \sigma_k)$ \cite{barabasi,krspoh}.
Pictorially one can interpret this mapping as shown in Fig.~\ref{fig9}.
Particle (hole) movement to the right (left) correspond to
local forward growth of the surface via particle deposition.
In this scenario a particle evaporation would correspond to
a particle (hole) movement to the left (right) which is not
allowed in the NaSch model. It is worth pointing out that
any quenched disorder in the rate of hopping between two
adjacent sites would correspond to {\it columnar} quenched
disorder in the growth rate for the surface \cite{barma}.

The surface growth model described above is known to be a 
discrete counterpart of continuum models of growing surfaces 
whose dynamics are governed by the so-called Kardar-Parisi-Zhang 
equation \cite{barabasi,krspoh}. Since the Kardar-Parisi-Zhang 
equation can be mapped onto the Burgers equation \cite{burgersbook} 
using the Cole-Hopf transformation \cite{barabasi}, it is not 
surprising that several features of vehicular traffic are 
described by the NaSch model at the microscopic level and by 
the noisy Burgers equation for the coarse-grained continuum of 
the fluid-dynamical theory \cite{Nagel96}.

\subsection{Limiting cases of the NaSch model} 

In spite of the fact that the deterministic limits $p=0$ and $p=1$
of the NaSch model do not capture some of the most essential features 
of vehicular traffic it may be instructive to examine these limits
to gain insight into the features of this simpler scenario. Another
limiting case which exhibits a surprisingly complex behaviour is
the case $v_{max}=\infty$.

\subsubsection{NaSch model in the deterministic limit $p = 0$}
\label{sec_p=0}

The NaSch model, a stochastic CA, becomes a deterministic CA in the 
limit $p = 0$. In this special case, the deterministic dynamical 
update rules of the model can be written as 
\begin{eqnarray}
v_n(t+1) & = & \min[v_{max}, v_n(t)+1, d_n(t)-1], \nonumber\\
 x_n(t+1) & = & x_n(t) + v_n(t+1) 
\end{eqnarray} 
which can lead to two types of steady states depending on the density
$c$ \cite{nh}. At low densities, the system can self-organize so that 
$d_n > v_{max}$ for all $n$ and, therefore, every vehicle can move 
with $v_{max}$, i.e., $v_n(t) = v_{max}$, giving rise to the corresponding
flux $cv_{max}$.  This steady-state is, however, possible only if
enough empty cells are available in front of every vehicle, i.e., for
$ c \leq c_m^{det} = (v_{max} + 1)^{-1}$ and the corresponding maximum
flux is $J_{max}^{det} = v_{max}/(v_{max} + 1)$. On the other hand,
for $c > c_m^{det}$, $d_n(t)-1 \leq \min[v_n(t)+1, v_{max}]$ and,
therefore, the relevant steady-states are characterized by $v_n(t) =
d_n(t)-1$, i.e., the flow is limited by the density of holes.  Since the
average distance-headway is $1/c-1$, the fundamental diagram in the
deterministic limit $p = 0$ of the NaSch model (for any arbitrary
$v_{max}$) is given by the {\it exact} expression 
\begin{equation}
J = \min(c v_{max},1 - c).
\end{equation}
Note that the result $v_n = 1/c - 1$ is identical with
Greenshields ansatz $v_n = 1/c - 1/c_{jam}$ if we identify
$c_{jam} = 1$.

\subsubsection{NaSch model in the deterministic limit $p = 1$}
\label{sec_p=1}

Aren't the properties of the NaSch model with maximum allowed 
speed $v_{max}$, in the deterministic limit $p = 1$, exactly 
identical to those of the same model with maximum allowed speed 
$v_{max} - 1$? Although this expectation may seem to be consistent 
with the observation that $J = 0$ for all $c$ in the special case 
$v_{max} = 1 = p$, the answer to the question posed above is: NO. 
To understand the subtle features of the deterministic limit 
$p = 1$ one has to consider $v_{max} > 1$. You can easily convince 
yourself that if, for example, $v_{max} = 2$, then, for $c > 1/3$, 
all stationary states correspond to $J = 0$ because at least one 
vehicle will have only one empty cell in front (i.e.\ $d_n=2$) and 
it will never succeed in moving forward. For $v_{max} = 2$ and 
$p = 1$, although there are stationary states corresponding to 
$J \neq 0$ for all $c \leq 1/3$, such states are metastable in the 
sense that any local external perturbation leads to complete breakdown 
of the flow. If the initial state is random, such metastable states 
cannot lead to non-zero $J$ because they have a vanishing weight in the 
thermodynamic limit. Hence, if $p = 1$, all random initial states 
lead to $J = 0$ in the stationary state of the NaSch model 
irrespective of $v_{max}$ and $c$!

\subsubsection{NaSch model in the limit $v_{max} = \infty$}

The limit $v_{max} = \infty$ has been introduced in 
\cite{sasvari}. One has to be aware that there are several
possible ways of performing this limit since only finite systems 
of length $L$ can be treated in computer simulations.
In \cite{sasvari} the case $v_{max}=L$ has been 
investigated\footnote{See also \cite{nh}, where the case $p=0$
was studied.}, but other limiting procedures are also possible, 
e.g.\ $v_{max}\propto L^\alpha$ with $\alpha > 0$ or even 
$v_{max}=\infty$ independent of the system size. In principle, 
these different limiting procedures could lead to different results, 
but up to now no systematic study has been performed. We therefore
restrict ourselves to the case $v_{max}=L$ studied in \cite{sasvari}.

\begin{figure}[ht]
\centerline{\epsfig{figure=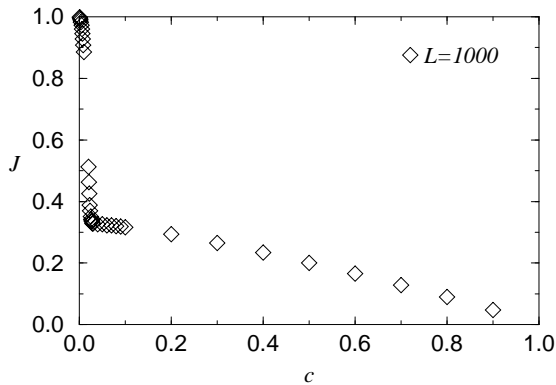,height=5.5cm}}
\caption{Fundamental diagram of the NaSch model with $v_{max}=L$ for
a system size of $L=1000$ (from \cite{sasvari}).}
\label{fig_infty}
\end{figure}

Surprisingly one finds that the fundamental diagram has a form
quite different from that of the case of a finite $v_{max}$.
The flow does not vanish in the limit $c\to 0$ since already one
single car produces a finite value of the flow, $J(c\to 0)=1$.
Due to the hindrance effect of other cars, $J(c)$ is a monotonically
decreasing function of the density $c$ (see Fig.~\ \ref{fig_infty}).
Another characteristic feature of the fundamental diagram is the
existance of a plateau at flow $J_P$ where the value $J_P$ depends
on the randomization $p$, but not on the system size $L$. The length
of the plateau, on the other hand, increases with $L$.

What is the microscopic structure of the stationary state leading to
such a fundamental diagram ? At low densities,  where flow $J$ is larger
than the plateau value $J_P$, the cars tend to be uniformly distributed
just as in the deterministic case $p=0$ (see Sec.\ \ref{sec_p=0}).
For densities in the plateau regime, however, one jam exists in the
system, whereas for higher densities there is more than one jam.
In the thermodynamic limit, one expects a phase transition at $c=0$
between a jamless phase with $J=1$ and a phase with one jam and flow
$J_P$ \cite{sasvari}. Increasing the density further, more jams
develop and the plateau ceases. Note that this behaviour is completely
different from the prediction of mean-field theory in that limit
\cite{ssni} (see Sec.\ \ref{Sec_naiveMF}) showing the importance 
of correlations.

\subsection{Analytical theories of the NaSch model with periodic 
boundary conditions} 
\label{sub_analyt}

CA are, by design, ideal for large scale computer simulations. 
However, proper interpretations of the numerical data obtained 
from computer simulations are not always quite straightforward 
because of the finite-size effects and "numerical noise". One 
cannot deny the importance of exact analytical results in providing 
a testing ground for the computer codes. On the other hand, the 
parallel updating makes exact analytical solution of CA models 
very difficult. Nevertheless, even in those situations where exact 
solutions are not possible, a combination of approximate analytical 
treatments and computer simulation often turns out out to be very 
powerful method of analysis of a problem. This approach has been 
quite successful in recent years in the studies of the NaSch model 
and its generalizations.

Before we proceed with the analytical theories in the non-deterministic 
NaSch model, we would like to point out that the fundamental 
diagram $J(c)$ is known exactly for arbitrary $v_{max}$ and $p$ 
in the two limits $c \rightarrow 0$ and $c \rightarrow 1$. In 
the former case, $J \simeq c v_F$ where $v_F = v_{max} - p$ is 
the free-flow velocity. On the other hand, in the latter case, 
$J \simeq (1-p)(1-c)$ as flow is determined by holes moving 
backwards at a speed $1-p$.

\subsubsection{Site-oriented naive mean-field theory for the NaSch model} 
\label{Sec_naiveMF}

In the "site-oriented" theories one describes the state of the 
finite system of length $L$ by completely specifying the state of 
each site, i.e., by the set $(\sigma_1, \sigma_2,...,\sigma_L)$ 
where $\sigma_j$ ($j = 1,2,...,L$) can, in principle, take 
$v_{max}+2$ values one of which represents an empty site while 
the remaining $v_{max}+1$ correspond to the $v_{max}+1$ possible 
values of the speed of the vehicle occupying the site $j$. 
In some of the analytical calculations of steady-state properties 
of the NaSch model one follows, for convenience, the sequence $2-3-4-1$, 
instead of $1-2-3-4$ of the stages of updating~\cite{ssni} as this 
merely shifts the starting step and, therefore, does not 
influence the steady-state properties of the model. The advantage 
of this new sequence is that, in a site-oriented theory, the 
variable $\sigma_j$ can now take $v_{max}+1$ values as none of 
the vehicles can have a speed $v=0$ at the end of the acceleration 
stage of the updating. 

Let us introduce the lattice gas variables $n(i;t)$ through the 
following definition: $n(i;t) = 0$ if the site labeled by $i$ is 
empty and $n(i;t) = 1$ if it is occupied by a vehicle (irrespective 
of the speed). Obviously, the space-average of $n(i;t)$ is the 
density of the vehicles, i.e., $\sum_i n(i;t)/L = c$.
Suppose, $c_v(i;t)$ is the probability that there is a vehicle 
with speed $v$ ($v = 0,1,2,...,v_{max}$) at the site $i$ at the 
time step $t$. Obviously, $c(i;t) = \sum_{v=0}^{v_{max}} c_v(i;t)$ 
is the probability that the site $i$ is occupied by a vehicle at 
the time step $t$ and $d(i;t) = 1 - c(i;t)$ is the corresponding 
probability that the site $i$ is empty at the time step $t$.

In the naive site-oriented mean-field (SOMF) approximation for 
the NaSch model one writes down the equations relating $c_v(i;t+1)$ 
($v = 1,...,v_{max}$) with the corresponding probabilities at 
time $t$ and, then, solves the equations in the steady-state (see 
the Appendix \ref{app_MF} for a detailed derivation of these equations for 
arbitrary $v_{max}$). In the simplest case of $v_{max} = 1$ and 
periodic boundary conditions one gets~\cite{ssni} 
\begin{eqnarray}
c_0(i;t+1) &=& c(i;t)c(i+1;t) + p c(i;t) d(i+1;t),
\label{mf-c0} \\
c_1(i;t+1) &=& q c(i-1;t) d(i;t).
\label{mf-c1} 
\end{eqnarray}
The equation (\ref{mf-c0}) expresses the simple fact that at the time 
step $t+1$ the speed of the vehicle at the $i$-th site can be   
zero either because the site $i+1$ was occupied at time $t$ or 
because of random deceleration (if the site $i+1$ was empty at 
time $t$). Similarly, the equation (\ref{mf-c1}) implies that the speed 
of the vehicle at the site $i$ can be $1$ at time $t+1$ if at 
the time step $t$ the site $i$ was empty while the site $i-1$   
was occupied by a vehicle which did not decelerate during the 
random deceleration stage of updating. 

In the steady state, $c_{v}(i,t)$ are independent of $t$.
Besides, if periodic boundary conditions are imposed, the
$i$-dependence of $c_{v}(i)$ also drops out in the 
translational-invariant steady-state.
Therefore, in the steady-state 
\begin{equation}
J = c_1 = q c (1-c) 
\label{eq-mfflux}
\end{equation}
It turns out \cite{ssni} that the naive SOMF underestimates the 
flux for all $v_{max}$. Curiously, if instead of parallel updating 
one uses the random sequential updating, the NaSch model with 
$v_{max} = 1$ reduces to the TASEP for which the equation 
(\ref{eq-mfflux}) is known to be the {\it exact} expression for the 
corresponding flux (see, e.g., \cite{ns})!

\subsubsection{Paradisical mean-field theory of the NaSch model}

\begin{figure}[ht]
\centerline{\psfig{figure=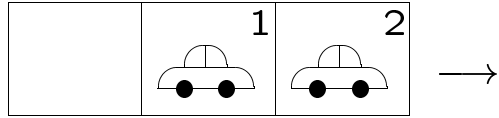,bbllx=50pt,bblly=420pt,bburx=550pt,bbury=540pt,height=3cm}}
\caption{A GoE state for the NaSch model with $v_{max} \geq 2$.}
\label{goe}
\end{figure}

What are the reasons for these differences arising from parallel 
updating and random sequential updating? There are "Garden of Eden" 
(GoE) \cite{Moore} states (dynamically forbidden states) ~\cite{ss98} 
of the NaSch model which cannot be reached by the parallel updating 
whereas no state is dynamically forbidden if the updating is done in 
a random sequential manner. For example, the configuration shown in 
Fig.~\ref{goe} is a GoE state{\footnote{The configuration shown in 
Fig.~\ref{carpic} is also a GoE state!}} because it could occur at 
time $t$ only if the two vehicles occupied the same cell simultaneously 
at time $t - 1$. 

\begin{figure}[ht]
\centerline{\psfig{figure=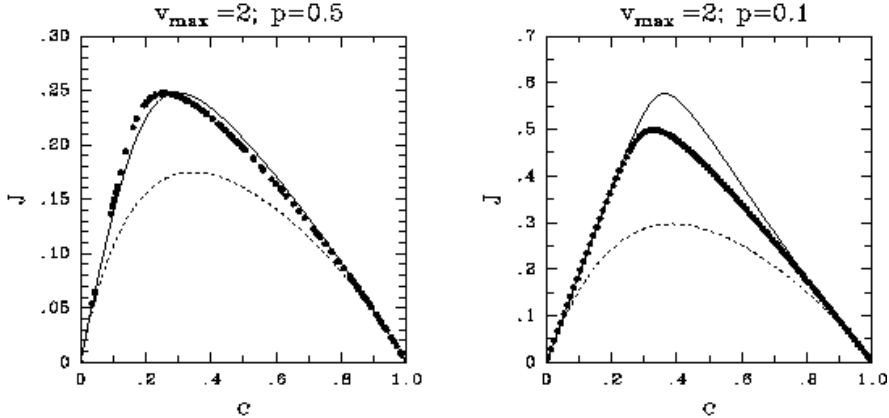,height=5.5cm}}
\caption{Fundamental diagram for $v_{max} =2$ and $p=0.5$ (left) and $p=0.1$
(right). Comparison of paradisical MFT (full line) with results from computer
simulations ($\bullet$) and the naive MFT (dotted line).}
\label{fund_pmf2}
\end{figure}

The naive SOMF theory, discussed in the preceding subsection, does 
not exclude the GoE states. On the other hand, results of the 
paradisical mean-field (pMF) theory are derived by repeating the 
calculations of the naive SOMF theory excluding all the GoE states 
from consideration. The exact expression, given in the next subsection, 
for the flux in the steady-state of the NaSch model with $v_{max} = 1$ 
is obtained in the pMF theory (see Appendix \ref{App_para} for detailed 
calculations), thereby indicating that the only source of 
correlation in this case is the parallel updating~\cite{ss98}. But, 
for $v_{max} > 1$, there are other sources of correlation because of 
which exclusion of the GoE states merely improves the naive SOMF 
estimate of the flux (Fig.~\ref{fund_pmf2}) but does not yield exact 
results \cite{ss98,shad99}. 

\subsubsection{Site-oriented cluster-theoretic approach to the 
NaSch Model} 
\label{Sec_cluster}

\begin{figure}[ht]
 \centerline{\psfig{figure=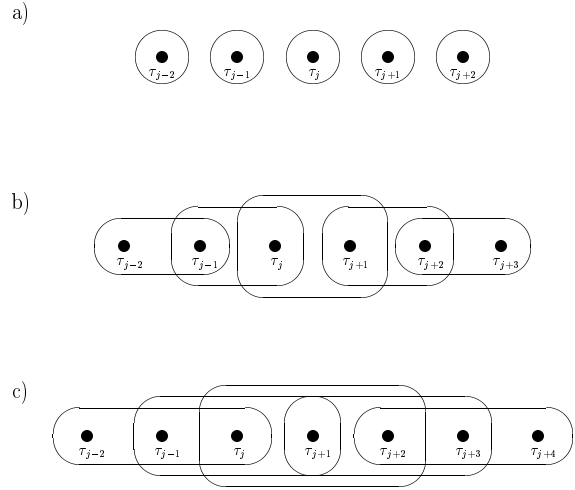,bbllx=80pt,bblly=285pt,bburx=560pt,bbury=740pt,height=8cm}}
\caption{\protect{Decomposition of a lattice into (a) 1-clusters
(b) 2-clusters and (c) 3-clusters in the SOMF theory. }}
\label{ncluster}
\end{figure}

The site-oriented cluster theoretic approach leads to a systematic 
improvement of the naive SOMF theory of the NaSch model. We define 
a $n$-cluster to be a collection of $n$ successive sites and denote 
the probability of finding an $n$-cluster in the state 
$(\sigma_1, \sigma_2,..., \sigma_n)$ in the steady-state of the 
system by the symbol $P_n(\sigma_1, \sigma_2,..., \sigma_n)$.  
In the general $n$-cluster approximation \cite{ssni}, one divides the 
lattice into "clusters" of length $n$ such that two neighbouring 
clusters have $n-1$ sites in common (see Fig.~\ref{ncluster}); an 
$n$-cluster is treated exactly and the cluster is coupled to the 
rest of the system in a self-consistent way, as we shall show in 
this subsection. Even without any calculation, one would expect 
that, for a given $v_{max}$, the $n$-cluster approximation should 
yield more accurate results with increasing $n$ and should give 
exact results in the limit $n \rightarrow \infty$. Fortunately, in 
the special case $v_{max} = 1$ exact results are obtained already 
for $n$ as small as $2$, i.e., the results of 2-cluster calculations 
are exact for $v_{max} = 1$ \cite{ssni}. 

Let us first explain the key concepts involved in the cluster theory.
It is straightforward to verify, for example, in the special case of 
$v_{max} = 1$, that the state of the 2-cluster $\sigma_i,\sigma_{i+1}$ 
at time $t+1$ depends on the state of the 4-cluster 
$(\tau_{i-1},\tau_i,\tau_{i+1},\tau_{i+2})$ at time $t$. In general, 
in the $n$-cluster approximation for an arbitrary $v_{max}$ 
one has to take into account the vehicles that can enter an $n$-cluster 
from one of the $v_{max}$ cells to its left and can leave it to occupy 
one of the $v_{max}$ cells to its right. Therefore, in general, the 
state $\sigma_j, \sigma_{j+1},...\sigma_{j+n-1}$ of an $n$-cluster 
at time $t+1$ depends on the state of a $n+2v_{max}$ cluster 
$\tau_{j-v_{max}},...\tau_j,\tau_{j+1},...,\tau_{j+n-1},...,
\tau_{j+n-1+v_{max}}$ at time $t$.
Therefore, in the special case of $v_{max} = 1$, the master equations 
\begin{eqnarray}
P_2(\sigma_i,\sigma_{i+1};t+1) = \sum_{\tau_j} 
W(\sigma_i,\sigma_{i+1}|\tau_{i-1},\tau_i,\tau_{i+1},\tau_{i+2}) 
P_4(\tau_{i-1},\tau_i,\tau_{i+1},\tau_{i+2};t) \nonumber\\
\label{mastereq}
\end{eqnarray}
governing the time evolution of the 2-cluster probabilities 
$P_2(\sigma_i,\sigma_{i+1})$ involve the 4-cluster probabilities for 
all those configurations $(\tau_{i-1},\tau_i,\tau_{i+1},\tau_{i+2};t)$ 
which can lead to the 2-cluster configuration $(\sigma_i, \sigma_{i+1};t+1)$ 
under consideration as well as the corresponding transition probabilities 
$W(\sigma_i,\sigma_{i+1}|\tau_{i-1},\tau_i,\tau_{i+1},\tau_{i+2})$.
Similarly, the master equation governing the time evolution of the 
4-cluster probabilities $P_4(\tau_{i-1},\tau_i,\tau_{i+1},\tau_{i+2})$
involve 6-cluster probabilities, and so on. In order to obtain a closed 
set of equations one has to truncate this hierarchy in an appropriate 
manner; in the $n$-cluster approximation one expresses the 
$(n+2v_{max})$-cluster probabilities in terms of products of $n$-cluster 
probabilities. 

The $n$-cluster approximation represented geometrically in 
Fig.~\ref{ncluster} for $n = 1$ can be expressed mathematically as 
\begin{equation}
P(\tau_{j-2}, \tau_{j-1}, \tau_j, \tau_{j+1}, \tau_{j+2}) 
= \prod_{i=\tau_{j-2}}^{\tau_{j+2}} P_1(\tau_i) 
\label{1cl-prod} 
\end{equation}
Thus, 1-cluster approximation is equivalent to the naive SOMF 
approximation. The $2$-cluster approximation represented geometrically 
in Fig.~\ref{ncluster} can be expressed mathematically as \cite{ssni} 
\begin{eqnarray}
P(\tau_{j-2}, \tau_{j-1}, \tau_j, \tau_{j+1}, \tau_{j+2}, \tau_{j+3}) 
&\propto& P_2(\tau_{j-2},\tau_{j-1}) P_2(\tau_{j-1},\tau_{j}) 
P_2(\tau_j, \tau_{j+1})\nonumber\\
& & \cdot P_2(\tau_{j+1},\tau_{j+2}) P_2(\tau_{j+2},\tau_{j+3}) 
\label{2cl-prod}
\end{eqnarray}
or, more precisely,
\begin{eqnarray}
P(\tau_{j-2}, \tau_{j-1}, \tau_j, \tau_{j+1}, \tau_{j+2}, \tau_{j+3}) 
&=& P_2(\tau_{j-2}|\underline{\tau_{j-1}}) 
P_2(\tau_{j-1}|\underline{\tau_{j}}) P_2(\tau_j, \tau_{j+1})
\nonumber\\
& &\cdot P_2(\underline{\tau_{j+1}}|\tau_{j+2}) 
P_2(\underline{\tau_{j+2}}|\tau_{j+3}) 
\label{2cl-prod2} 
\end{eqnarray}
where 
\begin{equation}
P_2(\tau_{j-1}|\underline{\tau_{j}}) 
= \frac{P_2(\tau_{j-1},\tau_{j})}{\sum_{\tau_{j-1}} 
P_2(\tau_{j-1},\tau_{j})} 
\end{equation}
are 2-cluster conditional probabilities.
Similarly, the 3-cluster approximation consists of the approximate 
factorization 
\begin{eqnarray}
&&P(\tau_{j-2}, \tau_{j-1}, \tau_j, \tau_{j+1}, \tau_{j+2}, 
\tau_{j+3}, \tau_{j+4}) 
= P_3(\tau_{j-2}|\underline{\tau_{j-1},\tau_{j}}) 
P_3(\tau_{j-1}|\underline{\tau_{j},\tau_{j+1}}) \nonumber\\
&&\quad\cdot P_3(\tau_j, \tau_{j+1}, \tau_{j+2}) 
P_3(\underline{\tau_{j+1}, \tau_{j+2}}|\tau_{j+3}) 
P_3(\underline{\tau_{j+2},\tau_{j+3}}|\tau_{j+4}).
\label{2cl-cond} 
\end{eqnarray}
Analoguous factorizations hold for an arbitrary number of sites
on the left-hand-side of (\ref{1cl-prod}),(\ref{2cl-prod2}) and 
(\ref{2cl-cond}).

Let us now illustrate the scheme of the cluster calculations for 
the NaSch model by carrying out the calculation for the simplest case, 
namely, the 2-cluster calculations for $v_{max} = 1$. For convenience, 
one follows the sequence $2-3-4-1$, instead of $1-2-3-4$ of the stages 
of updating so that, for $v_{max} = 1$, a two-state variable $\sigma$ 
is adequate to describe the state of a lattice site; $\sigma = 0, 1$ 
correspond, respectively, to an empty site and a site occupied by a 
vehicle with speed $1$. Thus, in the special case of $v_{max} = 1$
one would need only four $2$-cluster probabilities, namely, 
$P_2(0,0), P_2(1,0), P_2(0,1), P_2(1,1)$. Interestingly, the 
constraints
\begin{equation}
P_2(1,0) + P_2(1,1) = c
\label{constraint-1}
\end{equation}
and
\begin{equation}
P_2(0,0)+P_2(0,1) = 1- c
\label{constraint-2}
\end{equation}
together with the particle-hole symmetry
\begin{equation}
P_2(1,0) = P_2(0,1)
\label{constraint-3} 
\end{equation}
leave only one of the four $2$-cluster probabilities, say, $P_2(1,0)$,
as an independent variable which one needs to calculate by solving the 
corresponding master equation. For general $v_{max}$, the $n$-cluster 
approximation on the right hand side of the master equation leads to 
$(v_{max} + 1)^n$ nonlinear equations; the number of independent 
equations gets reduced by the so-called Kolmogorov consistency 
conditions \cite{gutowitz}. 

\begin{figure}[ht]
 \centerline{\psfig{figure=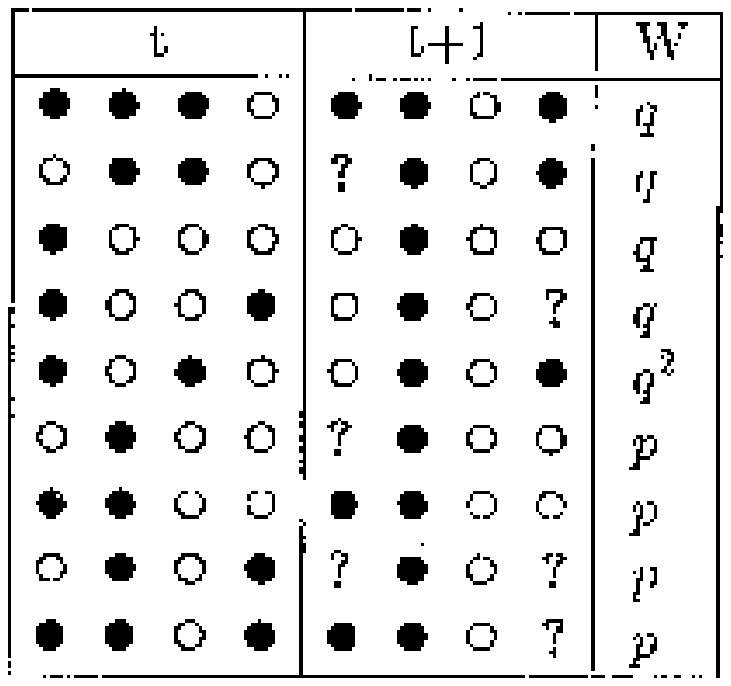,height=5cm}}
\caption{\protect{The four circles in the first two columns represent 
the states of the sites of a 4-cluster; the open and filled circles 
correspond, respectively, to empty sites and sites occupied by vehicles 
with speed $v = 1$.}}
\label{twocl_table}
\end{figure}

Using (\ref{2cl-prod2}) one factorizes the 4-cluster probabilities on
the right-hand-side of (\ref{mastereq}) for $P_2(1,0)$ in terms of
2-cluster conditional probabilities. In the first column of the table
in Fig.\ \ref{twocl_table} we list all those configurations
$(\tau_{i-1},\tau_i,\tau_{i+1},\tau_{i+2};t)$ which can lead to the
configurations, shown in the second column, which is the exhaustive
list of the 4-cluster configurations each having $\sigma_i =
1,\sigma_{i+1} = 0$; the corresponding transition probabilities
$W(1,0|\tau_{i-1},\tau_i,\tau_{i+1},\tau_{i+2})$ are given in the
third column.

Using the configurations at $t$ and $t+1$ as well as the corresponding
transition probabilities given in table in Fig.\ \ref{twocl_table} the
master equation (\ref{mastereq}) for $P_2(1,0)$ reduces to the
quadratic algebraic equation
\begin{equation}
q y^2 - y + c(1-c) = 0 
\label{quadratic}
\end{equation}
where we have used the shorthand notation $y = P_2(1,0)$. Solving this  
quadratic equation we get \cite{ssni} (see also \cite{wanghui})
\begin{equation}
P_2(1,0) = \frac{1}{2q}\left[1 - \sqrt{1- 4qc(1-c)}\right] 
\label{expr-10}
\end{equation} 
and, hence, $P_2(1,1), P_2(0,0), P_2(0,1)$ from the equations 
(\ref{constraint-1}),  (\ref{constraint-2}) and (\ref{constraint-3}). 
Moreover, the expression (\ref{expr-10}) establishes that 
 $P_2(1,0) \geq P_1(1) P_1(0) = c(1-c)$, which  indicates an 
effective particle-hole attraction (particle-particle repulsion) in 
the NaSch model with $v_{max} = 1$. Furthermore, from equation 
(\ref{expr-10}) one gets the expression  
\begin{equation}
J(c,p) = q P_2(1,0) = \frac{1}{2}\left[1 - \sqrt{1- 4qc(1-c)}\right] 
\label{fl-2cl}
\end{equation} 
which can be proved \cite{ssni} to be the {\it exact} expression 
for the corresponding flux. It is not difficult to carry out 2-cluster 
calculations for higher values of $v_{max}$, but one gets only 
approximate results for $v_{max} > 1$ \cite{ssni}. 

\begin{figure}[ht]
 \centerline{\psfig{figure=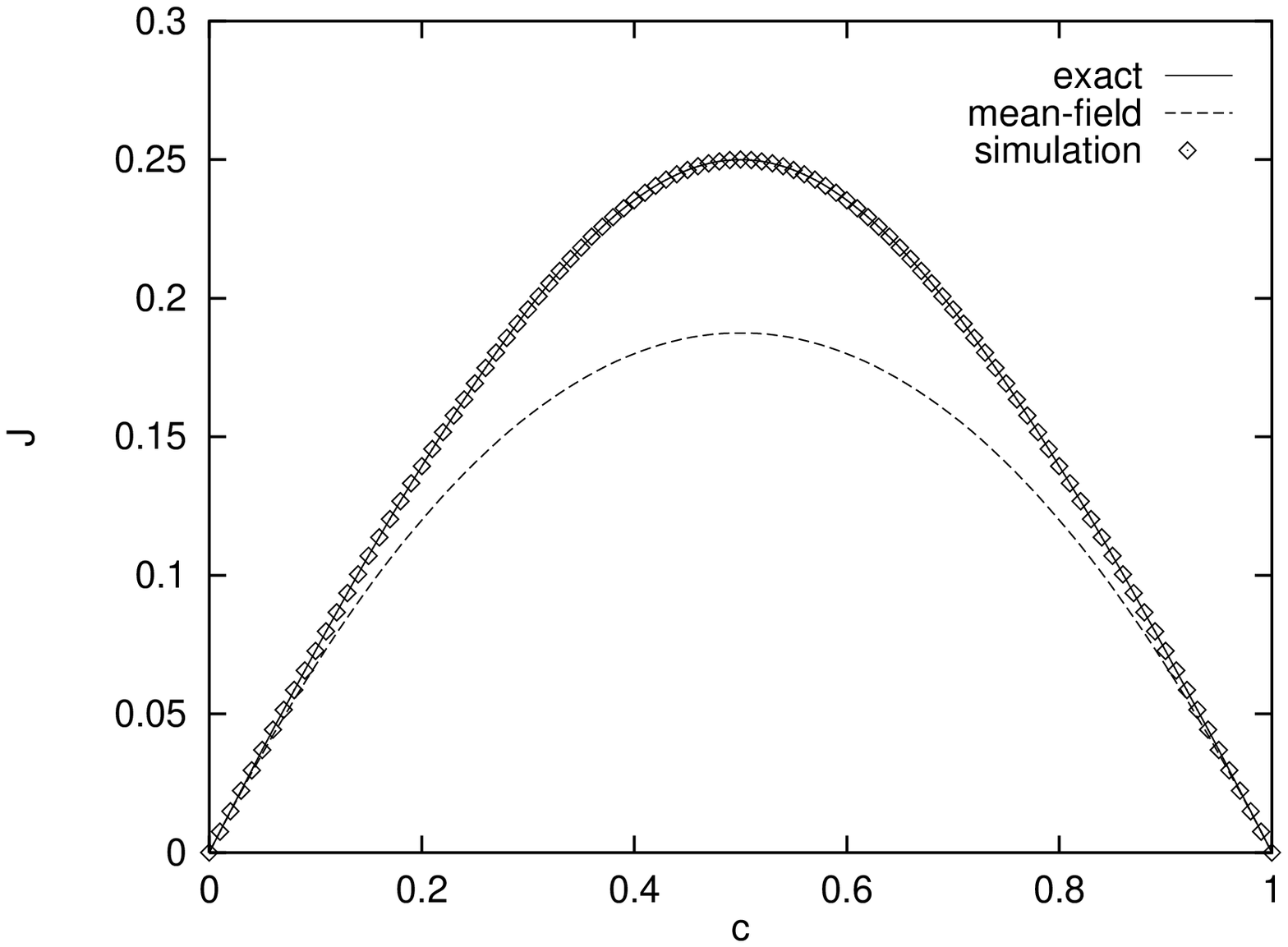,bbllx=50pt,bblly=50pt,bburx=550pt,bbury=400pt,height=5cm}
\psfig{figure=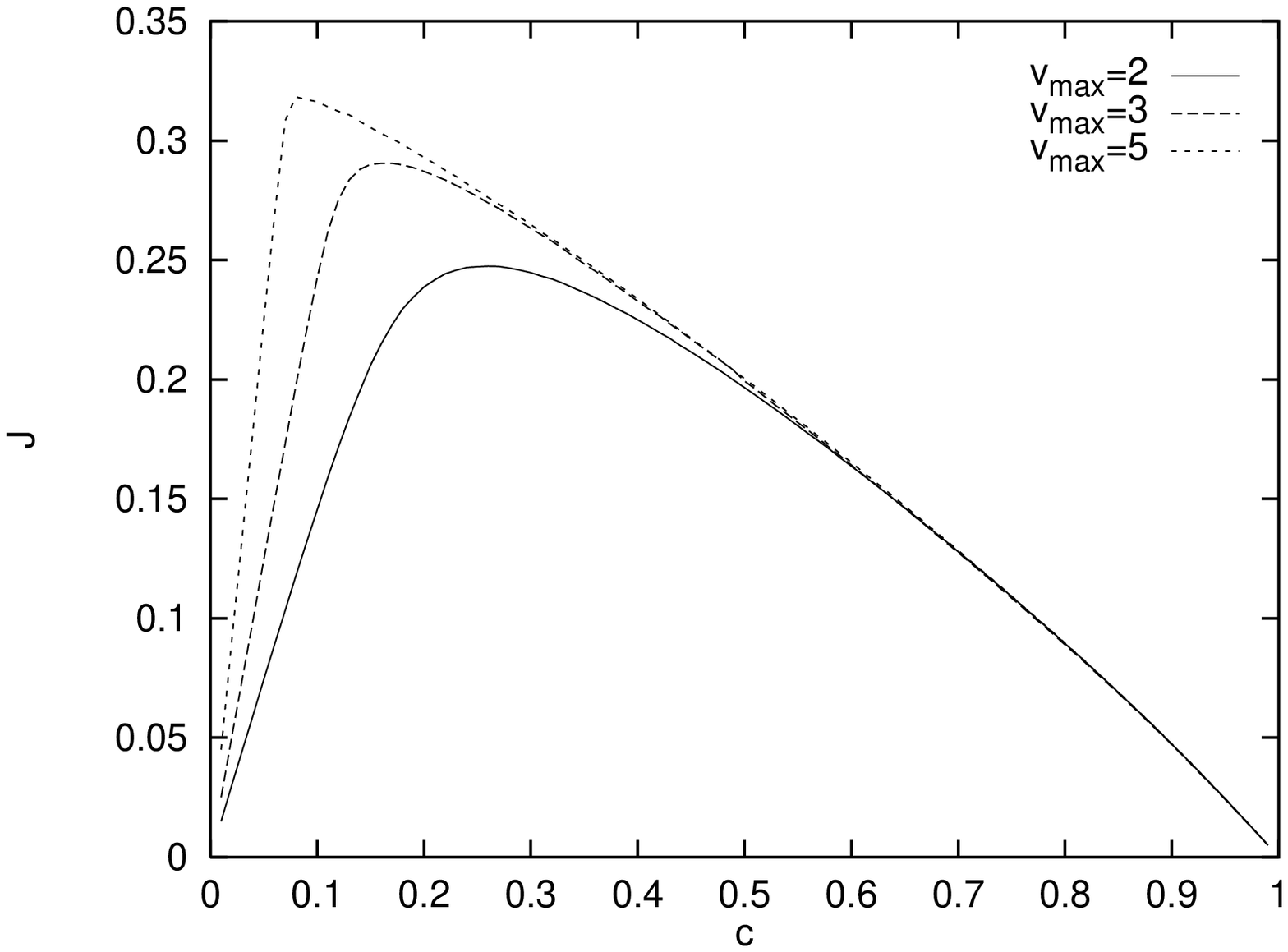,bbllx=50pt,bblly=50pt,bburx=550pt,bbury=400pt,height=5cm}}
\caption{\protect{The fundamental diagram in the NaSch model 
for (a) $v_{max} = 1$ and (b) $v_{max} > 1$, both for $p = 0.25$. 
The data for all $v_{max} > 1$ have been obtained through computer 
simulations.}}
\label{fund_ns235}
\end{figure}

\begin{figure}[ht]
\centerline{\psfig{figure=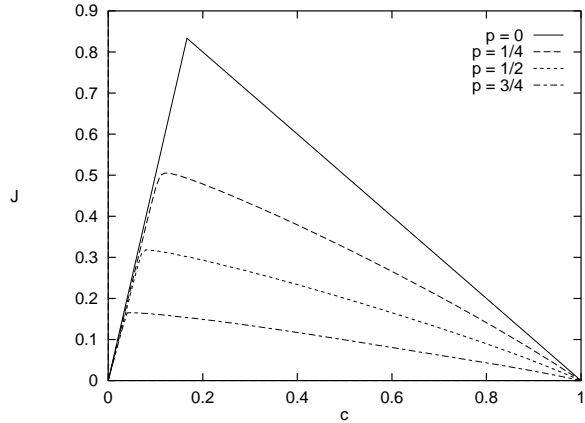,bbllx=40pt,bblly=20pt,bburx=550pt,bbury=400pt,height=6cm }}
\caption{\protect{The fundamental diagrams in the NaSch model 
with $v_{max} = 5$ are plotted for a few values of $p$. }}
\label{fund_nsp}
\end{figure}

An interesting feature of the expression (\ref{fl-2cl}) is that the 
flux is invariant under charge conjugation, i.e., under the operation 
$c \rightarrow (1-c)$ which interchanges particles and holes. Therefore, 
the fundamental diagram is symmetric about $c = 1/2$ when $v_{max} = 1$ 
(see Fig.~\ref{fund_ns235}(a)). Although this symmetry breaks down for all 
$v_{max} > 1$ (see Fig.~\ref{fund_ns235}(b)), the corresponding fundamental 
diagrams appear more realistic. Moreover, for given $p$, the magnitude 
of $c_m$ decreases with increasing $v_{max}$ as the higher is the 
$v_{max}$ the longer is the effective range of interaction of the 
vehicles (see Fig.~\ref{fund_ns235}). Furthermore, for $v_{max} = 1$, 
flux merely decreases with increasing $p$ (see eqn.~(\ref{fl-2cl})), 
but remains symmetric about $c = 1/2 =c_m$. On the other hand, for all 
$v_{max} > 1$, increasing $p$ not only leads to smaller flux but also 
lowers $c_m$ (Fig.~\ref{fund_nsp}). 

\begin{figure}[ht]
 \centerline{\psfig{figure=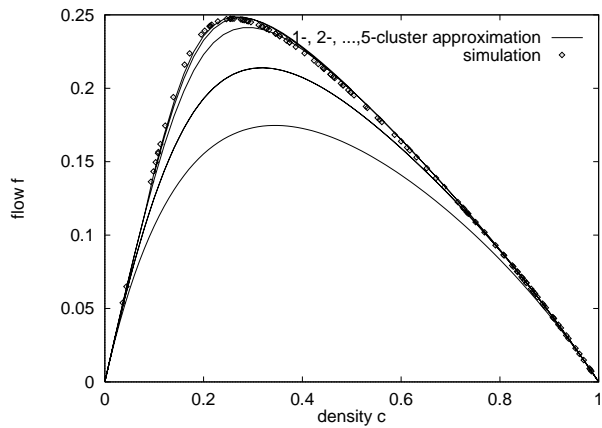,bbllx=10pt,bblly=200pt,bburx=520pt,bbury=550pt,height=6cm}}
\caption{\protect{The fundamental diagrams in the NaSch model 
with $v_{max} = 2$ in the $n$-cluster approximation ($n = 1,2,..,5$).}}
\label{fundns_high}
\end{figure}

For $v_{max} > 1$, one needs to carry out higher order cluster
calculations \cite{ssni,shad99} to get more accurate results than 
those obtained in the 2-cluster approximation. For $v_{max} = 2$, the
fundamental diagrams obtained from the $n$-cluster approximation ($n =
1,2,..,5$) are compared in Fig.~\ref{fundns_high} with the Monte Carlo
data. This comparison clearly establishes a rapid convergence with
increasing $n$ and already for $n = 4$ the difference between the
cluster calculation and MC data is extremely small. In \cite{shad99}
the cluster probabilities for $v_{max}=2$ have been obtained from 
computer simulations. The results suggest that the $n$-cluster
approximation for $n\geq 3$ becomes asymptotically exact in the 
limit $p\to 0$.

%

\subsubsection{Car-oriented mean-field theory of the NaSch model} 
\label{sec_comf}

In the "car-oriented" theories the state of the traffic system is 
described by specifying the positions and speeds of all the $N$ 
vehicles in the system \cite{ss97}. Suppose, $P_n(t)$ is the probability 
to find at time $t$ exactly $n$ empty sites immediately in front of a 
vehicle. Another auxiliary quantity, which turns out to be very 
convenient to use in several different calculations, is $g(t)$, 
the probability at time $t$ that a vehicle moves in the next time 
step. These two sets of quantities, namely, $P_n(t)$ and $g(t)$ 
are related. For example, in the NaSch model with $v_{max} = 1$,
a vehicle will move in the next time step if there is at least one
empty cell in front of it (probability $\sum_{n \geq 1} P_n(t)$) and
if it does not decelerate in the randomization step (probability $q$);
therefore, $g(t) = q[\sum_{n \geq 1} P_n(t)] = q [1 - P_0(t)]$. Hence
the flux $J(c,p)$ can be obtained from $J(c,p) = c g = cq[1-P_0]$.

The essence of the car-oriented mean-field (COMF) approximation \cite{ss97}
is to neglect the correlations between the gaps in front of the successive
cars\footnote{A similar approach, the so-called interparticle distribution
functions technique, is used for studying reaction-diffusion systems
\cite{ipdf}.}. The equations describing the time evolution of the 
probabilities $P_n(t)$, under this approximation (see Appendix 
\ref{App_COMF} for the derivation of these equations), can be solved 
in the steady-state using a generating function technique \cite{ss97}.  
Following this approach, in the special case $v_{max} = 1$, one recovers 
the exact expression (\ref{fl-2cl}) for $J(c,p) = cq[1-P_0]$.

For $v_{max} = 2$ one has to distinguish between $P_n(v=1)$ and 
$P_n(v=2)$. Moreover, one has to generalize the quantity $g$ to 
$g_{\alpha}$, the probability that the vehicle moves $\alpha$ 
cells ($\alpha = 1,2$) in the next time step. Applying the same 
generating function techniques as for $v_{max}=1$, one can also 
solve the coupled sets of steady-state equations for $P_n(v=1)$ 
and $P_n(v=2)$ for $v_{max} = 2$ but gets only approximate results 
\cite{ss97}. 

Interestingly, finite size of the system affects 
the equations for $v_{max} = 2$ in a much more dramatic way 
\cite{shad99} than those for $v_{max} = 1$ thereby revealing the 
intrinsic qualitative differences in the nature of correlations in 
the NaSch model for $v_{max}=1$ and $v_{max} >1$.

Comparisons with Monte Carlo simulations show that in contrast to the
3-cluster approximation for $v_{max}=2$ COMF does not become 
asymptotically exact in the limit $p\to 0$. This implies that even 
in this limit correlations between the headways are not negligible.
It is interesting, however, that for the fundamental diagram one
finds an excellent agreement between MC simulations and the predictions
of COMF \cite{ss97} for $p\to 0$. The reason is that in the deterministic
limit many configurations exist which produce the same flow. COMF
is not able to identify the dominating structures correctly, but
nevertheless can predict the correct current.

\subsubsection{Microscopic structure of the stationary state}
\label{sec_micro}

As we have seen MFT underestimates the flow in the stationary state
considerably. Deviations become larger for higher velocities $v_{max}$.
This shows the importance of correlations. As described above
a particle-hole attraction exist. Using the 2-cluster probabilities
for $v_{max}=1$ this attraction can be expressed in mathematical form 
as $P_2(1,0) > P_1(1)P_1(0)=c(1-c)$. 

For $v_{max}=1$ all improvements of MFT (2-Cluster, COMF and
pMFT) are exact. Here only correlations between neighbouring cells are
important. The fact that pMFT is exact shows that no 'true' correlations
exist. All correlations have their origin in the existence of GoE states. 
This also helps to understand why for random-sequential dynamics already 
MFT is exact for $v_{max}=1$ and the stationary state is uncorrelated.
The reason is simply that for random-sequential dynamics no GoE states
exist!

The situation changes for higher velocities $v_{max} > 1$. Here pMFT
is no longer exact\footnote{Note that for random-sequential dynamics
also MFT is no longer exact!}. Therefore 'true' correlations exist.
This corresponds to the observation made in \cite{ssni} that
the NaSch model shows a qualitatively different behaviour for $v_{max}=1$ 
and $v_{max} >1$. Furthermore it explains why so far the exact 
determination of the stationary state for $v_{max} > 1$ has not
been possible.

It is interesting to investigate how the microscopic structure of
the stationary state depends on the randomization $p$. 
For $p=0$ we have seen in Sec.\ \ref{sec_p=0} that for densities 
$c \leq 1/(v_{max} +1)$ the vehicles arrange themselves in such a way
that all headways are at least $v_{max}$. This is no longer possible
for larger densities, but still the vehicles have the tendency
to maximize their headway. Furthermore, for $p=0$ no spontaneous
formation of jams exists since overreactions are not possible.
The behaviour in this limit can be interpreted as coming from
a kind of "repulsive interaction" between the vehicles.

The behaviour for $p=1$ is a little bit different. Here we have
seen in Sec.\ \ref{sec_p=1} that metastable states with finite
flow exist for $c \leq 1/3$ and $v_{max}>1$.

For $0<p<1$ the microscopic structure interpolates between
these two limiting cases. This can be seen by analysing the
3-cluster probabilities obtained from Monte Carlo simulations
\cite{shad99}. For small $p$ the microscopic structure of the
stationary state is determined by the 'repulsive interactions'
between vehicles. With increasing $p$ one finds a tendency towards
phase separation into jammed and free-flow regions. A standing 
vehicle is able to induce a jam even at low densities since the
restart probability is small. The jams formed are typically not
compact, but of the form `.0.0.0.' since a vehicle approaching the
jam slows down in the randomization step with a rather high
probability. 

Concluding one might say that the microscopic structure for $0<p<1$
is determined by the competition of the two "fixed points" 
$p=0$ and $p=1$.

\subsection{Spatio-temporal organization of vehicles; is there a 
phase transition?}
\label{sec_transition}

The density $c_m$ corresponding to maximum flux is an obvious first 
candidate for a critical density separating the regimes of free-flow 
and congested flow in the NaSch model. We shall show in this 
subsection that this, indeed, is a critical point provided $p = 0$. 
However, in spite of strong indications that, probably, a noise-induced 
smearing of the transition takes place when $p \neq 0$, rigorous 
proofs are still lacking.

\noindent$\bullet${ Order parameter}

For a proper description of a phase transition one should introduce 
an appropriate order parameter which can distinguish the two phases 
because of its different qualitative behavior within the two phases 
\cite{stanley,goldenfeld}. 

A first candidate \cite{vdes} for the NaSch model would be the average 
fraction of vehicles at rest, i.e., with instantaneous speed $v = 0$. 
In the deterministic limit $p = 0$ this, indeed, serves the purpose 
of the order parameter for the sharp transition at $c_m^{det}$ from 
the free-flowing dynamical phase to the congested dynamical phase. 
But, in the general case of non-zero $p$, there is a non-vanishing 
probability that a vehicle comes to an instantaneous rest merely 
because of random braking even at extremely low density $c$; this 
probability is $p$ for $v_{max} = 1$ and decreases with increasing $c$. 

The next obvious choice would be \cite{eisen} the density of 
nearest-neighbor pairs in the stationary state
\begin{equation}
  \label{opdef}
   m = \frac{1}{T}\frac{1}{L}\sum_{t=1}^{T} \sum_{j=1}^{L} n_j n_{j+1},
\end{equation}
where, as defined earlier, $n_j=0$ for an empty cell and $n_j=1$ for 
a cell occupied by a vehicle (irrespective of its velocity). 
Because of the step 2 of the  
updating rule (deceleration due to other vehicles) $m$ gives the 
space-time-averaged density of those vehicles with velocity $0$ 
which had to brake due to the next vehicle ahead.

\begin{figure}[ht]
 \centerline{\psfig{figure=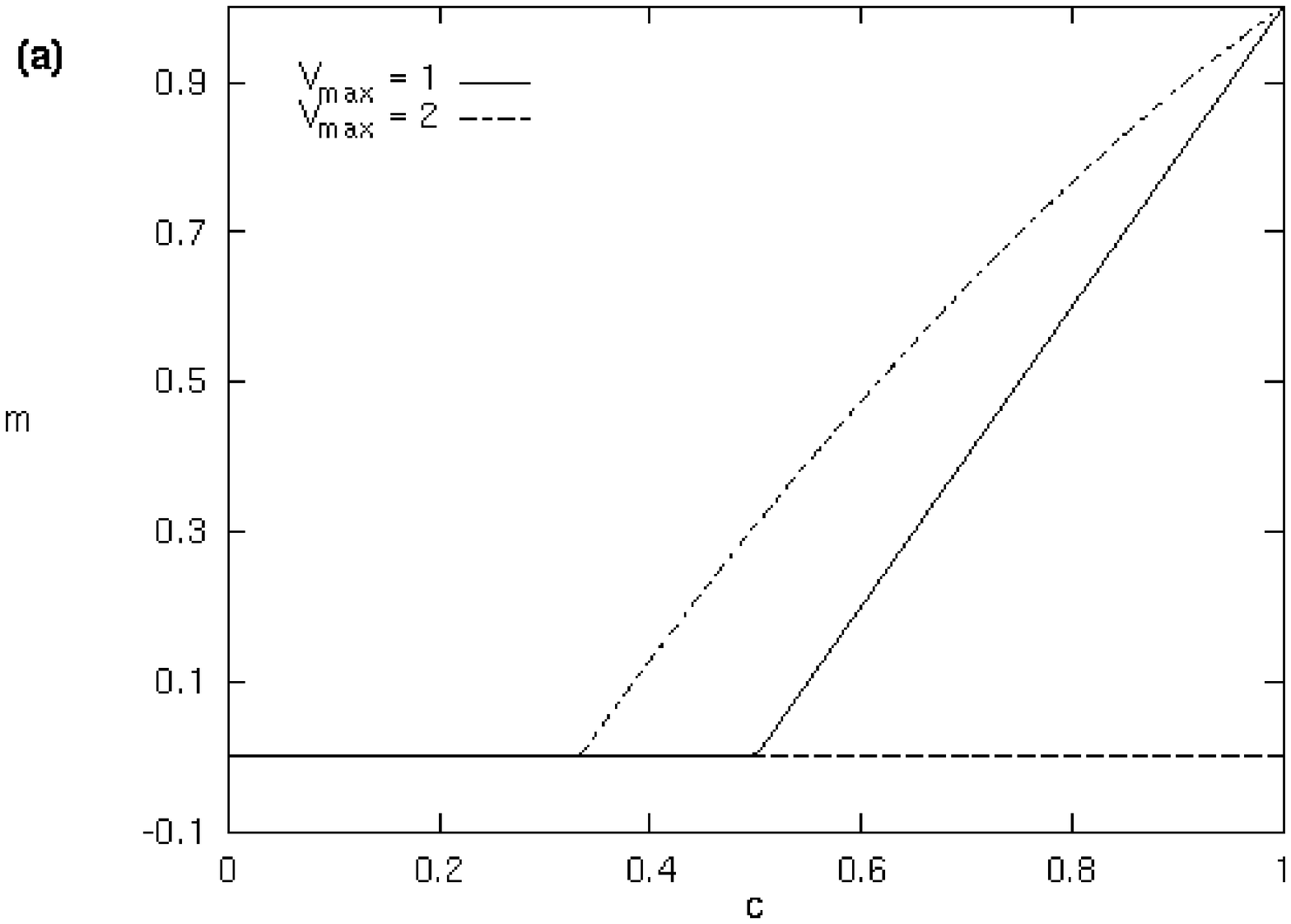,height=4.5cm}
\psfig{figure=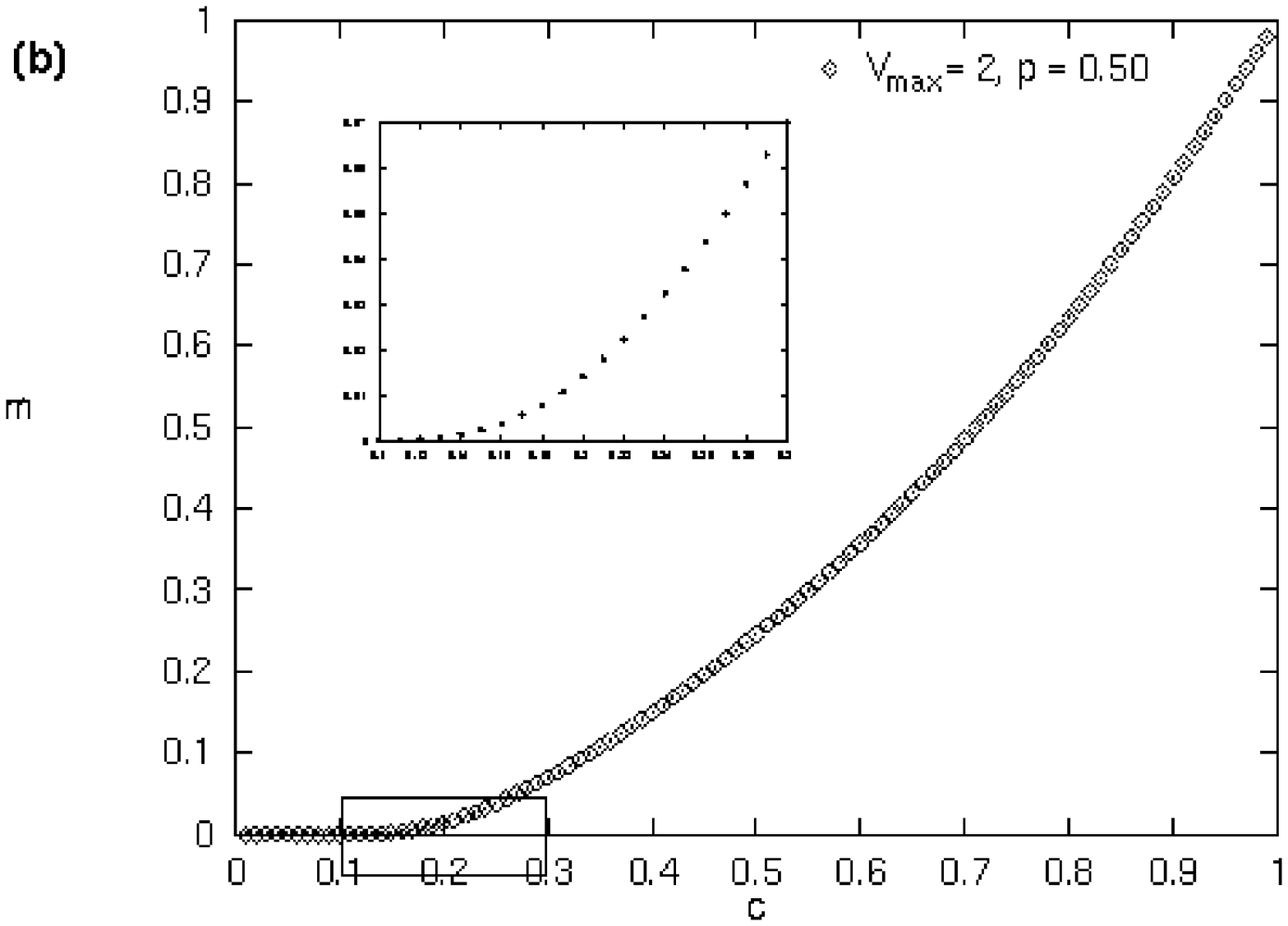,height=4.5cm}}
  \caption{\protect{The density-dependence of the order parameter 
      $m$ in (a) the deterministic limit $p = 0$ of the NaSch model 
      ($v_{max}= 1,2$) and (b) the NaSch model with $v_{max} = 2$, 
      $p = 0.5$. }}
\label{ordpar}
\end{figure}

The Fig.~\ref{ordpar}(a) shows that $m$ vanishes at $c_m^{det}$ if 
$p = 0$. On the other hand if $p \neq 0$, $m$ does not vanish even 
for $c < c_m$ although $m$ becomes rather small at small densities 
(see Fig.~\ref{ordpar}(b)). 

We now present a heuristic argument to point out why any quantity 
related to the fraction of jammed vehicles is non-zero at any 
density $c > 0$ and, hence, inadequate to serve as an order parameter 
\cite{krug99}. To slow down to $v_{max} - 2$ a vehicle must be 
hindered by one randomly braking vehicle in front. Similarly, to 
reach a speed $v_{max} - 3$ a vehicle must find two vehicles within 
the range of interaction, and so on. The probability for $n$ vehicles 
to be found in the close vicinity of a given vehicle is proportional 
to $c^n$. Therefore, the probability $P_v(c)$ of finding a vehicle 
with speed $v < v_{max} - 1$ is proportional to $c^{v_{max}-1-v}$ 
and, hence, even for $v = 0$, $P_v(c)$ is, in general, non-zero for 
all $c \rightarrow 0$. 

\noindent$\bullet${ Spatial correlations}

A striking feature of second-order phase transitions is the
occurrence of a diverging length scale at criticality and a 
corresponding algebraic decay of the correlation function 
\cite{stanley,goldenfeld}. Using lattice gas variables $n_j$, 
the equal-time density-density correlation function is defined by
\begin{equation}
  \label{korrdef}
  G(r) = \frac{1}{T} \frac{1}{L}\sum_{t=1}^{T} \sum_{j=1}^{L} 
  n_j n_{j+r} - c^2 .
\end{equation}
which measures the correlations in the density fluctuations  
that occur at the same time at two different points in space 
separated by a distance $r$.

Again it is very instructive to consider first the deterministic 
case $p=0$ (Fig.~{\ref{correl}(a)). Since, as argued before, there 
are exactly $v_{max}$ empty sites in front of each vehicle at 
$c = c_m^{det}$ the correlation function at $c = c_m^{det}$ is 
given by
\begin{equation}
   G(r) = \begin{cases}
c_m^{det}(1-c_m^{det})  &\text{for $r\equiv 0$~mod $(v_{max}+1)$,}\\ 
-(c_m^{det})^2          &\text{else.} 
\end{cases}
  \label{korrdet}
\end{equation}

\begin{figure}[ht]
\centerline{\epsfig{figure=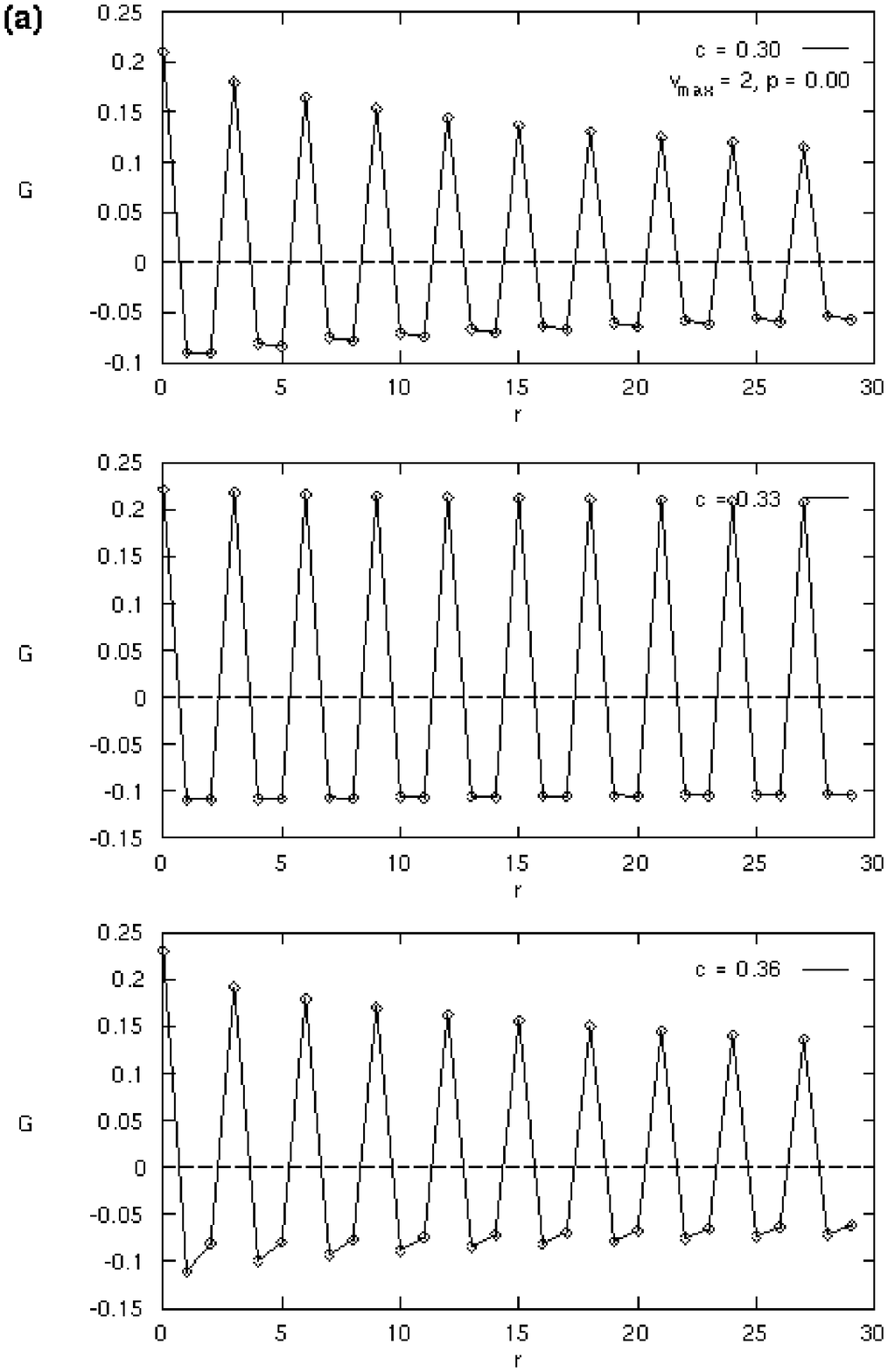,width=5cm}\epsfig{figure=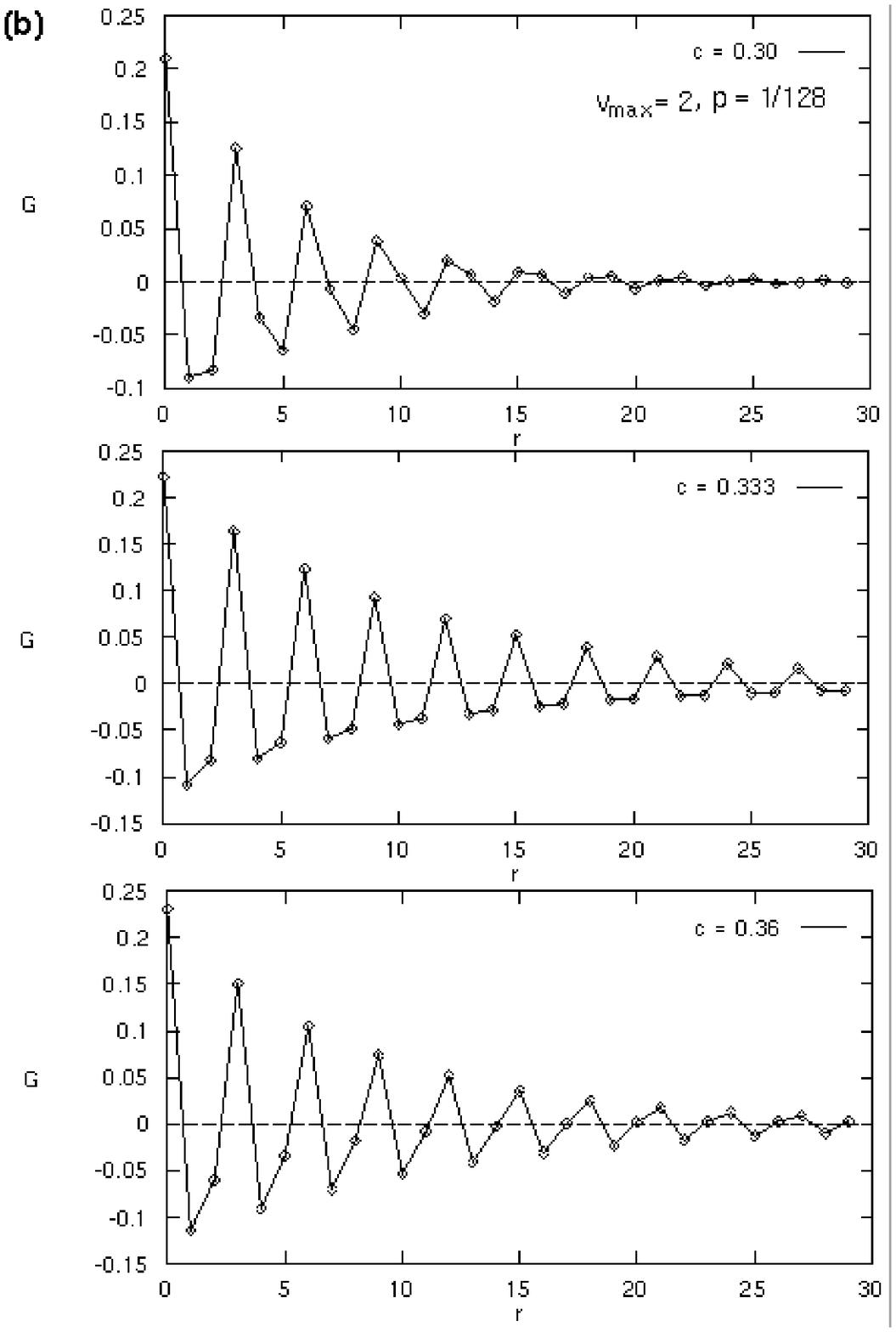,width=5cm}}
  \caption{\protect{The density-density correlation function $G(r)$ 
      (a) the deterministic limit $p = 0$ of the NaSch model 
      ($v_{max}= 2$) and (b) the NaSch model with 
      $v_{max} = 2$, $p = 1/128$. }}
\label{correl}
\end{figure}

For all $v_{max}$ the correlation function for small non-zero $p$ 
(Fig.~\ref{correl}(b)) has essentially the same structure as that 
for $p = 0$ (Fig.~\ref{correl}(a)) but the amplitude decays exponentially 
\cite{eisen} for all $c$. In the general case of non-vanishing $p$, 
the asymptotic behaviour ($r \rightarrow \infty$) of the correlation 
length $\xi$ can be obtained analytically \cite{shad99} only for 
$v_{max} = 1$. It turns out that, for given $p$, $\xi$ is maximum 
at $c = 1/2 = c_m$ and that $\xi(c=1/2) \propto p^{-1/2}$. Thus, 
for $v_{max} = 1$, $\xi$ diverges only for $p = 0$ but remains finite 
for all non-zero $p$.  For $v_{max} > 1$ the trend of variation of 
$\xi$ with $c$ (Fig.~\ref{corlength}(a)) in the vicinity of $c_m$ is 
the same as that for $v_{max} = 1$ \cite{eisen}. Moreover, for 
$v_{max} > 1$, the maximum value of the correlation length, 
$\xi_{max}$ plotted against $p$ (Fig.~\ref{corlength}(b)), is also 
consistent with the corresponding trend of variation for $v_{max} = 1$.  
Thus, the correlation function $G(r)$ gives a strong indication that 
the NaSch model exhibits a second order phase transition, at 
$c = c_m^{det}$, only for $p = 0$ but this transition is smeared 
out if $p \neq 0$. This noise-induced smearing of the phase transition 
in the NaSch model is very similar to the smearing of critical phenomena 
by finite-size effects. 
\begin{figure}[ht]
\centerline{\epsfig{figure=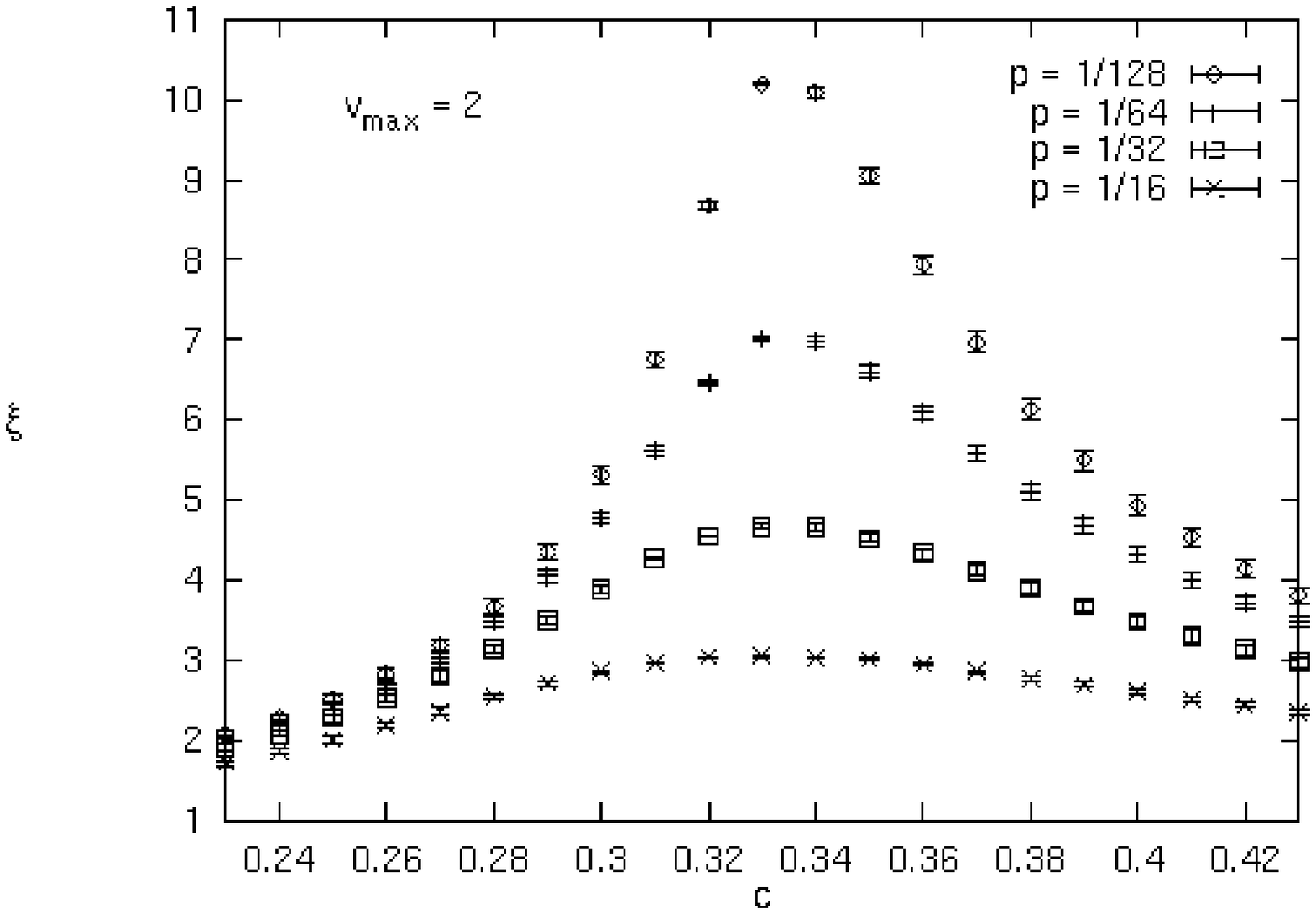,height=4.75cm}\quad
\epsfig{figure=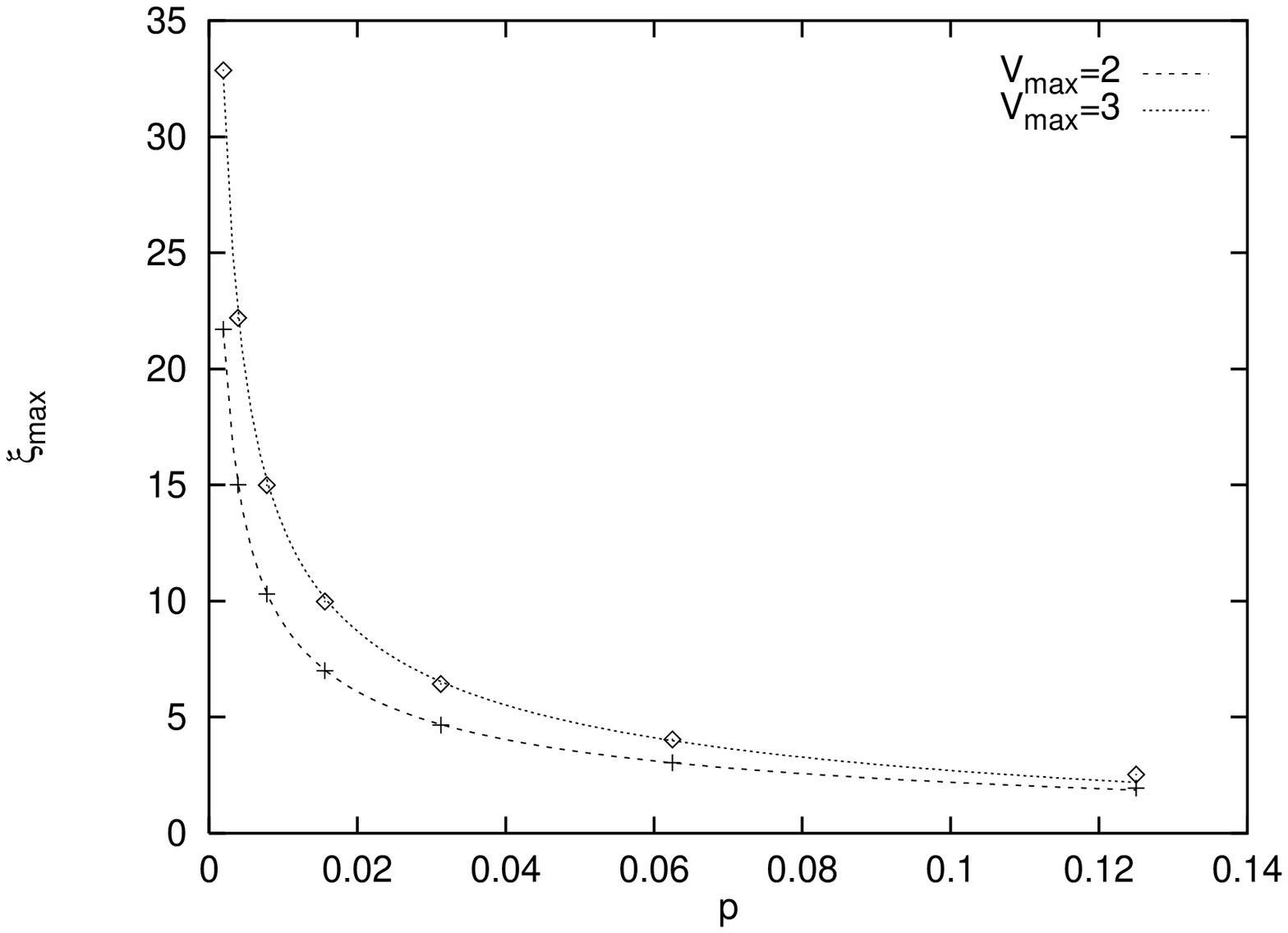,height=4.75cm}}
 \caption{\protect{The dependence of (a) $\xi(c,p)$ on $c$ for four 
                    different values of $p$ ($v_{max}= 2$) and (b) 
                    $\xi_{max}$ on $p$ ($v_{max} = 2,3$).}}
\label{corlength}
\end{figure}
%

\noindent$\bullet${ Distribution of lifetimes of jams}

Another quantity which should be able to give information about the
nature of the transition from free-flow to the jammed regime is
the distribution of lifetimes of jams. Following Nagel \cite{kn94}
each vehicle which has a velocity less than $v_{max}$ before the
randomization step will be considered jammed. This definition is 
motivated by the cruise-control limit (see Sec.\ \ref{sub_cruisesoc})
where it is more natural than in the NaSch model. One expects,
however, that the long-time behaviour of the lifetime distribution
is independent of the exact definition of a jam. The short-time
behaviour, on the other hand, might differ strongly, e.g.\ for
``compact jams'' where a jam is defined as a series of consecutive
standing vehicles without any empty cells in between.

\begin{figure}[ht]
 \centerline{\psfig{figure=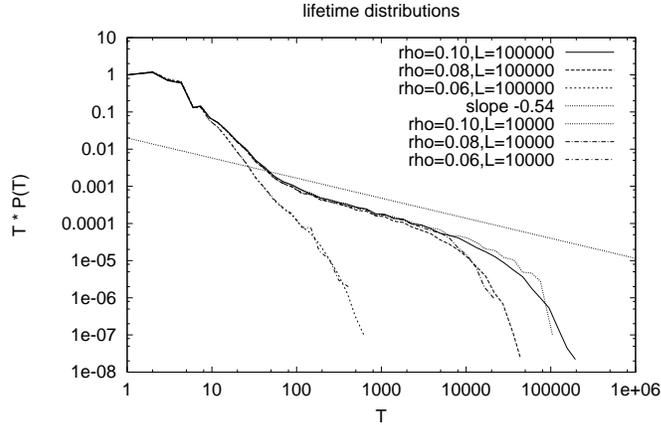,height=6cm }}
 \caption{\protect{Lifetime distribution in the NaSch model for
$v_{max}=5$ and $p=0.5$ and various densities below and above 
$c_m\approx 0.085$.}}
\label{fig_life}
\end{figure}

Fig.~\ref{fig_life} shows the results of Monte Carlo simulations for
the lifetime distribution in the NaSch model for different densities
near the transition region, $c \approx c_{m}=0.085\pm 0.005$
(for $v_{max}=5$, $p=0.5$), where $c_{m}$ is the density where
the flow is maximal. The most interesting feature of the lifetime
distribution is the existence of a cutoff near $\tau_c = 10000$.
It has been shown \cite{kn94} that this cutoff is neither a finite-size
nor a finite-time effect. For times smaller than $\tau_c$ a
scaling regime exists where the distribution decays algebraically.

\noindent$\bullet${ Dissolution of a megajam} 

Gerwinski and Krug \cite{krug99} tried to find an intuitive criterion
which allows the distinction of free-flow and jammed phases.
It is based on the investigation of jam dissolution times.
Starting from a megajam configuration, i.e.\ a block of $N$ consecutive
cells occupied by vehicles with the remaining $L-N$ cells being empty,
they determined the time until the jam\footnote{In \cite{krug99}
the same definition of a jam as in the previous point "Distribution 
of lifetimes of jams" (see \cite{kn94}) has been used.} has 
dissolved completely.

A simple estimate gives the density at which the lifetime is expected
to become infinite. Suppose that the jam dissolves with velocity
$v_J$. Since the first vehicle move freely with average velocity
$v_F=v_{max}-p$ it will reach the end of the jam at the same time
as the dissolution wave if the condition $(L-N)/v_F = N/v_J$ is
satisfied. The corresponding density is then given by
\begin{equation}
c^*=\frac{v_J}{v_J+v_F}=\frac{v_J}{v_J+v_{max}-p}.
\label{defcstar}
\end{equation}

For $v_{max}=1$ vehicles accelerate immediately to $v_{max}$. In this
case one has $v_J=q=1-p$. For higher velocities, $q=1-p$ is only an upper
bound for $v_J$. Inserting $v_J=1-p$ into (\ref{defcstar}) one therefore
obtains an upper bound for the density $c^*$. Taking into account
interactions between vehicles in the outflow region of the jam,
one can derive an effective acceleration rate $q_{eff}$, and thus
the jam dissolution velocity $v_J=q_{eff}$, as a function of $p$
\cite{krug99}. 

Computer simulations show a sharp increase of the lifetime near the
density $c^*$. It becomes ``infinite'', i.e.\ the jam does not 
dissolve within the measurement time, only at a higher density
$c_1^*$ which is considerably larger than the density $c_m$ of maximum
flow. At intermediate densities $c^* < c < c_1^*$ the jam does
not dissolve during the first lap, but later due to fluctuations of
the two ends of the jam. During this time other jams have usually
formed. All results found in \cite{krug99} are consistent with
the measurements of the lifetime distributions presented in the
previous point.

\noindent$\bullet${ Relaxation time}

A characteristic feature of a second order phase transition is the
divergence of the relaxation time at the transition point. 
For $p=0$ this has been studied first by Nagel and Herrmann
\cite{nh}. They found a maximum $\tau_{max}$ of the relaxation time
at the density $c=1/(v_{max}+1)$ which diverges in the thermodynamic
limit $L\to\infty$ as $\tau_{max}\propto L$. For finite $p$ the
behaviour of the relaxation time is more complicated.
For technical reasons Cs\'anyi and Kert\'{e}sz \cite{Csanyi}
made no direct measurements of the relaxation time, but used the 
following approach:
Starting from a random configuration of cars with velocity $v_j=0$ the
average velocity $\bar v(t)$ is measured at each time step $t$. 
For $t\rightarrow\infty$ the system reaches a stationary state 
with average velocity $\left < \bar v_{\infty} \right >$. The
relaxation time is characterised by the parameter \cite{Csanyi}
\begin{equation}
  \label{eq:relax}
  \tau= \int\limits_{0}^{\infty}\left[ {\rm min}\{v^{\ast}(t),\ 
  \langle \bar{v}_{\infty}\rangle\} - \langle \bar{v}(t)\rangle\right]\  
  dt \ . 
\end{equation}
$v^{\ast}(t)$ denotes the  average velocity  in the acceleration phase
$t\rightarrow 0$ for low vehicle density $\rho\rightarrow
0$. In this regime, due to the absence of interactions between the
vehicles, one has $v^{\ast}(t) = (1-p)t$.
Thus the relaxation time is obtained by summing up the deviations of the
average velocity $\langle \bar{v}(t)\rangle $ 
from the values of a system with one single vehicle which can move
without interactions with other cars ($\rho\rightarrow 0$).
Note that for a purely exponentially decaying quantity $v(t)=
v_\infty+C\exp(-t/\tau')$ the definition (\ref{eq:relax}) is proportional
to $\tau'$, i.e.\ the standard definition of the relaxation time.
One finds a maximum of the relaxation parameter near, but {\em below}, the
density of maximum flow for $p=0.25$ and ${v_{\mathit{max}}}=5$ 
(see \cite{Csanyi}).

\begin{figure}[ht]
\centerline{\epsfig{figure=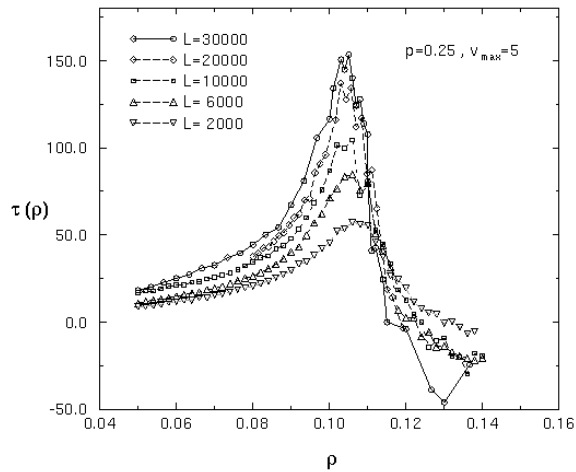,height=6.5cm}}
\caption{Relaxation parameter near the transition density for
$v_{max} = 5$ and  $p=0.25$.}
\label{fig_relax}
\end{figure}

The relaxation time (\ref{eq:relax}) shows interesting behavior
\cite{Csanyi,sasvari,eisen} which is difficult to interpret in terms 
of critical point phenomena. One finds a maximum of the
relaxation parameter near, but below, the density of maximal flow.
This maximum value increases with system size, but the width of the 
transition region does not seem to shrink \cite{sasvari}. Two
scenarios are possible: 1) The relaxation time converges to a large
but finite value for large system sizes beyond the present computer
power; 2) The relaxation time diverges for $L\to\infty$.
Scenario 1) appears to be more plausible in view of the finite lifetimes
of jams discussed above. Complicated interactions between jams could
in principle lead to a divergence. Keeping in mind the unusual scaling
behaviour of the width, this should occur in a finite interval, not 
at a special (critical) point.  

The interpretation of the parameter $\tau$ as a relaxation time
can be problematic. This can be seen clearly for $c>c_{transition}$,
where $\tau$ can become negative \cite{eisen}. Here it is possible 
that during relaxation the system can temporarily reach states with 
a higher average velocity than in the stationary state. This 
overreaction can be divided into two phases for $p>0$. Within the 
first few time steps small clusters which occur in the initial 
configuration vanish. The second phase is characterized by the 
growth of surviving jams. More and more cars get trapped into 
large jams and therefore the average flow decreases to its stationary 
value. This decrease causes negative values of $\tau$ at large 
densities.

\noindent$\bullet${ Distribution of distance-headways}

In order to get information on the spatial organization of the 
vehicles, one can calculate the distance-headway distribution 
${\mathcal{P}}_{dh}(\Delta x)$ 
by following either a site-oriented approach \cite{chow97a} or a 
car-oriented approach \cite{ss97} if $\Delta x_j = x_{j+1}-x_{j}$, 
i.e., if the number $\Delta x_j -1$ of empty lattice sites in front 
of the $j$-th vehicle is identified as the corresponding 
distance-headway. 

Stated precisely, ${\mathcal{P}}_{dh}(k)$ is the conditional probability 
of finding a string of $k$ empty sites in front of a site which is 
given to be occupied by a vehicle. A comparison between the naive 
mean-field expression
\begin{equation}
{\mathcal{P}}_{dh}^{mfa}(j) = c (1-c)^j
\label{eq-mfdh} 
\end{equation}
for the distance-headway distribution in the NaSch model with $v_{max}
= 1$ and the corresponding Monte Carlo data \cite{chow97a} reveals the
inadequacy of equation (\ref{eq-mfdh}) at very short distances which
indicates the existence of strong short-range correlations in the
NaSch model that are neglected by the mean-field treatment. This is
consistent with our earlier observation that there are particle-hole
effective short-range attraction in the NaSch model with $v_{max} =
1$. Again, this correlation disappears when a random sequential
updating is carried out! The exact distance-headway distribution in
the NaSch model with $v_{max} = 1$ is found to be \cite{chow97a,ss97}
\begin{equation}
{\mathcal{P}}_{dh}^{2c}(j) = \frac{y^2}{c(1-c)}
\left[1 - \frac{y}{(1 -c)}\right]^{j-1} \qquad (j=1,2,...)
\label{dh-2cl}
\end{equation}
where $y = P_2(1,0)$ is given by the equation (\ref{quadratic}).

\begin{figure}[ht]
 \centerline{\psfig{figure=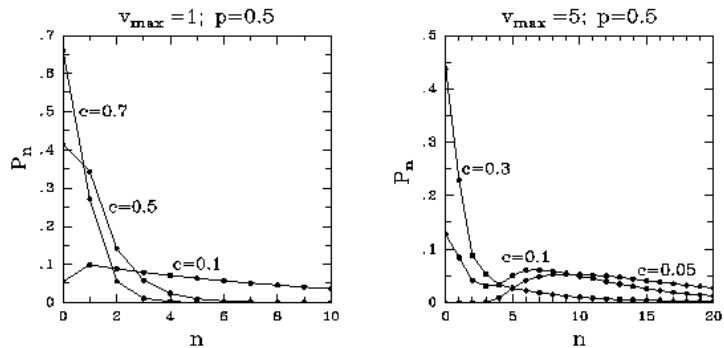,width=10cm }}
  \caption{\protect{Distributions of distance-headways in the NaSch 
model for $p = 0.5$ and different densities for (a) $v_{max} = 1$ and 
(b) $v_{max} = 5$. $n$ denotes the number of empty cells in front
of a vehicle and is related to the distance-headway by $n=\Delta x-1$.}}
\label{dist_head}
\end{figure}

For all $v_{max} > 1$, at moderately high densities,
${\mathcal{P}}_{dh}(\Delta x)$ exhibits two peaks, in contrast to a
single peak in the distance-headway distributions for $v_{max} = 1$ at
all densities (Fig.~\ref{dist_head}); the peak at $\Delta x = 1$ is
caused by the jammed vehicles while that at a larger $\Delta x$
corresponds to the most probable distance-headway in the free-flowing
regions.  At first sight, the simultaneous existence of free-flowing
and jammed regions may appear analogous to the coexistence of gaseous
and liquid phases of matter in equilibrium. In fact, when the
two-peaked structure of the distance-headway distribution was first
observed \cite{chow97a,krauss}, it was erroneously interpreted as a
manifestation of the coexistence of two dynamical phases, namely, the
free-flowing phase and the jammed phase. But, later works
\cite{chow98a} established that the analogy between the coexistence of
free-flowing and jammed regions in the NaSch model and the coexistence
of the gas and liquid phases of matter cannot be pushed too far
because the analogue of the gas-liquid interfacial tension is zero in
the NaSch model. Thus, one can not conclude that the NaSch model 
exhibits a first order dynamical phase transition.

\noindent$\bullet${ Distributions of jam sizes and gaps between jams} 

One can identify a string of $k$ successive stopped vehicles as a
jam of length $k$ (by definition, such jams are compact). Similarly,
when there are $k$ lattice sites between two successive jams, each
occupied by a moving vehicle or is vacant then we say that there is
a gap of length $k$ between the two successive jams. Analytical
expressions for the distributions of the jam sizes as well as of
the gaps between jams can be calculated for the NaSch model (and some 
of its extensions) using the 2-cluster approximation or or COMF
\cite{shad99,chow97a,chow99,as_tgf}. The expressions are exact in the
case $v_{max} = 1$ with periodic boundary conditions.
For higher velocities the results are only approximative. In COMF
the probability $C_k$ to find a jam of length $k$ is given by
\begin{equation}
C_{k}^{(COMF)}=(1-P_0)P_0^{k-1}
\label{COMFjamdist}
\end{equation}
whereas in the 2-cluster approach one finds
\begin{eqnarray}
C^{(2-cl)}_k &=& \frac{1}{{\mathcal{N}}_J}\, P(\underline{0}|1) 
    P(\underline{1}|1)^{k-2}P(\underline{1}|1) P(\underline{1}|0)
\qquad \qquad (k\geq 2), \nonumber\\
C^{(2-cl)}_1 &=& \frac{1}{{\mathcal{N}}_J}\sum_{v=1}^{v_{max}} 
    P(\underline{0}|v) P(\underline{v}|0)\ .
\label{2clustjamdist}
\end{eqnarray}
For the $n-$cluster approximation similar expressions can be derived.

Both distribitions (\ref{COMFjamdist}) and (\ref{2clustjamdist})
decay exponentially for large jam sizes. COMF always predicts a
monotonous distribution with $C_{k}^{(COMF)}\geq C_{k+1}^{(COMF)}$.
In contrast, the jam size distribution in the $n-$cluster approximation
can in principle exhibit a maximum at small jam sizes $1\leq k\leq n$.

\noindent$\bullet${ Distribution of time-headways} 

Since flux is equal to the inverse of the average time-headway, much
more detailed information is contained in the full distribution of the
time-headway than in the fundamental diagram. The time-headway
distribution contains information on the temporal organization of the
vehicles.

Suppose, ${\mathcal{P}}_m(t_1)$ is the probability that the following
vehicle takes time $t_1$ to reach the detector, moving from its
initial position where it was located when the leading vehicle just
left the detector site. Suppose, after reaching the detector site, the
following vehicle waits there for $\tau - t_1$ time steps, either
because of the presence of another vehicle in front of it or because
of its own random braking; the probability for this event is denoted
by $Q(\tau-t_1|t_1)$. The distribution ${\mathcal{P}}_{th}(\tau)$, of
the time-headway $\tau$, can be obtained from \cite{chow98a,ghosh}
\begin{equation}
{\mathcal{P}}_{th}(\tau) = \sum_{t_1=1}^{\tau-1} {\mathcal{P}}_m(t_1) 
Q(\tau-t_1|t_1) 
\label{eq-TH}
\end{equation}

Substituting the expressions for ${\mathcal{P}}_m(t_1)$ and 
$Q(\tau-t_1|t_1)$ for $v_{max} = 1$ in (\ref{eq-TH}) we, finally, 
get \cite{chow98a,ghosh} 
\begin{eqnarray}
 {\mathcal{P}}^{th}(t) = \frac{qy}{c-y}
 \left(1-\frac{qy}{c}\right)^{t-1} +  \frac{qy}{d-y}
 \left(1-\frac{qy}{d}\right)^{t-1} \nonumber\\
 - \left[\frac{qy}{c-y}+\frac{qy}{d-y}\right] p^{t-1} 
 - q^2(t-1)p^{t-2}.
\end{eqnarray}
where, $q = 1 - p$, $d = 1 - c$ and, for the given $c$ and $p$, $y$
can be obtained from equation (\ref{expr-10}). The expression is
plotted in Fig.~\ref{th}(a) for a few typical values of $c$ for a given
$p$. A few typical time-headway distributions in the NaSch model for
$v_{max} > 1$, obtained through computer simulation, are shown in
Fig.~\ref{th}(b).

\begin{figure}[ht]
\centerline{\epsfig{figure=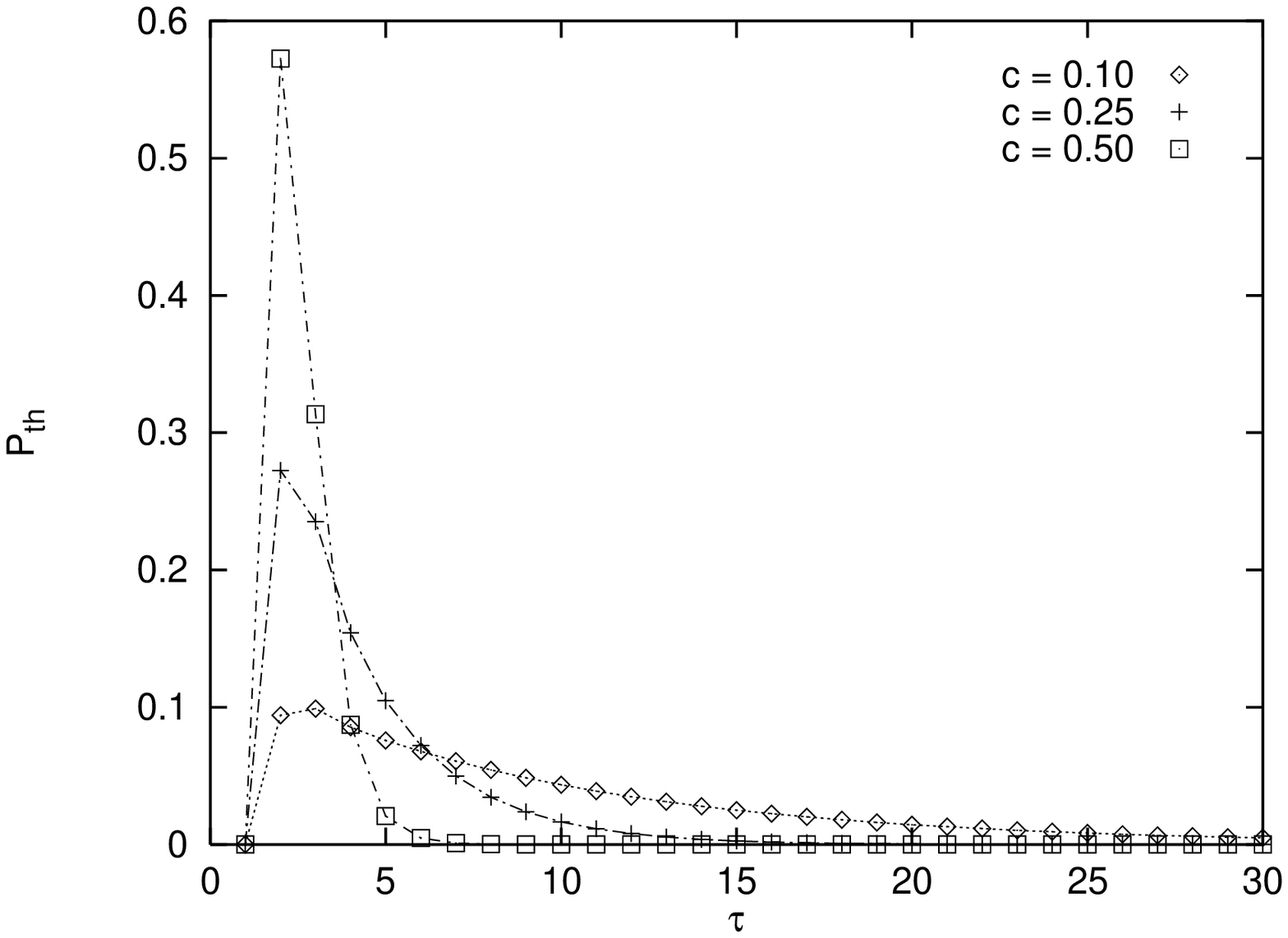,height=4.75cm}\qquad
\epsfig{figure=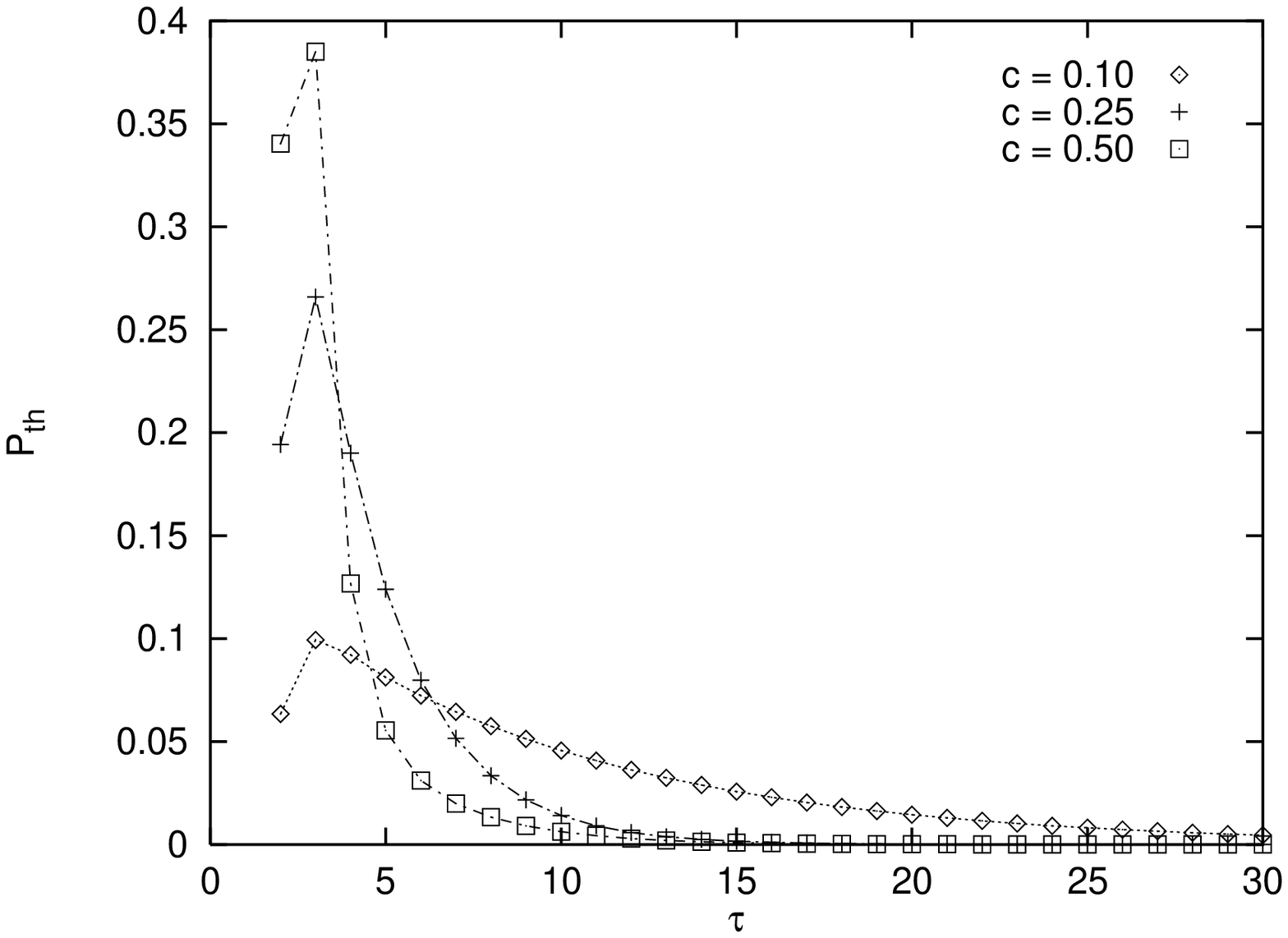,height=4.75cm}}
\caption{Time-headway distribution in the NaSch model with (a) 
$v_{max} = 1$ and (b) $v_{max} = 5$.}
\label{th}
\end{figure}

\noindent$\bullet${ Temporal correlations}

In order to probe the spatio-temporal correlations in the fluctuations 
of the occupation of the cells, one can study the space-time 
correlation function 
\begin{equation}
C(r,\tau) = \frac{1}{T} \frac{1}{L}
\sum_{j=1}^L \sum_{t=1}^T n_i(t) n_{j+r}(t+\tau) - c^2  
\end{equation}
which, by definition, vanishes in the absence of any correlation. 
In \cite{chebani} three different regimes have been distinguished.\\
{\em Free-flow} ($0<\rho\leq \rho_1$): ``Minijams'' occur which
resolve immediately. The correlation function shows anticorrelations
around propagating peak.\\
{\em Jamming} ($\rho_1<\rho\leq \rho_2$): Free-flow and jamming coexist,
i.e.\ jams with a finite lifetime and vehicles moving with $v_{max}$
occur. This coexistence is reflected in the behaviour of the 
correlation function which exhibits a double-peak structure.\\
{\em Superjamming} ($\rho_2<\rho\leq 1$): The whole system is congested.
Jamming waves are connected and form an infinite wave. As a consequence,
the propagating peak in the correlation function has disappeared.\\

Neubert et al.\ \cite{neublee} have introduced a special autocorrelation 
function of the density in order to study the velocity of jams. 
They have determined jam velocities for several variants of the
NaSch model which will be introducted in later sections.


\noindent$\bullet${ Structure factor} 

Structure factors are known to give valuable information about driven
systems \cite{sz}. For the NaSch model the static structure factor
\begin{equation}
S(k)=\frac{1}{L}\left\langle \left| \sum_{j=1}^L n_je^{ikj}
\right|\right\rangle
\end{equation}
has been investigated in \cite{LueSchUs}. Again $n_j$ denotes the
occupation number of cell $j$. Note that $S(k)$ is related to the
Fourier transform of the density-density correlation function
(\ref{korrdef}).

For all densities, $S(k)$ exhibits a maximum at $k_0 \approx 0.72$
which corresponds to the characteristic wavelength $\lambda_0=2\pi/k_0$
of the density fluctuations in the free-flow regime. For $v_{max}>1$
one finds $k_0(v_{max}+1)=const$.

In \cite{usadel} these investigations have been extended to the
dynamical structure factor in velocity-space,
\begin{equation}
  S_v(k,\omega) \; = \; \frac{1}{N\,T} \, \left\langle 
    \left| \sum_{n,t} \, v_{n}(t) \, e^{i (k n -
    \omega t) } \right|^2 \right\rangle ,
\label{eq:struc_fact_v}
\end{equation}
with $k=2\pi m_k/N$, $\omega=2\pi m_\omega/T$, where $N$ is the
number of vehicles and $m_k$ and $m_\omega$ are integers. $v_{n}(t)$
is the velocity of the $n$-th vehicle at time $t$.

Compared to the dynamical structure factor in real space, 
(\ref{eq:struc_fact_v}) has the advantage that the free-flow regime
only contributes white noise, $S_v(k,\omega)|_{free-flow}=const$.
Therefore it is easier to study jamming properties. It is found
in \cite{usadel} that $S_v(k,\omega)$ exhibits one ridge with negative
slope, corresponding to backward moving jams.
One finds that the velocity of the jams is a function of the
randomization parameter $p$ only. It is independent of the density $c$
and the maximal velocity $v_{max}$ \cite{usadel}. This is consistent
with results from a direct study of the autocorrelation function
\cite{neublee}\footnote{Measurements of the jam dissolution speed in
\cite{krug99}, however, show a decrease with increasing $v_{max}$
and saturation for large $v_{max}$.}.
Above a transition density, an algebraic behavior 
$S_v(k, \omega ) \bigl|_{\omega / k= v_{\rm j}} \; \sim \; k^{-\gamma}$ 
of the structure factor is found. 
This has been interpreted as an indication 
of critical behavior in \cite{usadel}. However, due to the 
difficulties involved in the calculation of (\ref{eq:struc_fact_v})
only relatively short times $T\leq 2048$ have been considered in
\cite{usadel}. This is much smaller than the cutoff found in the 
lifetime of jams (see the discussion above) and lies well in the 
region where an algebraic decay is found. Therefore the results for 
the dynamical structure factor (\ref{eq:struc_fact_v}) and the 
lifetime measurements are consistent, but the algebraic decay is not 
to be interpreted as an evidence for the existence of a critical point 
in the NaSch model. In order to see the cutoff, times $T > 10^4$ 
would have to be considered.


\subsection{Exact solution of the NaSch model with $v_{max}=1$ and open
boundary conditions}
\label{sub_openBC}

The analytical methods presented in Section \ref{sub_analyt} are well 
suited for the investigation of translationally-invariant stationary 
states which are achieved by imposing periodic boundary conditions.
For both practical and theoretical reasons sometimes different
boundary conditions are preferable. Imagine a situation where a
multilane road is reduced to one lane, e.g.\ due to road construction.
Such a bottleneck can be modeled by using a NaSch model with
open boundaries. The multilane part of the road acts as a particle
reservoir. If the first cell of the one-lane part is empty a car
is inserted here with probability $\alpha$.  At the other end
a car is removed from the last cell with probability $\beta$
(see Fig.\ \ref{asepdef}). These boundary conditions break the
translational invariance and in general one can expect stationary
states with a non-trivial density profile $\langle \tau_j \rangle$.
From a more theoretical point of view such models have been studied
intensively as prototypes of systems exhibiting so-called 
boundary-induced phase transitions \cite{krug91,henkel}. In contrast to
what one expects from experience with equilibrium systems, 
one-dimensional driven nonequilibrium systems can exhibit phase 
transitions, even when the interactions are short-ranged, just by 
'slightly' changing the boundary conditions.

\begin{figure}[ht]
\centerline{\epsfig{figure=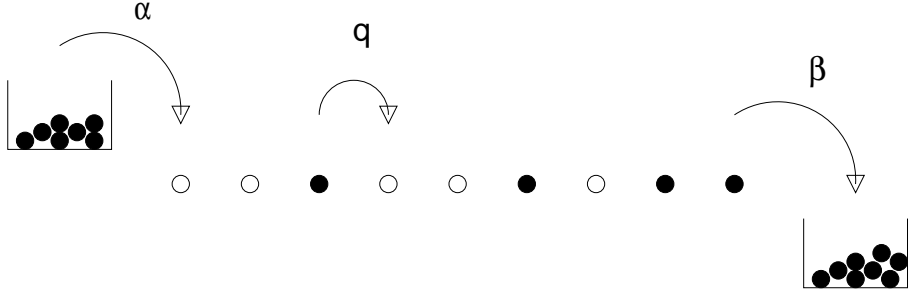,height=4cm}}
\caption{Definition of the ASEP with open boundary conditions.}
\label{asepdef}
\end{figure}

The probability distribution characterizing the steady state of 
the TASEP with parallel dynamics (i.e., NaSch model with 
$v_{max} = 1$) and open boundary conditions has been obtained 
recently in \cite{ERS} and \cite{degier} using
generalizations of techniques based on the {\em Matrix Product 
Ansatz (MPA)} (see Appendix \ref{App_MPA} for a more technical 
introduction). By varying the boundary rates $\alpha$ and 
$\beta$ one obtains a surprisingly rich phase diagram (see 
Fig.\ \ref{fig_asep}) which is qualitatively the same for all 
types of dynamics.  Three phases can be distinguished 
by the functional dependence of the current through the system on
the system parameters. In the low-density phase A ($\alpha < \beta,
\alpha_c(p)$) the current is independent of $\beta$. Here the current
is limited by the rate $\alpha$ which then dominates the
behaviour of the system. In the high-density phase B ($\beta < \alpha,
\beta_c(p)$) the behaviour is dominated by the output rate $\beta$ and 
the current is independent of $\alpha$. In the maximum current phase C
($\alpha>\alpha_c(p)$ and $\beta>\beta_c(p)$) the limiting factor for 
the current is the bulk rate\footnote{Note that conventionally
the hopping rate in the ASEP is denoted as $p$. Since in the NaSch
model $p$ is the braking probability the hopping rate in the ASEP (for
$v_{max}=1$) becomes $q=1-p$.} $q=1-p$. Here the current becomes 
independent of both $\alpha$ and $\beta$.\\

High- and low-density phase can be subdivided into two phases AI, AII
and BI and BII, respectively. These subphases can be distinguished by
the asymptotic behaviour of the density profiles at the boundaries.

\begin{figure}[t]
\centerline{\psfig{figure=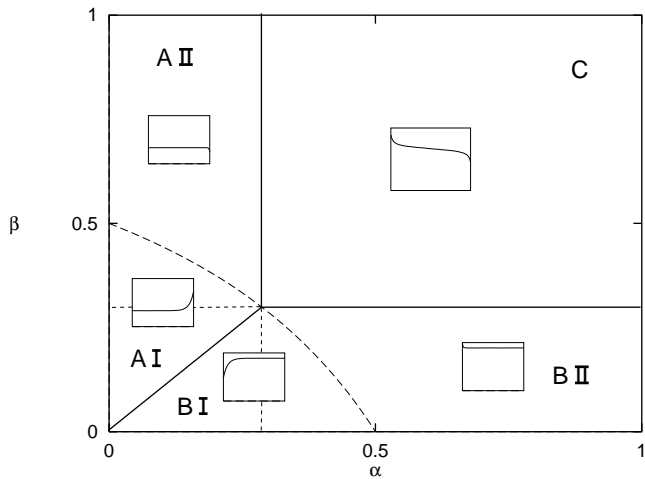,bbllx=75pt,bblly=40pt,bburx=560pt,bbury=410pt,height=7cm}}
\caption{Phase diagram of the ASEP for parallel dynamics. The inserts
show typical density profiles.}
\label{fig_asep}
\end{figure}

The transitions between the phases can be characterized by the
behaviour of two correlation lengths $\xi_\alpha$ and $\xi_\beta$
which only depend on $p$ and $\alpha$ or $\beta$. These lengths can be
obtained explicitly from the exact solution. Apart from $\xi_\alpha$
and $\xi_\beta$ also a third length $\xi^{-1}=|\xi_\alpha^{-1}
-\xi_\beta^{-1}|$ plays an important role.

The transition from AII (BII) to C is continuous with diverging
correlation length $\xi_\alpha$ ($\xi_\beta$). The transition from the
high- to the low-density phase is of first order. Here both
$\xi_\alpha$ and $\xi_\beta$ are finite, but $\xi$ diverges. On the
transition line one finds a linear density profile created by the
diffusion of a domain wall between a low-density region at the left
end of the chain and a high-density region at the right end.

For the case of parallel dynamics, i.e.\ the NaSch model with 
$v_{max}=1$, the currents in the three phases are given by
\cite{ERS,degier,rsss}
\begin{equation}
J_A=\frac{\alpha(q-\alpha)}{q-\alpha^2},\quad
J_B=\frac{\beta(q-\beta)}{q-\beta^2},\quad
J_C=\frac{1}{2}\left(1-\sqrt{p}\right).
\label{ASEPcurrent}
\end{equation}
The corresponding bulk densities\footnote{In the maximum current phase
no real bulk density can be defined due to the algebraic behaviour
of the density profile. $c_C$ is therefore just $\langle \tau_{L/2}
\rangle$.} are 
\begin{equation}
c_A=\frac{\alpha(1-\alpha)}{q-\alpha^2},\quad
c_B=\frac{q-\beta}{q-\beta^2},\quad
c_C=\frac{1}{2}.
\end{equation}
The phase boundaries are determined by the critical rates
\begin{equation}
\alpha_c(p)=\beta_c(p)=1-\sqrt{p}.
\label{alphacdef}
\end{equation}
In Fig.\ \ref{fig_asep} also the special line $(1-\alpha)(1-\beta)=p$
is indicated. Here the density profile is flat (i.e.\ constant). On
this line the exact solution can be obtained by the 2-cluster approach
of Sec.\ \ref{Sec_cluster} \cite{rsss}. Since it goes through all three phase
these results are sufficient to obtain exact analytic expressions,
e.g.\ for the currents, once the structure of the phase diagram is
established (e.g.\ by Monte Carlo simulations).

The stationary state of the ASEP can also be obtained for other 
types of updates (see App.\ \ref{App_updates}), e.g.\ 
random-sequential \cite{derrida93,schdom}, 
ordered-sequential \cite{rsss,rss,brankov3} and 
sublattice-parallel update \cite{gs93,hinrich,honecker,rsss}.
One finds that the phase diagram has 
the same basic structure for all updates \cite{rsss}. The functional
dependence of the currents, density profiles etc.\ on the model
parameters differs, however. For the important case of random-sequential
updating (\ref{ASEPcurrent})-(\ref{alphacdef}) have to be replaced by
\cite{derrida93,schdom}
\begin{alignat}{3}
J_A&=q\alpha(1-\alpha),&\qquad J_B&=q\beta(1-\beta),&\qquad 
J_C&=\frac{q}{4},\\
\rho_A&=\alpha,&\qquad \rho_B&=1-\beta,&\qquad \rho_C&=\frac{1}{2},\\
\alpha_c(p)&=\beta_c(p)=\frac{q}{2}.&\qquad  &\qquad &\qquad &\
\end{alignat}
Results for other updates can be found in \cite{rsss}. For a discussion
of the calculation of diffision constants and shock profiles we
refer to the reviews \cite{derr_review,derrida19} and references therein.

In \cite{krug91} the behaviour of the ASEP for $\beta=1$ was explained
by postulating a {\em maximal-current principle}. According to this
principle, independent of the details of the dynamics, the system tries
to maximize the stationary current $J$:
\begin{equation}
J=\max_{c\in[0,c_-]} J(c)
\label{maxcurrprinc}
\end{equation}
Here $J(c)$ is the fundamental diagram (for periodic boundary
conditions) and $c_-$ is the density at the left (input) boundary, 
i.e.\ $c_-=\alpha$ in the case described above.

In \cite{kolomeisky} a nice physical picture has been developed which
explains the structure of the phase diagram not only qualitatively,
but also (at least partially) quantitatively.
It is determined by the dynamics of a domain walls\footnote{A somewhat
related approach has been used to obtain an approximate solution
for the special case of parallel dynamics with deterministic
bulk dynamics ($p=0$) \cite{tilstra}.}. In nonequilibrium 
systems, a domain wall is an object connecting two possible stationary
states. 
The notion of domain walls in the ASEP can be illustrated in the
limit $\alpha L \ll 1$ and $\beta L \ll 1$ of small boundary rates.
At late times there will be a low-density region at the
left end of the chain and a high-density region at the right end,
with a domain wall in between.
Schematically this state can be depicted as $000011111$. 
For general values of the rates the wall not be sharp in general, but
spread over a few lattice sites.

For late times the dynamics of the system can then be interpreted
in terms of the motion of the domain wall. A particle entering the
system leads moves the wall one cell to left, and a particle leaving
the system moves it one cell to the right. Therefore the domain wall
performs a biased random walk with drift velocity $v_D=\beta-\alpha$
and diffusion coefficient $D=(\alpha + \beta)/2$.
For $\alpha < \beta$ the domain wall moves to the right until it reaches
the end of the system which is thereafter in the low-density stationary
state. For $\alpha > \beta$ the wall moves to the left until it reaches
the left end and the systems goes into the high-density stationary
state. In the case $\alpha = \beta$ there is no net drift in the position 
of the wall. It fluctuates with its rms displacement increasing with time
as $(Dt)^{1/2}$, i.e.\ it can be anywhere in the system resulting in 
a linear density profile.

In order to understand the case of general $\alpha$ and $\beta$ one
has to introduce a second kind of domain wall separating a maximum
current phase from the high-density phase. Since the maximal possible
flow for periodic boundary conditions is $J_{max}=1/4$ (for $p=0$ and
random-sequential update)
the dynamics for $\alpha=1/2$ is dominated by the overfeeding at the
left boundary. The injection rate could support a current larger than
$1/4$, but in the bulk it can not exceed this value. Therefore
at the left boundary a maximum current state is formed. If the particles
are not extracted fast enough at the right boundary a high-density
region will develop there. These two regions are separated by a new
kind of domain wall, the maximum current/high-density domain wall.
Schematically it can be represented as mmmm1111. Again this domain
wall performs a biased random walk.

In order to obtain more quantitative predictions one goes to a 
coarse-grained picture. Then it is useful to replace the boundary rates
$\alpha$ and $\beta$ by particle reservoirs with densities $c_-$ and
$c_+$. The continuity equation $\partial c /\partial t +
\partial J/\partial x =0$ in the continuum limit has traveling wave
solutions of the form $c(x-v_Dt)$ with the domain wall velocity
\begin{equation}
v_D=\frac{J_+-J_-}{c_+-c_-}
\end{equation}
which can be obtained by integration over the chain.

For the low-density/high-density domain wall one 
has\footnote{We assume $p=0$ and random-sequential dynamics.}
$c_+=1-\beta$, $J_+=J(c_+)=\beta(1-\beta)$ and $c_-=\alpha$, 
$J_-=J(c_-)=\alpha(1-\alpha)$. This gives indeed $v_D=\beta-\alpha$ 
which should be valid for $\alpha,\beta < 1/2$.
For the maximum current/high-density domain wall $c_-$ takes the
value $c_-=1/2$ so that $J_-=1/4$ and thus $v_D=\beta-1/2$.

The arguments described above can be generalized to any
process where the fundamental diagram $J(c)$ of the periodic system 
has only one maximum at a density $c^*$. For all currents $J<J(c^*)$ 
there exist two corresponding densities $c_1$ and $c_2$ with
$J(c_1)=J=J(c_2)$. For fundamental diagrams with more than one
maximum, more than two densities might exist for a given current
$J$. This implies the existence of a larger number of domain wall
types. The phase diagram of the open system than exhibits a larger
number of phases \cite{popkov}.
The maximal-current principle (\ref{maxcurrprinc}) for the TASEP
with $\beta=1$ is generalized to the {\em extremal-current principle}
\cite{popkov,gs}
\begin{equation}
J=\begin{cases}
\max_{c\in[c_+,c_-]} J(c)    &\text{for\ $c_- > c_+$},\\ 
\min_{c\in[c_-,c_+]} J(c)    &\text{for\ $c_- < c_+$}.
\end{cases}
\end{equation}

Since the above phenomenological picture does not depend on the 
microscopic details of the dynamics, it plausible that the phase 
diagrams for different updates are qualitatively the same.
Boundary-induced phase transitions have recently been observed 
\cite{psss} in measurements on a German motorway \cite{Neubert99}. 
One finds a first-order
nonequilibrium phase transition between a free-flow and a congested
phase. This transition is induced by the interplay between density
waves induced by an on-ramp and a shock wave moving on the motorway
\cite{psss}.


\section{Generalizations and extensions of the NaSch model} 
\label{Sec_gener}

As stated earlier, the NaSch model is a minimal model. The first 
obvious possible generalization would be to replace the {\it 
acceleration} stage of updating rule (U1) to 
$$
v_n \rightarrow \min(v_n+a_n,v_{max}) \eqno ({\rm U}1') 
$$
where, $a_n$, acceleration assigned to the $n$-th vehicle, 
need not be unity and, in general, may depend on $n$.
In the following subsections we consider more non-trivial 
generalizations and extensions of the NaSch model. 

\subsection{Single-lane highways}

In the next few subsubsections we shall demonstrate the rich variety 
of traffic phenomena that can be observed by appropriate modifications 
of the random braking. We have earlier mentioned in the context of 
empirical results that traffic flow exhibits metastability and the 
related hysteresis effects. Such phenomena have been observed in 
continuum formulations of "microscopic" models, i.e., in coupled-map 
lattice models \cite{krauss}. However, the NaSch model is too simple 
to account for these phenomena. We now briefly describe a few 
generalizations of the NaSch model, each of which is based on 
modifications of the braking rules of the original NaSch model; one 
common feature of all of these generalized models is that they show 
metastability and hysteresis. 

\begin{figure}[ht]
 \centerline{\epsfig{file=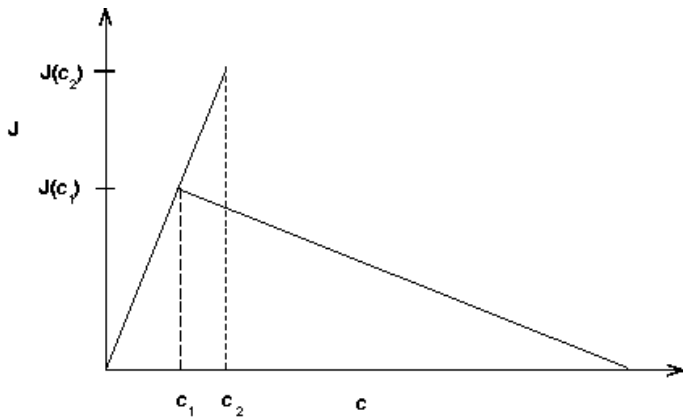,width=9cm}}
\caption{\protect{Metastability and hysteresis.}}
\label{hyssch}
\end{figure}

Before we begin our discussions on specific generalized versions 
of the NaSch model, which exhibit metastability, we make some 
general remarks. In the schematic stationary fundamental diagram 
of Fig.~\ref{hyssch}, the low density branch corresponds to 
homogeneous free-flow states, while the high density branch 
corresponds to configurations, where jammed states are present. 
Obviously, at densities $c_1 < c < c_2$, the flow depends 
non-uniquely on the global density. 

In order to establish the existence of meta-stable states 
one can follow two basic strategies. In the first method, 
the density of vehicles is changed adiabatically by adding or 
removing vehicles from the stationary state at a certain 
density. Starting in the jamming phase (large densities)
and removing vehicles, one obtains the lower branch of the 
hysteresis curve.  On the other hand, by adding vehicles to 
a free flowing state (low densities), one obtains the upper 
branch. This method is closely related to the experimental 
situation, where the occupancy of the road varies continuously.

The second method does not require changing the density. Instead 
one starts from two different {\it initial conditions}, the 
{\it mega-jam} and the {\it homogeneous} state. The mega-jam 
consists of one large, compact cluster of standing vehicles. 
In the homogeneous state, vehicles are distributed periodically 
with same constant gap between successive vehicles (with one 
larger gap for incommensurate densities). Then, for $c > c_1$ 
the homogeneous initialization leads to a free-flow state, while 
the mega-jam initialization leads to the jammed high-density 
states. 

\subsubsection{Cruise-control limit and self-organized criticality}
\label{sub_cruisesoc}

In the cruise-control limit of the NaSch model \cite{paczus}  
vehicles moving with their desired velocity $v_{max}$ are not 
subject to noise. This is exactly the effect of a cruise-control 
which automatically keeps the velocity constant at a desired 
value. In this model the acceleration, deceleration (due to 
other vehicles) and movement stages of updating are identical 
to those in the general case of the NaSch model; however, the 
randomization step is applied only to vehicles which have a 
velocity $v < v_{max}$ after step 2 of the update rule. We can 
express this more formally by recasting the {\it randomization} 
stage of the update rules in the NaSch model as follows:\\
$$
v_n \rightarrow \max(0, v-1)
$$ 
with probability 
\begin{equation}
p =
\begin{cases}
p_{v_{max}}   & \text{if $v = v_{max}$,}\\
  p & \text{if $v < v_{max}$,}
\end{cases}
\label{ccl_p}
\end{equation}
where $v$ is the velocity of the vehicle at the end of the step 2 
of the update rule, i.e., after deceleration due to blocking by 
other vehicles. In the original formulation of the NaSch model 
$p_{v_{max}} = p$. On the other hand, the cruise-control limit corresponds 
to $p_{v_{max}} \rightarrow 0, p \neq 0$. 

For $p_{v_{max}} \ll 1$, at sufficiently low densities, all the vehicles move 
deterministically with the velocity $v_{max}$; this deterministic 
motion is, however, interrupted by small perturbations at a 
vanishingly small rate. Consequently, the system gets enough 
time to relax back to the state corresponding to the deterministic 
algorithm before it is perturbed again. This effectively separates 
completely the time scales for perturbing the system and the 
response of the system. 

\begin{figure}[ht]
\centerline{\epsfig{file=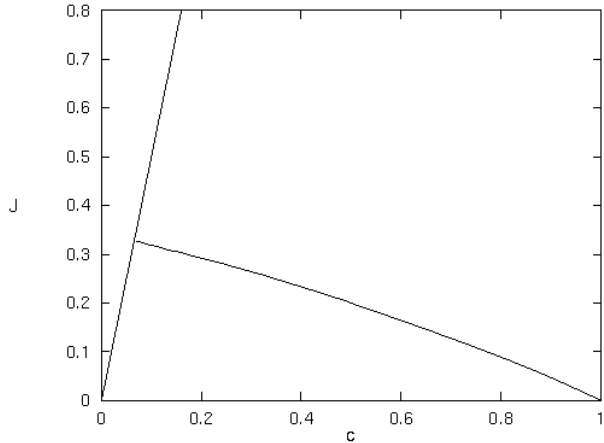,width=8cm}}
\caption{\protect{Fundamental diagram in the cruise-control limit of 
the NaSch model for $v_{max}=5$ and $p_{v_{max}}=0$, $p=0.5$.}} 
\label{ccfund}
\end{figure}

First, let us consider the periodic boundary conditions which 
is easier to treat than than the open boundary conditions. In 
this model, a sharp transition from the "free-flowing" dynamical 
phase to the "congested" phase takes place at a critical 
concentration $c_*(p,v_{max})$  which depends on $p$ as well 
as $v_{max}$ and, for all $p \neq 0$, $c_*(p,v_{max})$ is 
smaller than $c_m^{det} = 1/(v_{max}+1)$. For a given $v_{max}$, 
$c_*(p)$ increases with decreasing $p$ and, in the deterministic 
limit $p \rightarrow 0$, $c_*(0,v_{max}) \rightarrow c_m^{det}$.

In this model a jam is defined to consist of vehicles all of 
which have their instantaneous velocities smaller than $v_{max}$. 
For all $c < c_*$, jams present in the initial configuration 
eventually disappear and in the jam-free stationary state every 
vehicle moves with the velocity $v_{max}$. Therefore, in the 
density regime $c < c_*$ the flux increases linearly with 
density following $J = c v_{max}$, just like that in the 
deterministic limit $p = 0$ of the NaSch model (Fig.~\ref{ccfund}). 
But, unlike the deterministic limit $p = 0$, the cruise-control 
limit of the NaSch model exhibits metastability in the interval 
$c_* < c < c_m^{det}$. In this context, the metastability means 
that, in the interval $c_*< c < c_m^{det}$, on appropriate 
initialization, the system can reach apparently steady states 
where no jam appears and where the fluxes are higher than 
$J(c_*)$; but, perturbations of such a "metastable" state creates 
long-lived jams thereby reducing the flux to a level consistent 
with the stable branch of the fundamental diagram. At all $c > c_*$ 
jams present in the initial configuration never disappear 
completely and, in this density regime, the stable steady-state 
is a mixture of laminar flow regions and jams. The long-lived 
jams lower the flux beyond $c_*$ and the flux decreases linearly 
with density (Fig.~\ref{ccfund}).

Let us assume that at densities slightly above $c_*$, only one 
jam of length $L_{jam}$ containing $N_{jam}$ vehicles exists 
in the system. Then, because of the periodic boundary conditions, 
the total number of vehicles $N$ is conserved and, hence, 
\begin{equation}
N = c_{jam} L_{jam} + c_{out} (L - L_{jam}) 
\label{np1}
\end{equation}
where $c_{jam} = N_{jam}/L_{jam}$ and $c_{out} =
(N-N_{jam})/(L-L_{jam})$ are the densities of the vehicles in the jam
and in the outflow region, respectively. Dividing both sides of
(\ref{np1}) by $L$ we get
\begin{equation}
  c = c_{jam} \frac{L_{jam}}{L} + c_{out} \left(1 -
    \frac{L_{jam}}{L}\right)
\label{np2}
\end{equation}
Since in the cruise-control limit of the NaSch model $L_{jam}$ 
must vanish as $c \rightarrow c_*$, we conclude \cite{paczus} 
that we must have $c_{out} = c_*$, i.e., {\it the average density 
in the outflow region of a jam is equal to the critical density} 
$c_*$.

In order to study the traffic at the critical point of the 
cruise-control limit, Nagel and Paczuski \cite{paczus} used a 
a special boundary condition which enables the system to 
select automatically the state of maximum throughput, i.e, 
the system exhibits {\it self-organized criticality}. This 
special boundary condition consists of an infinite jam  from 
$- \infty$ to $0$ (i.e., at the left boundary) while the 
right boundary is open. Vehicles emerge from the infinite 
jam in a jerky way, before attaining the velocity $v_{max}$. 
Far away from the infinite jam all vehicles move with the 
same velocity $v_{max}$. In order to show that the state 
selected this way is "critical" \cite{paczus} we perturb a 
vehicle, far downstream from the infinite jam, slightly by 
reducing its velocity from $v_{max}$ to $v_{max} - 1$. This 
particular vehicle initiates a chain reaction and gives 
rise to a jam if the following vehicle is sufficiently close 
to it although it itself accelerates and, eventually, attains 
$v_{max}$. This phantom jam has a time-dependent size $n(t)$, 
measured by the number of vehicles $n$ in this jam at time
$t$ and it has a lifetime $\tau_{life}$. The statistics of 
this features of the phantom jam can be obtained by repeating 
the computer experiment sufficiently large number of times; 
sometimes the phantom jam created is small and has a short 
lifetime and sometimes it is large and has quite long lifetime. 
Interestingly, the characteristic quantities like, for example, 
the distributions of the sizes of the jams, lifetimes of the 
jams, etc. do, indeed, exhibit power-laws which are hall mark 
of the self-organized criticality \cite{bak,ddhar}. E.g. the
branching behaviour of the jams gives rise to intermittent
dynamics with a $1/f$ power law spectrum \cite{paczus}.
$1/f$ noise in real traffic has been discovered by Musha and
Higuchi \cite{musha}. They recorded transit times of vehicles
passing underneath a bridge. The corresponding power spectral
density of the flow fluctuations shows $1/f$ behaviour at 
low frequencies.

The exponents associated with the various power laws in the 
cruise-control limit of the NaSch model can be calculated 
analytically, at least for $v_{max} = 1$, by utilizing a 
formal relation with one-dimensional unbiased random walk.  
If $v_{max} = 1$, all the vehicles in the jams have velocity 
$v = 0$. Moreover, the jams are compact so that the number 
of vehicles in a jam is identical to its spatial extent. 
The probability distribution $P(n;t)$ for the number of 
vehicles $n$ in such a jam, at time $t$, is determined by 
the following equation:
\begin{equation}
P(n;t+1) = (1-r_{in}-r_{out}) P(n;t) + r_{in} P(n-1;t) 
+ r_{out} P(n+1;t) 
\label{soc1}
\end{equation}
where the phenomenological parameters $r_{in}$ and $r_{out}$ 
are the rates of incoming and outgoing vehicles. Of course, 
$r_{in}$ depends on the density of the vehicles behind the jam. 
For large $n$ and $t$, taking the continuum limit of the 
equation (\ref{soc1}) and expanding we get 
\begin{equation}
\frac{\partial P}{\partial t} = (r_{out}-r_{in}) 
\frac{\partial P}{\partial n} + \frac{r_{in}+r_{out}}{2} 
\frac{\partial ^2 P}{\partial n^2} 
\label{soc2}
\end{equation}
If $r_{in} > r_{out}$ the jams would grow for ever. On the other 
hand, the jams would shrink, and eventually disappear, if 
$r_{in} < r_{out}$.  If $r_{in} = r_{out}$, the first term on the 
right hand side of the equation (\ref{soc2}) vanishes and the 
resulting equation governing the time evolution of $P(n;t)$ is 
identical to that of the probability of finding, at time $t$, an 
unbiased one-dimensional random walker at a distance $n$ which 
was initially at the origin. Thus, when $r_{in} = r_{out}$, 
the jams exhibit large ("critical") fluctuations which can be 
characterized by critical exponents. Using this formal mapping 
onto unbiased random walk, we find 
(a) that the mean size of jam at time $t$ corresponds to the 
mean displacement of the random walker from the origin after 
time interval $t$ and (b) that the lifetime of a jam corresponds 
to the time taken by the random walker to return to the origin 
for the first time. Hence, using the well known results from the 
theory of random walks \cite{feller,weiss}, we get 
\begin{equation}
n(t) \propto t^{1/2}, \qquad\text{and}\qquad
P(\tau_{life}) \propto \tau_{life}^{-3/2} .
\end{equation}
It turns out that the power-law exhibited by the size of the 
jams, the distributions of the lifetimes, etc. are not restricted 
merely to the special case $v_{max} = 1$ of the cruise-control 
limit but is also shown by the corresponding computer simulation 
data also for arbitrary $v_{max}$. The power-law distributions of 
$P(\tau_{life})$ in the cruise-control limit of the NaSch model 
is in sharp contrast with the exponential distribution observed 
in the NaSch model \cite{kn94}. Thus, in the cruise-control 
limit of the NaSch model, the large jams are fractal \cite{mandel} 
in the sense that there are smaller sub-jams inside larger jams, 
ad infinitum. In other words, in between sub-jams, there are 
holes of all sizes. 

\subsubsection{Slow-to-start rules, metastability and hysteresis} 
\label{Sec_s2s}

The slow-to-start rules can lead not only to metastability and, 
consequently, hysteresis, but also to phase separated states 
at high densities, as we now show. 

\noindent ${\bullet}$ Takayasu-Takayasu slow-to-start rule

Takayasu and Takayasu (TT) \cite{tt} were the first to suggest a CA
model with a slow-to-start rule. Here, we investigate the
generalization, as suggested in \cite{sss2s}, of the original
slow-to-start rule.  According to this generalized version, a standing
vehicle (i.e., a vehicle with the instantaneous velocity $v = 0$) with
exactly one empty cell in front accelerates with probability $q_t =
1-p_t$, while all other vehicles accelerate deterministically. The
other steps of the update rule (U2-U4) of the NaSch model are kept
unchanged.

As in the case of the NaSch model, it is instructive to consider first
the deterministic limits of the TT model \cite{tt,fis2s}. The TT model
reduces to the NaSch model in the limit $p_t = 0$. What happens in
the other deterministic limit, namely, $p_t = 1$? In the latter
deterministic limit, a stopped vehicle can move only if there are at
least two empty cells in front \cite{tt}. Obviously, completely blocked
states exist for densities $c \geq 0.5$, where the number of empty
cells in front of each vehicle is smaller than two. However, in the
region $0.5\le c \lesssim 0.66$ the number of blocked configurations
is very small compared to the total number of configurations and
states with a finite flow exist. Precisely at $c = 0.5$, there are
only two blocked states and the time to reach these states diverges
exponentially with the system size.

The fundamental diagram for the TT model with $v_{max} = 1$ has been
derived analytically by carrying out (approximate) 2-cluster
calculations in the site-oriented approach \cite{sss2s}. But, the
fundamental diagrams of the TT model for all $v_{max} > 1$ have been
obtained so far only numerically by carrying out computer simulations
(see Fig.~\ref{taka_fund5}).

\begin{figure}[ht]
\centerline{\psfig{figure=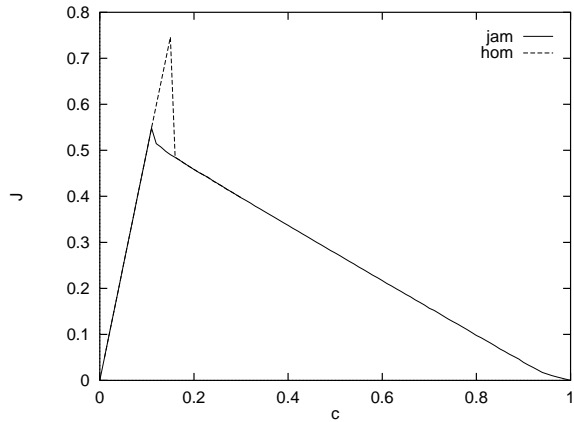,bbllx=50pt,bblly=50pt,bburx=550pt,bbury=400pt,height=5.5cm}}
\caption{\protect{The fundamental diagram in the TT model ($v_{max} = 5$,
$p = 0.01, p_t = 0.75, L = 10000$) obtained using two different initial 
conditions, namely, a completely jammed state (jam) and a homogeneous 
state (hom) and averaging over 100000 sweeps through the lattice. }}
\label{taka_fund5}
\end{figure}

Comparing these fundamental diagrams with the corresponding ones for
the NaSch model ($p_t = 0$), we find the following effects of the TT
slow-to-start rule: (i) for a given density $c$, the flux $J(c)$ is
smaller in the TT model as compared to that in the NaSch model; (ii)
the particle-hole symmetry is not exhibited by the TT model for any
$v_{max}$ (not even for $v_{max} = 1$) and (iii) the TT model exhibits
metastability and hysteresis which are absent in the NaSch model. Note
that the mechanism for meta-stability in the case $p_t=1$ is
different from that for the metastability observed for $0 < p_t < 1$.

Because of the slow-to-start rules, the separations between the
vehicles coming out of a jam are larger than those between the
vehicles coming out a jam in the NaSch model. Since the density far
downstream is smaller than the density of maximum flow, the vehicles
can propagate freely in the low density regions of the lattice where
spontaneous formation of jams is highly unlikely, if the parameter $p$
is sufficiently small.  Therefore, the phase-separated steady-states
at high global densities consist of a macroscopic jam and a
macroscopic free-flow regime both of which coexist simultaneously
(Fig.~\ref{xttt}).

\begin{figure}[ht]
\centerline{\epsfig{figure=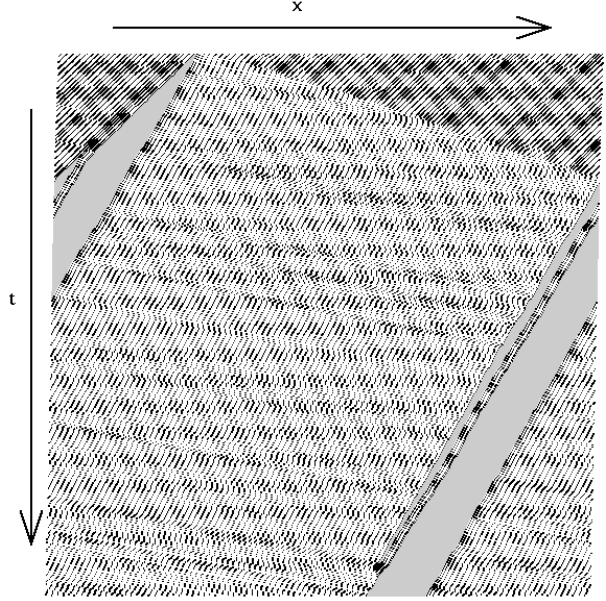,width=8cm}}
\caption{\protect{Typical space-time diagram of the TT model
with $v_{max} = 5$ and $c = 0.20$, $p = 0.01$ and $p_t = 0.75$.
Each horizontal row of dots represents the instantaneous positions of
the vehicles moving towards right while the successive rows of dots
represent the positions of the same vehicles at the successive time 
steps.}}
\label{xttt}
\end{figure}

\noindent${\bullet}$ The BJH model of slow-to-start rule

Benjamin, Johnson and Hui (BJH) \cite{bjh} modified the updating rules
of the NaSch model by introducing an extra step where their
"slow-to-start" rule is implemented; this slow-to-start rule is
different from that introduced by TT \cite{tt}. According the BJH
"slow-to-start" rule , the vehicles which had to brake due to the next
vehicle ahead will move on the next opportunity only with probability
$1-p_s$. The steps of the update rules can be stated as follows:\\ 

\noindent {\it Step 1: Acceleration.} 
$v_n \rightarrow \min(v_n+1,v_{max})$.

\noindent{\it Step 2: Slow-to-start rule}:
If $flag = 1$, then $v_n \rightarrow 0$ with probability $p_s$.

\noindent{\it Step 3: Blockage (due to other vehicles).}
$v_n \rightarrow \min(v_n,d_n-1)$ and, then,\\ ~~ $flag = 1$ if $v_n = 0$,
else $flag = 0$.

\noindent{\it Step 4: Randomization.} $v_n \rightarrow \max(v_n-1,0)$
with probability $p$.

\noindent{\it Step 5: Vehicle movement.}
$x_n \rightarrow  x_n + v_n$.\\

Here $flag$ is a label distinguishing vehicles which have to obey the
slow-to-start rule ($flag = 1$) from those which do not have to
($flag = 0$).

Obviously, for $p_s=0$ the above rules reduce to those of the NaSch
model.  The slow-to-start rule of the TT model is a `spatial' rule. In
contrast, the BJH slow-to-start rule requires `memory', i.e.\ it is a
`temporal' rule depending on the number of trials and not on the free
space available in front of the vehicle. The fundamental diagram of
the BJH model (Fig.~\ref{bjh_fund5}) clearly shows the existence of
metastable states and, therefore, expected to exhibit hysteresis
effects \cite{chow99}.  But, in the special case of $v_{max} =1$, for 
which approximate analytical calculations can be carried out \cite{sss2s},
no meta-stable states exist.

\begin{figure}[ht]
\centerline{\psfig{figure=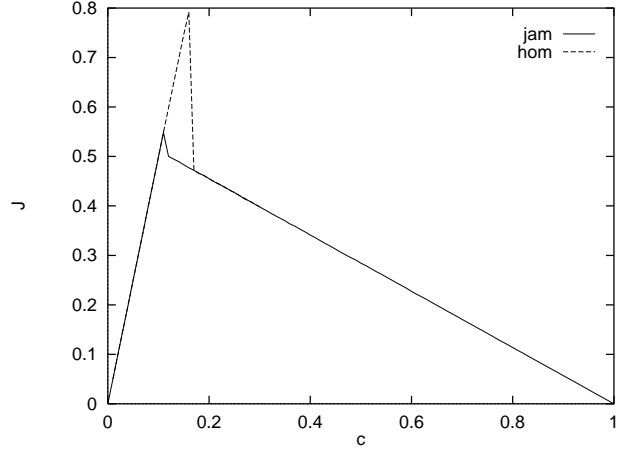,bbllx=50pt,bblly=50pt,bburx=555pt,bbury=405pt,height=6cm}}
\caption{\protect{The fundamental diagram in the BJH model 
($v_{max} = 5$, $p = 0.01, p_s = 0.75$) obtained using two different 
initial conditions, namely, a completely jammed state (jam) and a 
homogeneous state (hom). }}
\label{bjh_fund5}
\end{figure}

Since for all $v_{max} > 1$ in the BJH model, just as in the TT model,
the outflow from a jam is smaller than the maximal flow, the
phase-separated steady-states at high global densities consist of a
macroscopic jam and a macroscopic free-flow regime both of which
coexist simultaneously (Fig.~\ref{xtbjh}) \cite{chow99}.

\begin{figure}[ht]
 \centerline{\psfig{figure=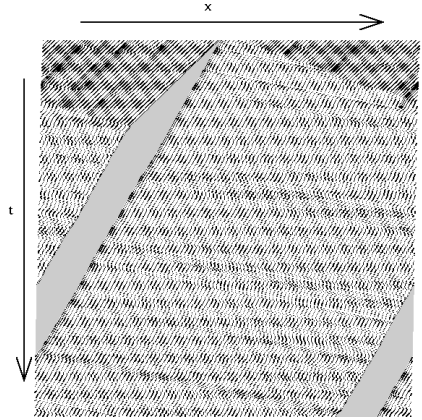,height=6cm}}
\caption{\protect{Typical space-time diagram of the BJH model
with $v_{max} = 5$ and $c = 0.20$, $p = 0.01$ and $p_s = 0.75$.
Each horizontal row of dots represents the instantaneous positions of
the vehicles moving towards right while the successive rows of dots
represent the positions of the same vehicles at the successive time steps. }}
\label{xtbjh}
\end{figure}

However, the macroscopic jam is not compact.  The typical size of the
macroscopic free-flow regime can be estimated by measuring the
distribution of the gaps between the successive jams \cite{chow99}.  
A peak occurs in this distribution for headways of the order of the 
system size (see the inset of the right part of Fig.~\ref{BJH_dist}).  
The position of the peak indicates the typical size of the macroscopic 
free-flow regime.

\begin{figure}[!h]
\centerline{\epsfig{figure=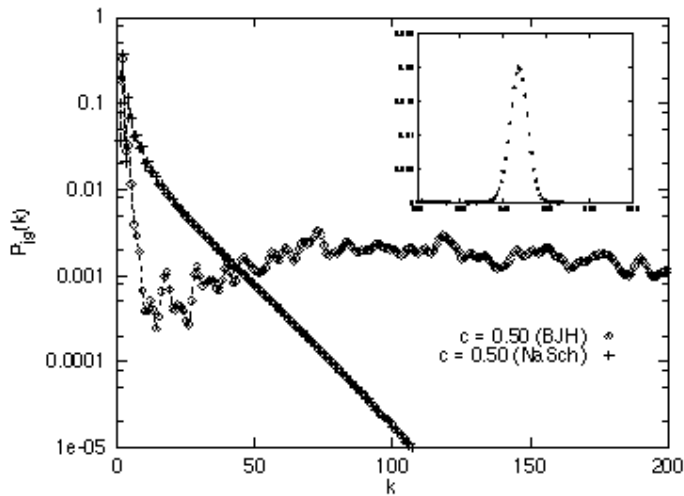,width=10cm}}
  \caption{Distribution of gaps between jams in the BJH and the NaSch 
model for $v_{max}=5$, $p=0.05$, $p_s=0.5$ and density $c=0.5$.}
\label{BJH_dist}
\end{figure}

\noindent${\bullet}$ The NaSch model with a velocity-dependent
slow-to-start rule

Although the NaSch model does not exhibit metastable states and
hysteresis, a simple generalization exists which is able to 
reproduce these effects. It is the so-called 
Velocity-Dependent-Randomization (VDR) model \cite{barlovic}.
Here, in contrast to the original NaSch model, the randomization
parameter depends on the velocity of the vehicle, $p=p(v)$.
The rules (see Sec.~\ref{sec_NaSch}) are supplemented by a new
rule,

\noindent {\it Step 0: Determination of the randomization parameter.} 
The randomization parameter used in step 3 for the $n$-th vehicle 
is given by $p=p(v_n(t))$.

This new step has to be carried out before the acceleration step 1.
The randomization parameter used in step 3 depends on the velocity
$v_n(t)$ of the $n-$th vehicle after the previous timestep.
In order to implement a simple slow-to-start rule one chooses
\cite{barlovic}
\begin{equation}
p(v)=\begin{cases}
p_{0}   & \text{for\ $v = 0$,}\\
p       & \text{for\ $v > 0$,}
\end{cases}
\end{equation}
with $p_0>p$. This means that vehicles which have been standing in
the previous timestep have a higher probability $p_0$ of braking
in the randomization step than moving vehicles.

The rules of the VDR model can be recast in a form similar to those
of the BJH model. We define a label $flag$ which distinguishes between 
vehicles which have to obey the slow-to-start rule ($flag = 1$) from 
those which do not have to ($flag = 0$). $flag = 1$ if $v_n = 0$ at 
the beginning of a time step, else $flag = 0$.
Explicitly, the update rules are as follows:\\ 

\noindent {\it Step 1: Acceleration.} 
$v_n \rightarrow \min(v_n+1,v_{max})$.

\noindent{\it Step 2: Blockage (due to other vehicles).}
$v_n \rightarrow \min(v_n,d_n-1)$,

\noindent{\it Step 3: Randomization.} 
$v_n \rightarrow \max(v_n-1,0)$ with probability $p_0$ if $flag = 1$.
else, $v_n \rightarrow \max(v_n-1,0)$ with probability $p$.

\noindent{\it Step 4: Vehicle movement.}
$x_n \rightarrow x_n + v_n$.\\ 

Let us compare this VDR model with the cruise-control limit of the
NaSch model. The vehicles with velocity $v = v_{max}$ (at the end of
the step 2) are treated deterministically in the cruise-control limit
whereas in the VDR model velocities of all those with the velocity 
$v > 0$ (just before the step 3) are updated stochastically, but 
using different values of the braking parameter. 

Typical fundamental diagrams look like the one shown in 
Fig.~\ref{vdr_fund} where, over a certain interval of $c$, $J(c)$ can
take one of the two values depending on the initial state and, 
therefore, exhibit metastability.

\begin{figure}[ht]
\centerline{\psfig{figure=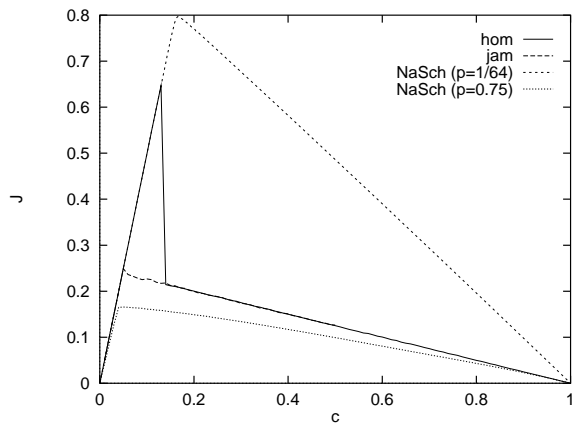,bbllx=50pt,bblly=50pt,bburx=550pt,bbury=400pt,height=5.5cm}}
\caption{\protect{The fundamental diagram in the NaSch model 
with a velocity-dependent slow-to-start rule ($v_{max} = 5$, 
$p_0 = 0.75, p = 1/64$) obtained using two different initial conditions, 
namely, a completely jammed state (jam) and a homogeneous state (hom). }}
\label{vdr_fund}
\end{figure}

Moreover, typical space-time diagrams of the VDR model (see
Fig.\ref{xtvdr}) clearly demonstrate that metastable homogeneous
states have a lifetime after which their decay leads to a phase
separated steady state. The microscopic structure of these
phase-separated high-density states is qualitative similar to those
observed in the high-density regimes of the TT and BJH models but
differs drastically from those found in the NaSch model.

\begin{figure}[ht]
 \centerline{\psfig{figure=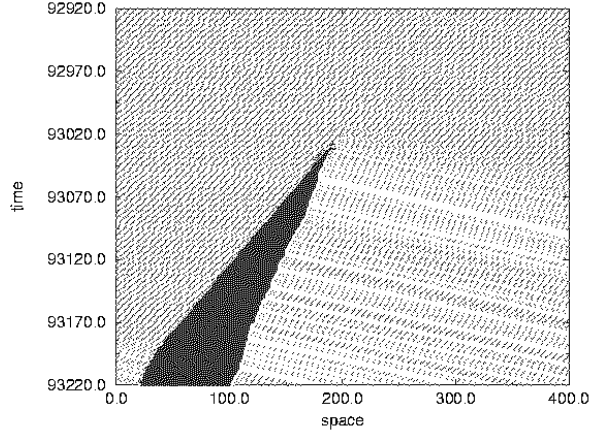,height=6cm}}
\caption{\protect{Typical space-time diagram of the VDR model
    with $v_{max} = 5$ and $c = 0.20$, $p = 0.01$ and $p_0 = 0.75$.
    Each horizontal row of dots represents the instantaneous positions
    of the vehicles moving towards right while the successive rows of
    dots represent the positions of the same vehicles at the
    successive time steps. }}
\label{xtvdr}
\end{figure}

It is instructive to compare the fundamental diagram of the VDR model
with those of the corresponding NaSch models. We now present a simple
derivation of the fundamental diagram of the VDR model on the basis of
heuristic arguments utilizing the observed structures of the
steady-states. For small densities $c \ll 1$ there are no slow
vehicles in the VDR model since interactions between vehicles are
extremely rare. In this regime every vehicle can move with the
free-flow velocity $v_{f} = (1-p) v_{max} + p (v_{max} - 1) = v_{max}
- p$ and, therefore, the flux is given by
\begin{equation}
  \label{jhom}
  J_{hom}(c) = c(v_{max}-p)
\end{equation}

which is identical to the NaSch model with randomization $p$.  On the
other hand, for densities close to $c=1$, the vehicles are likely to
have velocities $v=0$ or $v=1$ only and, therefore, the random braking
is dominated by $p_0$, rather than $p$, while the flow is determined
by the movement of the holes.  Hence, for large densities, i.e., $1-c
\ll 1$, the flow is given by $J(c) \approx (1-p_0)(1-c)$ which
corresponds to the NaSch model with randomization $p_0$. This
expression for flux in the high-density regime can also be derived as
follows.  In the phase-separated state the vehicles are expected to
move with the velocity $v_f = v_{max} - p$ in the free-flow region.
Neglecting interactions between vehicles in the free-flowing region
(which is justified because of the corresponding low density), the
average distance of two consecutive vehicles in the free-flow region
is given by $\Delta x = c_f^{-1} = T_w v_{f}+1$ where the average
waiting time $T_w$ of the first vehicle at the head of the megajam is
given by $T_w = \frac{1}{1-p_0}$.  In other words, the density in the
free-flow regime $c_f$ is determined by the average waiting time $T_w$
and $v_f$. Now suppose that $N_J$ and $N_F$ are the number of vehicles
in the megajam and free-flowing regions, respectively. Using the
normalization $L = N_J + N_F \Delta x$ we find that for the density $c
= \frac{N_F+N_J}{L}$, the flux $J_{sep}(c)$ is given by $J_{sep}(c) =
\frac{N_F}{L}(v_{max} - p)$ and, hence,
\begin{equation}
  \label{jjam}
  J_{sep}(c) = (1-p_0)(1-c).
\end{equation}

Obviously, $c_f$ is precisely the lower branching density $c_1$,
because for densities below $c_f$ the jam-length is zero.  It should
be noted that the heuristic arguments presented above remain valid for
$p_0 \gg p$ and $v_{max} >1$. The condition $p\ll 1$ guarantees that
the jams are compact in that limit.  In the case $v_{max}=1$, vehicles
can stop spontaneously, even in the free-flow regime and these vehicles
might initiate a jam.  This is the basic reason why hysteresis is
usually not observed for $v_{max}=1$.

\begin{figure}[ht]
\centerline{\epsfig{figure=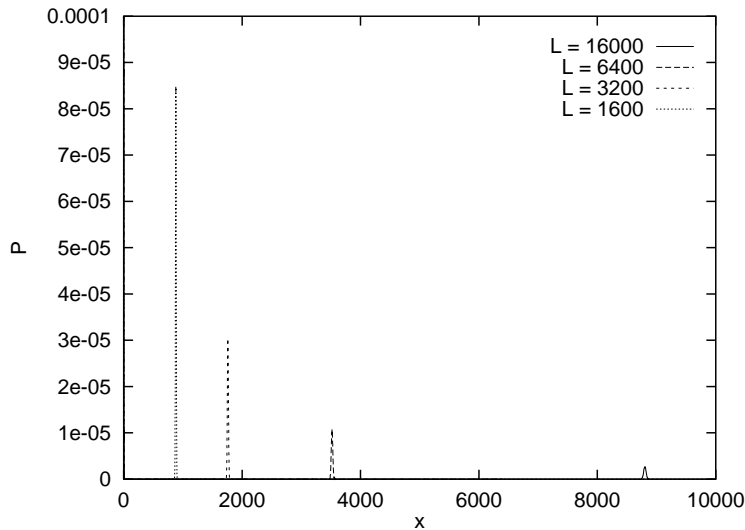,width=10cm}}
  \caption{\protect{Peaks of the jam-gap distribution at macroscopic
      distances.}}
\label{peak}
\end{figure}

Analogous to the BJH model phase separation can be directly
identified using the results of the jam-gap distribution.
Fig.~\ref{peak} shows that the size of the free-flow regime 
is proportional to the system size.

The results for the slow-to-start models discussed above have been
obtained by computer simulations of periodic systems of finite
length. It was shown that the fundamental diagram which is sketched 
in Fig.~\ref{hyssch} is generic for all models under consideration. 
Now it is self-evident to ask what kind of stationary states are 
realised in the thermodynamic limit $L \to \infty$. The simulation 
results indicate that  $\Delta c = c_1 - c_2$ decreases with 
larger system sizes and is expected to vanish
for $L\to \infty$, i.e. the jammed branch is stable in that limit. This
is readily understood if one analyses the typical configurations
which lead to an emerging jam or, vice versa, the
mechanism of the dissolution of a jam. Jams emerge if overreactions of
drivers lead to a chain reaction. This is possible in dense regions of
the free-flow state where the gap between the vehicles is not larger than
$v_{max}$. Obviously the probability to find large platoons of 
vehicles driving with small spatial headways increases with the 
system size (for fixed density).
In addition to that the jammed states are  phase separated, i.e. the size 
of the jam is of the order of the system size. During a simulation run 
the size of the jam fluctuates due to the stochastic movement and 
acceleration of the vehicles. Jams can
dissolve if the amplitude of these fluctuations are of the order of
the length of the jam, which is impossible in the thermodynamic limit.

Therefore the non-unique behaviour of the fundamental
diagram is only observable if finite system sizes are considered or if
the vehicles move deterministically in the free flow regime. Nevertheless
the results discussed above are highly relevant for practical
purposes, because the hysteresis effects have been observed at
realistic system sizes (e.g. $L=10000$ corresponds to a highway of
length $75~km$).

\subsubsection{Flow-optimization and meta-stable states}

Hysteresis effects and meta-stable states are not only of theoretical
interest, but also have interesting applications. From the previous
discussion of the slow-to-start models it is evident, that one can
optimize the maximum flow, if the homogeneous state is stabilized by
controlling the density so that it never exceeds $c_2$. This strategy
was followed in minimizing frequent jams in the Lincoln- and the
Holland-Tunnels in New York. Before the traffic lights were installed
at the entrance of the tunnels jams used to form spontaneously within
the tunnel because (a) the vehicle density used to be sufficiently
high and (b) the drivers used to drive more carefully inside the
tunnel thereby giving rise to stronger fluctuations which caused the
jams. But, the traffic lights installed at the entrance of the tunnels
do not allow the density to exceed $c_2$ and, consequently, jams are
not formed spontaneously by the decay of any metastable high-density
state.

\begin{figure}[ht]
\begin{center}
\epsfxsize=\columnwidth\epsfbox{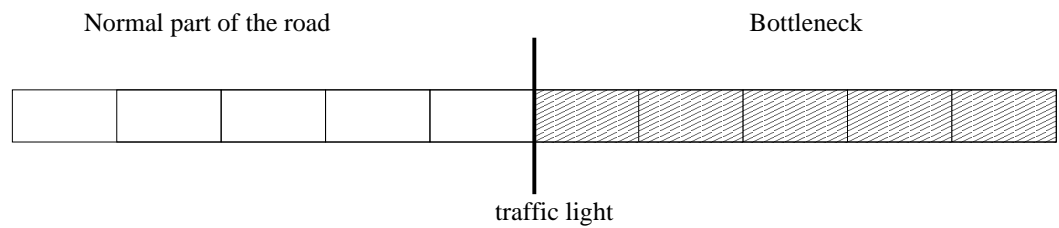}
\caption{A schematic representation of a tunnel on a highway. 
Larger values of the braking parameters are used for the right part, 
which mimics the tunnel.} 
\label{lincoln}
\end{center}
\end{figure}

One can mimic the situation of Lincoln and Holland-tunnels within the
framework of the CA models in the following way
\cite{barlovic,bsss,santen}.  The tunnel is considered as part of the
road, where the braking probabilities 
$p^t,p^t_0$ are higher compared to the remaining part of the
lattice ($p,p_0$, see Fig.~\ref{lincoln}).
Therefore, if one allows for an uncontrolled
inflow of the vehicles, jams typically appear inside the ``tunnel'' and
the system capacity is governed by $p^t_0$.

\begin{figure}[h]
\begin{center}
\centerline{\epsfig{file=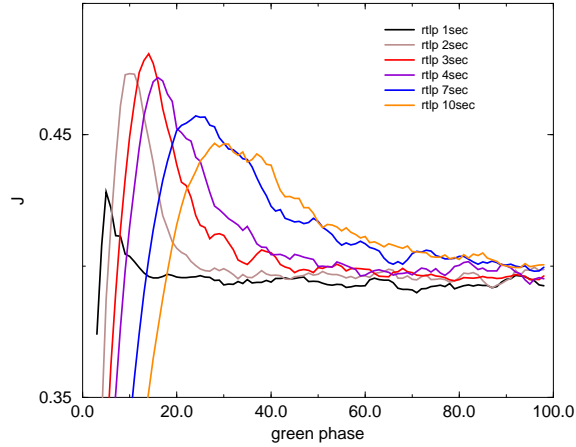,height=7cm}}
\caption{Average flow for different combinations of red-/green-signal
periods for a system of length $L=1000$ and $v_{max}=5$. Inside the
tunnel $p^t=0.15$, $p_0^t=0.60$ and outside $p=0.01$, $p_0=0.20$
has been used.} 
\label{lincolncurrent}
\end{center}
\end{figure}

The situation differs drastically if traffic lights are implemented
\cite{barlovic,bsss,santen}. As shown in Fig.~\ref{lincolncurrent}, 
a considerable increase of the maximum capacity can be achieved for 
an optimal combination of the red-/green-signal periods. The gain of  
capacity obtained for the optimal intervals of the signal is of the 
same order as for the realistic examples \cite{greenbergdaou}.


\subsection{Multi-lane highways}
\label{sub_multi}

For a realistic description of traffic on highways the idealized
single-lane models must be generalized to develop CA models of
multi-lane traffic; the main ingredient required for this
generalization being the {\it lane-changing rules}. Several attempts
have been made so far in this direction
\cite{nagatani96,rickert,wagner97,chow97b,nagel98,santen99,zhanghu,awazu2}.  
The lane changing rules for two-lane traffic can be symmetric or
asymmetric
with respect to the lanes. 
Similarly, if there are two 
(or more) different types of vehicles (say, cars and trucks) with
two different $v_{max}$, the lane-changing rule can be symmetric or
asymmetric with respect to the vehicles.

In general, the update in the two-lane models is divided into two
sub-steps: in one sub-step, the vehicles may change lanes in parallel
following the lane-changing rules and in the other sub-step each
vehicle may move forward effectively as in the single-lane NaSch
model. Drivers must find some incentive in changing the lane. Two
obvious incentives are (a) the situation on the other lane is more
convenient for driving, and (b) the need to make a turn in near
future. Two general prerequisites have to be fulfilled in order to
initiate a lane change: first, there must be an incentive and second,
the safety rules must be fulfilled \cite{sparmann}. Lane changing rules 
according to this scheme have been introduced by Rickert et al. 
\cite{rickert}. They suggested that vehicles are allowed to change 
the lane if the following four criteria are satisfied:

\begin{description}
\item{({\bf C1})\ \ \ } $gap(i) < l$,
\item{({\bf C2})\ \ \ } $gap_o(i) > l_o$,
\item{({\bf C3})\ \ \ } $gap_{o,back}(i) > l_{o,back}$, 
\item{({\bf C4})\ \ \ } $rand() < p_{c}$.
\end{description}

Here $gap(i)$ and $gap_o(i)$ are the gaps in front of vehicle $i$ on the
own lane and the other lane\footnote{The gap on the other lane is defined
in the same way as the gap on the own lane by imagining that the
vehicle occupies the site parallel to its current position.},
respectively. $gap_{o,back}(i)$ is the gap on the other lane to the
next vehicle {\em behind}. $l$, $l_{o}$, $l_{o,back}$ and $p_{c}$
are parameters specifying the rule and $rand()$ is a random number
in the interval $[0,1]$.

The first rule {\bf C1} represents the incentive criterion, i.e. if the
gap $gap(i)$ in front of the vehicle is not sufficiently large vehicles want 
to change the lane. Typical choices of the parameter $l$ are given by 
$l=\min(v+1,v_{max})$. This choice of the minimal headway ensures that
vehicles driving in a slow platoon try to change the lane if
possible. In the next rule {\bf C2} it is checked if the situation
on the other lane is indeed more convenient. This motivates the choice
$l=l_o$. The third rule {\bf C3} avoids too small distances 
to following vehicles on the other lane. Rickert and coworkers suggested
$l_{o,back} = v_{max}$. It is also important to perform lane changing
stochastically. Even if the incentive and safety criteria are fulfilled
a lane change is performed only with probability $p_c$ ({\bf C4}). 
This avoids, at least partially, so called ping-pong lane
changes, i.e. multiple lane-changes of vehicles in consecutive
timesteps\footnote{This artifact of the parallel update was
already pointed out by Nagatani \cite{nagatani96}, who simulated a
two-lane system with $v_{max} =1 $.}. 
Already 
implementations of the NaSch model using the basic lane-changing
rules revealed quite realistic results.
Nevertheless several variants of the basic rules
have been developed in order to improve the realism. A large number of
lane changing rules considered in the literature have been tabulated and 
compared by Nagel et al.~\cite{nagel98} (see Fig.~\ref{multlane_table}).

\begin{figure}[ht]
\begin{center}
\epsfig{file=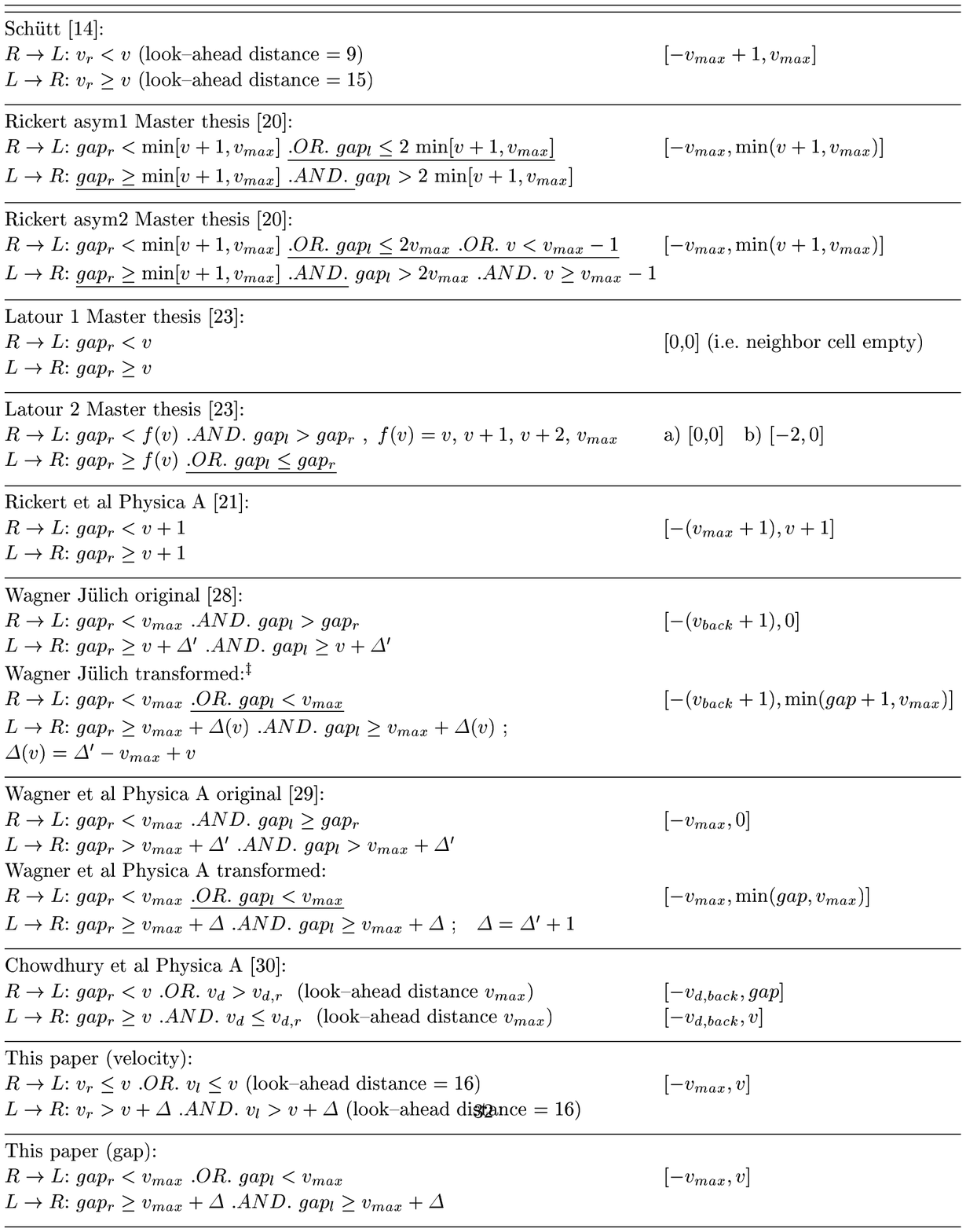,width=10cm}
\caption{The table gives an overview over different choices of the lane
changing rules discussed in the literature. The numbers of the
references correspond to those in \cite{nagel98}. The first column shows
different incentive criteria and the second column denotes the
corresponding minimal gap on the other lane. In general the incentive
criteria can be different for a change from left to right ($L\to R$ )
and ($R \to L$) (from \cite{nagel98}).} 
\label{multlane_table}
\end{center}
\end{figure}

The lane changing rules for two-lane traffic can be symmetric or 
asymmetric with respect to the lanes \cite{rickert}. If symmetric lane
changing rules are applied the rules  do not depend
on the direction of the lane changing maneuver. In contrast also
asymmetric lane changing rules have been considered. Lane changing
rules can be asymmetric in two ways. First it is possible that it is
preferred to drive on the right lane at low densities. This behavior
can be implemented simply by leaving out the first rule for a change
from the left to the right lane. Second it is also possible that
it is even forbidden to overtake a vehicle on the right lane, e.g. on
german highways. Then 
the single lane dynamics on the right lane
depends on the configuration on the left lane. These examples show
the flexibility of the CA 
approaches. Moreover the simulations also
show that the details of the lane changing rules may lead to
considerable changes of the model results \cite{wagner97,nagel98}.

In multi-lane traffic it is of particular interest to investigate
systems with different types of vehicles. For CA models this has been
done first by Chowdhury et al. \cite{chow97b} who simulated a periodic 
two-lane system with slow and fast vehicles, i.e. vehicles with different
$v_{max}$. The simulation results have been shown that already for
small densities the fast vehicles take on the average the free-flow
velocity of the slow vehicles, even if only a small  fraction of  slow
vehicles have been considered. Analogous results have been obtained by
Helbing and Huberman \cite{helbhub} who used a different CA model for
the in-lane update (see Sec.~\ref{sub_CAOVM}
for the definition of the model).  In addition to that Nagel et al.
\cite{nagel98} have been shown that for a suitable choice of the
lane changing rules and different types of vehicles even the
phenomenon of ''lane inversion'' which has been observed at german
highways can be reproduced\footnote{On german highways the left lane 
is considered for overtaking vehicles only. Therefore, at low 
densities, the right lane is used more often. Surprisingly, at 
higher densities not simply a balancing of the lane usage has been 
observed, but for densities close to the optimum flow the left lane 
is even higher frequented.}. The results
discussed so far show the strong influence of slow vehicles in multi-lane
systems. They fit fairly well the empirical results,
which show an alignment of the speeds on different lanes and and of
different types of vehicles. Nevertheless recent simulation results of 
Knospe et al. \cite{santen99} indicate that the influence of slow vehicles 
seems to be overestimated 
by the multi-lane variants of the NaSch model. In particular
for symmetric lane changing rules even a single slow vehicle can dominate
the dynamics close to  the optimal value of the flow. In order to
weaken the effect of slow vehicles they suggested to consider anticipation
effects, i.e. the driver estimates the velocity of the vehicle in the 
next timestep \cite{santen99}. 

Another interesting quantity to look at is the frequency of lane
changes at different densities. Here the simulation results show that
close to the density of maximal flow the number of lane changing
maneuvers drastically decreases if the small values of the braking noise 
are considered in CA models where the velocities of vehicles are solely
determined by the distance to the vehicle ahead. This is due to the
fact that for homogeneous states at high densities no sufficiently
large gaps exist. 
For larger values of the braking noise large density fluctuations 
are observable. Therefore the local minimum of the lane-changing
frequency is not found for larger values of $p$. 

In general the simulation results show that some generic multi-lane
effects can be pointed out. First of all the maximal performance of
multi-lane systems is slightly increased compared to corresponding 
single-lane network. In addition, slow vehicles lead to an alignment 
of velocities of different type of vehicles already at low densities 
which is confirmed by empirical observation. This effect is quite 
robust for different choices of the CA model as well as 
for different lane changing rules. It can be weakened most efficiently
if anticipation effects 
are applied. The details of
the lane changing rules, however, may have strong influence on the
lane usage characteristics.


\subsection{Bidirectional traffic}
\label{sub_direct}

Simon and Gutowitz \cite{SimGut} have introduced a two-lane CA model
where the vehicles move in opposite directions. Passing may be 
allowed on one or on both lanes. It is only attempted if there is
a chance to complete the pass. Therefore drivers measure the local
density, i.e.\ the density of vehicles in front that have to be
passed. If it is sufficiently low, a pass will be attempted. This means
that at high global densities the lanes are effectively decoupled since
only very few passes will occur. 

In principle, three types of jams can occur on a bidirectional road:
1) Spontaneous jamming and start-and-stop waves on one of the lanes;
2) jams caused by drivers who try to pass but can not return to their
home lane since there is not enough space and 3) ``super jams'' 
when an adjacent pair of drivers tries to pass simultaneously. These
super jams halt traffic on both lanes and can be prevented by
breaking the symmetry between the lanes.

The precise rules of the CA are in the same spirit as the rules for
multilane traffic described in the previous subsection \ref{sub_multi}. 
First, the situation on the own lane is
examined. If the motion is hindered by another vehicle (moving in the
same direction), a pass is attempted. This will only be initiated
if the safety criteria are satisfied: 1) The gap on the other lane
has to be sufficiently large, and 2) the number of vehicles to be passed
has to be small. Even if these criteria are satisfied a lane change
occurs only with probability $p_{change}$. After this lane changing
step the vehicles move forward similar to the dynamics of the NaSch
model. There are, however, important differences. Passing vehicles
never decelerate randomly. In order to break the symmetry between
the two lanes moving vehicles which are on their home lane and see
oncoming traffic decelerate deterministically by one unit. This rule
prevents the occurance of a super jam. 

The results of \cite{SimGut} show the expected behaviour, namely that
passing makes traffic more fluid. Start-stop waves are surpressed if the 
density is not too large. The improvement of the flow on one lane 
compared to the the one-lane model depends on the density of
vehicles on the other lane. It is maximal for very small densities
($c\to 0$) on the passing lane. If the density on the other lane
is small ($c<0.25$) the flow may be lower than in the one-lane model
since passing oncoming vehicles create an additional hindrance.
For large densities on at least on of the lanes there is little
difference between the one- and two-lane models.

Lee et al.\ \cite{leepopkov} have proposed a toy model for
bidirectional traffic based on a multispecies generalization of
the ASEP. Here no passing is allowed. Instead oncoming traffic
on the opposite lane reduces the hopping rates of the vehicles.
The dynamics on each lane is given by that of the ASEP with
random-sequential update and $v_{max}=1$, but the hopping rate
from an occupied cell $j$ to an empty cell $j+1$ on lane 1
depends on the occupancy of cell $j+1$ on the opposite lane (lane 2).
When this cell is empty, vehicles hop with rate 1, otherwise
with rate $1/\beta$. On lane 2 vehicles move in the opposite direction
and the hopping rate from cell $j+1$ to cell $j$ depends on the 
occupancy of cell $j$ on lane 1. It is given by $\gamma$ when
this cell is empty and by $\gamma/\beta$ if it is occupied.

For $\gamma<1$ the uninfluenced hopping rate on lane 2 is smaller
than that of lane 1. The vehicles on lane 2 might therefore be
interpreted as trucks. The interlane interaction parameter $\beta$
can be interpreted as a kind of road narrowness. For $\beta=1$
vehicles are not slowed down by oncoming traffic. This corresponds
to a highway with divider. The case $\beta \to 0$ corresponds to
a narrow road being completely blocked by the oncoming traffic.

The behaviour of the model with only one truck is rather similar
to that of the NaSch model with quenched disorder 
(see Sec.\ \ref{Sec_disorder}). For $\beta>\beta_c$ the system
segregates into two phases, a high-density phase in front of the
truck and a low-density phase behind it.

By forbidding trucks and cars to occupy parallel cell $j$ 
simultaneously the model can be mapped onto an exactly solvable
2-species variant of the ASEP.
Using the matrix-product Ansatz (see App.\ \ref{App_MPA}) many 
steady-state properties for the single-truck case can be obtained 
exactly.
Two phase can be distinguished: A low-density phase for $c\beta<1$
and a jammed phase for $c\beta>1$ where $c$ is the density of
vehicles on lane 1. In contrast to the case of a fixed defect
site (see Sec.\ \ref{sec_bottle}) only one critical density
$c_{crit}=1/\beta$ exists since the particle-hole symmetry is broken.

Generalizations of this model to other updates and higher velocities
can be found in \cite{foolad}.

\section{Effects of quenched disorder on traffic} 
\label{Sec_disorder}

\subsection{Randomness in the braking probability of drivers and 
Bose-Einstein-like condensation}
\label{sub_BoseEin}

We have seen how modifications of the random braking probability or
the rule(s) for random braking in the NaSch model can give rise to a
rich variety of physical phenomena, e.g., self-organized criticality,
metastability and hysteresis, etc. Now we consider the effects of
quenched randomness in the random braking probability $p$, i.e., we
study the effects of assigning randomly different time-independent
braking probabilities $p_i$ to different drivers $i$ in the NaSch
model. Such "quenched" (i.e., time-independent) randomness in the
random braking in the NaSch model can lead to exotic phenomena
\cite{ktitarev,santen99} which are reminiscent of "Bose-Einstein-like
condensation" in the TASEP where particle-hopping rates are
quenched random variables \cite{kf,evans}.  Various aspects of these
phenomena have been thoroughly reviewed by Krug \cite{krug98} and,
therefore, we'll restrict our discussion to only the essential points.

Let us first consider the special case of the NaSch model with
$v_{max} = 1$. As explained earlier, this model reduces to the TASEP
if the parallel updating is replaced by random sequential updating
scheme. If the same hopping probability $q$ is assigned to every
particle except one for which the the hopping probability is $q' < q$,
then the single "impurity" particle is the slowest moving one. The
faster particles can be allowed to overtake the slow one at a non-zero
rate \cite{derrida19,mallik96}; however, if this rate of overtaking
vanishes the slow particle will give rise to a platoon of particles
behind it. This phenomenon is very similar to the formation of 
{\it platoons} of vehicles in a traffic behind the slow vehicles 
(e.g., trucks).

Here we are interested in a more general situation of quenched
"disorder" in the form of a distribution of intrinsic hopping
probabilities of the vehicles in the system rather than that of the
single "defect" particle. In such situations random initial conditions
can lead to the formation of platoons if (a) slow particles are
sufficiently rare and (b) if the density of vehicles is sufficiently
low. Following their formation, starting from a random initial
condition, the platoons grow through coalescence. The coarsening of
the platoons has been investigated in the same spirit in which
coarsening of domains (the so-called Oswald ripening) is monitored
while studying spinodal decomposition in, for example, binary 
alloys \cite{gunton}.
Suppose, $\xi(t)$ is the typical platoon size at time $t$. Starting
from a homogeneous spatial distribution of the vehicles, $\xi(t)$ can
be monitored as a function of time $t$ to find out the law of "growth"
of the size of the platoons.

Before describing the effects of the quenched randomness in the
hopping probabilities on the steady-states of the TASEP and the NaSch
model, we consider an even simpler model of platoon formation
\cite{ben-naim,nagaplatoon} which was developed using the language of 
aggregation phenomena. In this model an initial velocity $v_j$ is 
assigned to each vehicle $j$, drawn randomly from a continuous 
probability density $f(v)$. The particles then move {\it ballistically} 
along a line and
coalesce whenever a faster vehicle catches up with a slower one in
front. It has been found that $\xi(t)$ increases indefinitely
according to the power law
\begin{equation}
\xi(t) \sim t^{(n+1)/(n+2)} 
\label{platoon}
\end{equation}
where the exponent $n$ characterizes the behaviour of $f(v)$ in the
vicinity of the minimal velocity $v_{min}$, i.e., $f(v) \sim A (v -
v_{min})^n$ as $v \rightarrow v_{min}$ with some positive constant
$A$. An attempt has been made to develop a coarse-grained description
of this phenomenon \cite{emme}.

It has been shown \cite{kf,evans} that if quenched random hopping
probabilities are assigned to each particle in the TASEP, there are
small gaps between particles in the high-density congested phase but
in the inhomogeneous low-density phase there is a macroscopically
large empty region in front of the slowest particle (i.e., the
particle with smallest hopping probability) behind which a platoon is
formed. The phase transition from the low-density inhomogeneous phase
(which consists of a macroscopic free region and a platoon) to the
high-density congested phase is, in many respects, analogous to the
Bose-Einstein transition.

In order to see this analogy, let us imagine that the empty sites are
bosons and the state of a boson is determined by which particle it is
immediately in front of. In the language of the ideal Bose gas, in the
high-density phase the bosons are thinly spread over all the states.
On the other hand, in the low-density phase there is a finite fraction
of the empty sites are condensed in front of the slowest particle in
such "Bose-Einstein-like condensed" state. The steady-state velocity
of the particles is the analogue of the fugacity of the ideal Bose
gas. What makes the system interesting is the fact that the platoon
appears at low-density rather than at high density of the vehicles.

The Bose-Einstein-like-condensation in the TASEP with quenched random
hopping probabilities of the individual particles survives when the
random sequential updating is replaced by parallel
updating \cite{evans}. Finally, it is worth emphasizing that, the
analogy with the ideal Bose gas is only formal as the empty sites in
the TASEP are not non-interacting quantum particles.

The qualitative features of the dynamical phases and phase transitions
observed in the NaSch model with random braking probabilities, for
$v_{max} = 1$ as well as for larger $v_{max}$, are very similar to
those described above for the TASEP with random hopping
probabilities \cite{ktitarev}. Typical snapshots of the system at three
different stages of evolution from a random initial state are shown in
Fig.~\ref{xtplatoon}.

\begin{figure}[ht]
 \centerline{\epsfig{figure=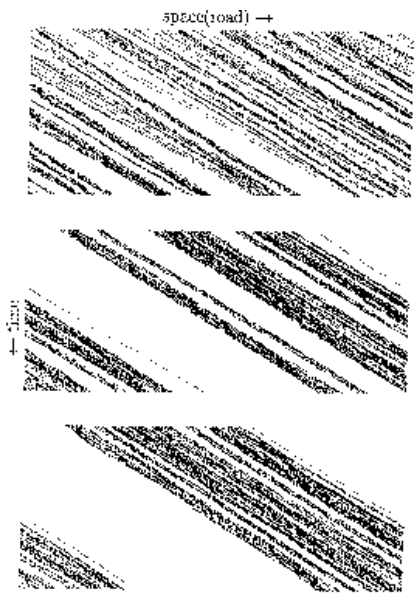,height=8cm}}
\caption{\protect{Typical space-time diagram of the NaSch model
    with $v_{max} = 2$ and $c = 0.1$ and random braking
    probabilities.  Each horizontal row of dots represents the
    instantaneous positions of the vehicles moving towards right while
    the successive rows of dots represent the positions of the same
    vehicles at the successive timesteps. }}
\label{xtplatoon}
\end{figure}

The typical size of the platoons $\xi(t)$ can be computed directly
\cite{ktitarev} by computing the correlation function (\ref{korrdef})
and identifying the separation $r = R_0$ of the first zero-crossing of
this correlation as $\xi(t)$. Following this procedure, it has been
observed that $\xi(t)$ follows the {\it power law} (\ref{platoon})
when the distribution of the random braking probabilities is given by
$P(p) = 2^n (n+1)(\frac{1}{2} - p)^n$.

\subsection{Random $v_{max}$} 

The two important parameters of the NaSch model are $p$ and $v_{max}$.
In the preceding subsection we have seen the effects of randomizing
$p$ assigning the same $v_{max}$ to all the vehicles. In this
subsection, on the other hand, we investigate the effects of
randomizing $v_{max}$, assigning a non-random constant $p$ to every
driver.

The simplest possible model to investigate the effects of quenched
randomness in $v_{max}$ is that considered by Ben-Naim et al. 
\cite{ben-naim} which was discussed to motivate the phenomenon of platoon
formation in the preceding subsection.

In order to model traffic consisting of two different types of
vehicles, say, for example, cars and trucks, of which a fraction
$f_{fast}$ are intrinsically fast (say, cars) while the remaining
fraction $1-f_{fast}$ are intrinsically slow (say, trucks), Chowdhury
et al.\ \cite{chow97b} assigned a higher $v_{max}$ (e.g., $v_{max}=5$)
to a fraction $f_{fast}$ of vehicles chosen randomly while the
remaining fraction $1-f_{fast}$ were assigned a lower $v_{max}$ (e.g.,
$v_{max} = 3$). As the density of the vehicles increases, the vehicles
with higher $v_{max}$ find it more difficult to change lane in order
pass a vehicle with lower $v_{max}$ ahead of it in the same lane. This
leads to the formation of "coherent moving blocks" of vehicles each of
which is led by a vehicle of lower $v_{max}$ \cite{helbhub}. Two main
causes of traffic accidents, namely, differences in vehicles speeds
and lane changes, are reduced considerably in this state thereby
making this state of traffic much safer. It is worth mentioning
that even a small number of slow vehicles in 2-lane models, where
overtaking is possible, can have a drastic effect. For details we 
refer to \cite{santen99} and the discussion in Sec.~\ref{sub_multi}.

\subsection{Randomly placed bottlenecks on the roads and the maximum 
flux principle} 
\label{sec_bottle}

So far we have investigated the effects of two different types of
quenched randomness both of which were associated with the vehicles
(i.e., particles). We now consider the effects of yet another type of
quenched randomness which is associated with the road (i.e., lattice).

In order to anticipate the effects of such randomness associated with
the highway, let us begin with the simplest possible caricature of
traffic with a "point defect" \cite{chunghui}: a single "impurity"
(or, "defect") site in the deterministic limit $p = 0$ of the NaSch
model with $v_{max} = 1$. In this model, vehicles move forward, in
parallel, by one lattice spacing if the corresponding site in front is
empty; each vehicle takes $T_{imp}$ ($> 1$) timesteps to cross the
"impurity" site but only one time step to cross a normal site when the
next site is empty. The impurity sites acts like a {\it blockage} for
all $T_{imp} > 1$. As explained in section \ref{sec_p=0}, in the
absence of the impurity, $J = c$ for $0 < c \leq 1/2$ and $J = 1-c$
for $1/2 < c \leq 1$. Note that, if the impurity is present,
$1/T_{imp}$ vehicle passes through the impurity site per unit time.
Therefore, the bottleneck created by the impurity introduces an upper
cut-off of the flux, viz., $1/T_{imp}$.  Obviously, $J = c <
1/T_{imp}$ so long as $c < c_1 = 1/T_{imp}$.  Similarly, $J = 1-c <
1/T_{imp}$ for $c > c_2$ where $c_2 = 1- c_1$. In the density interval
$c_1 < c < 1-c_1$, the bottleneck at the impurity is the flow-limiting
factor and, hence, in this regime, $J = 1/T_{imp}$ is independent of
$c$. Thus, in the simple caricature of traffic under consideration one
would expect the flux to vary with density following the relation
\begin{equation}
J =
\begin{cases} 
c         & {\rm if} \quad  0 < c \leq c_1,\\
1/T_{imp} & {\rm if} \quad c_1 < c \leq c_2,\\
1-c       & {\rm if} \quad c_2 < c \leq 1,
\end{cases}
\label{nsimp}
\end{equation}
where 
\begin{equation} 
c_1 = \frac{1}{1+(\Delta t)_{imp}} \quad\quad \text{and} \quad\quad c_2 
= \frac{(\Delta t)_{imp}}{1+(\Delta t)_{imp}}, 
\label{c1c2}
\end{equation}
and $T_{imp} = 1 + (\Delta t)_{imp}$ such that $(\Delta t)_{imp} = 0$
for the normal sites but $(\Delta t)_{imp} > 0$ for the impurity site.

The fundamental diagram, obtained numerically through computer
simulation of the NaSch model with a single defect and non-zero $p$
(Fig.~\ref{dasep_fund}) is in qualitative agreement with those of
the fundamental diagram (\ref{nsimp}). The qualitative features of the
fundamental diagram in Fig.~\ref{dasep_fund} are also similar to
those of the TASEP with a single defect \cite{lebo} where the hopping
probability $q$ is smaller than that at all the normal sites. Equation
(\ref{nsimp}) also indicates that the larger is $(\Delta t)_{imp}$ the
lower is the maximum flux $1/[1+(\Delta t)_{imp}]$ and the wider is
the interval $c_1 \leq c \leq c_2$ over which the flux remains
constant.

\begin{figure}[ht]
\centerline{\epsfig{figure=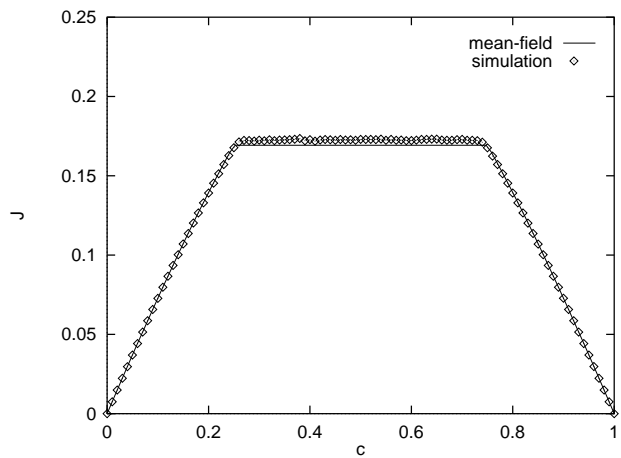,height=6cm}}
\caption{\protect{Fundamental diagram of the NaSch model with 
    $v_{max} = 1$ and a blockage site located at the site 1. The
    hopping probability in the bulk is given by $q = 0.75$ and at the
    defect by $q = 0.25$. }}
\label{dasep_fund}
\end{figure}
\begin{figure}[ht]
\centerline{\epsfig{figure=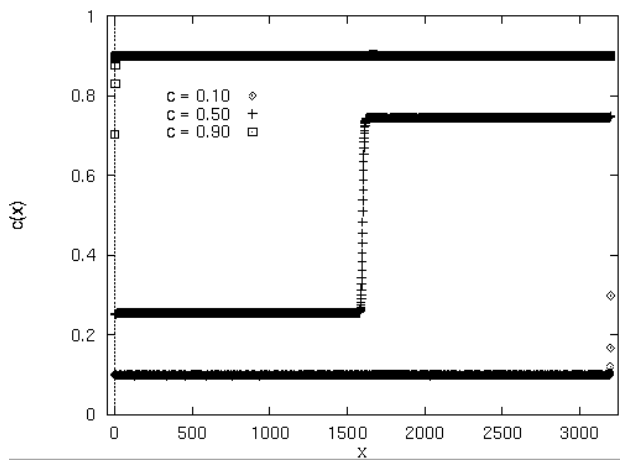,height=6cm}}
\caption{\protect{Density profiles for different values of $c$ 
    in the NaSch model with $v_{max} =1$ and a blockage site located
    at the site 1. The hopping probability in the bulk is given by $q
    = 0.75$ and at the blockage site by $q = 0.25$. }}
\label{dasep_prof}
\end{figure}

What makes the problem of a single "point defect" nontrivial is that,
over the interval $c_1 \leq c \leq c_2$ of the density of the
vehicles, where the $J$ is maximum and independent of $c$, the 
{\it localized blockage} has {\it global} effects whereby the traffic
exhibits {\it macroscopic} phase segregation into high-density and
low-density regions. Evidence for such macroscopic phase segregation
can be obtained directly from the density profiles (see
Fig.~\ref{dasep_prof}).  Fig.~\ref{dasep_prof} implies that so long as
$c < c_1$ the particles will not pile up but a local increase of
density will compensate for the reduced local velocity at the blockage
so that the flux around the blockage is identical to that far from it.
However, if the global density exceeds $c_1$, the particles pile up
during the transient period leading to the phase-segregated steady
state. Because of the particle-hole symmetry the phase-segregation
does not take place if the particle density exceeds $c_2$.

We now develop a semi-phenomenological theory\footnote{A microscopic
approach for deterministic dynamics can be found in \cite{tilstra}.} 
for the NaSch model with $v_{max} = 1$, non-zero $p$ and a single 
"impurity" site {\it assuming} the steady-state to be phase-segregated, 
as demonstrated by computer simulation (Fig.~\ref{dasep_prof}).  
Naturally, this theory cannot {\it explain} the underlying mechanism 
that gives rise to the phase-segregated structure of the steady-state. 
But, as we shall see soon, it provides a good estimate of the flux in 
the phase-segregated regime.  Our calculations are based on arguments
similar to those suggested originally by Janowsky and Lebowitz
\cite{lebo} in the context of TASEP with a single defect.

Using equation (\ref{fl-2cl}), the flux in the high-density and
low-density regions, far from their interface, are given by $J_{h} =
\frac{1}{2}\left(1 - \sqrt{1 - 4 q c_h(1-c_h)}\right)$ and 
$J_{\ell}=\frac{1}{2}\left(1 -\sqrt{1-4qc_{\ell}(1-c_{\ell})}\right)$ 
and that across the defect bond is given
by $J_{def} \simeq \frac{1}{2}\left(1-\sqrt{1- 4 q_d c_h(1-c_{\ell})}
\right)$.  Since,
in the steady state, the flux is same across the entire system, we
must have $q c_h (1-c_h) = q c_{\ell} (1-c_{\ell})$ and, hence,
\begin{equation}
c_h = c_{\ell} \quad \text{or} \quad c_h = 1 - c_{\ell}. 
\end{equation}
The condition $c_h = c_{\ell}$ is satisfied by the uniform density
profiles whereas the condition $c_h = 1 - c_{\ell}$ is satisfied by
the phase-segregated density profile (see Fig.~\ref{dasep_prof}).
Moreover, using the condition $J_h = J_{def} = J_{\ell}$ we get
\begin{equation}
c_h (1-c_h) = c_{\ell} (1-c_{\ell}) \simeq r c_h (1-c_{\ell})   
\label{janlebo}
\end{equation}
where $r = q_d/q < 1$ may be interpreted as the "transmission
probability" or "permeability" of the blockage. From the (approximate)
equations (\ref{janlebo}) we get
\begin{equation}
c_h \simeq \frac{1}{r+1} = \frac{p}{p+p_d} \quad \text{and} 
\quad c_{\ell} \simeq \frac{r}{r+1} = \frac{p_d}{p+p_d}
\label{density}
\end{equation}
and, hence, 
\begin{equation}
J = \frac{1}{2} \left[1 - \sqrt{1 - \frac{4 q r}{(1+r)^2}} \right] 
\label{leboflux}
\end{equation}
The estimate (\ref{leboflux}) is in good agreement with the numerical
data (Fig.~\ref{dasep_fund}) obtained from computer simulation
\cite{santen}. Moreover, the estimates $c_{\ell}$ and $c_h$ are also
in good agreement with $c_{\ell}$ and $c_h$, respectively, in 
Fig.~\ref{dasep_prof}.

Note that $c_{\ell}$ and $c_h$ depend only on $r$ and are independent
of $c$. Moreover, the estimates (\ref{density}) of $c_{\ell}$ and
$c_h$ are in excellent agreement with $c_1$ and $c_2$, respectively,
in Fig.~\ref{dasep_fund}. At first sight, these two results may appear
surprising and counter-intuitive.  But, we'll now show that these are
related to the mechanism of the phase-segregation. Conservation of the
vehicles demand that
\begin{equation}
c L = c_h h + c_{\ell} = c_h h + c_{\ell} (L - h) 
\end{equation} 
where $h$ and $\ell = L - h$ are the lengths of the high-density 
and low-density regions, respectively. Thus, 
\begin{equation}
\frac{h}{L} = \frac{c - c_{\ell}}{c_h - c_{\ell}} = \frac{c(1+r)-r}{1-r} 
\label{h_thick}
\end{equation}
The equation (\ref{h_thick}) shows that $h/L \rightarrow 0$ as $c
\rightarrow c_{\ell}$ and $h/L \rightarrow 1$ as $c \rightarrow c_h$.
Therefore, keeping $r$ fixed as the density is increased beyond $c_1 =
c_{\ell}$, the densities of the two regions remain fixed but the
high-density region grows thicker at the cost of the length of the
low-density region as more and more vehicles pile up and, eventually,
at $c = c_2 = c_h$ the low-density region occupies a vanishingly small
fraction of the total length of the system signaling the disappearance
of the phase segregation. Interestingly, recasting the expressions for
$c_h$ and $c_{\ell}$ as $c_{\ell} = 1/[1+(\Delta t)_{imp}]$ and $c_h =
(\Delta t)_{imp}/[1+(\Delta t)_{imp}]$, where $(\Delta t)_{imp} =
1/r$, we find close formal analogies with $c_1$ and $c_2$,
respectively, in equation (\ref{c1c2}) \cite{yukawa94}.

Sch\"utz \cite{schutz93} considered a TASEP with sublattice-parallel
update (see App.\ \ref{App_updates}) where the motion of the particles
is {\it deterministic} (i.e., $q=1$) everywhere except at a defect
site where they move with the probability $q_d < 1$ (i.e., $r = q_d
<1$).  Exact solution is possible through a mapping on a 6-vertex
model. Later, a solution using the matrix-product Ansatz was presented
in \cite{hhss}. Except for minor differences, the qualitative features 
of the results do not differ from the corresponding approximate results
obtained for $q_d \neq 1$ \cite{lebo}. Qualitatively similar phase
segregation phenomena have also been observed in a related model
\cite{leepopkov}.  

The qualitative features of the fundamental diagram do not change
significantly if the "point-like" defect (or, impurity) is replaced by
an "extended" defect \cite{santen}, i.e. a few consecutive defect 
sites. However, with increasing length of the defect, the maximum value 
of the flux decreases monotonically and approaches the maximum flow of 
the homogeneous system where the hopping probability associated with each
of the bonds is identical to that associated with the defects in our
model (Fig.~\ref{extdefect}). From Figs.~\ref{extdefect} and
\ref{extdef_prof} we conclude that the monotonic decrease of the
flow with increasing length of the extended defects, leads to a larger
difference $c_h - c_{\ell}$ between the densities of the high-density
and the low-density regions of the phase-segregated steady-state.

\begin{figure}[!h]
\centerline{\epsfig{figure=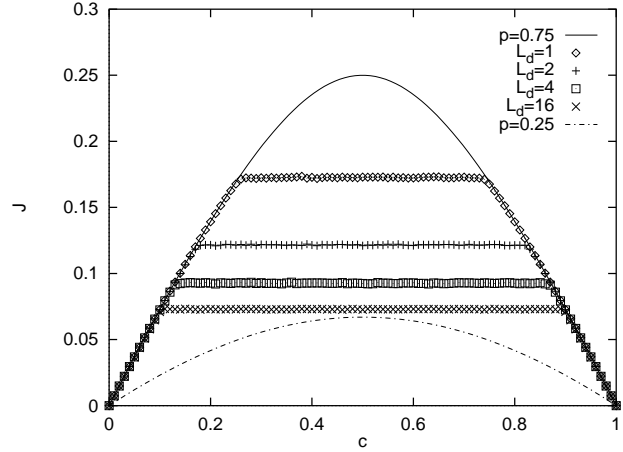,height=6cm}}
  \caption{Fundamental diagram for the NaSch model with $v_{max} = 1$
    and defects of different length $L_d$. Again $p=0.75$ and
    $p_d=0.25$ is chosen. For comparison the fundamental diagram of
    the ``fast'' (p=0.75) and ``slow'' homogeneous systems are shown.}
\label{extdefect}
\end{figure}
\begin{figure}[!h]
\centerline{\epsfig{figure=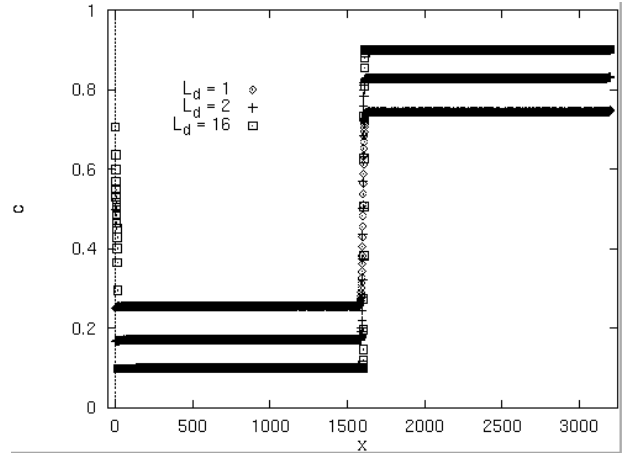,height=6cm}}
  \caption{Density profiles for different lengths of the defect in 
    the NaSch model with $v_{max} = 1$. Again $p = 0.75$ and $p =
    0.25$ are used and the average density is given by $c=0.50$. The
    defect extended over the first $L_d$ sites of the system.}
\label{extdef_prof}
\end{figure}

Next, instead of a single point-like or extended "defect", let us
consider the more general case of quenched "disorder" in the NaSch
model with $v_{max} = 1$ where the quenched random hopping
probabilities $q_{j,j+1} = 1 - p_{j,j+1}$ are chosen independently
from some probability distribution $P(q)$, for the hopping from the
cell $j$ to the cell $j + 1$ ($j = 1,2,...,L$).  For a given
realization of the disordered system, every vehicle hopping from a
given cell $i$ to the next cell $i+1$ must hop with the same
probability $q_{i,i+1}$ and a given vehicle hops across different
bonds, in general, with different probabilities assigned to these
bonds as it moves forward with time. A similar generalization of the
TASEP has also been studied \cite{barma}.  We shall refer to this model
as disordered TASEP (i.e., DTASEP).

Suppose, $q$ are chosen from the binary distribution
\begin{equation}
P(q_{j,j+1} = q_d) = f, \qquad P(q_{j,j+1} = q) = 1 - f 
\label{bin_dist}
\end{equation}
i.e., a fraction $f$ of the bonds have a permeability $r < 1$ while
the remaining fraction $1-f$ have unit permeability.  A mean-field
theory has been developed \cite{barma} (see Appendix \ref{App_solvMF}
for details) for computing the fundamental diagram of the DTASEP. The
flux in this model has interesting symmetry properties under the
operations of "charge conjugation" (which interchanges particles and
holes), "parity" (which interchanges forward and backward hopping
rates on each bond and "time reversal" (which reverses the direction
of the current) \cite{barma,goldstein}.

The quenched disorder in these "disordered" models can be 
viewed as "point-like impurities" distributed randomly over 
the lattice. But, the qualitative features of the fundamental 
diagram of DTASEP are similar to those observed for a single 
point-like defect and those for a single extended defect. 
Although the random distribution of the point-like impurities 
leads to a "rough" density profile for all densities, in an 
intermediate regime of density, phase-segregated steady-sates 
with macroscopic high- and low-density regions have been identified. 

What is the underlying mechanism for the "macroscopic" phase
segregation in all the models DTASEP \cite{barma} ? Let us
denote the stretches of bonds with permeability $1$ by $X$ and the
stretches of bonds with permeability $r$ by $Y$. The two parabolas in
Fig.~\ref{tripathyfig} are the two steady-state fundamental diagrams for
the two pure reference systems consisting of all $X$ and all $Y$,
respectively. Since the flux must be spatially constant in the
steady-state, the possible densities are given by the four
intersections of the line $J = J_0$ with the two parabolas. If the
average density is less (greater) than $1/2$ then the two possible
densities are $c_1$ and $c_2$ ($c_3$ and $c_4$). The variation of
density between $c_1$ and $c_2$ (or $c_3$ and $c_4$) in the $X$ and
$Y$ stretches is merely micro phase-segregation while, on a
macroscopic scale, the density remains uniform. For simplicity, we
assume that the density in each stretch of like bonds is uniform. The
global density of the system is approximately $c \simeq (1-f)
c_{1,4}(J_0) + f c_{2,3}(J_0)$ where $f$ is given by equation 
(\ref{bin_dist}). However, as the density increases the
flux also increases till it attains the maximum allowed flux of the
pure system consisting of all $Y$ (this happens at a global density
smaller than $1/2$). What happens when the density increases further?
According to the "maximum current principle" \cite{krug91}, no further
increase of the flux is possible and the excess density is taken care
of by increasing the density in some of the $X$ stretches from $c_1$
to $c_4$ (or, vice versa if $c > 1/2$). This conversion takes place
adjacent to the largest stretch of $Y$ bonds where the density also
changes from $c_2$ to $c_3$ (or, vice versa if $c > 1/2$) to
accommodate the additional particles added to the system. This leads
to the macroscopic phase segregation as the system consists of two
macroscopic regions of two different mean densities- one with lower
densities $c_1,c_2$ in the $X$ and $Y$ stretches and the other with
the higher densities $c_3,c_4$ in the $X$ and $Y$ stretches.

\begin{figure}[!h]
\centerline{\epsfig{figure=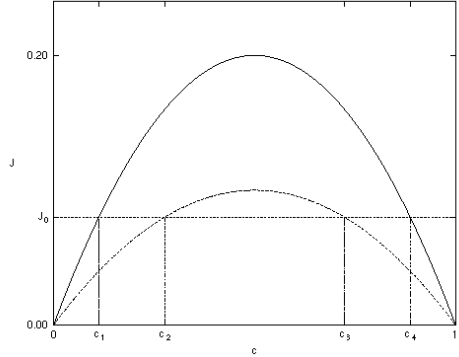,height=5cm}}
  \caption{Origin of phase separation in DTASEP. Shown are the fundamental
diagrams of two reference non-disordered TASEPs.}
\label{tripathyfig}
\end{figure}

It is not difficult to generalize the DTASEP to disordered NaSch
(DNaSch) model by replacing the random sequential updating by parallel
updating. However, we face subtle conceptual difficulties in extending
DNaSch model further to arbitrary $v_{max}$; if the position of a
vehicle at time $t+1$ is decided by $q$ at its current position at
time $t$, it may be forced to move $v$ sites downstream by hopping
over sites of even smaller $q$ if $v > 1$.

Some alternative parameterization of the defect (or, disordered) sites
in the NaSch model for arbitrary $v_{max}$ have also been suggested.
In \cite{knospe,santen} localized defects have been investigated
where the randomization parameter $p_d$ is larger than in the
rest of the system. Csahok and Vicsek \cite{cv} have considered the
blockages as sites with a "permeability" smaller than unity whereas the
permeability of all the other sites is unity. This effectively reduces
$v_{max}$ while the vehicle is at a blockage. On the other hand,
Emmerich and Rank \cite{er1,chow98a} considered a model of where the
velocity of every vehicle in the region occupied by the blockage (or,
more appropriately, hindrance) at the time step $t+1$ is half of that
at time $t$, i.e., $v_n(t+1) = v_n(t)/2$ is the $n$-th vehicle is
located within the hindrance region.
Some effects of static hindrances on vehicular traffic have also been
investigated following alternative approaches, e.g., car-following
theory \cite{nagacf}.
From a practical point of view, ramps have effects very similar to
those of a static defect. For the NaSch model this has been 
investigated in \cite{DSSZ}.

\section{Other CA models of highway traffic}
\label{Sec_other}

All the CA models of highway traffic described so far are basically
generalizations of the minimal CA model proposed originally by Nagel
and Schreckenberg \cite{ns}. We now describe a few other alternative
minimal CA models and the interesting features of the corresponding
results.

\subsection{Fukui-Ishibashi model}
\label{sec_FI}

The update rules of the Fukui and Ishibashi (FI) \cite{fukuishi}
model of single-lane highway traffic are as follows:\\ 

If $v_{max}$ or more sites in front of the $n$-th vehicle is empty at the
time step $t$, then it has a probability $1-p$ to move forward by $v_{max}$
sites and a probability $p$ to move forward by $v_{max}-1$ sites in the
time step $t+1$. However, if only $d$ sites ($d < v_{max}$) in front of
the $n$-th vehicle are empty at time $t$ then it moves by $d$ sites
in the next time step.  Since no site can be occupied simultaneously
by more than one vehicle, a vehicle must not move forward in the time
step $t+1$ if the site immediately in front of it is occupied by a
vehicle at the time step $t$. The model becomes deterministic
in both the limits $p= 0$ and $p = 1$.

The FI model differs from the NaSch model in two respects: (a) the
increase of speed of the vehicles is not necessarily gradual and (b)
the stochastic delay applies only to high-speed vehicles. The FI
model, obviously, reduces to the NaSch model if $v_{max} = 1$. A
site-oriented mean-field theory \cite{wang98} and a car-oriented
mean-field theory \cite{wang97,wangpre} for the FI model have been
developed for arbitrary $v_{max}$ and $p$. Note that the FI model is
equivalent to the deterministic CA rule $184$ (in the notation of
Wolfram~\cite{wolfram}) in the limit $v_{max} = 1, p = 0$. 
Generalizations of the deterministic limit $p = 0$ of the FI model 
have also been proposed~\cite{fuksboc,fuks}.

According to the classification of Sec.\ \ref{sub_KWG}, the
FI model belongs to class I, i.e.\ the high-acceleration limit where
no spontaneous jamming exists. In \cite{as_tgf} an alternative
high-acceleration variant has been proposed.
Here only the acceleration step $(U1)$ of the NaSch model is changed 
to $v\to v_{max}$, i.e.\ all vehicles accelerate immediately to the
maximal possible velocity. The other update steps of the NaSch model
are left unchanged. In contrast to the FI model, all vehicles
are subject to the randomization step. The behaviour of this variant
is therefore similar to that of the NaSch model, e.g.\ one finds
spontaneous jam formation.
 
\subsection{Galilei-invariant vehicle-vehicle interaction and 
metastability}

In the NaSch model it is postulated that the gap between a pair of
successive vehicles is adjusted according to the velocity of the
leading vehicle alone. In contrast, often in real traffic, drivers
tend to adjust the gap in front taking into account the {\it difference}
between the velocity of their own vehicle and that of the leading
vehicle.  The latter aspect of real traffic is captured by a recent
model developed by Werth, Froese and Wolf (WFW) \cite{wolf,wfw}.

If both the following vehicle and the leading vehicle move with
constant acceleration $b$, then a collision between the two can be
avoided provided
\begin{equation}
gap + \ell(v_{LV}) \geq v_{FV} \tau_r + \ell(v_{FV}) 
\label{coll1}
\end{equation}
where $v_{LV}$ and $v_{FV}$ are the velocities of the leading
vehicle and the following vehicle, respectively,
$\tau_r$ is the reaction time of the following vehicle and  
\begin{equation}
\ell(v) = \frac{v^2}{2b}
\label{eq-101} 
\end{equation}
is the distance covered by a vehicle with initial velocity $v$ 
before it comes to a stop by moving with a constant deceleration 
$b$. Using equation (\ref{eq-101}) the condition (\ref{coll1})
for avoiding collision can be written as 
\begin{equation}
gap \geq v_{FV} \tau_r + \frac{{\bar v}}{b}(v_{FV} - v_{LV}) 
\label{coll2}
\end{equation}
where ${\bar v} = (v_{FV} + v_{LV})/2$ if the average velocity of the
pair of vehicles under consideration. Therefore, a {\it sufficient}
condition for avoiding a collision is
\begin{equation} 
gap \geq 
\begin{cases}
v_{FV} \tau_r                                   
&\text{for $v_{FV} \leq v_{LV}$,}\cr
v_{FV} \tau_r + \frac{v_{max}}{b}(v_{FV}-v_{LV}) 
& \text{for $v_{FV} >
  v_{LV}$.}
\end{cases}
\label{suff_cond}
\end{equation}
In the limiting case $b \rightarrow \infty$ the sufficient condition
(\ref{suff_cond}) reduces to $gap \geq v_{FV} \tau_r$ which is
identical to the form of vehicle-vehicle interaction in the NaSch
model if one chooses $\tau_r$ as the unit of time.  In the opposite
limit $\tau = 0$, the sufficient condition (\ref{suff_cond}) reduces
to
\begin{equation} 
  gap \geq
\begin{cases} 
  0 & \text{for $v_{FV} \leq v_{LV}$,}\cr
  \frac{v_{max}}{b}(v_{FV}-v_{LV}) & \text{for $v_{FV} > v_{LV}$.}
\end{cases}
\label{suff_cond2}
\end{equation}
Since this type of vehicle-vehicle interaction involves the difference
of the velocities $v_{FV} - v_{LV}$, it is clearly invariant under a
Galilean transformation and, hence, the name.  The vehicle-vehicle
interactions in real traffic may be somewhere in between the two
limiting cases of NaSch model and the Galilei-invariant model.

Suppose the indices $n-1$ and $n$ label the leading vehicle and the
following vehicle, respectively, of a pair. The update rules suggested
by WFW \cite{wfw} for implementing the Galilei-invariant
vehicle-vehicle interaction are as follows:\\ 

\noindent {\it Step 1: Acceleration.} 
$$
v^{(1)}_n = \min(v_n+1,v_{max}) 
$$

\noindent{\it Step 2: Deceleration (due to other vehicles).} 
$$
v^{(2)}_n = \min(v^{(1)}_n,d_n-1+v_{n-1}) 
$$

\noindent{\it Step 3: Randomization.}  
$$
v^{(3)}_n {\stackrel{p}{=}}\ \max(v^{(2)}_n-1,0) \quad\quad
\text{with probability $p$}
$$

\noindent{\it Step 4: Deceleration (due to other vehicles).} 
$$
v^{(4)}_n = \min(v^{(3)}_n,d_n-1+v^{(4)}_{n-1}) 
$$

\noindent{\it Step 5: Vehicle movement.}
\vspace{-0.5cm}
\begin{eqnarray*}
x_n &=& x_n + v^{(4)}_n,\\
v_n &=& v^{(4)}_n.
\end{eqnarray*}

Thus, the rule for {\it deceleration (due to other vehicles)}
is applied twice. Step 4 makes sure collisions are avoided. Since
also the new velocity $v^{(4)}_{n-1}$ of the preceding car enters,
this step can not be performed in parallel for all cars. Instead
it is performed sequentially, but the final configuration is
independent of the starting point of this sequential updating.
Step 4 has then to be applied twice in order to determine all
velocities $v^{(4)}_n$ consistently.

The rules as given above define the {\em retarded} version of the
Galilei-invariant model. In the {\em non-retarded} version, in step 2
$v_{n-1}$ is replaced by the new velocity $v^{(2)}_{n-1}$.
To determine $v^{(2)}_{n}$ consistently for all cars, step 2 has
then to be iterated $v_{max}-1$ times.

The most interesting feature of the Galilei-invariant model is that
its fundamental diagram has a metastable branch although its update
scheme involve neither cruise-control nor slow-to-start rules.
The mechanism leading to the existence of metastable states
is different from the models with slow-to-start rules 
(see Sec.~\ref{Sec_s2s}). The outflow
from jams is the same as in the NaSch model since it is independent 
of the interaction between vehicles. However, due to the inclusion
of anticipation effects (i.e.\ the driver knows the velocity of
the preceding vehicle) the free-flow is less sensitive to fluctuations.

\subsection{CA versions of the optimal-velocity model} 
\label{sub_CAOVM}

The traffic jams appear spontaneously in both the OV models and 
the CA models. However, in the OV models spontaneous formation 
of the jams are caused by the {\it non-linearity} of the dynamical 
equations whereas in the CA models it is triggered primarily by 
the {\it stochasticity} of the update "rules". The mechanism for 
the spontaneous formation of jams in real traffic may be a 
combination of these two. 

In the OV model, the control of velocity is given by the control 
of acceleration through the OV function which gives the optimal 
velocity for the current distance-headway. Thus, unlike the CA 
models like the NaSch model, the vehicles in the OV models get 
an opportunity to avoid crash without any need to exert unphysically 
large deceleration. In fact, collision of vehicles may take place 
in naive discretization of the dynamical equations for the OV 
models unless special care is taken in the discretization process
(see below). 

In the following we present several CA model analogues of the 
OV model that have been proposed by different authors.
In principle, the NaSch model is also an OV model, but with a
linear OV function, $v(d)=\min[d-1,v_{max}]$. The first attempt
to generalize this relation is due to Emmerich and Rank \cite{er2}.
The update rules of their model are as follows:

\noindent{\it Step 1: Find largest gap.}\\
Find the vehicle with the largest gap to the next vehicle ahead.

\noindent{\it Step 2: Acceleration.}
$$
v_n \to \min(v_{max},v_n+1)
$$

\noindent{\it Step 3: Deceleration due to other vehicles.}
$$
v_n \to M_{d_n-1,v_n} \quad\quad \text{if $d_n-1 \leq v_{max}$},
$$
i.e.\ a vehicle with velocity $j$ and $i$ empty cells in of it
(i.e.\ a gap $d_n=i+1$) reduces its velocity
to $M_{i,j}$ ($0\leq i,j\leq v_{max}$).

\noindent{\it Step 4: Randomization.}
$$
v_n\to \max(v_n-1,0) \quad\quad \text{with probability $p$}
$$

\noindent{\it Step 5: Vehicle movement.}
$$
x_n\to x_n+v_n
$$

\noindent{\it Step 6: Next vehicle.}\\
Repeat steps 2-5 for the next vehicle behind, i.e.\ proceed
in the direction opposite to the motion of the vehicles.

For the NaSch model with $v_{max}=5$ the matrix $M_{i,j}$ is given by
\begin{equation}
M^{(NaSch)}_{i,j}=
\begin{pmatrix}
0 & 0 & 0 & 0 & 0 & 0\\
0 & 1 & 1 & 1 & 1 & 1\\
0 & 1 & 2 & 2 & 2 & 2\\
0 & 1 & 2 & 3 & 3 & 3\\
0 & 1 & 2 & 3 & 4 & 4\\
0 & 1 & 2 & 3 & 4 & 5
\end{pmatrix}
\end{equation}
A general matrix $M_{i,j}$ has to satisfy certain conditions (e.g.\
$M_{i,j}\leq \min(i,j)$ and $M_{i,j}\leq M_{i,k}$ for $j\leq k$) 
to guarantee e.g.\ the absence of collisions in the model.
In order to model the fact that faster vehicles keep a relatively
larger headway to the preceding vehicle, Emmerich and Rank suggested
the following matrix:
\begin{equation}
M^{(ER)}_{i,j}=
\begin{pmatrix}
0 & 0 & 0 & 0 & 0 & 0\\
0 & 1 & 1 & 1 & 1 & 1\\
0 & 1 & 2 & 2 & 2 & 2\\
0 & 1 & 2 & 2 & 3 & 3\\
0 & 1 & 2 & 3 & 3 & 4\\
0 & 1 & 2 & 3 & 4 & 4
\end{pmatrix}
\end{equation}

Using a parallel update scheme, the model shows unrealistic behaviour
in the free-flow regime, especially in the deterministic limit $p\ll 1$.
Here the fundamental diagram is non-monotonic \cite{unpub}, as can
be seen from a simple example for $p=0$. At density $c=1/7$, the 
stationary state is of the form $5......5......5......$ where
numbers denote the velocity of vehicles and '.' an empty cell.
At density $c=1/6$, on the other hand, the stationary state is
$4.....4.....4.....$. Comparing the corresponding flows, one finds
$J(c=1/7)=5/7 > J(c=1/6)=2/3$. At density $c=1/5$, the stationary
state is $4....4....4....$ with flow $J(c=1/5)=4/5$ which is again
larger than the flow at $c=1/6$ and corresponds to the maximal
possible flow. This kind of behaviour persists even in the 
presence of randomness ($p>0$) \cite{unpub}.
In order to circumvent this problem, Emmerich and Rank had to 
introduce a special kind of ordered-sequential update, where
first the vehicle with the largest gap ahead is updated. Then, the 
position of the next vehicle upstream is updated, and so on, using 
periodic boundary conditions.

Emmerich and Rank also investigated more general rules where even
for gaps larger than $v_{max}$ the velocity of the vehicles is
reduced to a value $v < v_{max}$.


Later, a similar model has been proposed by Helbing and Schreckenberg 
\cite{helbshrek}. It is closer to the spirit of the original
optimal-velocity model (see Sec.\ \ref{sec_OVM}).

\noindent{\it Step 1: Vehicle movement.}
$$
x_n \to x_n +v_n(t)
$$
\noindent{\it Step 2: Acceleration.}
$$
 v_n'(t+1) = v_n(t) + \Big\lfloor \lambda [ V_{opt}(d_n({t})) - v_n(t) ] 
\Big\rfloor \, ,\nonumber\\
$$
\noindent{\it Step 3: Randomization.}
$$
v_n(t+1) = v_n'(t+1) - 
\begin{cases}
1 & \text{with probability $p$ \ \ \ (if $v_n'(t+1) > 0$)}\\
0 & \text{otherwise.} 
\end{cases} 
$$

Here $\lfloor y \rfloor$ denotes the floor function, i.e.\
the largest integer $i\le y$. In \cite{helbshrek} various
optimal velocity functions $V_{opt}(d)$ have been used in order
to fit experimental data. The simplest, but unrealistic, choice
was $V_{opt}(d)=\min(d,v_{max})$ where $d$ is the distance-headway.
The parameter $\lambda$ corresponds formally to the sensitivity
parameter in the OV model where it determines the timescale of
relaxation towards the stationary fundamental diagram. 
Such an interpretation is not possible for discrete time models.
Here the main effect of the parameter $\lambda$ is a rescaling
of the OV function.

The naive discretization of the OV function produces some undesirable
features of the model, e.g.\ the flow corresponding to the OV function
is non-monotonic in the free-flow region. Furthermore one finds -- for
certain initial conditions -- a breakdown of the flow at a finite
density $c_*<1$. This indicates a lack of robustness of model against
small modifications of the rules.

For $\lambda <1$ the model is not intrinsically collision-free 
\cite{ssprep}, in contrast to most other models discussed in this 
review. Problems occur e.g.\ when fast vehicles approach the end
of a jam. For the $V_{opt}(d)=\min(d,v_{max})$ and the 'realistic'
choice $\lambda =0.77$ \cite{helbshrek} collisions can occur for
$v_{max}\geq 5$ \cite{ssprep}. For $\lambda > 1$, on the other hand,
a backward motion of vehicles is possible. For a given OV function 
it is possible to derive conditions for the parameter $\lambda$
which ensure the absence of collisions and backward motion \cite{ssprep}.

Nagatani \cite{NagOV} has suggested a CA model which combines the OV
idea with the TASEP. It is the discrete analogue of the simplified
OV model (\ref{eq:simpleOV}) presented in Sec.~\ref{sec_OVM}.
Here the $n$-th vehicle moves ahead with
probability $v_n(t)$ where $v_n(t)$ is interpreted as velocity.
This velocity is obtained by integration of the OV function
\begin{equation}
\ddot{x}_n(t)=
\begin{cases}
a  & \text{\ \ for $\Delta x_n(t) \geq \Delta x_c$}\\
-a & \text{\ \ for $\Delta x_n(t) < \Delta x_c$} 
\end{cases} 
\end{equation}
where $a>0$, $\Delta x_n(t)=x_{n+1}(t)-x_n(t)$ and $\Delta x_c$ is a 
safety distance. Furthermore the velocity is restricted to the interval
$0\leq v_n(t)\leq v_{max}\leq 1$.


\subsection{CA from ultra-discretization}

Several CA which can be interpreted as traffic models have
been derived using the so-called ultra-discretization method (UDM)
\cite{tokihiro}.
This approach allows to establish a direct connection between certain
differential equations and CA. The problem in the derivation of CA
from differential equations lies in the discretization of the 'state'
(or dependent) variable. E.g.\ in a numerical treatment, only space 
and time variables are discretized.

The basic procedure of the UDM is as follows: 1) Start from a
nonlinear wave equation, e.g.\ the KdV equation or Burgers'
equation; 2) Discretize space- and time-variables in a standard
way to obtain the discrete analogue of the wave equation which
is still continuous in the state variable $u_j(t)$;
3) The discrete analogue is now ultra-discretized. Defining
$U_j(t)=\epsilon \ln(u_j(t))$ (where $\epsilon$ depends on the
discretization $\Delta x$ and $\Delta t$ of space and time) one
can use the identity $\lim_{\epsilon \to 0} \epsilon \log 
(e^{A/\epsilon}+e^{B/\epsilon}+\cdots)=\max[A,B,\ldots]$
to derive the CA analogue fo the nonlinear wave equation.

By applying the UDM to Burgers' equation $v_t=2vv_x+v_{xx}$ one
obtains the $(L+1)$-state, deterministic CA \cite{nishinari,takahashi}.
\begin{equation}
n_j(t+1)=n_j(t)+\min[M,n_{j-1}(t),L-n_j(t)]
  -\min[M,n_{j}(t),L-n_{j+1}(t)].
\label{eq_BCAdef}
\end{equation}
$n_j(t)$ is the occupation number of cell $j$ at time $t$. In contrast
to most other CA discussed in this review multiple occupations of
cells are allowed. The maximum number of particles which can occupy
the same cell is given by $L$, i.e.\ $0\leq n_j(t)\leq L$. The model
defined by (\ref{eq_BCAdef}) might therefore by interpreted as a
simple model for a highway with $L$ lanes where the effects of lane 
changes are completely neglected. In \cite{takahashi} it has also been
suggested to interprete $n_j(t)/L$ for large $L$ as a coarse-grained 
density. The parameter $M$ denotes the maximum number of vehicles
that can move from cell $j$ to cell $j+1$.

For $M\geq L=1$ the model reduces to the rule-184 CA, i.e.\ the
NaSch model with $v_{max}=1$ and $p=0$. In the case $L\leq 2M$ the
fundamental diagram looks similar to that of rule-184 CA, but it
is 'degenerate' in the sense that qualitative different stationary 
states with the same flow exist.
For $L>2M$ the fundamental diagram the fundamental diagram resembles 
that of rule-184 CA with a blockage site (see Sec.\ \ref{sec_bottle},
especially Fig.\ \ref{dasep_fund}).
In the region $M/L \leq c \leq (L-M)/L$ the flow takes the constant
value $M/L$, i.e.\ the parameter $M$ can be interpreted as a flow
limiter.

In \cite{takahashi} also a generalization of (\ref{eq_BCAdef}) to higher
velocities $v_{max}>1$ has been suggested. For $v_{max}=2$ the
generalized update rule is given by
\begin{eqnarray}
n_j(t+1)&=&n_j(t)+a_{j-2}(t)-a_j(t)\nonumber\\
&+&\min[b_{j-1}(t)-a_{j-1}(t),L-n_j(t)-a_{j-2}(t)]\nonumber\\
&-&\min[b_{j}(t)-a_{j}(t),L-n_{j+1}(t)-a_{j-1}(t)].
\label{eq_gBCA}
\end{eqnarray}
Here $b_j(t)=\min[n_j(t),L-n_{j+1}(t)]$ is the maximum number of 
vehicles at site $j$ that can move and 
$a_j(t)=\min[n_j(t),L-n_{j+1}(t),L-n_{j+2}(t)]$ is the number
of vehicles that move two cells forward. The idea behind this
dynamics is that first the vehicles try to move two cells forward.
This is only possible if the two cells ahead are not fully 
occupied, i.e.\ $L-n_{j+1}(t)>0$ and $L-n_{j+2}(t)>0$.
Then $\min[b_j(t)-a_j(t),L-n_{j+1}(t)-a_{j-1}(t)]$ vehicles
move forward one cell.

The model defined by (\ref{eq_gBCA}) can be considered as a
generalization of the Fukui-Ishibashi model (see Sec.\ \ref{sec_FI})
to which it reduces for $L=1$. Note that the flow limiter $M$
has been dropped in the extended model.

The fundamental diagram of the model (\ref{eq_gBCA}) has a structure
similar to that shown in Fig.~\ref{hyssch}, i.e.\ states of high flow
exist. Due to the higher velocity the degeneracy found in the
model (\ref{eq_BCAdef}) is lifted. In the case $L=2$, the $c=1/2$
states $\ldots 1111\ldots$ and $\ldots 2020 \ldots$ are degenerate
in the simple model.
In the $v_{max}=2$ case, however, in the state $\ldots 1111\ldots$ the
vehicles can increase their velocity, while in the state
$\ldots 2020 \ldots$ they can not due to the presence of
the fully-occupied cells '2'. The high-flow states are unstable 
against local perturbations.

Another CA model obtained by ultra-discretization of the
modified KdV-equation has been suggested in \cite{nagaov4}.
It is a second-order CA since the configuration at time $t+1$ does not
only depend on the configuration at time $t$, but also on the 
previous one at time $t-1$. The update rule for the position $x_j(t)$
of vehicle $j$ is given by
\begin{equation}
x_j(t+2)=x_j(t)+\Delta x_j(t)+\max[0,\Delta x_j(t+1)-M]
-\max[0,\Delta x_j(t+1)-L]
\end{equation}
where $\Delta x_j(t)=x_{j+1}(t)-x_j(t)$ is the gap and $M$ 
and $L$ are constants.

Although this model is deterministic it exhibits start-stop waves 
similar to those found in the NaSch model \cite{nagaov4}. It
seems that a second-order deterministic CA might produce 
effects similar to those of noise in a first-order stochastic CA.
One should note, however, that the rules allow vehicles to move
backwards.
Other models obtained using the UDM have been discussed in 
\cite{otherultra}.


\section{Cellular automata models of city traffic and road networks}
\label{Sec_city}

\subsection{Biham-Middleton-Levine model and its generalizations}

In the BML model \cite{bml}, each of the sites of a square lattice
represent the crossing of a east-west street and a north-south street.
All the streets parallel to the $\hat{x}$-direction of a Cartesian
coordinate system are assumed to allow only {\it single-lane}
east-bound traffic while all those parallel to the $\hat{y}$-direction
allow only single-lane north-bound traffic. Let us represent the
east-bound (north-bound) vehicles by an arrow pointing towards east
(north).  In the initial state of the system, vehicles are {\it
randomly} distributed among the streets. The states of east-bound
vehicles are updated in parallel at every odd discrete time step
whereas those of the north-bound vehicles are updated in parallel at
every even discrete time step following a rule which is a simple
extension of the TASEP: a vehicle moves forward by one lattice
spacing if and only if the site in front is empty, otherwise the
vehicle does not move at that time step.

Thus, the BML model is also a driven lattice gas model where each of
the sites can be in one of the three possible states: either empty or
occupied by an arrow $\uparrow$ or $\rightarrow$.  Note that the
parallel update rules of the BML model is fully {\it deterministic}
and, therefore, it may also be regarded as a deterministic CA. The
randomness arises in this model only from the {\it random initial
  conditions} \cite{fuishjp96}.  Suppose, $N_{\rightarrow}$ and
$N_{\uparrow}$ are the numbers of the east-bound and north-bound
vehicles, respectively, in the initial state of the system. If
periodic boundary conditions are imposed in all directions, the number
of vehicles in {\it every street} is conserved since no turning of the
vehicles are allowed by the updating rules. In a finite $L \times L$
system the densities of the east-bound and north-bound vehicles are
given by $c_{\rightarrow} = N_{\rightarrow}/L^2$ and $c_{\uparrow} =
N_{\uparrow}/L^2$, respectively, while the global density of the
vehicles is $c = c_{\rightarrow} + c_{\uparrow}$.

Computer simulations of the BML model with periodic boundary
conditions demonstrate that a {\it first order} phase transition takes
place at a finite non-vanishing density $c_*$, where the average
velocity of the vehicles vanishes discontinuously signaling complete
jamming; this jamming arises from the mutual blocking ("grid-locking")
of the flows of east-bound and north-bound traffic at various
different crossings (see \cite{bmlopen} for the corresponding results
of the BML model with {\it open} boundary conditions).

At concentrations just above $c_*$, in the jammed phase, all the
vehicles together form a {\it single} cluster which is stretched along
the diagonal connecting the {\it south-west} to the {\it north-east}
of the system. In other words, the lowest-density jammed
configurations consist of a single diagonal band where the
$\rightarrow$ and $\uparrow$ occupy nearest-neighbour sites on the
band in a zigzag manner. With further increase of
density more and more vehicles get attached to the band in the form of
off-diagonal branches and the infinite cluster of the jammed vehicles
looks more and more random. Thus, in general, a
typical infinite cluster of the jammed vehicles consists of a
"backbone" and "dangling vehicles" which are the analogs of the
"backbone" and the "dangling ends" of the infinite percolation
clusters in the usual site/bond percolation \cite{stauaha}. However,
in contrast to the infinite percolation cluster in the usual random
site/bond percolation, the infinite spanning cluster of vehicles in
the BML model emerges from the self-organization of the system.
Nevertheless, concepts borrowed from percolation theory have been used
to characterize the structure of the infinite cluster of jammed
vehicles in the BML model at $c > c_*$ \cite{tadaki,gupta}.  The
distribution of the waiting times of the vehicles at the signals
(i.e., at the lattice sites) has also been investigated through
computer simulations \cite{wait1,wait2}.

%

\subsubsection{Poor man's mean-field estimates for the BML model}

If one is not interested in detailed information on the "structure"
of the dynamical phases, one can get a mean-field estimate of $c_*$ by
carrying out a back-of-the-envelope calculation.  Suppose,
$c_{\rightarrow}$ and $c_{\uparrow}$ denote the average densities
while $v_{\rightarrow}$ and $v_{\uparrow}$ denote the {\it average}
speeds of the east-bound and north-bound vehicles, respectively. In
order to estimate $c_*$ one has to take into account interaction of
the east-bound (north-bound) vehicles not only with the north-bound
(east-bound) vehicles \cite{nagajp93} but also, in a self-consistent
manner, with other east-bound (north-bound) vehicles \cite{wfh96a}.
Following the arguments of Appendix \ref{App_microBML}, one can show
that in the symmetric case $c_{\rightarrow} = c_{\uparrow} = c$, a
self-consistency equation for the speed is
\begin{equation}
v_{\rightarrow} = v_{\uparrow} = v = \frac{1}{2}\left[1+ \frac{c}{2}
+ \sqrt{\left(1 + \frac{c}{2}\right)^2 - 4 c}\right]
\label{mf-bml}
\end{equation}
for $c < c_*$. The critical density $c_*$ is determined by the
condition that at $c \geq c_*$ the equation (\ref{mf-bml}) does not
give a real solution. Hence we get $c = c_* = 6 - \sqrt{32} \simeq
0.343$ which, in spite of the approximations made, is surprisingly
close to the corresponding numerical estimate obtained from computer
simulation \cite{bml}.  Moreover, the mean-field estimate $c_* \simeq
0.343$ is also consistent with the more rigorous 
result\footnote{Ishibashi and Fukui \cite{Ishi2d} claimed that
  complete jamming can occur in the BML model only for $c = 1$.
  However, a plausible flaw in their arguments was pointed out by 
  Chau et al.~\cite{chau1}. In \cite{shi} it has been argued that
  $c_*\propto L^{-0.14}$, i.e.\ $c_*=0$ in the thermodynamic limit.} 
\cite{chau1,chau2} that $c_*$ is strictly less than $1/2$. The BML
model in three dimension, although not relevant for vehicular traffic,
has also been studied \cite{chau3}.

\subsubsection{Mean-field theory of the BML model}

Recall that the occupation variables $n(i;t)$ in the NaSch model
describe the state of occupation of the sites $i$ ($i = 1,2,...$) by
the vehicles on the one-dimensional highway. A scheme for a truly
microscopic analysis of the BML model begins \cite{mmcb95} by
introducing the corresponding generalized occupation variables
$n_{\uparrow}(x,y;t)$ and $n_{\rightarrow}(x,y;t)$, which describe the
state of occupation of the sites $(x,y)$ by the north-bound and
east-bound vehicles, respectively, on the two-dimensional
street-network. The analogs of the equations (\ref{mf-c0}) and
(\ref{mf-c1}) (see Appendix \ref{App_microBML} for details) have
analogous physical interpretations. As usual, in the naive SOMF
approximation one neglects the correlations between the occupations of
different sites \cite{mmcb95}. Since none of the sites is allowed to be
occupied by more than one particle at a time, Pauli operators may be
used to develop an analytical theory of vehicular traffic
\cite{kaulke} but one should keep in mind that the particles
representing the vehicles are purely classical and the system does not
have any quantum mechanical characteristics.

\subsubsection{Generalizations and extensions of the BML model}

The BML model has been generalized and extended to take into account
several realistic features of traffic in cities.

\noindent$\bullet$ {\bf Asymmetric distribution of the vehicles:}

Suppose the vehicles are distributed asymmetrically among the
east-bound and north-bound streets \cite{nagajp93}, i.e.,
$c_{\rightarrow} \neq c_{\uparrow}$. For convenience, let us write
$c_{\uparrow} = c f_a$ and $c_{\rightarrow} = c (1-f_a)$ where $f_a$
is the fraction of the vehicles moving towards north.  Clearly $f_a =
1/2$ corresponds to the symmetric case $c_{\rightarrow} = c_{\uparrow}
= c/2$. On the other hand, $f_a = 0$ ($f_a = 1$) correspond to the
extreme asymmetric case where all the vehicles are east-bound
(north-bound). Obviously, the absence of grid-locking in the extreme
limits $f_a = 0$ and $f_a = 1$ rules out the possibility of BML-like
complete jamming transition, i.e., $c_* = 1$ for both $f_a = 0$ and
$f_a = 1$. Moreover, $c_*$ decreases with decreasing asymmetry in the
distribution of the vehicles; $c_*$ is the smallest for $f_a = 1/2$,
i.e., symmetric distribution of the vehicles.  These results can be
presented graphically by plotting the curve $c_*(f_a)$ in the phase
diagram on the $c-f_a$ plane \cite{nagajp93}.

\noindent$\bullet$ {\bf Unequal maximum velocities:}

In the BML model both east-bound and north-bound vehicles can move by
a maximum of one lattice spacing at a time and, therefore, the average
speeds of both types of vehicles can never exceed unity. On the other
hand, recall that in the Fukui-Ishibashi model \cite{fukuishi} of
highway traffic vehicles can move up to a maximum of $M$ lattice sites
at a time and, hence, have average velocities larger than unity.
Incorporating similar high-speed vehicles Fukui et al.\ \cite{fukubml}
generalized the BML model where the east-bound vehicles are allowed tom
ove by $M$ sites at a single time step while the north-bound vehicles
can move by only one lattice spacing.

\noindent$\bullet$ {\bf Overpasses or two-level crossings:}

The BML model has been extended to take into account the effects of
overpasses (or two-level crossings) ~\cite{nagapr93}. A fraction $f_o$
of the lattice sites in the BML model are randomly identified as
overpasses each of which can accommodate up to a maximum of two
vehicles simultaneously. The overpasses weaken the grid-locking in the
BML model. Therefore, $c_*$ is expected to increase with increasing
$f_o$. Besides, $c_*$ is expected to be unity if $f_o = 1$. Naturally,
we address the question: does jamming disappear (i.e., $c_*$ is unity)
only at $f_o = 1$ or for even smaller values of $f_o$? In order to
answer this question we extend the self-consistent mean-field
arguments, which led to the equation (\ref{mf-bml}), incorporating the
effects of the overpasses thereby getting the generalized
self-consistency equation \cite{wfh96b}
\begin{equation}
v_{\rightarrow} = v_{\uparrow} = v = \frac{1}{2}\left[1 
+ \frac{1-f_o}{2}~c + \sqrt{\left(1 + \frac{1-f_o}{2}~c \right)^2 
- 4 (1-f_o)c}\right]
\label{mf-bmlop}
\end{equation}
in the symmetric case $c_{\rightarrow} = c_{\uparrow} = c$.  The
equation (\ref{mf-bmlop}) reduces to the equation (\ref{mf-bml}) in
the limit $f_o = 0$. It predicts that, if $f_o \neq 0$, a moving phase
exists in the BML model with overpasses for vehicle densities $c \leq
c_* = (6 - \sqrt{32})/(1-f_o)$ and that the jammed phase disappears
altogether (i.e., $c_*$ becomes unity) for $f_o \geq 1 - 0.343 =
0.657$, an underestimate when compared with the corresponding computer
simulations. These results can be presented graphically by drawing the
curve $c_*(f_o)$ in the mean-field phase diagram on the $c-f_o$ plane
not only for the symmetric distribution (i.e., for $f = 1/2$) but also
for asymmetric distributions of the vehicles among the east-bound and
north-bound streets \cite{wfh96b}.

\noindent$\bullet$ {\bf Faulty traffic lights:}

The effects of faulty traffic lights have been modeled by generalizing
the BML as follows \cite{chung95}: a fraction $f_{tl}$ of the sites
are identified randomly as the locations of the faulty traffic lights.
Both the north-bound vehicles currently south of the faulty traffic
lights and the east-bound vehicles currently west of a faulty traffic
light are allowed to hop onto the {\it empty} crossings where the
faulty traffic lights are located, irrespective of whether the
corresponding time step of updating is odd or even. If an east-bound
vehicle and a north-bound vehicle simultaneously attempt to enter the
same crossing, where the faulty traffic light is located, then only
one of then is allowed to enter that crossing by selecting randomly.

Since a north-bound (east-bound) vehicle will be able to move forward
although only east-bound (north-bound) vehicles would have moved
forward had $f_{tl}$ been zero, the average velocity of the vehicles
is expected to increase with increasing $f_{tl}$. However, with the
increase of $f_{tl}$ there is an increasing likelihood that an
east-bound (north-bound) vehicle would be blocked by a north-bound
(east-bound) vehicle at a faulty traffic light.  Thus, the increasing
density of faulty traffic lights increases the effect of grid-locking
thereby decreasing $c_*$.

\noindent$\bullet$ {\bf Static hindrances:}

The BML model has been extended to incorporate the effects of static
hindrances or road blocks (e.g., vehicles crashed in traffic
accident), i.e., stagnant points \cite{acci,gu}. A vehicle, which stays
at a normal site for only one time step before attempting to move out
of it, stays at a point-like blockage for $T_p$ time steps before
attempting to move out of it. Obviously, the longer is $T_p$, the
lower is the corresponding $c_*$. The time-dependent phenomenon of
spreading of the jam from the blockage site during the approach of the
system to its jammed steady-state configurations has also been
investigated \cite{acci,acci2}.
 
\noindent$\bullet$ {\bf Stagnant street:}

Let us consider the effects of a stagnant street, where the local
density $c_s$ of the vehicles is initially higher than that in the
other streets \cite{stag}, on the traffic flow in the BML model. The
stagnant street, effectively, acts like a "line-defect", rather than a
"point-defect". However, in contrast to the static roadblocks, a
stagnant street offers a time-dependent hindrance to the vehicles
moving in the perpendicular streets. As intuitively expected, the
jamming transition has been found to occur at a lower global density
when the local density in the stagnant street is higher.

\noindent$\bullet$ {\bf Independent turning of the vehicles:}

Let us now generalize the BML model by assigning a {\it trend} or
preferred direction of motion, $W_n(x,y)$, to each vehicle $n$ ($n =
1,2,...$) located at the site $x,y$. According to this definition, the
vehicle $n$, located at $x,y$ jumps to the next site towards {\it
  east} with probability $W_n(x,y)$ while $1-W_n(x,y)$ is the
corresponding probability that it hops to the next site towards {\it
  north} \cite{turn}. In this generalized model vehicles can take a
turn but the processes of turning from east-bound (north-bound) to
north-bound (east-bound) streets is {\it stochastic}.  For simplicity,
suppose, $N/2$ vehicles are assigned $W_n(x,y) = \gamma$ while the
remaining $N/2$ vehicles are assigned $W_n(x,y) = 1-\gamma$ where $0
\leq \gamma \leq 1/2$; this implies that $N/2$ vehicles move
preferentially east-ward whereas the remaining $N/2$ vehicles move
preferentially towards north. In the limit $\gamma = 0$ no vehicle can
turn and we recover the original BML model with deterministic update
rules. The most dramatic effect of the stochastic turning is that the
discontinuous jump $\Delta \langle v\rangle$ of the average velocity
$\langle v\rangle$ decreases with increasing $\gamma$ and, eventually,
the first order jamming transition ends at a critical point where
$\Delta \langle v\rangle$ just vanishes.

\noindent$\bullet$ {\bf Jam-avoiding turn and drive:}

In the model of turning considered in references \cite{turn} a vehicle
turns stochastically independently of the other vehicles. In real
traffic, however, a vehicle is likely to turn if its forward movement
is blocked by other vehicles ahead of it in the same street.
Therefore, let us now consider a model \cite{jamavoidt} where an
east-bound (north-bound) vehicle turns north (east) with probability
$p_{turn}$ if blocked by another vehicle in front of it. Computer
simulations of this model shows that $c_*(p_{turn})$ increases with
increasing $p_{turn}$. In a slightly different model \cite{jamavoidd},
on being blocked by a vehicle in front, an east-bound (north-bound)
vehicle hops with probability $p_{ja}$ to the next east-bound
(north-bound) street towards north (east).

\noindent$\bullet$ {\bf A single north-bound street cutting across
  east-bound streets}

Let us now consider a special situation where only one north-bound
street exists which cuts all of the $L$ equispaced mutually parallel
east-bound streets of length $L$ \cite{nagaseno}.  
This situation can be modeled as
a $L \times L$ square lattice and each cell on the north-bound street
can be in one of the three allowed states, namely, either empty or
occupied by a $\rightarrow$ or a $\uparrow$. But, in contrast to the
BML model, there are {\it two} allowed states for each cell 
(outside the crossings) on the east-bound street; 
these can be either empty or occupied by only
$\rightarrow$.  Since no grid-locking is possible with only one
north-bound street, complete jamming occurs trivially in this case
only if each of the cells either on the east-bound streets or on the
north-bound street or on all the streets are occupied simultaneously
by vehicles.  Nevertheless, at any finite nonvanishing density, the
crossings of the east-bound and north-bound streets act like
hindrances with finite non-vanishing permeability for the flow of the
east-bound traffic. Obviously, the higher is the density of the
north-bound vehicles, the lower is the permeability and the stronger
is the rate-limiting factor of the bottleneck. A comparison of this
problem with that of highway traffic in the presence of static
hindrances \cite{lebo,barma} explains not only why the flux along the
east-bound streets exhibits a flat plateau over an intermediate
density regime but also why the plateau appears at lower values of the
flux with increasing density of the north-bound vehicles
\cite{nagaseno}. A mean-field theory for the macroscopic phase
segregation in this model has been developed by appropriately
modifying that for the similar phenomenon in the one-dimensional
models of highway traffic in the presence of static hindrances.  It is
worth emphasizing here that in \cite{nagaseno} each site on the
east-bound streets has been interpreted as a cell, which can
accommodate one vehicle at a time, rather than as a crossing of the
east-bound street with a north-bound street. Conceptually, this is an
extension of the BML model. Finally, we briefly mention that in
\cite{IshiF2street} the phase diagram of a system consisting of 
one east-bound and one north-bound street with one crossing has been
investigated.

\noindent$\bullet$ {\bf Green-wave synchronization} 

\begin{figure}[ht]
\begin{center}
\psfig{figure=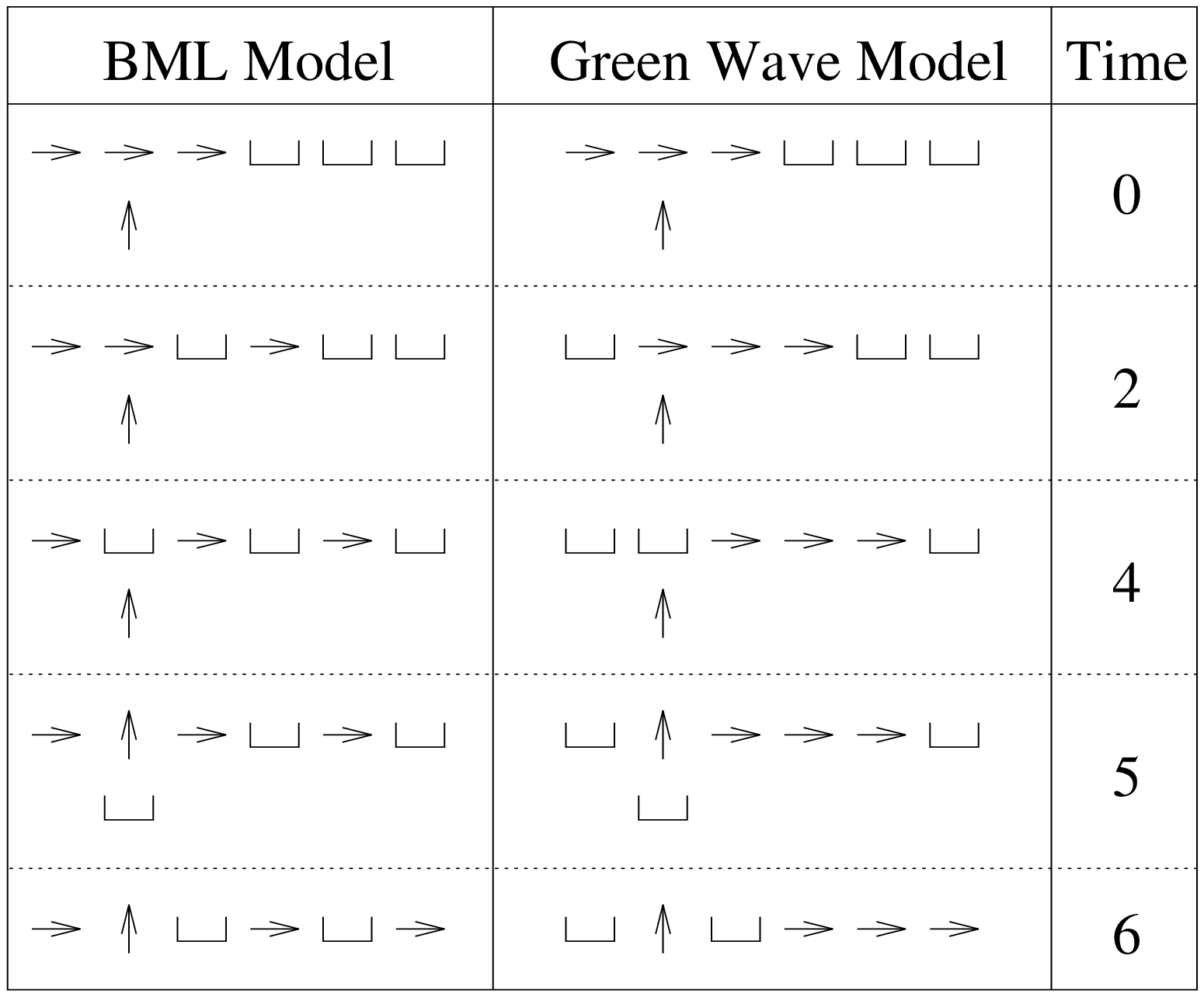,bbllx=65pt,bblly=190pt,bburx=540pt,bbury=600pt,width=10cm}
\caption{Comparison of the update procedure in the BML model (left)
and the GW model (right) (from \cite{greenwave}).}
\label{bml_gwm.fig}
\end{center}
\end{figure}

Often the traffic lights along the main streets in cities are
synchronized to allow continuous flow; this is usually referred to as
"green-wave" synchronization. A green-wave (GW) model has been
developed \cite{greenwave} by replacing the parallel updating scheme
of the BML model by an updating scheme which is partly
backward-ordered sequential (see Appendix \ref{App_updates} for a
general explanation of this update scheme).  At odd time steps, an
east-bound vehicle moves by one lattice spacing if the target site was
empty at the end of the previous time step or has become empty in the
current time step (this is possible because of the backward-ordered
sequential updating at every time-step). Similarly the positions of
the north-bound vehicles are updated at every even time step.  The
main difference between the BML model and the GW model (see
Fig.~\ref{bml_gwm.fig}), arising from the different updating schemes,
is that in the GW model vehicles move as "convoys" (a cluster of
vehicles with no empty cell between them) thereby mimicking the
effects of green-wave synchronization of the traffic lights in real
traffic. The jamming transition in the GW model has been investigated
by a combination of a mean-field argument and numerical input from
computer simulations \cite{greenwave}.

\noindent$\bullet$ {\bf More realistic description of streets and
  junctions}

At first sight the BML model may appear very unrealistic because the
vehicles seem to hop from one crossing to the next. However, it may
not appear so unrealistic if each unit of discrete time interval in
the BML model is interpreted as the time for which the traffic lights
remain green (or red) before switching red (or green) simultaneously
in a synchronized manner, and over that time scale each vehicle, which
faces a green signal, gets an opportunity to move from $j$-th crossing
to the $(j+1)$-th (or, more generally \cite{fukubml}, to the $(j+r)$-th
where $r > 1$).

In the original version of the BML model the vehicles are located on
the lattice sites which are identified as the crossings. Brunnet and
Goncalves \cite{brugon} considered a modified version where, instead,
the vehicles are located on the bonds and, therefore, never block the
flow of vehicles in the transverse direction. Consequently, in this
version of the CA model of city traffic jams of only {\it finite}
sizes can form and these jams have {\it finite} lifetime after which
they disappear while new jams may appear elsewhere in the system; an
infinitely long-lived jam spanning the entire system is possible only
in the trivial limit $c=1$.  In contrast, Horiguchi and Sakakibara
\cite{horiguchi} generalized the BML by replacing each of the bonds
connecting the nearest-neighbour lattice sites by a bond {\it
  decorated} with an extra lattice site in between. 
In \cite{horiguchi2} a generalization to $s$ extra lattice sites
between crossings has been presented. 
However, the vehicles are still allowed to hop forward by only one 
lattice spacing in the model of Horiguchi and Sakakibara.
Moreover, generalizing the rules for turning of the vehicles in
Ref.~\cite{turn}, Horiguchi and Sakakibara also allowed probabilistic
turning of the vehicles in their model. The model exhibits a
transition from the flowing phase to a completely jammed phase.

The streets in the original BML model were assumed to allow only
one-way traffic. This restriction was relaxed in a more realistic
model proposed by Freund and P\"oschel \cite{freund} which allows
both-way traffic on all the streets. Thus, each east-west
(north-south) street is implicitly assumed to consist of two lanes one
of which allows east-bound (north-bound) traffic while the other
allows west-bound (south-bound) traffic.  Moreover, each site is
assumed to represent a crossing of a east-west street and a
north-south street where four numbers associated with the site denote
the number of vehicles coming from the four nearest-neighbour
crossings (i.e., from north, south, east and west) and queued up at
the crossing under consideration. So, in this extended version of the
BML model, each site can accommodate at most $4Q$ particles if each of
the four queues of vehicles associated with it is allowed to grow to a
maximum length $Q$.

In the model proposed by Freund and P\"oschel \cite{freund},
initially, each of the vehicles is assigned a site selected at random,
the queue to which it belongs (i.e., whether it is approaching the
crossing from north,south, east or west) and the desired direction
(i.e., left, right or straight) for its intended motion at the next
time step. At each discrete time step a vehicle is allowed to move
forward in its desired direction of motion by one lattice spacing
provided (a) it is at the front of the queue in its present location
and (b) there are fewer than $Q$ vehicles queued up at the next crossing
in the same desired direction of motion. Once a vehicle moves to the
next crossing it finds a place at the end of the corresponding queue
at the new crossing while the vehicles in the queue it left behind are
moved "closer to the crossing" by one position (by mere relabeling as
no physical movement of the vehicles in the queue takes place
explicitly). Various reasonable choices for the rule which determines
the desired direction of each vehicle at every time step have also
been considered.

The finite space of the streets in between successive crossings do not
appear {\it explicitly} in the extension of the BML model suggested by
Freund and P\"oschel \cite{freund} although it is more realistic than
the BML model because it {\it implicitly} takes into account the
possibility of formation of queues by vehicles approaching one
crossing from another. Chopard et al.  \cite{chopard} have developed a
more realistic CA model of city traffic where the stretches of the
streets in between successive crossings appear explicitly. In this
model also each of the streets consist of two lanes which allow
oppositely directed traffic. The rule for implementing the motion of
the vehicles at the crossing is formulated assuming a {\it rotary} to
be located at each crossing. Depending on the details of the rules to
be followed by the vehicles at the rotary, the system can exhibit a
variety of phenomena. For example, the flow can be metastable at all
densities if each of the vehicles on the rotary is required to stop
till the destination cell becomes available for occupation
\cite{chopard}. Moreover, the bottleneck created by the vehicles on
the rotaries at the junctions can lead to a plateau in the fundamental
diagram which is analogous to that caused by a static hindrances on a
highway \cite{lebo,barma}.

\subsection{Marriage of the NaSch model and the BML model; a "unified" CA
model of city traffic} 

If one wants to develop a more detailed "fine-grained" description of
city traffic than that provided by the BML model then one must first
decorate each bond \cite{nagaseno,horiguchi} with $D-1$ ($D > 1$) sites
to represent $D-1$ cells in between each pair of successive crossings
\cite{nagaseno,chopard} thereby modeling each segment of the streets
in between successive crossings in the same manner in which the entire
highway is modelled in the NaSch model. Then, one can follow the
prescriptions of the NaSch model for describing the positions, speeds
and accelerations of the vehicles \cite{chopard,simon} as well as for
taking into account the interactions among the vehicles moving along
the same street. Moreover, one should flip the color of the signal
periodically at regular interval of $T$ ($T \gg 1$) time steps where,
during each unit of the discrete time interval every vehicle facing
green signal should get an opportunity to move forward from one cell
to the next. Such a CA model of traffic in cities has, indeed, been
proposed very recently ~\cite{cs,csprep} where the rules of updating 
have been formulated in such a way that, (a) a vehicle approaching a
crossing can keep moving, even when the signal is red, until it
reaches a site immediately in front of which there is either a halting
vehicle or a crossing; and (b) no grid-locking would occur in the
absence of random braking.

Let us model the network of the streets as a $N \times N$ square
lattice. For simplicity, let us assume that the streets parallel to
$\hat{x}$ and $\hat{y}$ axes allow only single-lane east-bound and
north-bound traffic, respectively, as in the original formulation of
the BML model. Next, we install a signal at every site of this $N
\times N$ square lattice where each of the sites represents a crossing
of two mutually perpendicular streets.  We assume that the separation
between any two successive crossings on every street consists of $D$
cells so that the total number of cells on every street is $L = N
\times D$. Each of these cells can be either empty or occupied by at
most one single vehicle at a time. Because of these cells, the network
of the streets can be viewed as a decorated lattice. However, unlike
the BML model \cite{bml}, which corresponds to $D = 1$, and the model
of Horiguchi and Sakakibara \cite{horiguchi}, which corresponds to $D
= 2$, $D ~(< L)$ in this model is to be treated as a parameter.  Note
that $D$ introduces a new length scale into the problem.

The signals are synchronized in such a way that all the signals remain
green for the east-bound vehicles (and simultaneously, red for the
north-bound vehicles) for a time interval $T$ and then,
simultaneously, all the signals turn red for the east-bound vehicles
(and, green for the north-bound vehicles) for the next $T$ timesteps. 
Clearly, the parameter $T$
introduces a new time scale into the problem. Thus, in contrast to the
BML model, the forward movement of the individual vehicles over
smaller distances during shorter time intervals are described
explicitly in this "unified" model.

If no turning of the vehicles is allowed, as in the original BML
model, the total number of vehicles on each street is determined by
the initial condition, and does not change with time because of the
periodic boundary conditions.
Following the prescription of the NaSch model, we allow the speed $v$ of
each vehicle to take one of the $v_{max}+1$ {\it integer} values
$v=0,1,...,v_{max}$.  Suppose, $v_n$ is the speed of the $n$-th
vehicle at time $t$ while moving either towards east or towards north.
At each {\it discrete time} step $t \rightarrow t+1$, the arrangement
of $N$ vehicles is updated {\it in parallel} according to the
following "rules":

\begin{figure}[hbt]
\epsfxsize=\columnwidth\epsfbox{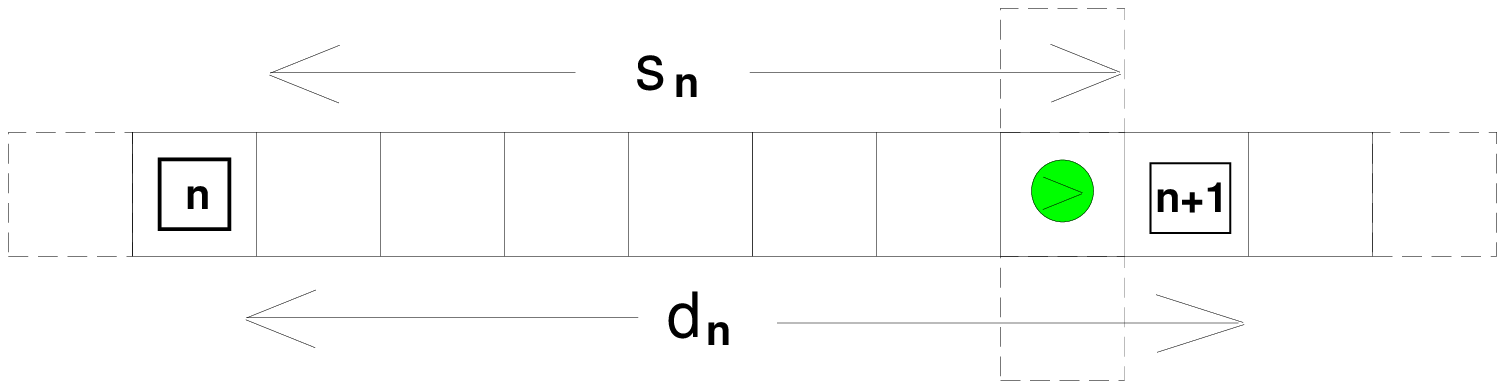}
\caption{A segment of a east-west street, where the $n$-th and
  the $(n+1)$-th vehicles are located, is shown schematically together
  with one of the nearest signals.}
\label{jamconf}
\end{figure}

\noindent {\it Step 1: Acceleration.}\\
$v_n \rightarrow \min(v_n+1,v_{max})$.

\noindent{\it Step 2: Deceleration (due to other vehicles or signal).}\\
Suppose, $d_n$ is the gap in between the $n$-th vehicle and the
vehicle in front of it, and $s_n$ is the distance between the same
$n$-th vehicle and the closest signal in front of it 
(see Fig.~\ref{jamconf}).
\begin{description}
\item{{\sl Case I}}: The signal is {\bf red} for the $n$-th vehicle under
consideration:\\
$v_n \rightarrow \min(v_n,d_n-1,s_n-1)$.
\item{{\sl Case II}}: The signal is {\bf green} for the $n$-th vehicle under
consideration:\\
Suppose, $\tau$ is the number of the remaining time steps before the
signal turns red. Now there are two possibilities in this case:\\
$(i)$ When $d_n \leq s_n$, then $v_n \rightarrow \min(v_n,d_n-1)$.
The motivation for this choice comes from the fact that, when
$d_n \leq s_n$, the hindrance effect comes from the leading vehicle.\\
$(ii)$ When $d_n > s_n$, then $v_n \rightarrow \min(v_n,d_n - 1)$
if $\min(v_n,d_n-1) \times \tau > s_n$;
else $v_n \rightarrow \min(v_n,s_n-1)$.
The motivation for this choice comes from the fact that, when
$d_n > s_n$, the speed of the $n$-th vehicle at the next time step
depends on whether or not the vehicle can cross the crossing in
front before the signal for it turns red.
\end{description}
\noindent{\it Step 3: Randomization.}\\
$v_n \rightarrow \max(v_n-1,0)$ with probability $p$ ($0 \leq p \leq 1$);
$p$, the random deceleration probability, is identical for all
the vehicles and does not change during the updating.

\noindent{\it Step 4: Vehicle movement.} \\
For the east-bound vehicles, $x_n \rightarrow  x_n + v_n$
while for the north-bound vehicles, $y_n \rightarrow  y_n + v_n$.

The rule in case II of step 2 can be simplified without
changing the overall behaviour of the model \cite{csprep}:

\noindent{\sl Case II':}\\
If the signal turns to red in the next timestep ($\tau=1$):\\             
$v_n \rightarrow \min ( v_n, s_n -1, d_n-1)$\\
else\\
$v_n \rightarrow \min ( v_n, d_n-1)$.

These rules are not merely a combination of the rules proposed 
by Biham et al.~\cite{bml} and those introduced by Nagel and
Schreckenberg \cite{ns} but also involve some modifications.
For example, unlike all the earlier BML-type models, a vehicle 
approaching a crossing can keep moving, even when the signal is 
red, until it reaches a site immediately in front of which there 
is either a halting vehicle or a crossing. Moreover, if $p=0$ 
every east-bound (north-bound) vehicle can adjust speed in the 
deceleration stage so as not to block the north-bound (east-bound) 
traffic when the signal is red for the east-bound (north-bound) 
vehicles.

Initially, we put $N_{\rightarrow}$ and $N_{\uparrow}$ vehicles  
at random positions on the east-bound and north-bound streets, 
respectively. We update the positions and velocities of the 
vehicles in parallel following the rules mentioned above.
After the initial transients die down, at every time step $t$, 
we compute the average velocities $\langle v_x(t)\rangle$ and 
$\langle v_y(t)\rangle$ of the east-bound and north-bound 
vehicles, respectively.

\begin{figure}[hbt]
\centerline{\psfig{figure=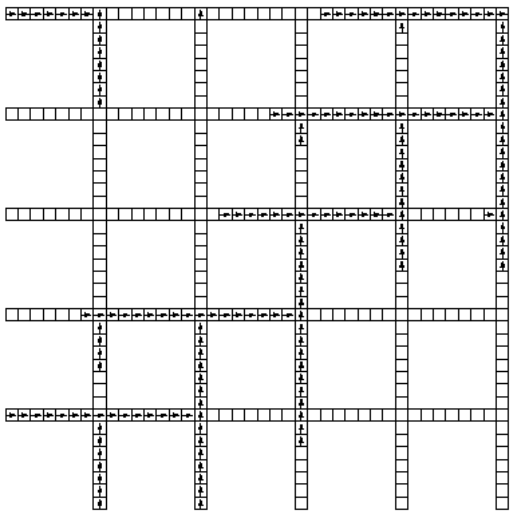,height=7cm}}
\caption{A typical jammed configuration of the vehicles ($N=5$, $D=8$).
The east-bound and north-bound vehicles are represented by the
symbols $\rightarrow$ and $\uparrow$, respectively.}
\label{fig8}
\end{figure}

A phase transition from the "free-flowing" dynamical phase to the
completely "jammed" phase takes place in this model at a vehicle
density $c_*(D)$. The dependence on the dynamical parameters $p$,
$v_{max}$ and $T$ is not clear yet \cite{csprep}. The data obtained 
so far from computer simulations do not conclusively rule out 
the possibility that the transition density only depends on the
structure of the underlying lattice, similar to the percolation
transition \cite{stauaha}, and is independent of the 
dynamical parameters. The intrinsic
stochasticity of the dynamics, which triggers the onset of jamming, is
similar to that in the NaSch model, while the phenomenon of complete
jamming through self-organization as well as the final jammed
configurations (see Fig.~\ref{fig8}) are similar to those in the BML
model.  The variations of $\langle v_x \rangle$ and $\langle v_y
\rangle$ with time as well as with $c$, $D$, $T$ and $p$ in the
flowing phase are certainly more realistic that in the BML model
\cite{cs}.

\begin{figure}[hbt]
\centerline{\epsfig{figure=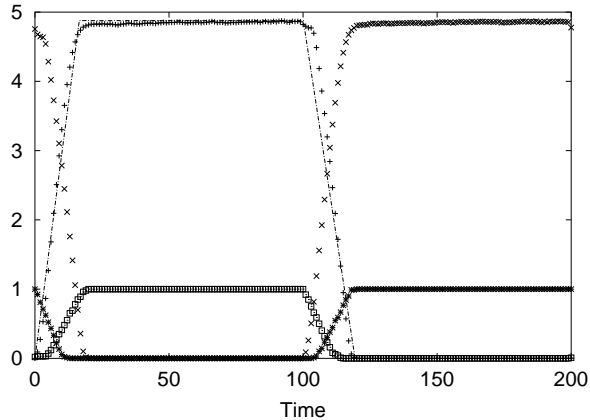,width=8cm}}
\caption{Time-dependence of average speeds of vehicles. The symbols
$+$, $\times$, $\ast$ and $\Box$ correspond, respectively, to
$\langle v_x\rangle$, $\langle v_y\rangle$, and the 
fractions of vehicles with instantaneous speed $v = 0$, 
$f_{x0}$ and $f_{y0}$, respectively. 
The common parameters are $v_{max} = 5, p = 0.1, D = 100$,
$T = 100$ and $c = 0.1$. The continuous line has been obtained from 
heuristic arguments given in \cite{cs}.}
\label{v5d100_t100}
\end{figure}

The "unified" model has been formulated intentionally to keep 
it as simple as possible and at the same time capture some of 
interesting features of the NaSch model as well as the BML model. 
We believe that this model can be generalized (i) to allow 
traffic flow both ways on each street which may consist of 
more than one lane, (ii) to make more realistic rules for the 
right-of-the-way at the crossings and turning of the vehicles, 
(iii) to implement different types of synchronization or 
staggering of traffic lights \cite{simon} (including green-wave), 
etc.

\subsection{Practical applications of the models of vehicular traffic; 
On-line simulation of traffic networks}

A large fraction of the available resources are spent by the
governments, particularly in the industrialized developed
countries, to construct more highways and other infrastructural
facilities related to transportation.
The car-following models, the coupled-map lattice models as well 
as the CA models have been used for computer simulation with a 
hope to utilize the results for on-line traffic control.
For {\it planning and design} of the transportation network
\cite{nagel99}, for example, in a metropolitan area
\cite{duisburg,dallas,geneva}, one needs much more than just
{\it micro-simulation} of how vehicles move on an idealized linear or
square lattice under a specified set of vehicle-vehicle and
road-vehicle interactions. For such a simulation, to begin with,
one needs to specify the roads (including the number of lanes,
ramps, bottlenecks, etc.) and their intersections. Then, times
and places of the activities, e.g., working, shopping, etc.,
of individual drivers are planned.  Micro-simulations are
carried out for all possible different routes to execute
these plans; the results give informations on the efficiency
of the different routes and these informations are utilized in
the designing of the transportation network \cite{nagel99}. 
Some socio-economic questions as well as questions on the 
environmental impacts of the planned transportation infrastructure 
also need to be addressed during such planning and design.
For a thorough discussion of these aspects we refer to the
recent review by Nagel et al. \cite{nagel99}.

\section{Some related systems, models and phenomena} 
\label{Sec_related}

In this section we will briefly present some stochastic models
and phenomena which are somehow related to the main topic of this
review. For some models the relation to traffic flow is rather obvious,
e.g.\ there are toy models which share some features with traffic
models, but which can be solved exactly. We also point out similarities
in the description of other phenomena, e.g.\ granular flow or
surface growth. Similarities also exist to systems from solid state
physics, namely ionic conductors. This point will not discussed
here, instead we refer to refs.\ \cite{ionic,trimper}.
Finally we like to mention that there are some resemblances
between the CA models of vehicular traffic and the CA models of driven
diffusive Frenkel-Kontorova-type systems \cite{fktype}.

\subsection{Generalizations of the TASEP}

The TASEP is probably one of the best studied models in nonequilibrium
physics. Using powerful methods like MPA or Bethe Ansatz recently
it became possible to calculate not only simple expectation values
in the stationary states, but also more complicated quantities
like diffusion constants or the large-deviation function. For
a more detailed discussion and a list of references we refer to the
recent reviews on the TASEP \cite{derr_review,derrida19} (see also
\cite{gs}).
Several variants of the TASEP have been proposed. Most of those 
preserve the exact solvability. 
In the following we will discuss briefly some of the variants and
generalizations of the TASEP discussed so far in the literature.

\noindent ${\bullet}$ {\bf Partially asymmetric exclusion process}\\
An obvious generalization of the TASEP is to allow hopping processes 
in both directions \cite{partASEP}. Here only results for the
random-sequential update exist, since for parallel dynamics ambiguities
in the updating appear when two particles attempt to hop onto the 
same site. One finds that the phase diagram looks essentially like
that of the TASEP with three different phases (see Sec.\ \ref{sub_openBC}).
Recently it has been shown \cite{sasa99}, however, that the high- and 
low-density phases can be divided into three subphases (AI, AII, AIII 
and BI, BII, BIII) instead of two in the TASEP case. 
Again the phase boundaries of these subphases are determined by
the behaviour of the density profiles and the corresponding correlation
lengths.

\noindent ${\bullet}$ {\bf Multispecies models I}\\
Karimipour and collaborators \cite{karimi} have developed 
a multispecies generalization of the TASEP which retains 
the solvability by MPA. It is similar to the disordered
model discussed in Sec.\ \ref{sub_BoseEin} where each particle 
is characterized by a hopping rate (also called ``velocity'' in
this context) $v_j$ ($j=1,\ldots,N_s$), i.e.\
there are $N_s$ different ``species''' of particles.
In contrast to the models with quenched disorder discussed in 
Sec.\ \ref{Sec_disorder}, however, overtaking of particles is
possible, i.e.\ the ordering of the particles is not fixed.\\
More specifically the dynamics of the model is defined as follows:
A particle is chosen at random (random-sequential update). If the
particle is of species $j$ and the cell to its right is empty, it 
hops there with rate $v_j$. If the cell is occupied by a particle
of species $l$ and $v_l < v_j$, then they interchange there positions 
with rate $v_j-v_l$. This means that a fast particle can overtake
a slower one with a rate proportional to their velocity difference.

\noindent ${\bullet}$ {\bf Multispecies models II}\\
In several papers multispecies generalizations of the ASEP  have 
been suggested which exhibit phase separation and spontaneous 
symmetry breaking.

Arndt et al.\ \cite{arndt} considered a system of positive and negative
charged particles diffusing on a ring in opposite directions.
Positive particles move to an empty right neighbour and negative
particles move to an empty left neighbour with the same rate $\lambda$.
Furthermore positive and negative particles on neighbouring sites can
exchange their positions. The process $-+ \to +-$ occurs with rate 1,
and the inverse process $+- \to -+$ with a different rate $q$. For
equal densities of positive and negative particles the system
exhibits three phases. For $q<1$ in the thermodynamic limit,
the system organizes itself into configurations consisting of
blocks of the type $000\cdots +++ \cdots ---\cdots$
The dynamics out of theses states is extremely slow.
Translational invariance is broken and the current vanishes.
This phase is called 'pure phase'. For $1<q<q_c$ the system is in the
'mixed phase' which consists of two coexisting phase, the dense
phase and the fluid phase. The dense phase where no vacancies exist
covers a macroscopic region which shrinks to 0 for $q\to q_c$.
It is remarkable that the current for $1<q<q_c$ takes the value
$J=(q-1)/4$ independent of the total density and the rate $\lambda$.
For $q>q_c$ the fluid phase extends through the whole system. There is 
no charge separation and density profiles are uniform. 

Similar results have been found by Evans et al.\ \cite{evansmulti}
in a slightly different model. The dynamics of their 3-species
model\footnote{One species may be interpreted as a vacant site.}
is given by
\begin{equation}
\begin{picture}(130,20)(0,28)
\unitlength=1.0pt
\put(-80,28){$AB$}
\put(-50,26) {$\longleftarrow$}
\put(-43,18) {\footnotesize $1$}
\put(-50,30) {$\longrightarrow$}
\put(-43,39) {\footnotesize $q$}
\put(-20,28){$BA$,}
\put(20,28){$BC$}
\put(50,26) {$\longleftarrow$}
\put(57,18) {\footnotesize $1$}
\put(50,30) {$\longrightarrow$}
\put(57,39) {\footnotesize $q$}
\put(80,28){$CB$,}
\put(120,28){$CA$}
\put(150,26) {$\longleftarrow$}
\put(157,18) {\footnotesize $1$}
\put(150,30) {$\longrightarrow$}
\put(157,39) {\footnotesize $q$}
\put(180,28){$AC$,}
\end{picture}
\end{equation}
i.e.\ the rates are cyclic in $A$, $B$ and $C$ and the numbers $N_A$, 
$N_B$ and $N_C$ of particles of each species is conserved.

In the case $N_A=N_B=N_C$ the dynamics satisfies the detailed balance
condition with respect to a Hamiltonian with long-range asymmetric
interactions. Stationary states are of the form 
$A\cdots AB\cdots BC\cdots C$ and exhibit phase separation, i.e.\
for large separations $r$ the two-point function satisfies
$\lim_{L\to\infty} \left[\langle A_1A_r\rangle -\langle A_1\rangle
\langle A_r\rangle\right] > 0$.


\subsection{Surface growth, KPZ equation and Bethe Ansatz}

In Sec.\ \ref{sub_growth} we have explained how the NaSch model
with $v_{max}=1$ can be mapped onto a stochastic surface growth
model. This connection can be employed to calculate several
properties of the noisy Burgers and KPZ equation exactly.

Gwa and Spohn \cite{gwa} used the Bethe Ansatz (see e.g.\ \cite{BetheA})
to determine the spectrum of the stochastic Hamiltionian (see 
App.\ \ref{App_MPA})
\begin{equation}
\mathcal{H}=-\frac{1}{4} \sum_{j=1}^L\left[\pauli_j\cdot\pauli_{j+1}
-1+i\epsilon\left(\sigma^x_j\sigma^y_{j+1}-\sigma^y_j\sigma^x_{j+1}
\right)\right]
\end{equation}
corresponding to the ASEP with random-sequential updating and
periodic boundary conditions. $\pauli_j=(\sigma^x_j,\sigma^y_j,\sigma^z_j)$
are the standard Pauli matrices at site $j$ and $\epsilon$ is the
asymmetry of the hopping rates, $q_{right}=\half(1+\epsilon)$ and
$q_{left}=\half(1-\epsilon)$.

The ``ground state'' of $\mathcal{H}$ has eigenvalue $0$ and is $L$-fold
degenerate. For a fixed number $N$ of up-spins\footnote{i.e.\ a fixed 
number $N$ of particles} every configuration has equal weight in the
ground state. In order to determine the dynamical scaling exponent
of the noisy Burgers and KPZ equations, Gwa and Spohn investigated
the finite-size behaviour of the energy gap of $\mathcal{H}$. Since
$\mathcal{H}$ is non-hermitian its spectrum is complex. 
The first excited state is then defined as the eigenvalue with
the smallest (positive) real part $E_{gap}$.
In \cite{gwa} it was shown that $E_{gap}\propto L^{-3/2}$ for 
$\epsilon=1$ and $N=L/2$. This implies that the dynamical exponent $z$
for the stationary correlations of the KPZ equation is given by 
$z=3/2$. The dynamical exponent relates temporal and spatial
scaling behaviour on large scale.
Generalizations and related results can be found in
\cite{gwagener}.

By an extension of the Bethe Ansatz method of Gwa and Spohn, Derrida
and Lebowitz \cite{DerrLeb} calculated the large deviation function
(LDF) of the time-averaged current of the TASEP. The LDF is related to
the total displacement $Y(t)$, i.e.\ the total number of hops to the
right minus the total number of hops to the left between time $0$ and
time $t$. In the corresponding growth model $Y(t)$ is the total number of
particles deposited until time $t$. The LDF is then defined as
\begin{equation}
f(y)=\lim_{t\to\infty}\frac{1}{t}\ln\left[Prob\left(\frac{Y(t)}{t}=
\bar{v}+y\right)\right]
\end{equation}
where $\bar{v}=\lim_{t\to\infty}\langle Y(t)/t =\epsilon c(1-c)L^2/(L-1)$
is the mean current for a ring of finite size $L$ and density $c=N/L$.
The results of \cite{DerrLeb} have been extended and generalized in
\cite{DerrLebgen}.

Apart from the treatment of finite systems, the BA can also be
used to solve the master equation for an infinite system with a finite
number of particles \cite{gunterBA}. This allows e.g.\ to study the
collective diffusion of two single particles.


\subsection{Protein synthesis}

You must have noticed in the earlier sections that some of the 
models of traffic are non-trivial generalizations or extensions of 
the TASEP, the simplest of the {\it driven-dissipative} systems 
which are of current interest in nonequilibrium statistical 
mechanics \cite{sz,gs,vp}. Some similarities between these systems 
and a dynamical model of protein synthesis have been pointed out 
\cite{shubio,gs}.

In a simplified picture of the mechanism of biopolymerisation
ribosomes read the genetic information encoded in triplets of
base pairs. They attach to one end of a messenger-RNA and then
move along the chain after adding a monomer to a biopolymer attached
to the ribosome. The type of monomer added depends on the genetic
information read by the ribosome. When the ribosome reaches the other
end of the m-RNA the biopolymer is fully synthesized and the ribosome
is released.

MacDonald et al. \cite{macd} have described the kinetics of this 
process using an ASEP-type model. The m-RNA is represented by a
chain of $L$ sites where each site corresponds to one triplet of
base pairs. The ribosome is given by a hard-core particle covering
$r$ neighbouring sites ($r\approx 20-30$) which moves forward
by one site with rate $q$. At the beginning of the chain particles
are added with rate $\alpha q$ and at the end they are released with
rate $\beta q$. In the idealized case $r=1$ this is exactly the 
TASEP of Sections \ref{Sec_asep} and \ref{sub_openBC}. The
relevant case for the experiments is $\alpha=\beta < 1/2$. The 
exact solution of the TASEP allowed for an explanation of many
aspects of the experiments \cite{shubio}.


\subsection{Granular flow}

Another quasi-one-dimensional driven-dissipative system, which is also
receiving wide attention of physicists in recent years, is the
granular material flowing through a pipe \cite{proc1,proc2}.
Since the fascinating phenomena (e.g.\ size segregation, convection,
standing waves, localized excitations) found in granular materials
have been subject of several excellent reviews \cite{granRevs} we
discuss only briefly the similarities between the clustering
of vehicles on a highway and particle-particle (and particle-cluster) 
aggregation process \cite{similar,gavrilov,ben-naim}.

Obviously both highway traffic and granular flow through a pipe may
be described as quasi-onedimensional systems consisting of discrete
elements (vehicles, grains). The dynamics of these elements is
determined by an intricate interplay between a driving force (driver,
gravitation) and dissipation (braking, inelastic scattering processes).
These similarities already show that both systems can be described
by similar approaches. One important difference between traffic flow
and granular flow exist, however. In granular flow density waves can
move both in and against the direction of the flow whereas in traffic
flow they only move backwards.

An important success of a description of granular flow using the
optimal velocity model (Sec.~\ref{sec_OVM}) is the explanation of
the experimentally observed $f^{-4/3}$-behaviour of the power 
spectrum \cite{moriyama}.


\subsection{The bus route model}

The bus route model (BRM) \cite{cates} has been formulated as a 
one-dimensional lattice with periodic boundary conditions. 
The sites represent cells, each of which may be thought of as a 
bus stop and are labeled by an index $i$ ($i = 1,2,...,L$)
\cite{cates}. Two binary variables $\sigma_i$ and $\tau_i$ are 
assigned to each cell $i$: (i) If the cell $i$ is occupied by a 
bus then $\sigma_i = 1$; otherwise $\sigma_i = 0$. (ii) If cell 
$i$ has passengers waiting for a bus then $\tau_i = 1$; otherwise 
$\tau_i = 0$. Since a cell cannot have simultaneously a bus and 
waiting passengers, let us impose the condition that a cell 
cannot have both $\sigma_i = 1$ and $\tau_i = 1$ simultaneously.  
Each bus is assumed to hop from one stop to the next. 

\begin{figure}[ht]
\centerline{\epsfig{figure=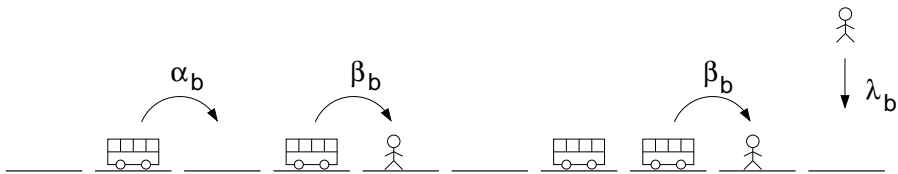,height=2.2cm}}
\caption{The bus-route model (from \cite{cates}).}
\label{bus}
\end{figure}

Next, let us specify the update rules (see Fig.~{\ref{bus}):  
a cell $i$ is picked 
up at random. Then, (i) if $\sigma_i = 0$ and $\tau_i = 0$ (i.e, 
cell $i$ contains neither a bus nor waiting passengers), then 
$\tau \rightarrow 1$ with probability $\lambda$, where $\lambda$ 
is the probability of arrival of passenger(s) at the bus stop.   
(ii) If $\sigma_i = 1$ (i.e., there is a bus at the cell $i$) 
and $\sigma_{i+1} = 0$, then the hopping rate $\mu$ of the bus 
is defined as follows: (a) if $\tau_{i+1} = 0$, then 
$\mu = \alpha_b$ but (b) if $\tau_{i+1} = 1$, then $\mu = \beta_b$,
where $\alpha_b$ is the hopping rate of a bus onto a stop which 
has no waiting passengers and $\beta_b$ is the hopping rate onto 
a stop with waiting passenger(s). Generally, $\beta_b < \alpha_b$, 
which reflects the fact that a bus has to slow down when it 
has to pick up passengers. We can set $\alpha_b = 1$ without 
loss of generality. When a bus hops onto a stop $i$ with waiting 
passengers $\tau_i$ is reset to zero as the bus takes all the 
passengers. Note that the density of buses $c = N/L$ in a conserved 
quantity whereas that of the passengers is not.

An ideal situation in this bus-route model would be one where 
the buses are evenly distributed over the route so that each 
bus picks up roughly the same number of passengers.  However, 
because of some fluctuation, a bus may be delayed and, consequently, 
the gap between it and its predecessor will be larger than the 
average gap. Therefore, this bus has to pick up more passengers 
than what a bus would do on the average, because during the 
period of delay more passengers would be waiting for it and, as a 
result, it would get further delayed. On the other hand, the 
following bus has to pick up fewer passengers than what a bus 
would do on the average and, therefore, it would catch up with 
the delayed bus from behind. The slowly moving delayed bus 
would slow down the buses behind it thereby, eventually, creating 
a jam. In other words, once a larger-than-average gap opens up 
between two successive buses, the gap is likely grow further 
and the steady state in a finite system would consist of a 
single jam of buses and one large gap. This is very similar 
to the Bose-Einstein-condensation-like phenomenon we have 
observed earlier in particle-hopping models with slow impurities.
On the basis of heuristic arguments and mean-field approximation 
it has been argued \cite{cates} that this model exhibits a 
true phase transition from an inhomogeneous low-density phase 
to a homogeneous (but congested) high-density phase only in 
the limit $\lambda \rightarrow 0$.
Finally we mention that the BRM with parallel dynamics has been 
recently been studied in \cite{ChowDes} where also its connection
with the NaSch model has been elucidated.

\subsection{Mobile directional impurities}

We have considered the effects of random hopping probabilities, 
assigned either to the lattice sites \cite{lebo,barma,santen} 
or to the particles \cite{ktitarev,kf,evans} in the TASEP and 
in the NaSch-type models of vehicular traffic, on the nature of 
the corresponding steady states as well as their approach to 
the steady-states starting from random initial conditions. 
Toroczkai and Zia \cite{toroczkai} solved analytically a model 
with one "mobile directional impurity"; this model is also an 
extension of the ASEP. In this model, $N$ particles, labeled 
by integers $1$ to $N$ (from left to right), occupy the sites 
of a one-dimensional lattice of length $N+1$ where periodic 
boundary conditions are applied. Thus, there is a single "hole" 
(i.e., empty site) in this model. The shifting of the hole from 
the {\it site} in between the particles $n$ and $n+1$ to the 
{\it site} in between particles $n+1$ and $n+2$ is described 
by the statement "hole jump from {\it position} $n$ to $n+1$". 
In the absence of any impurity, the hole at position $n$ can 
exchange position with either the particle on its left 
(with probability $W_{n-1,n}$) or the particle on its right 
(with probability $W_{n+1,n}$). These probabilities are arbitrary 
and direction-dependent (i.e., in general, $W_{n-1,n} \neq W_{n+1,n}$) 
but time-independent. Note that the hopping probabilities of 
the hole is determined by the {\it particles} (rather than the 
lattice {\it sites}) in front and behind it. The general case, 
where $W_{n-1,n} \neq W_{n+1,n}$ and the probabilities $W_{n-1,n}$ 
as well as $W_{n+1,n}$ for different $n$ are chosen randomly, 
is referred to as the {\it random} {\it asymmetric} case. 
As the hole wanders, the string of particles also shifts as a whole. 
However, this system is translationally invariant in the sense 
that the jump rate of any particular particle-hole pair is 
independent of its location on the lattice. 

The "directional impurity" is introduced by identifying a 
specific bond (whose position is fixed with respect to the 
lattice) as a defect bond such that the time-independent rates 
of particle-hole exchanges across it are fixed at, say, $q$ 
and $q'$ irrespective of the particle-hole pair involved. 
In other words, when the hole is in between the particles $n$ 
and $n+1$, the hopping probability to the right (left) is always 
$W_{n+1,n}$ ($W_{n-1,n}$), except when the defect bond is 
involved. This model can be represented as in Fig.~\ref{tz}
where the defect bond is shown as a kink; the motivation for 
such a kink came from an earlier model \cite{duke} of gel 
electrophoresis \cite{gel}. 

\begin{figure}[ht] 
\centerline{\epsfig{figure=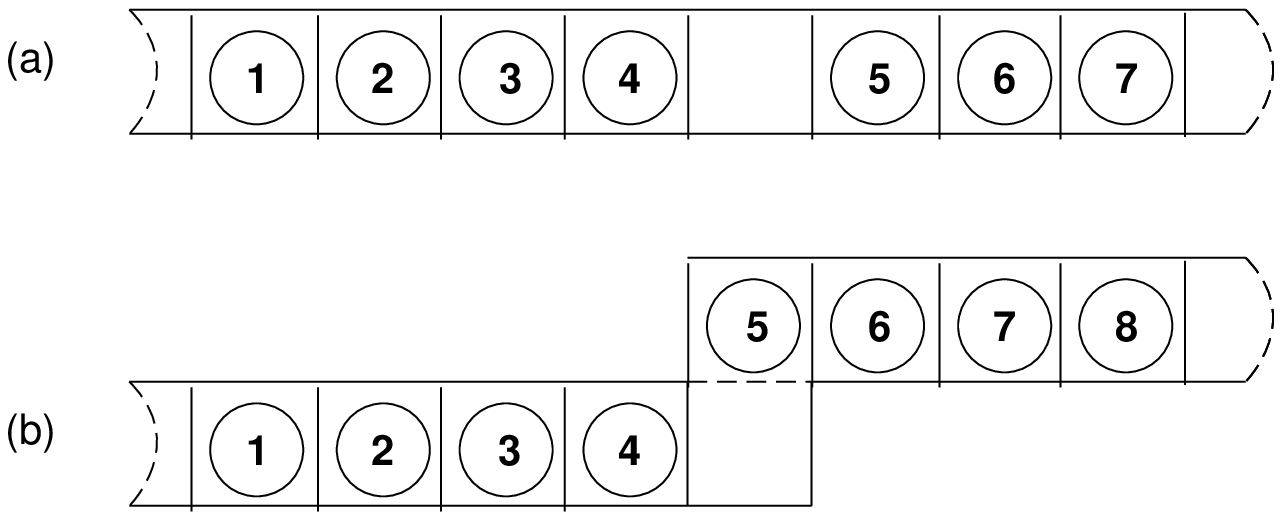,height=4cm}}
\caption{(a) The translationally-invariant Toroczkai-Zia model 
without any defect bond. (b) The translational invariance is 
broken by the defect bond (represented by the kink denoted by 
the dashed line).}
\label{tz}
\end{figure}

More specifically, suppose, the probabilities for the exchange of the
hole with a particle across the defect bond are $q$ if the hole moved
upward and $q'$ if the hole moved downward, {\it independent} of the
particle involved. In the so-called {\it pure} limit, we have, for all
$n$, $W_{n+1,n} = W_{\rightarrow}$ and $W_{n-1,n} = W_{\leftarrow}$.
In this limit, the hole can also be regarded as a particle undergoing
biased diffusion (if $W_{\rightarrow} \neq W_{\leftarrow}$) everywhere
except across across a specific defect bond. If we make the further
assumption that no backward motion of any particle is allowed, then
this model reduces to the model of TASEP with a single blockage
\cite{lebo}.


\subsection{Computer networks}

Inspired by the recent success of the methods of statistical 
mechanics outside the traditional domain of physics, tools 
of statistical mechanics have also been applied to analyze 
fundamental properties of information traffic on the international 
network of computers (Internet) \cite{csabai,taka,vandewalle,ohira}. 
Messages in the form of information packets are continuously 
being emitted from the hundreds of millions of host computers 
and transported to their destination computers through this 
network. Each of these packets is relayed through the so-called 
routers on its way. The routers can deal with the packets one 
by one. Each router has a finite buffer where the arriving packets 
get queued up and forwarded one by one from the head of the 
remaining queue to their respective next destinations. Since 
packets run with the velocity of light through the cables, 
information congestion does not take place inside the cables. 
It is the routers which give rise to the information congestion 
on the internet. Measuring  the fluctuations in the round-trip 
time taken by a message on the internet (using the {\it ping} 
command of the UNIX operating system), $1/f$-like power spectrum 
has been observed \cite{csabai,taka}. 

In the square lattice model of a computer network developed by 
Ohira and Sawatari \cite{ohira} information packets are generated 
at the sites on the boundary at a rate $\lambda$ with the 
corresponding destination addresses chosen randomly from among 
the boundary sites. The packets can form queues of unlimited 
length at the the inner nodes, which act as routers of the 
network. At every time step, the packets from the heads of 
the queues at the routers are forwarded to the tail ends of 
the queues at the next router. Both deterministic and 
probabilistic strategies have been considered for selecting 
the next router to which the individual packets are to be 
forwarded. On reaching their individual destinations the packets 
die. The average number of time steps  between the birth and 
death of a packet is referred to as the average lifetime of a 
packet. Computing the average lifetime as a function of 
the birth rate $\lambda$ of the packets, Ohira and Sawatari 
\cite{ohira} observed a transition from a low-congestion phase 
to a high-congestion phase at a non-zero finite value $\lambda_*$. 


\section{Summary and conclusion} 
\label{Sec_summ}

In this section we summarize our conclusions regarding the 
current status of understanding of the statistical physics 
of vehicular traffic. We also speculate on the future trends 
of research in this area.

As we stated in the beginnning, one of the main aims of basic 
research on vehicular traffic, from the point of view of 
statistical physics, is to understand the nature of the 
steady-states of the system. We have summarized the empirical 
evidences available at present in support of the occurrence of 
three distinct dynamical phases, namely, those corresponding 
to {\it free-flow}, {\it synchronized flow} and {\it stop-and-go} 
traffic. Our critical review of the theoretical works has made 
it clear that, at present, the physical mechanisms at the 
"microscopic" level, which give rise to the synchronized traffic, 
are not as well understood as those responsible for the free-flow 
and stop-and-go traffic. There are, however, strong indications
that for a complete theory, which would account for all these 
three phases, one must take into account not only the vehicles 
on a given stretch of the highway but also on the on- and 
off-ramps. 

We have explored the possibility of transitions from one dynamical 
phase to another in the NaSch model (and TASEP) with periodic 
boundary conditions. Moreover, we have also presented the generic 
phase diagram of the TASEP with open boundary conditions and 
explained the notion of boundary-induced phase transitions in 
such one-dimensional driven-diffusive lattice gases which are far 
from equilibrium. Furthermore, we have found that while some models 
exhibit first-order phase transitions, some others exhibit second 
order phase transitions and the signatures of "criticality" while 
in some rare situations, e.g., in the cruise-control limit of the 
NaSch model, the system is found to exhibit even "self-organized" 
criticality.

In the beginning, we stated that one of the aims of basic research 
on vehicular traffic is to understand the nature of the dynamical 
fluctuations around the steady-states. The time-dependent correlations 
functions and the distributions of the relaxation times have been 
computed for the NaSch model and some other models but the general 
questions of the validity of dynamic scaling and dynamic universality 
classes have not been addressed. Another aim of the statistical 
mechanical approach to vehicular traffic, as we stated also in the 
beginning, is to investigate how the system evolves from initial 
states which are far from the corresponding steady-state. The 
phenomenon of "coarsening" of the platoons of vehicles during 
evolution from random initial states have been studied in some 
models. But, the questions of "universality", if any, of the growth 
exponents have not been addressed so far. Metastable states have 
been observed in several CA models. But, to our knowledge, the 
mechanisms of spontaneous decay of such states (analogue of 
homogeneous nucleation) has not been investigated so far. Besides, 
to our knowledge, so far it has not been possible to develope any 
powerfull analytical technique for calculating the dynamical 
properties of the traffic models. 

While stating the aims of basic research on vehicular traffic, we 
also mentioned the need to understand the effects of quenched 
disorder on the steady-states as well as on the dynamical properties 
of the systems. We have seen that the randomization of the hopping 
probability of the vehicles can lead to some exotic platooning 
phenomena which are close analogs of the "Bose-Einstein-condensation". 
An alternative prescription for introducing quenched disorder into 
the traffic models is to install random bottlenecks on the road and 
assign a time-independent hopping probability (or, equivalently, a 
"permeability") for hopping across bonds in such locations. It 
has been found that such localized bottlenecks can lead to global 
phase-segregation.

The NaSch model is the most extensively studied minimal CA model
of vehicular traffic on idealized single-lane highways. We have 
explained the conceptual framework, and illustrated the use of 
the mathematical formalism, of the cluster-theoretic analytical 
calculations for the NaSch model. This formalism, which yields 
exact results for the NaSch model with $v_{max} = 1$, gives quite 
accurrate estimates of various quantities of interest as long as 
$v_{max}$ is not too large. It would be desirable to develope a
new formalism to carry out {\it exact} analytical calculations for 
higher velocities too. Here also the limit $v_{max} = \infty$
is interesting since it shows a rather peculiar behaviour.

In addition to the detailed discussions on the NaSch model and 
its various generalizations, we have also mentioned briefly some 
other alternative CA models of single-lane highway traffic, 
e.g., the VDR model, the Fukui-Ishibashi model, the Werth-Froese-Wolf 
model, etc. so that one can appreciate the ongoing efforts to formulate 
the most satisfactory minimal model. 

From the point of view of practical applications, modelling 
vehicular traffic on {\it multi-lane} highways are more relevant 
than that on idealized single-lane highways which are, nevertheless, 
interesting from the point of fundamental understanding of truly 
non-equilibrium phenomena in driven-diffusive lattice gases.
At present, there are several different alternative prescriptions 
for formulating the CA rules for lane-changing of the vehicles on 
multi-lane highways. But, in order to pick out the most appropriate 
one from among these CA models theorists would require input from 
careful further observations of the phenomenon of lane-changing 
on real multi-lane highways. Empirical observations may also 
indicate modifications or extensions of the CA rules necessary for 
more realistic modelling of the multi-lane traffic.

The generalizations of the CA models of traffic on idealized 
single-lane highways to those on multi-lane highways may be 
regarded as extensions of one-dimensional model chains to
one-dimensional strips. The BML model of vehicular traffic in 
cities may be regarded as a further generalizations of these 
models from one-dimensional chains to two-dimensional lattices, 
or further, to decorated lattices. A few different CA rules have 
been considered so far for taking into account the effects of the 
traffic lights at the crossings of streets in such idealized 
street networks. We have emphasized the intrinsic differences 
between percolation clusters and the cluster of jammed vehicles 
in the BML (and similar) models in spite of some apparent 
similarities between them.

We have focused attention mainly on the progress made in the
recent years using "particle-hopping" models, formulated in
terms of {\it cellular automata}, and compared these with
several other similar systems. Although this may be a slightly
biased overview (as all reviews usually are) of the theory
of vehicular traffic, we have also discussed the main ideas
behind all the major approaches including the fluid-dynamical,
gas-kinetic and car-following theories of vehicular traffic.  
At present the relationships between different approaches of 
modelling have not explored in great detail. It would be very 
useful if the phenomenological parameters of the macroscopic 
theories can be estimated by utilizing the mathematical formulae 
relating these with those of the "microscopic" models. 

It is now quite clear that, in order to make significant further 
progress, we not only need more realistic models and better 
techniques of calculation but we also need more detailed and 
accurate empirical data from real traffic on highways as well 
as more careful re-analysis of the existing data in the light 
of recently developed concepts. So far as the observation of 
the real traffic is concerned, a lot can be learnt from a 
systematic analysis of aereal pictures or video photographs. 
Alternatively, as the second best choice, a series of counting 
loops along the highway can give more insight, by providing 
detailed information on, for example, time-headways (flux), 
velocity and local density.

If you are a critical thinker (or a pragmatist) you may ask: 
"armed with the theoretical tools at our disposal now, can 
we {\it predict} the occurrence of a traffic jam at a 
specific place on a given highway (or street) at a particular 
instant of time"? This question sounds similar to questions 
often asked in the context of some other interdisciplinary 
topics of current research in the area of complex systems, 
e.g., "can we predict an earthquake", or, "can we predict 
a stock market crash"? Of course, we know that, at present, 
the best we can hope for is to {\it predict} (if at all 
possible) probabilities of occurrences of all these phenomena. 
But, we must admit that, we have a long way to go before we 
come even close to this goal. Nevertheless, we hope you have 
enjoyed the fascinating twists and turns of the way we have 
covered so far. Our endeveour will be more successful if your 
interest has been stimulated by the intellectual challenges 
posed by the open problems and if you are willing to uncover 
the current mysteries as well as anticipating new surprises 
that may lie ahead. We are just at the beginning of a long road!

\vspace{1cm}
\noindent
{\em "The volume of vehicular traffic in the past several years 
has rapidly outstripped the capacities of the nation's highways. 
It has become increasingly necessary to understand the dynamics 
of traffic flow and obtain a mathematical description of the 
process"} - H. Greenberg (1959)


\vspace{2cm}

\noindent{\bf Acknowledgements:} It is our pleasure to thank 
P. Arndt, R. Barlovic, M. Barma, J.G. Brankov, E. Brockfeld,
G. Diedrich, B. Eisenbl\"atter, J. Esser, K. Ghosh, N. Ito, J. Kert\'esz, 
K. Klauck, W. Knospe, S. Krauss, J. Krug, D. Ktitarev, A. Majumdar, K. Nagel, 
L. Neubert, A. Pasupathy, V. Popkov, V.B. Priezzhev, N. Rajewsky, 
M. Schreckenberg, G. Sch\"utz, S. Sinha, D. Stauffer, R.B. Stinchcombe, 
Y. Sugiyama, P. Wagner, D.E. Wolf and J. Zittartz 
for enjoyable collaborations the results of some of which
have been reviewed here, for useful discussions and for critical
comments as well as suggestions on a preliminary draft of this review.
One of us (DC) acknowledges warm hospitality of ICTP, Trieste during
the preparation of this manuscript. This work is supported by SFB341
K\"oln-Aachen-J\"ulich.

\vspace{1cm}

\newpage 


\begin{appendix}
\section{Definition of update orders}
\label{App_updates}

A dynamical model is not fully defined just by its local transition
rules. In addition one has to specify the order in which the rules
are applied to the different particles, i.e.\ the update ordering
(sometimes also called 'dynamics'). This is an essential part of
the definition of the model since the transient and even the stationary
state may differ dramatically \cite{lecaer,HubGlance}.

For the NaSch model one uses a {\em parallel update} scheme
\cite{ssni}  where the rules are applied to all particles (i.e.,
vehicles) at the {\em same} time. This kind of ordering is
sometimes also called {\em synchronous} updating.

Among the various types of {\em asynchronous} update schemes most
frequently the the so-called {\em random-sequential} update is used.
Here one picks one particle at random and applies the transition
rules to it. Then one makes another random choice (which can also
be the same particle again) and so on. This update is sometimes
called {\em continuous} since it can be described by a master
equation in continuous time.

Apart from the parallel update there are other updates which are
{\em discrete} in time. We just mention the {\em ordered-sequential}
updates. Here one starts by applying the transition rules to one
particle. After that the rules are applied to the other particles
in a fixed order, e.g.\ one might continue with the next particle
ahead of the first one ({\em forward-ordered}) or the next particle
behind it ({\em backward-ordered}). We would like to point out that,
in principle, one has to distinguish two different types of
ordered-sequential updates which one could name {\em site-ordered-sequential}
and {\em particle-ordered-sequential}, respectively.
In contrast to the particle-ordered-sequential update described above,
in the site-ordered-sequential update the rules are applied to
{\em all} sites consecutively. This might have a strong effect,
since a particle might move to next cell ahead which then is
updated next (for the forward-particle-ordered-sequential update).
Then this particle might move again and so on. This is different from
the particle-ordered-sequential case where a particle at most moves
once during a sweep through the lattice.
As an example consider the extreme case of the NaSch mode with $v_{max}=1$
where only one particle is present which moves with probability $q=1$
to an empty cell in front. This particle
will move through the whole lattice during one sweep!
By looking at a lattice with two particles, one can
already see that the two different updates might introduce rather different
correlations. Starting with particles separated by $d$ empty sites, in the
site-ordered-sequential update the left particle will move to the right until
it reaches the right particle, which then starts to move. On the other hand,
in the case of particle-ordered-sequential update the particles will stay
always $d$ or $d-1$ sites apart. For general values of $q$ the situation
is similar.

There are several other updates which can be defined. We refer to the
literature (see e.g.\ \cite{rsss,lecaer,HubGlance,ChoiHub,Birgitt,Blok})
for a comparison of different update procedures.
The parallel update usually produces the strongest correlations and is
used for traffic simulations \cite{ssni}. Note that the
forward-particle-ordered-sequential update is almost identical to the
parallel update. In the case of periodic boundary conditions a difference
only occurs during the update of the last particle. In the forward-ordered
case the particle in front of it (i.e.\ the first particle) might already have
moved since it has been updated earlier. Although this difference appears
to be minor it can have a large effect. The difference between parallel
and forward-particle-ordered update can be viewed as a dynamical defect.

\section{TASEP} 
\label{App_TASEP}

This simple model of driven systems of interacting particles is
one of the most exhaustively studied prototype models in 
nonequilibrium statistical mechanics~\cite{sz,gs,spohn,spit,lige}. 
This model can be divided into four classes on the basis of the 
boundary conditions and the update scheme for the implementation 
of the dynamics. In this appendix we consider the TASEP with only 
random-sequential dynamics. 

Let us consider the TASEP with periodic boundary conditions and
random-sequential dynamics. Since only two states, namely empty and
occupied, are allowed for each site we can use a two-state variable
$n_i$ to denote the state of the $i$-th site where $n_i = 0$ if the
$i$-th site is empty and $n_i = 1$ if the $i$-th site is occupied. For
any given initial configuration $\{n_i(0)\}$, we can write the
equations governing the time evolution of $\langle n_i(t)\rangle$ (and
all the correlation functions) by taking into account all the
processes during the elementary time interval $dt$. It is not
difficult to establish that
\begin{equation}
  n_i(t+dt) = \begin{cases} n_i(t) & \text{with probability $1-2dt$}
    \\ n_{i-1}(t)+n_i(t)-n_{i-1}(t)n_i(t) &\text{with probability
      $dt$}\\ n_i(t)n_{i+1}(t) &\text{with probability $dt$}
          \end{cases}
\end{equation} 
\begin{equation}
\frac{dn_i}{dt} = n_{i-1}(1-n_i) - n_i(1-n_{i+1})
\label{eq-asep1}
\end{equation}
which, upon averaging over the history between times $0$ and $t$, 
leads to the equation 
\begin{equation}
\frac{d\langle n_i\rangle}{dt} = \langle n_{i-1}\rangle 
- \langle n_i\rangle - \langle n_{i-1}n_i\rangle 
+ \langle n_i n_{i+1}\rangle
\label{eq-asep2}
\end{equation}
for $\langle n_i\rangle$, the average occupation of the $i$-th site. Note 
that the equation for $\langle n_i\rangle$ involves two-site correlations. 
Similarly, it is straightforward to see that the equations for 
the two-site correlations involve three-site correlations and so 
on. Thus, the problem is an intrinsically $N$-body problem! 
The probability distribution for this system in the steady-state 
is given by \cite{derr_review} 
\begin{equation}
P_{steady-state}(\{n_i\}) = \frac{N! (L-N)!}{L!} 
\end{equation} 
where $L$ and $N$ refer to the total number of sites and the total
number of particles, respectively. From this distribution it follows
that $\langle n_i\rangle = N/L$ and $\langle v\rangle =
\frac{L-N}{L-1}$ which lead to $\langle n_i\rangle = c$ and $\langle
v\rangle = (1-c)$ in the thermodynamic limit.

The stationary state of the TASEP with open boundary conditions 
and random-sequential dynamics has been determined exactly using 
the so-called {\em Matrix Product Ansatz (MPA)} (see Appendix 
F for a more technical introduction) in \cite{derrida93} 
and in \cite{schdom} using recursion relations. This solution
has been generalized to different types of discrete dynamics in
\cite{hinrich,honecker,rsss}. The solution for parallel dynamics
was obtained recently in \cite{ERS} and \cite{degier} using
generalizations of the MPA technique.

\section{Naive site-oriented mean-field treatment 
  of the NaSch model}
\label{app_MF}

Suppose, $c_{v}(i,t) \equiv $ Probability that there is a
vehicle with speed $v$ ($v = 0,1,2,...,v_{max}$) at
the site $i$ at the time step $t$. Then, obviously,
$c(i,t)= \sum_{j=0}^{v_{max}} c_j(i,t) \equiv$ Probability that
the site $i$ is occupied by a vehicle at the time step $t$ and
$d(i,t) = 1 - c(i,t)$ is the corresponding probability that the
site $i$ is empty at the time step $t$. Using the definition
$$ J(c,p) = \sum_{v=1}^{v_{max}} v c_{v}. $$
for the flux $J(c,p)$ one can get the mean-field fundamental
diagram for the given $p$ provided one can get $c_{v}$ in
the mean-field approximation.

\noindent{\it Step I}:\ \ Acceleration stage ($t \rightarrow t_1$)
\begin{eqnarray}
c_0(i,t_1) &=& 0,\label{stepIeq1}\\
c_{v}(i,t_1) &=& c_{v-1}(i,t), \qquad\qquad\qquad (0 < v < v_{max})\\
c_{v_{max}}(i,t_1) &=& c_{v_{max}}(i,t) + c_{v_{max}-1}(i,t) 
\end{eqnarray}

\noindent{\it Step II}:\ \  Deceleration stage ($t_1 \rightarrow t_2$)
\begin{eqnarray}
c_0(i,t_2) &=& c_0(i,t_1) + c(i+1,t_1) \sum_{v=1}^{v_{max}} 
   c_v(i,t_1) \\
c_{v}(i,t_2) &=& c(i+v+1,t_1)\prod_{j=1}^{v} d(i+j,t_1) 
\sum_{v'=v+1}^{v_{max}}c_{v'}(i,t_1)\nonumber\\
&&+ c_{v}(i,t_1)\prod_{j=1}^{v} d(i+j,t_1), 
\qquad\qquad (0 < v < v_{max})\\
c_{v_{max}}(i,t_2) &=& \prod_{j=1}^{v_{max}} d(i+j,t_1) 
c_{v_{max}}(i,t_1) 
\end{eqnarray}

\noindent{\it Step III}:\ \  Randomization stage ($t_2 \rightarrow t_3$)
\begin{eqnarray}
c_0(i,t_3) &=& c_0(i,t_2) + p c_1(i,t_2)\\
c_{v}(i,t_3) &=& q c_{v}(i,t_2) + p c_{v+1}(i,t_2), 
\qquad\qquad (0 < v < v_{max})\\
c_{v_{max}}(i,t_3) &=& q c_{v_{max}}(i,t_2) 
\end{eqnarray}

\noindent{\it Step IV}:\ \  Movement stage ($t_3 \rightarrow t+1$)
\begin{equation}
c_{v}(i,t+1) = c_{v}(i-v,t_3), \qquad\qquad (0 \leq v \leq v_{max})
\label{stepIVeq}
\end{equation}

In the special case $v_{max} = 1$ the equations 
(\ref{stepIeq1}-\ref{stepIVeq}) get
simplified and, hence, we get
\begin{eqnarray}
c_0(i,t+1) &=& c(i,t)c(i+1,t) + p c(i,t) d(i+1,t),\\ 
c_1(i,t+1) &=& q c(i-1,t) d(i,t).
\end{eqnarray}
Similarly, in the case of $v_{max} = 2$, one gets 
\begin{eqnarray}
c_0(i,t+1) &=& [c(i,t)+p d(i,t)]c_0(i,t) 
+ [1 + p d(i,t)] c(i,t)[c_1(i,t)+c_2(i,t)],\nonumber\\
&&\\
c_1(i,t+1) &=& d(i,t)q c_0(i,t) + d(i,t)[q c(i,t)+p d(i,t)] 
[c_1(i,t) + c_2(i,t)],\nonumber\\
&&\\
c_2(i,t+1) &=& q d^2(i,t)[c_1(i,t)+c_2(i,t)].
\end{eqnarray}

\section{Paradisical mean-field theory} 
\label{App_para}

For $v_{max}=1$ the question, whether a state is a GoE state or not, 
can be decided locally by investigating just nearest-neighbour configurations.
By analysing the update rules one finds that all states containing the 
local configurations $(0,1)$ or $(1,1)$, i.e.\ configurations where a 
moving vehicle is directly followed by another car, are GoE states. 
This is not possible as can be seen by looking at the previous configurations.
The momentary velocity gives the number of cells that the car moved
in the previous timestep. In both configurations the first car
moved one cell. Therefore it is immediately clear that $(0,1)$ is a
GoE state since otherwise there would have been a doubly-occupied
cell before the last timestep. The configuration $(1,1)$ is also
not possible since both cars must have occupied neighbouring cells
before the last timestep too. Therefore, according to step 2, the
second car could not move.

The MFT equations (\ref{mf-c0}) and (\ref{mf-c1}) have to be modified
to take into account the existence of GoE states. 
In general, one has to follow the procedure outlined in 
Appendix \ref{app_MF}.
A quicker way to derive the paradisical mean-field (pMF) equations
is to analyse the MF equations (\ref{mf-c0}) and (\ref{mf-c1}).
In (\ref{mf-c0}) the contribution $c(i;t)c(i+1;t)$ appears.
Since we know that site $i+1$ can never be occupied by a car with
velocity $1$ if site $i$ is not empty, this contribution has to
be modified to $c(i;t)c_0(i+1;t)$ in pMFT. All other contributions
are left unchanged compared to MFT.

Due to this modification and the corresponding reduction of the
configuration space the normalization $c_0+c_1=c$ is no longer 
satisfied automatically.
Therefore a normalization constant ${\mathcal{N}}$ has to be introduced.
The final equations for a homogeneous stationary state are than given by
\begin{eqnarray}
c_0&=&{\mathcal{N}} (c_0+pd)c ,\label{pmf10}\\
c_1&=&{\mathcal{N}} qcd ,\label{pmf11}
\end{eqnarray}
with the normalization
\begin{equation}
{\mathcal{N}} =\frac{1}{c_0+d}.
\end{equation}
Since $c_0+c_1=c$ we have only one independent variable for fixed density
$c$, e.g.\ $c_1$. Solving (\ref{pmf10}), (\ref{pmf11}) for $c_1$ we obtain
\begin{equation}
c_1 = \frac{1}{2}\left(1-\sqrt{1-4q(1-c)c}\right).
\end{equation}
The flow is given by $f(c)=c_1$ and we recover the exact solution 
for the case $v_{max}=1$.

In the case $v_{max}=2$ more GoE states exist. In order to identify 
these it is helpful to note that the rules step 1 -- step 4 imply 
$d_j(t)=d_{j}(t-1) +v_{j+1}(t)-v_j(t)$ and therefore
\begin{eqnarray}
d_j(t)&\geq& v_{j+1}(t)-v_j(t),\label{distcond1}\\
v_j(t)&\leq& d_{j}(t-1)\label{distcond2}
\end{eqnarray}
The second inequality (\ref{distcond2}) is a consequence of step 2.

In the following we list the elementary GoE states, i.e.\ the 
local configurations which are dynamically 
forbidden (cars move from left to right): 
\begin{eqnarray}
(0,1),\quad (0,2),\quad (1,2),\quad (0,\bullet,2),
\label{goe1}\\
(1,1),\quad (2,1),\quad (2,2),\quad (1,\bullet,2),\quad (2,\bullet,2),
\label{goe2}\\
(0,\bullet,\bullet,2).\label{goe3}
\end{eqnarray}
The numbers give the velocity of a vehicle in an occupied cell and
$\bullet$ denotes an empty cell. 

The elementary GoE states in (\ref{goe1}) violate the inequality 
(\ref{distcond1}), and the configurations in (\ref{goe2}) violate 
(\ref{distcond2}). The state in (\ref{goe3}) is a second order GoE state. 
Going one step back in time leads to a first order GoE state since
$(0,\bullet,\bullet,2)$ must have evolved from $(0,v)$ (with $v=1$ or $v=2$).

Again we can derive the pMF equations by modifying the method 
for the derivation of the MFT.
Taking into account only the first order GoE states (\ref{goe1}) and 
(\ref{goe2}) one obtains the following pMF equations:
\begin{eqnarray}
c_0 &=& {\mathcal{N}} \left[ c_0c+pd(c_0+c_1c)\right],\label{c0eq}\\
c_1 &=& {\mathcal{N}} \left[ pd^2(c_1+c_2)+qd(c_0+c_1c)\right],\label{c1eq}\\
c_2 &=& {\mathcal{N}} qd^2(c_1+c_2)\label{c2eq}.
\end{eqnarray}
The normalization ${\mathcal{N}}$ ensures $c_0+c_1+c_2=c$ and is given
explicitly by
\begin{equation}
{\mathcal{N}} = \frac{1}{c_0+dc_1+d^2c_2}=\frac{1}{c_0+d(1-c_2)}.
\end{equation}

These equations have been analysed in \cite{ss98}.
After expressing $c_2$ through $c_0$ and $c$ by
\begin{equation}
c_2 = \frac{1}{2d}\left(c_0+d-\sqrt{(c_0+d)^2-4qd^3(c-c_0)}\right).
\label{c2}
\end{equation}
Inserting this result into (\ref{c0eq}) we obtain a cubic equation
which determines $c_0$ in terms of the parameters $c$ and $p$ 
\cite{ss98}. Results for different values of $p$ are shown
in Fig.\ \ref{fund_pmf2}. These results are only slightly modified
when also the second order GoE state is taken into account \cite{ss98}.

\section{Equations of car-oriented theory of NaSch 
Model and COMF approximation} 
\label{App_COMF}

In terms of $p$, $q$, $g$ and $\bar{g}(t) = 1 - g(t)$, the equations
describing the time evolution of the probabilities $P_n(t)$ for
the NaSch model with $v_{max} = 1$ are given by
\begin{eqnarray}
P_0(t+1) &=& \bar{g}(t)[P_0(t) + q P_1(t)],\\
P_1(t+1) &=& g(t) P_0(t) + [qg(t)+p\bar{g}(t)]P_1(t) 
       + q \bar{g}(t) P_2(t),\\
P_n(t+1) &=& p g(t) P_{n-1}(t) + [qg(t)+p\bar{g}(t)]P_n(t) 
       + q \bar{g}(t) P_{n+1}(t) \quad (n \geq 2).\nonumber\\
&&
\end{eqnarray}

It is worth mentioning here that, for the NaSch model with $v_{max} = 1$,
the 2-cluster probabilities $P_2(\sigma_i,\sigma_j)$ of the the SOMF
theory are related to the probabilities $P_n$ of the COMF theory
through
\begin{eqnarray}
P_2(1,1) &=& c P_0,\\
P_2(1,0) &=& c (1-c) P_1,\\
P_2(0,0) &=& (1-c) \frac{P_{n+1}}{P_n} \qquad\qquad  (n \geq 1)
\end{eqnarray}

\section{The matrix-product Ansatz for stochastic systems}
\label{App_MPA}

For the stochastic systems considered here the time-evolution of
the probability $P(\btau ,t)$ to find the system in the configuration
$\btau=(\tau_1,\ldots, \tau_L)$ is determined by the master equation. 
For random-sequential dynamics it has the form
\begin{equation}
\frac{\partial P(\btau,t)}{\partial t} = \sum_{\bttau} 
w(\bttau\to\btau) P(\bttau,t) - \sum_{\bttau} w(\btau\to\bttau)
P(\btau,t)
\label{master}
\end{equation}
with transition rates $w(\bttau\to\btau)$ from state $\bttau$ 
to state $\btau$. Eq.\ (\ref{master}) can be rewritten in the
form of a Schr\"odinger equation in imaginary time \cite{alca},
\begin{equation}
\label{schroed}
\frac{\partial}{\partial t} |P(t)\rangle =-\mathcal{H} \, |P(t)\rangle ,
\end{equation}
with the state vector $|P(t)\rangle=\sum_{\btau} P(\btau,t)|\btau\rangle$. 
The vectors $|\btau \rangle=|\tau_1,...,\tau_L \rangle$ corresponding
to the configurations $\btau$ form an orthonormal basis of the 
configuration space. The {\em stochastic Hamiltonian} $\mathcal{H}$
is defined through its matrix elements
\begin{equation}
\label{hamm1}
\langle \btau|\mathcal{H}|\bttau\rangle=-w(\bttau \rightarrow \btau),
\qquad
\langle \btau|\mathcal{H}|\btau\rangle=\sum_{\btau \ne \bttau}
w(\btau \rightarrow \bttau)\quad (\btau\neq\bttau).
\end{equation}
The stationary state of the stochastic process corresponds to the
eigenvector $|P_0\rangle$ of $\mathcal{H}$ with eigenvalue $0$.

For discrete-time dynamics the master equation takes the form
\begin{equation}
P(\btau,t+1) = 
    \sum_{\bttau} W(\bttau\to\btau) P(\bttau,t) 
\label{master_diskret}
\end{equation}
where $W(\bttau\to\btau)=w(\bttau\to\btau)\cdot\Delta t$ are 
transition probabilities.
This can be rewritten as
\begin{equation}
|P(t+1)\rangle =\mathcal{T} |P(t)\rangle.
\label{master_diskret2}
\end{equation}
Here the stationary state corresponds to the eigenvector
$|P_0\rangle$ of the transfer matrix $\mathcal{T}$ with eigenvalue $1$.

A very powerful method for the determination of stationary solutions
of the master equation is the so-called matrix-product Ansatz (MPA).
For a system with open boundaries the weights $P(\btau)$ in the
stationary state can be written in the form
\begin{equation}
\label{mpgweights}
P(\tau_1,\ldots,\tau_L)=\frac{1}{Z_L}\langle W|\prod_{j=1}^L\left[
\tau_jD+(1-\tau_j)E\right]|V\rangle.
\end{equation}
For periodic boundary condition the MPA takes the form
\begin{equation}
\label{mpgweights2}
P(\tau_1,\ldots,\tau_L)={\rm Tr} \left( \prod_{j=1}^L\left[
\tau_jD+(1-\tau_j)E\right]\right).
\end{equation}
For simplicity we have assumed a two-state system where e.g.\
$\tau_j=0$ corresponds to an empty cell $j$ and $\tau_j=1$ to an
occupied cell. $Z_L$ is a normalization constant that can be
calculated as $Z_L=\langle W|C^L|V\rangle$.
In (\ref{mpgweights}), (\ref{mpgweights2}) $E$ and $D$ are matrices and
$\langle W|$ and $|V\rangle$ are vectors characterizing the
boundary conditions. The explicit form of these quantities has to be 
determined from the condition that (\ref{mpgweights}) or
(\ref{mpgweights2}) solves the
master equation. This leads in general to a algebraic relations between
the matrices $E$ and $D$ and the boundary vectors $\langle W|$ and 
$|V\rangle$. Once one these have been determined one has a simple
recipe for determining $P(\tau_1,\ldots,\tau_L)$: First, translate
the configuration $\tau_1,\ldots,\tau_L$ into a product of matrices 
by identifying each empty cell ($\tau_j=0$) with a factor $E$ and each
occupied cell ($\tau_j=1$) with $D$. In that way the configuration 
$011001\cdots$ corresponds to the product $EDDEED\cdots=ED^2E^2D\cdots$. 
The weight of the configuration is then just the matrix element
with the vectors $\langle w|$ and $|v\rangle$.

A simple example is the ASEP discussed in Sect.\ \ref{Sec_asep}.
Here these quantities have to satisfy
\begin{eqnarray}
p\, DE & = & D+E,\label{derr1}\\
\alpha \langle W| E & = & \langle W| ,\label{derr2} \\ \beta
D|V\rangle & = & |V\rangle .\label{derr3}
\end{eqnarray}
If one is able to find explicit representations for this algebra
one can determine in principle all expectation values in the stationary
state exactly. For (\ref{derr1})--(\ref{derr3}) one can show that
all representations are infinite-dimensional \cite{derrida93}.
Only on the line $\alpha+\beta=p$ one-dimensional representations 
(with $E,D$ and $\langle W|$, $|V\rangle$ being real numbers) are 
possible\footnote{This can be seen easily from 
(\ref{derr1})--(\ref{derr3}).}.

At this point it might appear that the MPA works only in very special
cases. However, it can be shown that the stationary state of one-dimensional
stochastic processes is generically of matrix-product form 
\cite{krebs,rajschreck,klauck}\footnote{We have to mention here that
these results up to now do not include the case of parallel dynamics.}.
Even if it is not straightforward to find general representations
of the resulting algebras, one can at least search systematically for
finite-dimensional representations on special lines in the parameter
space of the model. Furthermore, since the mathematical structure
of the stationary state is known it is sometimes possible to derive
rather general results. As an example in \cite{rajschreck} interesting
relations between expectation values for ordered-sequential and
sublattice-parallel dynamics have been derived.

For a more detailed description of the MPA for different types of
dynamics and its relation with the MPA technique for
quantum-mechanical spin systems \cite{ksz} we refer to \cite{rsss}. A
review of the treatment of the ASEP using the MPA is given in
\cite{derr_review}. 
The MPA has also been extended to treat the full dynamics, not 
only the stationary state \cite{stinchsch}. Using time-dependent
matrices $D(t)$ and $E(t)$ one obtains the Bethe Ansatz equations
for the corresponding stochastic Hamiltonian 
\cite{stinchsch,schuetzZ60,SasaWad}.

\section{Two schemes for solving the mean-field 
approximation of the DTASEP}
\label{App_solvMF}

A mean-field approximation scheme for this model has also been
developed \cite{barma}. The time-averaged steady-state current
$J_{j,j+1}$ in the bond $(j,j+1)$ is given by $J_{j,j+1} = q_{j,j+1}
\langle n_j (1-n_{j+1})\rangle$. In the mean-field approximation,
$\langle n_j (1-n_{j+1})\rangle = \langle n_j\rangle\langle
(1-n_{j+1})\rangle$ and, hence,
\begin{equation}
J = J_{j,j+1} = q_{j,j+1} c_j (1 - c_{j+1})
\label{mf-barma}
\end{equation}
where $c_j = \langle n_j\rangle$. In order to calculate the steady-state
flux $J$ as a function of the mean density $c$ of the particles,
Tripathy and Barma \cite{barma} used two different iteration
schemes based on the equation (\ref{mf-barma}). \\
(i) {\it Constant-current iteration scheme:} In this scheme,
for a given system length $L$ and a fixed flux $J = J_0$,
one starts with some value of $c_1$ and, computes all the
other $c_j$ $(j > 1)$ using the equation (\ref{mf-barma}), i.e.,
\begin{equation}
c_{j+1} = 1 - \frac{J_0}{q_{j,j+1} c_j}, \quad j = 1,2,...,L
\end{equation}
together with the periodic boundary condition $c_{j+L} = c_j$.
If the iteration converges, i.e., one gets all the site densities
in the physically acceptable range $[0,1]$, one accepts the
average of these final site densities to be the global mean
density of the particles corresponding to the flux $J_0$.
(ii) {\it Constant-density iteration scheme:} In this scheme,
for a given system length $L$ and fixed global average density
$c$, one begins by assigning the site densities $0 \leq c_j(0) \leq 1$
to the lattice sites subject to the global constraint
$\frac{1}{L} \sum_j c_j(0) = c$. Then, the site densities are updated
in parallel according to
\begin{equation}
c_j(t+1) = c_j(t) + J_{j-1,j}(t) - J_{j,j+1}(t), \quad j = 1,2,...,L
\end{equation}
which follows from the equation (\ref{mf-barma}). It is
straightforward to verify that the this iteration scheme
keeps the average global density $c$ unchanged at every step
of updating and hence the name. After sufficient number of
iterations the set of densities converge to a set $\{c_j\}$
and the flux on each bond converge to the steady-state flux
$J_0$.

\section{Self-consistent equations for $v_x$ and $v_y$ in the
mean-field approximation of the BML model}
\label{App_selfBML}

Suppose, $v_x$ and $v_y$ denote the {\it average} speeds
of east-bound and north-bound vehicles, respectively.
Then, on the average, an east-bound vehicle spends a time
$1/v_x$ at a site whereas a north-bound vehicle spends
a time $1/v_y$ at a site. The north-bound vehicles lead
to a reduction of the speed of the east-bound vehicles
by $n_y/v_y$. Moreover, because of the hindrance of the
east-bound vehicles by other east-bound vehicles ahead
of it there will be further reduction of the speed of
the east-bound vehicles by $n_x[\frac{1}{v_x}-1]$. Furthermore,
if the density of the overpasses is $f_o$, then 
\begin{equation}
v_x = 1 - (1-f_o)\left[\frac{n_y}{v_y} + n_x\left(
\frac{1}{v_x}-1\right)\right]
\label{eqvx}
\end{equation}
Similarly, the corresponding equation for $v_y$ is given by
\begin{equation}
v_y = 1 - (1-f_o)\left[\frac{n_x}{v_x} 
+ n_y\left(\frac{1}{v_y}-1\right)\right]
\label{eqvy}
\end{equation}
In the special case $n_x = n_y = n/2$ both the equations (\ref{eqvx})
and (\ref{eqvy}) reduce to the form
\begin{equation}
v = 1 - (1-f_o)\left[\frac{1}{v} - \frac{1}{2}\right]n
\label{quadratico}
\end{equation}
where $v_x = v_y = v$. The solution of the quadratic equation
(\ref{quadratico}) for $v$ is
\begin{equation}
v = \frac{1}{2}\left[1 + \frac{1-f_o}{2}n 
+ \sqrt{\left(1 + \frac{1-f_o}{2}n \right)^2 
- 4 (1-f_o)n}\right]
\end{equation}

\section{Derivation of the equations in the microscopic 
theory of the BML model} 
\label{App_microBML}

By definition, $n_{\uparrow}(x,y;t)$ ($n_{\rightarrow}(x,y:t)$) is
unity if the site $(x,y)$ is occupied at time $t$ by a north-bound
(east-bound) vehicle and zero if the site $(x,y)$ is not occupied by a
north-bound (east-bound) vehicle. Normalization requires
$n_{\uparrow}(x,y;t) + n_{\rightarrow}(x,y;t) = 1 - n_{empty}(x,y;t)$
where the two-state variable $n_{empty}(x,y;t)$ is unity if, at time
$t$, the site $(x,y)$ is empty and zero if the site $(x,y)$ is not
empty. In order to describe the state of the signals at time $t$, one
also defines a two-state variable $S(t)$: $S(t) = 1 (0)$ if the signal
is green (red) for the vehicles under consideration. The space-average
of $n_{\uparrow}(x,y;t)$ and $n_{\rightarrow}(x,y;t)$ are
$c_{\rightarrow}(t)$ and $c_{\uparrow}(t)$, respectively.  Besides,
the time-average of $S(t)$ is $1/2$.

The updating rules of the BML model lead to the equations
\begin{eqnarray}
&&n_{\rightarrow}(x,y;t+1) = n_{\rightarrow}(x,y;t)
\left[n_{\uparrow}(x+1,y;t) + n_{\rightarrow}(x+1,y;t)\right] 
S(t)\nonumber\\
&&+ n_{\rightarrow}(x-1,y;t)\left[1- n_{\uparrow}(x,y;t) 
- n_{\rightarrow}(x,y;t) \right] S(t)
+ n_{\rightarrow}(x,y;t) [1 - S(t)]\nonumber\\
&&
\label{bml-micro}
\end{eqnarray}
The first term on the right hand side of equation (\ref{bml-micro})
describes the situation when the east-bound vehicle, which was at
the site $(x,y)$ at time $t$, finds a green signal but cannot move
because the next site towards east is occupied by another
vehicle. The second term on the right hand side of (\ref{bml-micro})
corresponds to the situation where the east-bound vehicle,
which was at the site $(x-1,y)$ at time $t$, finds a green
signal and moves to the next site towards east, which was empty.
The last term on the right hand side of equation (\ref{bml-micro})
arises from the possibility that the east-bound vehicle, which was
at $(x,y)$ at time $t$, could not move because of a red signal,
irrespective of the state of occupation of the next site towards east.
Following the similar arguments, one can also write down the
corresponding equation for the north-bound vehicles.
\end{appendix}

\newpage


\end{document}